\documentclass{aa}
\usepackage{graphics,epsfig,txfonts}
\usepackage[section]{placeins}
\usepackage{natbib}
\bibpunct{(}{)}{;}{a}{}{,}

\newcommand{\ergs}[1]{$\rm{\times 10^{#1}}$ erg s$\rm{^{-1}}$}

\newcommand{\ltsima}{$\buildrel < \over \sim$}
\newcommand{\lsim}{\lower.5ex\hbox{\ltsima}}
\newcommand{\gtsima}{$\buildrel > \over \sim$}
\newcommand{\gsim}{\lower.5ex\hbox{\gtsima}}
\def\rahour{\hbox{\ensuremath{^{\rm h}}}}
\def\ramin{\hbox{\ensuremath{^{\rm m}}}}
\def\rasec{\hbox{\ensuremath{^{\rm s}}}}

\usepackage{color}

\begin{document}
 
\title{X-ray source variability study of the M\,31 central field using \emph{Chandra} HRC-I
\thanks{The primary source catalogue table (Table 1) and the variability table (Table 2) will only be available in electronic form at the CDS (http://cds.u-strasbg.fr/).}}

\author{      F.~Hofmann\inst{1}
     \and     W.~Pietsch\inst{1}
     \and     M.~Henze\inst{1,2}
     \and     F.~Haberl\inst{1}
     \and     R.~Sturm\inst{1}
     \and     M.~Della~Valle\inst{3,4}
     \and     D.~H.~Hartmann\inst{5}
     \and     D.~Hatzidimitriou\inst{6,7}
       }

\titlerunning{X-ray source variability in M\,31}
\authorrunning{Hofmann et al.}

\institute{Max-Planck-Institut f\"ur extraterrestrische Physik, Giessenbachstra{\ss}e, 85748 Garching, Germany
        \and European Space Astronomy Centre, ESA, P.O. Box 78, 28691 Villanueva de la Ca\~nada, Madrid, Spain 
	\and INAF-Napoli, Osservatorio Astronomico di Capodimonte, Salita Moiariello 16, 80131 Napoli, Italy 
	\and International Centre for Relativistic Astrophysics, Piazzale della Repubblica 2, 65122 Pescara, Italy 
	\and Department of Physics and Astronomy, Clemson University, Clemson, SC 29634-0978, USA
        \and Department of Astrophysics, Astronomy and Mechanics, Faculty of Physics, University of Athens, Panepistimiopolis, GR15784 Zografos, Athens, Greece 
	\and Foundation for Research and Technology Hellas, IESL, Greece}


 \abstract{The central field of the Andromeda galaxy (M\,31) has been monitored, using the \emph{Chandra} HRC-I detector (about 0.1-10\,keV energy range) from 2006 to 2012 with the main aim to detect X-rays from optical novae. We present a systematic analysis of all X-ray sources found in the 41 nova monitoring observations, along with 23 M\,31 central field HRC-I observations available from the \emph{Chandra} data archive starting in December 1999.}
          {Based on these observations, we studied the X-ray long-term variability of the source population and especially of X-ray binaries in M\,31.}
          {We created a catalogue of sources, detected in the 64 available observations, which add up to a total exposure of about 1 Ms. To study the variability, we developed a processing pipeline to derive long-term \emph{Chandra} HRC-I light curves for each source over the 13 years of observations. In the merged images we also searched for extended X-ray sources.}
          {We present a point-source catalogue, containing 318 X-ray sources with detailed long-term variability information. 28 of which are published for the first time. The spatial and temporal resolution of the catalogue allows us to classify 115 X-ray binary candidates showing high X-ray variability or even outbursts in addition to 14 globular cluster X-ray binary candidates showing no significant variability. The analysis may suggest, that outburst sources are less frequent in globular clusters than in the field of M\,31. We detected 7 supernova remnants, one of which is a new candidate and in addition resolved the first X-rays from a known radio supernova remnant. Besides 33 known optical nova/X-ray source correlations, we also discovered one previously unknown super-soft X-ray outburst and several new nova candidates.}
          {The catalogue contains a large sample of detailed long-term X-ray light curves in the M\,31 central field, which helps to understand the X-ray population of our neighbouring spiral galaxy M\,31.}

 
\maketitle

\section{Introduction}
\label{sec:introduction}
\begin{figure}
  \resizebox{\hsize}{!}{\includegraphics[angle=0,clip=]{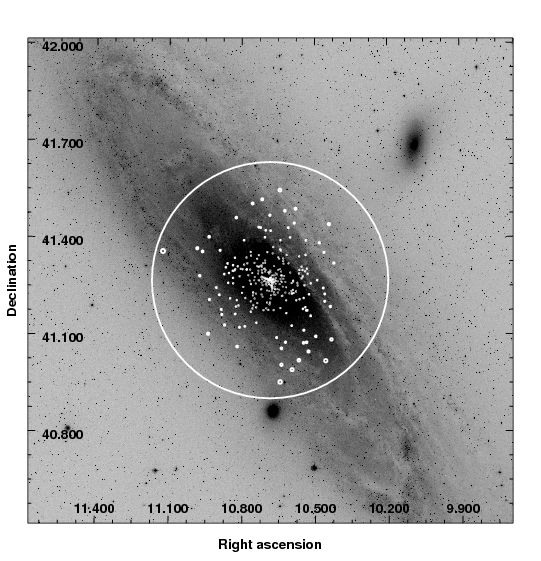}}
  \caption{
    Optical image of the Andromeda Galaxy (DSS2 blue) with X-ray source regions from the \emph{Chandra} HRC-I catalogue ($\sim 50\%$ encircled count fraction radius) and a circle marking the outer rim of the detector FOV (image obtained from the NASA \emph{Sky View} website).}
  \label{fig:andromeda}
\end{figure}

\addtocounter{table}{+2}

M\,31 is an ideal target for X-ray observations since it is the closest spiral galaxy and similar to our own galaxy. The outside view and the distance of M\,31 enable us to monitor several hundred X-ray sources in just one telescope pointing. The Einstein X-ray observatory \citep[e.g.][]{1991ApJ...382...82T} and the ROSAT satellite \citep{1993ApJ...410..615P, 1997A&A...317..328S, 2001A&A...373...63S} gave a first insight into the X-ray population of M\,31, followed by the instruments of \emph{Chandra} \citep[e.g.][]{2002ApJ...577..738K, 2002ApJ...578..114K, 2004ApJ...609..735W, 2007A&A...468...49V} and \emph{XMM-Newton} \citep[e.g.][]{2005A&A...434..483P, 2008A&A...480..599S, 2011A&A...534A..55S} with better spatial resolution and deeper observations.\\
Several individual X-ray sources in M\,31 have been discussed in previous publications (e.g. sources showing X-ray dips, pulsations, bursts or ultra-luminous X-ray outbursts, for references see section \ref{sec:discussion} and Table 1). In addition many source classes have been studied. The main class of M\,31 X-ray sources are X-ray binaries (XRB), in which a compact object like a white dwarf, a neutron star or a black hole accretes matter from a companion star. Such sources can show highly variable or transient behaviour \citep[e.g.][]{2001A&A...378..800O, 2006ApJ...643..356W}. The subclass of globular cluster X-ray sources (GlC), which are mostly low-mass XRBs (LMXBs) is covered in several papers \citep[e.g.][]{ 2002ApJ...570..618D, 2004ApJ...616..821T, 2009MNRAS.392L..55P, 2012ApJ...757...40B}. A further subclass of XRBs are defined by their super-soft X-ray spectrum (SSS) \citep[e.g.][]{2004ApJ...610..247D}. Most of these SSSs have been found to be associated with optical novae \citep{2005A&A...442..879P, 2007A&A...465..375P, 2006ApJ...643..844O}.
Another class of X-ray sources are supernova remnants (SNR), which can be spatially resolved in M\,31 using X-ray, optical or radio observations \citep[][and references therein]{2002ApJ...580L.125K, 2012A&A...544A.144S}.\\
Beginning in 1999 the local group galaxy M\,31 (Andromeda Galaxy) has been observed regularly with the \emph{Chandra} X-ray Observatory. The satellite High Resolution Mirror Assembly \citep[HRMA, see][]{1988SSRv...47...47W}, was frequently pointed at the centre of M\,31, which allows us to examine the long term evolution of the intensity of sources within the central field of the galaxy. We used 64 observations of the \emph{Chandra} High Resolution Camera in imaging mode \citep[HRC-I, see][]{1997SPIE.3114...11M} from 1999 to 2012. The single \emph{Chandra} HRC-I observation with the longest exposure time is discussed by \citet{2002ApJ...578..114K}. \citet{2004ApJ...609..735W} discuss snapshots, covering the whole galaxy until 2002.\\
The observations from 2006 to 2012 are part of an extensive monitoring program (41 observations with about 20 ks exposure), to investigate X-ray sources associated with optical novae. These observations have been searched for novae by \citet{2010A&A...523A..89H} and \citet{2011A&A...533A..52H}. We now use these monitoring observations, along with all previous archival HRC-I observations to systematically analyse the time variability of all sources, detected in the M\,31 central field with \emph{Chandra} HRC-I.\\
We used the HRC-I detector, which has a wide field of view (32{\arcmin} x 32{\arcmin}) and a very high angular resolution (below 1{\arcsec}) for sources close to the pointing of the telescope\footnote{http://cxc.harvard.edu/proposer/POG/html/HRC.html}. The dense HRC-I monitoring of M\,31 allows us to create the deepest (about 1 Ms of total exposure) catalogue of X-ray sources at the centre of M\,31 and provide a detailed long-term variability study for each source. Figure \ref{fig:andromeda} shows an image of the digitized sky survey 2 (DSS2) taken with a blue filter and the approximate HRC-I field of view (FOV). 

\section{Observations and data reduction}
\label{sec:observations}

As mentioned earlier, there are 64 \emph{Chandra} HRC-I observations to date, pointed to within 3{\arcmin} from the centre of M\,31 \citep[RA: 00{\rahour} 42{\ramin} 44.33{\rasec}, Dec: +41{\degr} 16{\arcmin} 07.50{\arcsec}, J2000,][]{2006AJ....131.1163S}. We excluded two $\sim$1 ks observations with high background count rates (ObsID 00270 and 01569), which would not have contributed new sources to the catalogue and would have deteriorated the signal to noise ratio, when merging the individual observations. Table \ref{tab:cat} contains information on all HRC-I observations used. The observations from 1999 to 2002 mostly have an exposure of $\sim$1.2 ks (one with 2.5 ks and one with 5.2 ks), except for ObsID 1912 (46.7 ks), which was used by \citet{2002ApJ...578..114K} for his investigation. From 2004-12-06 till 2005-02-21 there were six HRC-I observations of the M\,31 centre, with exposures between $\sim$20 and $\sim$50 ks to monitor the variability of M\,31*, the source at the M\,31 nucleus. The period from 2006 to 2012 is covered by the $\sim$20 ks HRC-I observations of the \emph{XMM-Newton}/\emph{Chandra} nova monitoring program and additional \emph{Chandra} guaranteed time observations of the Max-Planck-Institut f\"ur extraterrestrische Physik (MPE). In this period we have rather regularly spaced data points with about the same exposure. This monitoring delivers the best input for the time variability studies.

\subsection{Source detection}
\label{sec:detection}

The data for each observation were reprocessed using the \emph{Chandra} Interactive Analysis of Observations software package \citep[CIAO,][]{2006SPIE.6270E..60F} version 4.4 and the \emph{Chandra} Calibration Database \citep[CalDB,][]{2007ChNew..14...33G} version 4.4.7, to receive a level 2 event list according to the \emph{Chandra} Standard Data Processing (SDP). For each observation we created a light curve of the whole detector and removed events at times of high background from the event file (good time interval (GTI) filtering) using the Munich Image Data Analysis System (MIDAS) software package provided by the European Southern Observatory (ESO). We only used events in the pulse invariant (PI) channel range from 48-293, which according to the CIAO science threads increases the contrast between sources and background\footnote{http://cxc.harvard.edu/ciao/threads/hrci\_bg\_spectra/}.\\
As the point spread function (PSF) of the \emph{Chandra} HRMA causes higher source-blurring with increasing off-axis angle, we used five images with different resolution \citep[similar to][]{2002ApJ...578..114K}. This reduces computation time, since the detection does not have to run on the entire detector field of view at highest resolution. The highest resolution image covers 200{\arcsec}$\times$200{\arcsec} with 0.13{\arcsec} pixel size (HRC-I detector resolution). The second, third and fourth image cover 400{\arcsec}$\times$400{\arcsec}, 800{\arcsec}$\times$800{\arcsec} and 1400{\arcsec}$\times$1400{\arcsec}, with a binning of 2$\times$2, 4$\times$4 and 8$\times$8 detector pixels. The fifth image covers the entire detector and has a 16$\times$16 binning.\\
The size of the PSF increases by a factor of $\rm{\sim30}$ from the centre to an off-axis angle of $\rm{\sim15 \arcmin}$. where the effective exposure decreases to about $\rm{70\%}$. This means that the spatial resolution and the sensitivity of the catalogue depends on the off-axis angle of a source.\\
Using CIAO tools we created an exposure map for each image, correcting for vignetting effects of the HRMA and for the efficiency of the detector. After testing various background models (by subtracting background maps from images and inspecting residuals) we found that the \emph{XMM-Newton} Science Analysis System \citep[SAS,][]{2004ASPC..314..759G} task {\tt esplinemap} delivered the best results for modelling background contributions and the diffuse emission of M\,31. To separate point-source contributions, we did a detection run with the CIAO tool {\tt wavdetect} with the sensitivity parameter $sigthresh = \rm{10^{-4}}$. We chose a slightly higher sensitivity than recommended in the CIAO science threads\footnote{http://cxc.harvard.edu/ciao/threads/wavdetect/} to avoid missing any detections. Later we removed detections below a significance of 3$\rm{\sigma}$. To avoid spurious detections at the detector edges, we used a detection mask, to exclude areas where the effective exposure is below $\rm{50\%}$ of the maximum exposure, due to vignetting effects of the mirrors and detector efficiency. The scales of the wavelets were optimized to detect point sources in the relevant region of each image. The {\tt esplinemap} tool used the {\tt wavdetect} output to remove sources above a likelihood of $\rm{5\sigma}$ from the image and then created a background map, fitted to the re-binned image with 16 spline nodes.\\
With these background maps as input we then ran {\tt wavdetect} again on every image, with the same parameters as in the first run, to create a final source list for each image.\\
We merged the output source lists of the five images. The lists cover regions from radius 0-95{\arcsec}, 90-200{\arcsec}, 190-400{\arcsec}, 380-700{\arcsec} and 665-1315{\arcsec}. Sources that were found in two images in an overlap-region of two lists were taken from the higher resolution image.\\
Based on the {\tt wavdetect} count rates, we calculated the 0.2-10 keV flux, corrected for Galactic foreground absorption \citep[$\rm{N_{H} = 6.6 \cdot 10^{20} ~cm^{-2}}$,][]{1992ApJS...79...77S} toward M\,31 and used a power-law spectrum with a photon index of $\rm{\Gamma = 1.7}$. For the luminosity calculation, we assumed a distance of 780 kpc to M\,31 \citep{1998ApJ...503L.131S} and isotropic radiation, which has been used in many previous X-ray studies. Although there are more recent papers on the distance to M\,31 \citep[e.g.][$\rm{752 \pm 27 ~kpc}$]{2012ApJ...745..156R}.\\
To improve the astrometry of individual observations we compared the source-list positions with the \citet{2002ApJ...578..114K} catalogue, which is based on the longest individual observation and then corrected systematic offsets. The \citet{2002ApJ...578..114K} catalogue has been astrometrically corrected after cross correlation with the Two Micron All-Sky Survey catalogue \citep[2MASS, see][]{2006AJ....131.1163S}. After this correction, the mean angular distance to the \citet{2002ApJ...578..114K} sources was 0.23{\arcsec}. Since \citet{2002ApJ...578..114K} states, that systematic uncertainties are about 0.15{\arcsec}, we estimated a conservative systematic $\rm{1\sigma}$ uncertainty of 0.38{\arcsec} for the catalogue.

\subsection{Merged image}
\label{sec:merge}

\begin{figure}
  \centering
  \resizebox{\hsize}{!}{\includegraphics[angle=0,clip=]{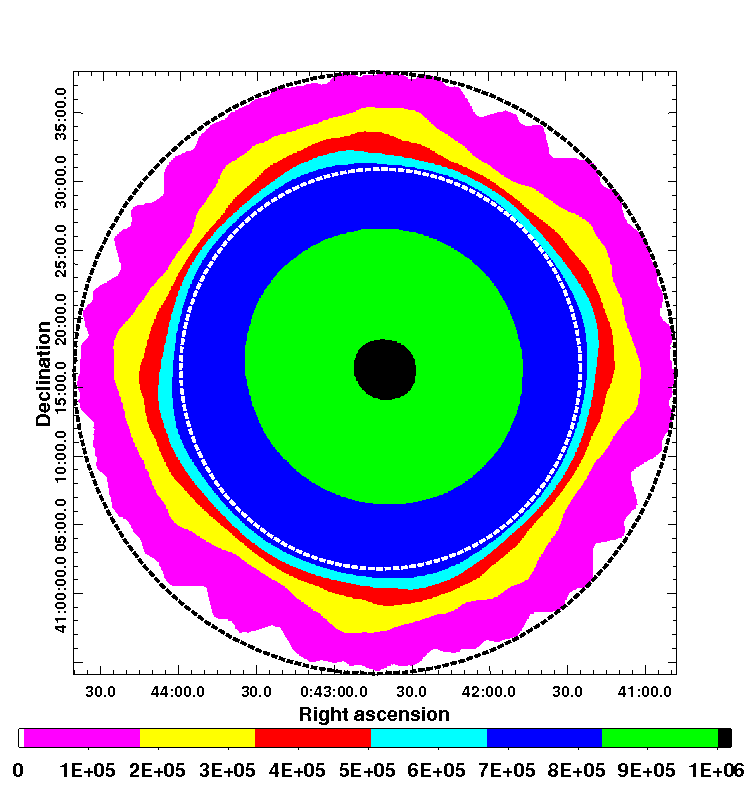}}
  \caption{Colour image of the merged HRC-I exposure maps, showing the entire detector FOV. The dashed black circle marks the outer limit of the catalogue (see Fig. \ref{fig:andromeda}). The dashed white circle includes the approximate area with $\rm{>70 \%}$ effective exposure. The colour bar shows the effective exposure time value (in seconds). The innermost black region marks the approximate 1 Ms effective exposure region.}
  \label{fig:exp}
\end{figure}

We merged the individual observations to find possible faint sources, that have not been detected in the individual observations.\\
We corrected for the systematic offset against the \citet{2002ApJ...578..114K} catalogue by re-projecting the events for each observation. We again produced images at the five resolutions (see \ref{sec:detection}), but aligned to the coordinates of the centre of M\,31 instead of the centre of the detector. Thereby, we ensure that a given pixel coordinate in one image corresponds to the same RA/Dec and pixel coordinate in the images of each observation. Therefore, we were able to add up the images pixel by pixel, instead of merging the event lists, which would have become too large to handle for some of the programs and would have used up too much computational resources.\\
The images and exposure maps of the individual observations were merged. We created new background images, accepting only merged areas, for which the effective exposure is above $\rm{70 \%}$ of the maximum. This was necessary, because the corners of the detector fields do not overlap further off-axis in the merged image, as the roll angle of the satellite changed during the course of the observations, leading to a fast drop in exposure (see Fig. \ref{fig:exp}).\\
For these images, we ran the detection pipeline, described in section \ref{sec:detection}.\\
The maximum exposure time of the merged image of all 62 observations, we used, adds up to 1.02 Ms in the central area (see Fig. \ref{fig:high}, \ref{fig:mid}, \ref{fig:low}).\\

\subsection{Spectral information}
\label{sec:HR}
\begin{table*}
\caption[]{Photon count distribution observed with \emph{Chandra} ACIS in the 0.2 keV to 10 keV photon energy range.}
\begin{center}
\begin{tabular}{rllrrr}
\hline\hline\noalign{\smallskip}
\multicolumn{1}{l}{No. \tablefootmark{a}} &
\multicolumn{1}{l}{RA(J2000)} &
\multicolumn{1}{l}{Dec(J2000)} &
\multicolumn{1}{l}{Observation \tablefootmark{b}} &
\multicolumn{1}{l}{Photon counts (ct)} &
\multicolumn{1}{l}{} \\
\multicolumn{1}{l}{} &
\multicolumn{1}{l}{} &
\multicolumn{1}{l}{} &
\multicolumn{1}{l}{} &
\multicolumn{1}{l}{$\rm{0.2 keV \leq ct \leq 1 keV}$} &
\multicolumn{1}{l}{$\rm{1 keV < ct \leq 10 keV}$} \\
\noalign{\smallskip}\hline\noalign{\smallskip}
 47 & 00:42:21.29 & +41:15:52.6 & ACIS-I, ObsID 11275, 2009-11-11 &  57 &  50 \\
 65 & 00:42:29.08 & +41:15:47.6 & ACIS-I, ObsID  9523, 2008-09-01 &  14 &  21 \\
127 & 00:42:41.46 & +41:16:17.9 & ACIS-I, ObsID  7064, 2006-12-04 & 271 & 413 \\
128 & 00:42:41.61 & +41:14:37.2 & ACIS-I, ObsID  7064, 2006-12-04 &  87 & 190 \\
154 & 00:42:43.80 & +41:16:12.7 & ACIS-I, ObsID  9524, 2008-10-13 &  28 &  42 \\
159 & 00:42:43.93 & +41:16:10.8 & ACIS-I, ObsID  9524, 2008-10-13 &  19 &  20 \\
161 & 00:42:43.99 & +41:16:37.1 & ACIS-I, ObsID 12974, 2011-09-28 &  23 &  15 \\
162 & 00:42:44.12 & +41:16:04.3 & ACIS-I, ObsID  7138, 2006-06-09 &  18 &   9 \\
180 & 00:42:45.10 & +41:15:42.8 & ACIS-S, ObsID 14197, 2011-09-01 & 108 & 151 \\
189 & 00:42:46.42 & +41:16:10.2 & ACIS-I, ObsID 11279, 2010-03-05 &  41 &  62 \\
222 & 00:42:51.79 & +41:17:27.3 & ACIS-I, ObsID  7064, 2006-12-04 & 141 & 120 \\
244 & 00:42:55.99 & +41:17:21.5 & ACIS-I, ObsID 12161, 2010-11-16 &   6 &   4 \\
247 & 00:42:56.97 & +41:20:05.1 & ACIS-I, ObsID 12160, 2010-10-19 &  48 &  49 \\
259 & 00:43:01.15 & +41:13:17.7 & ACIS-I, ObsID  7064, 2006-12-04 &  12 &  22 \\
\noalign{\smallskip}\hline
\end{tabular}
\tablefoot{
\tablefoottext{a}{Source number in the catalogue.}
\tablefoottext{b}{Observation, from which the spectrum was estimated (instrument, observation number (ObsID), date).}
}
\end{center}
\label{tab:HR}
\end{table*}
In order to distinguish between XRBs with harder spectrum or SSSs with super-soft X-ray spectrum, we checked the hardness ratios (HR) of correlating X-ray sources from previous catalogues (see section \ref{sec:corr}). In some cases we also checked hardness information from previous studies on individual sources, in which data from the \emph{XMM-Newton} European photon imaging camera \citep[EPIC,][]{2001A&A...365L..18S,2003SPIE.4851..169T} detectors, the \emph{Swift} X-ray telescope \citep[XRT,][]{2004ApJ...611.1005G} or the \emph{Chandra} Advanced CCD Imaging Spectrometer \citep[ACIS,][]{2003SPIE.4851...28G} were analysed. Where we could not obtain spectral information from the identification with known X-ray sources, we extracted hardness information from one adjacent detection with \emph{Chandra} ACIS, obtained from the \emph{Chandra} data archive\footnote{http://cxc.harvard.edu/cda/}, although there were many more observations available in most cases. We used observations with the ACIS-I or ACIS-S CCD array.\\
We extracted ACIS source counts in the 0.2 - 1 keV and 1 - 10 keV band (see Table \ref{tab:HR}). The counts were extracted from regions with $\rm{\sim 95\%}$ encircled count fraction PSF size and were not corrected for background contributions. For the ACIS-I observations the background was negligible. In the ACIS-S observation (ObsID 14197, $\rm{\sim 37.5}$ ks exposure) we extracted 9 counts in the background. The obtained count ratio given in Table \ref{tab:HR} can be used to clearly distinguish hard X-ray spectra from super-soft spectra. In all cases the source counts in the bands were of similar order, indicating a hard spectrum. For a SSS spectrum most of the counts would be expected in the 0.2 - 1 keV band. SSS selection criteria have been discussed by \citet{2002ApJ...577..738K} for \emph{Chandra} ACIS and by \citet{2005A&A...434..483P} for \emph{XMM-Newton} EPIC.\\
For two outburst sources (No. 116 and No. 300, see section \ref{sec:xrb}), which were only observed with \emph{Chandra} HRC-I, we give a rough estimate of hardness based on HRC-I HRs, as has been successfully demonstrated by \citet{2007A&A...465..375P}.\\
The \emph{Chandra} observatory proposers guide\footnote{http://cxc.harvard.edu/proposer/POG/html/chap7.html\#tth\_chAp7} suggests this procedure.

\section{Analysis and results}
\label{sec:analyses}

\subsection{Catalogue compilation}
\label{sec:catalogue}

\begin{figure}
  \centering
  \resizebox{\hsize}{!}{\includegraphics[angle=0,clip=]{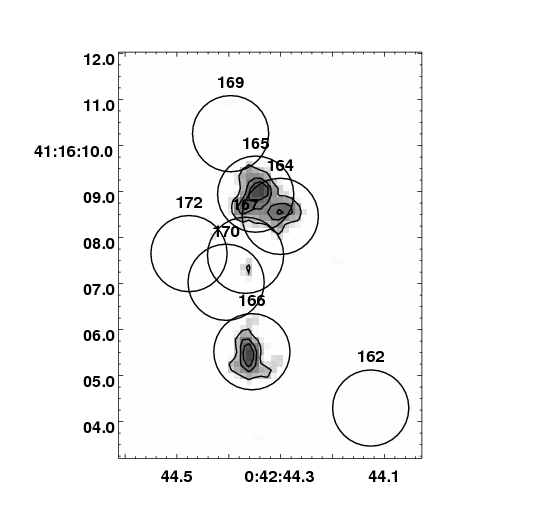}}
  \caption{Logarithmically scaled contour image resolving the M\,31 nucleus (ObsID 10684). Coordinate system in RA and Dec. Image smoothed with Gaussian function of 2 pixel FWHM and contours are chosen at 0.5, 1.1 and 1.5 counts per pixel to illustrate the separation of the two sources at the core of M\,31 (double core east No. 165 and west No. 164). No. 162, 169, 170, 172 were not active in this observation. The region size is the $\rm{90\%}$ encircled count fraction PSF.}
  \label{fig:nucleus}
\end{figure}

To produce the source catalogue from the detection lists of the individual and merged observations (for sources with significance above $\rm{3 \sigma}$), we matched positions of sources in each observation, using a script implemented in {\tt Python}\footnote{http://www.python.org/}. We consecutively merged the detection lists of the individual and merged observations, by adding sources, which were not correlating with any source position in the preliminary merged list within a radius of two times the $\rm{50 \%}$ encircled count fraction PSF size. This conservatively exceeds the $\rm{3 \sigma}$ positional errors given by {\tt wavdetect} and in addition accounts for systematic effects, which are caused mainly by the strong degradation of the PSF size with increasing off-axis angle. While building this list we excluded sources, which were only detected in one observation and had a source significance below $\rm{5 \sigma}$. The parameters adopted for each source are derived from the most significant detection with {\tt wavdetect}. We then removed sources with a maximum source significance below $\rm{4 \sigma}$ and screened the list. During the screening, we removed one duplicate of a source, spurious detections around very bright sources, faint and far off-axis sources with bad positions and false sources found at the detector edges with unrealistic wavelet scales. Overall there were 11369 individual detections with {\tt wavdetect}. After the merge and after applying the $\rm{> 3 \sigma}$ cut, there were 384 sources in the catalogue. The further significance cuts and the screening reduced the number of sources from 384 to the final 318.\\
As the detection was optimised to find point sources, we also removed known supernova remnants \citep[SNRs included in][]{2012A&A...544A.144S} from the catalogue, if the source was split into multiple spurious detections, due to the extent of the SNR. We will discuss them separately in section \ref{sec:snr}. The {\tt wavdetect} tool did not find the correct position for the far off axis neighbouring sources No. 20, 28 in the catalogue (with the merging parameters), so we determined the detection properties from an observation (HRC-I ObsID 5925) where one of the sources was too faint to be detected. The properties of No. 18 and 21 were also determined from single detections (HRC-I ObsID 5926 and 5927) since they were too close together to be resolved during the automatic compilation of the catalogue. We also added the positions for the double nucleus of the galaxy (No. 164 and 165 see Fig. \ref{fig:nucleus}), from ObsID 10684, where the nucleus is resolved.\\
The final catalogue includes 318 sources, 20 of which are only detected in the merged images. It is sorted ascending in right ascension (J2000). From this point onwards we shall refer to specific sources in the catalogue with their source number (see SRC\_ID column in section \ref{sec:col}) and ``No.'' in front.\\
It should be noted, that the systematic positional errors increase, the further off-axis a source is located as the PSF of the \emph{Chandra} HRMA degrades with increasing off-axis angle. Figures \ref{fig:high}, \ref{fig:mid}, \ref{fig:low} show the source regions of the HRC-I catalogue plotted over the merged images at different resolutions. The sources are marked by circles with $\rm{\sim 90\%}$ encircled count fraction PSF radius for 1 keV photons at the source position.

\subsection{Correlations}
\label{sec:corr}

The catalogue was correlated with the X-ray catalogues, listed below (abbreviation, reference and assumed $\rm{68\%}$ confidence errors) to identify previous detections and classifications of sources:\\
\begin{itemize}
     \item KGP2002: \cite{2002ApJ...577..738K}, individual errors given
     \item Ka2002: \cite{2002ApJ...578..114K}, 0.35{\arcsec} error
     \item PFH2005: \cite{2005A&A...434..483P}, individual errors given
     \item SHL2001: \cite{2001A&A...373...63S}, individual errors given
     \item SHP97: \cite{1997A&A...317..328S}, individual errors given
     \item SPH2008: \cite{2008A&A...480..599S}, individual errors given
     \item SPH2011: \cite{2011A&A...534A..55S}, individual errors given
     \item VG2007: \cite{ 2007A&A...468...49V}, 0.4{\arcsec} error
     \item WGK2004: \cite{2004ApJ...609..735W}, 0.5{\arcsec} error
     \item WNG2006: \cite{2006ApJ...643..356W}, 0.5{\arcsec} error
     \item HPH2011: \cite{2011A&A...533A..52H}, 0.4{\arcsec} error
     \item PHS2007: \cite{2007A&A...465..375P}, individual errors given
     \item \emph{Chandra} Source Catalog (CSC): \cite{2010ApJS..189...37E}, individual errors given
     \item D2002: \cite{2002ApJ...570..618D}, 1{\arcsec} error
     \item DKG2004: \cite{2004ApJ...610..247D}, 1{\arcsec} error
     \item O2006: \cite{2006ApJ...643..844O}, 2{\arcsec} error
     \item OBT2001: \cite{2001A&A...378..800O}, 1{\arcsec} error
     \item SBK2009: \cite{2009A&A...495..733S}, 3.1{\arcsec} error
     \item TF91: \cite{1991ApJ...382...82T}, 1.5{\arcsec} error
     \item TP2004: \cite{2004ApJ...616..821T}, 1.5{\arcsec} error
     \item TP2006: \cite{2006ApJ...645..277T}, 1{\arcsec} error
\end{itemize}
We considered two source positions as correlated, if their angular distance (d) from each other meets the criterion
\begin{equation}\label{eq:corr}
 \rm{d < 3.439 \cdot \sqrt{{\sigma_1}^2 + {\sigma_2}^2}}
\end{equation}
where $\rm{\sigma_{1,2}}$ are the $\rm{68\%}$ confidence errors of the positions, of detection 1 and 2, respectively.
Assuming that the probability distribution of a source position follows a Rayleigh distribution, the above formula (\ref{eq:corr}) results in $\rm{99.73\%}$ completeness of the correlation \citep[i.e. the probability, that we will detect all real counterparts, see e.g.][]{2009A&A...493..339W}.\\
The catalogue was also compared to source catalogues in other wavelengths. In the near infrared we used the 2MASS catalogue \citep[][]{2006AJ....131.1163S}, while in the optical we used the LGGS catalogue \citep[][]{2006AJ....131.2478M}, for which we visually inspected the images by creating finding charts for all the X-ray sources as described in section \ref{sec:oc}. For the comparison with radio sources, we used the \citet{1990ApJS...72..761B} catalogue (hereafter B90), based on Very Large Array (VLA) radio observations. The classification criteria are discussed in section \ref{sec:discussion} and summarized in Table \ref{tab:classes}. In order to find out which sources are associated with globular clusters in M\,31 we compared the catalogue positions with globular clusters from \citet{2007A&A...471..127G} (and checked classifications in the latest online version of the Revised Bologna Catalog (RBCv5, August 2012)). To identify X-ray sources with novae in M\,31, we correlated the catalogue with optical nova positions, obtained from the M\,31 nova monitoring project\footnote{http://www.mpe.mpg.de/$\sim$m31novae/opt/m31/index.php}.\\

\subsection{Variability}
\label{sec:variability}
\begin{figure}
  \centering
  \resizebox{\hsize}{!}{\includegraphics[angle=0,clip=]{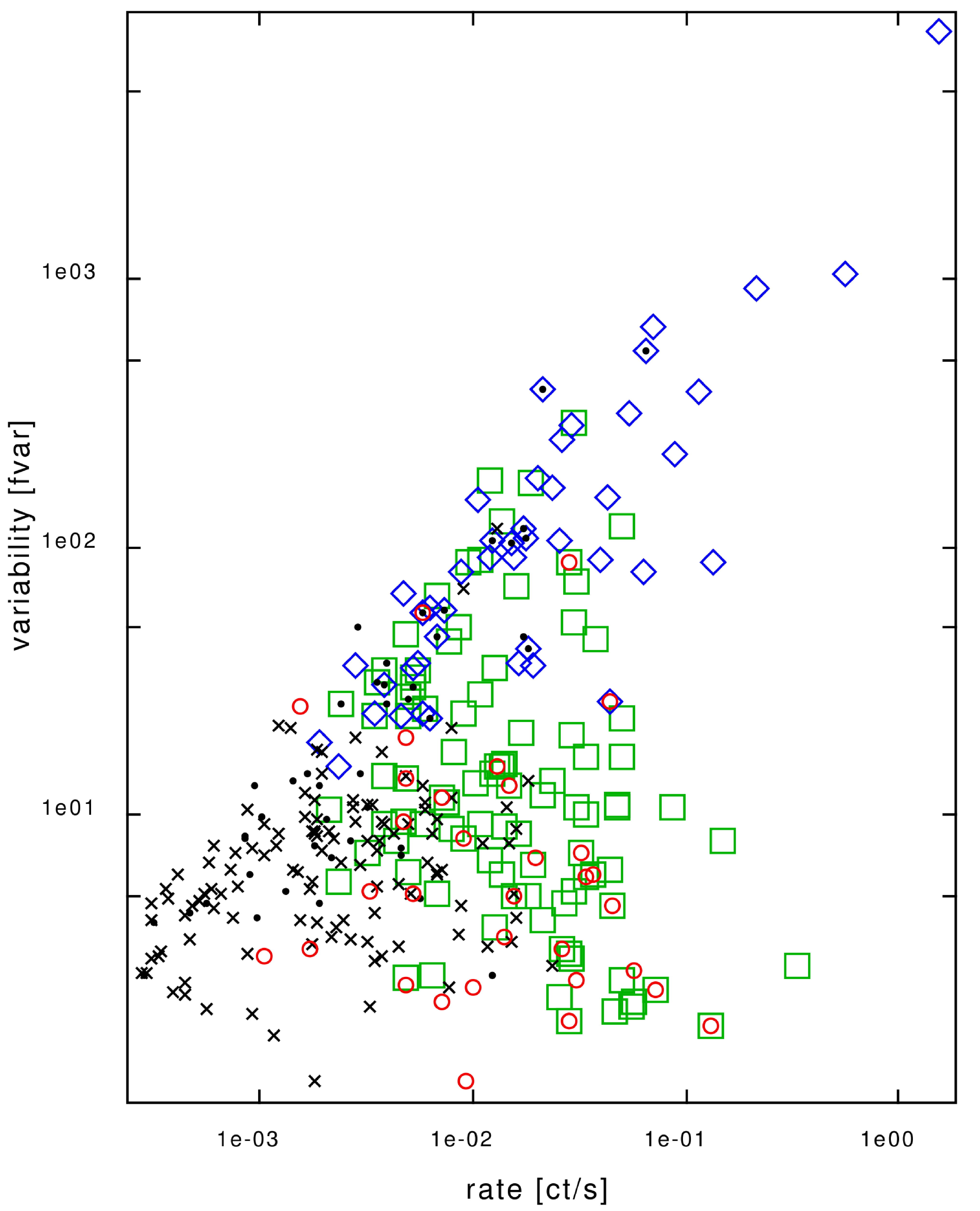}}
  \caption{Population plot showing the source variability ratio fvar versus the maximum photon count rate of a source. Blue diamonds are OBs, green squares are HVs, red circles are GlCs, black dots are novae and black crosses are the rest of the sources.}
  \label{fig:fvar}
\end{figure}
\begin{figure*}
    \centering
    \includegraphics[width=\linewidth]{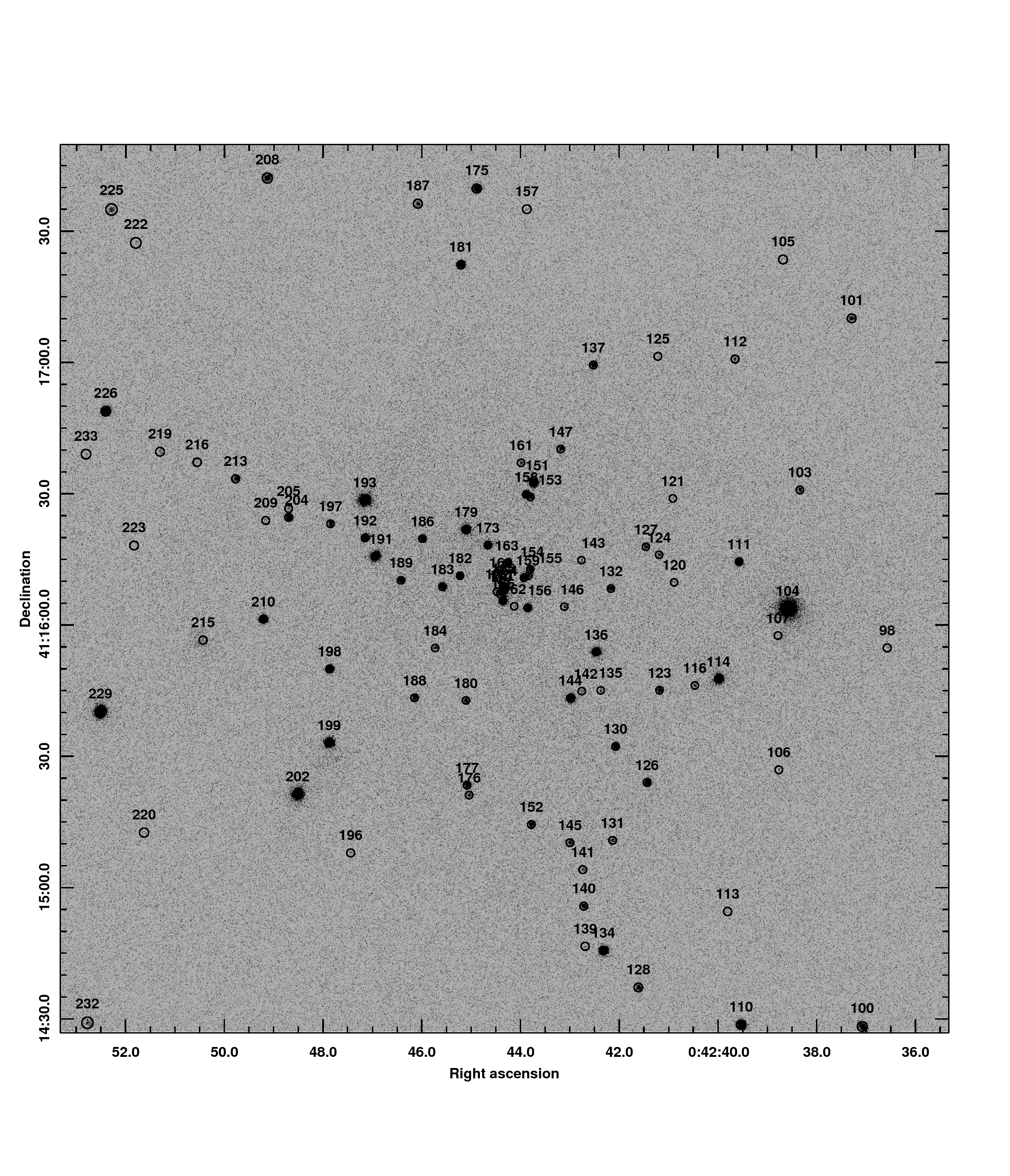}
  \caption{HRC-I merged high resolution image (0.13\arcsec \space pixels) with source positions and source No. from the catalogue. The region size is the $\rm{90\%}$ encircled count fraction PSF.}
  \label{fig:high}
\end{figure*}
\begin{figure*}
    \centering
    \includegraphics[width=\linewidth]{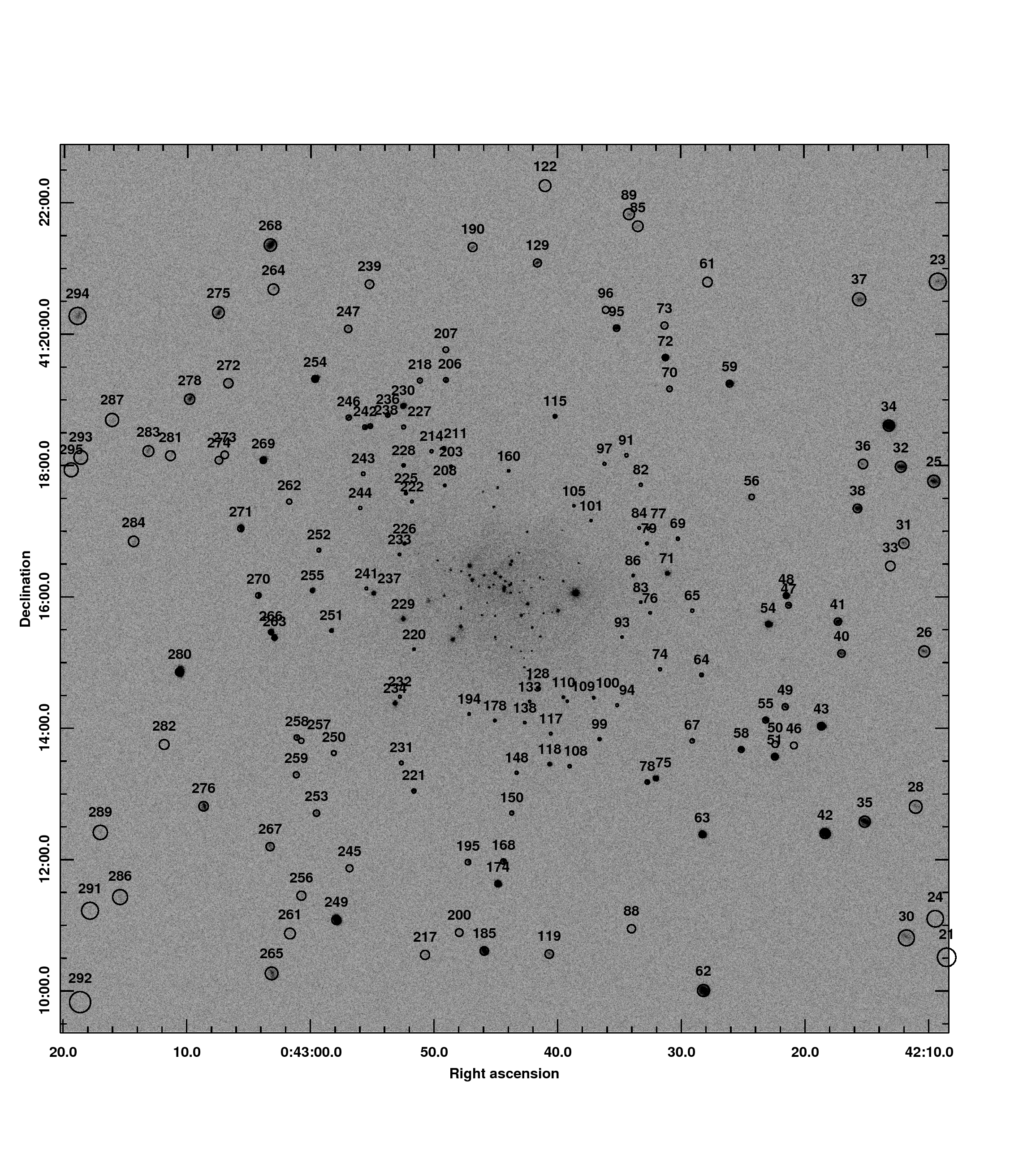}
  \caption{HRC-I merged middle resolution image (0.52\arcsec \space pixels) with source positions and source No. from the catalogue. The region size is the $\rm{90\%}$ encircled count fraction PSF.}
  \label{fig:mid}
\end{figure*}
\begin{figure*}
    \centering
    \includegraphics[width=\linewidth]{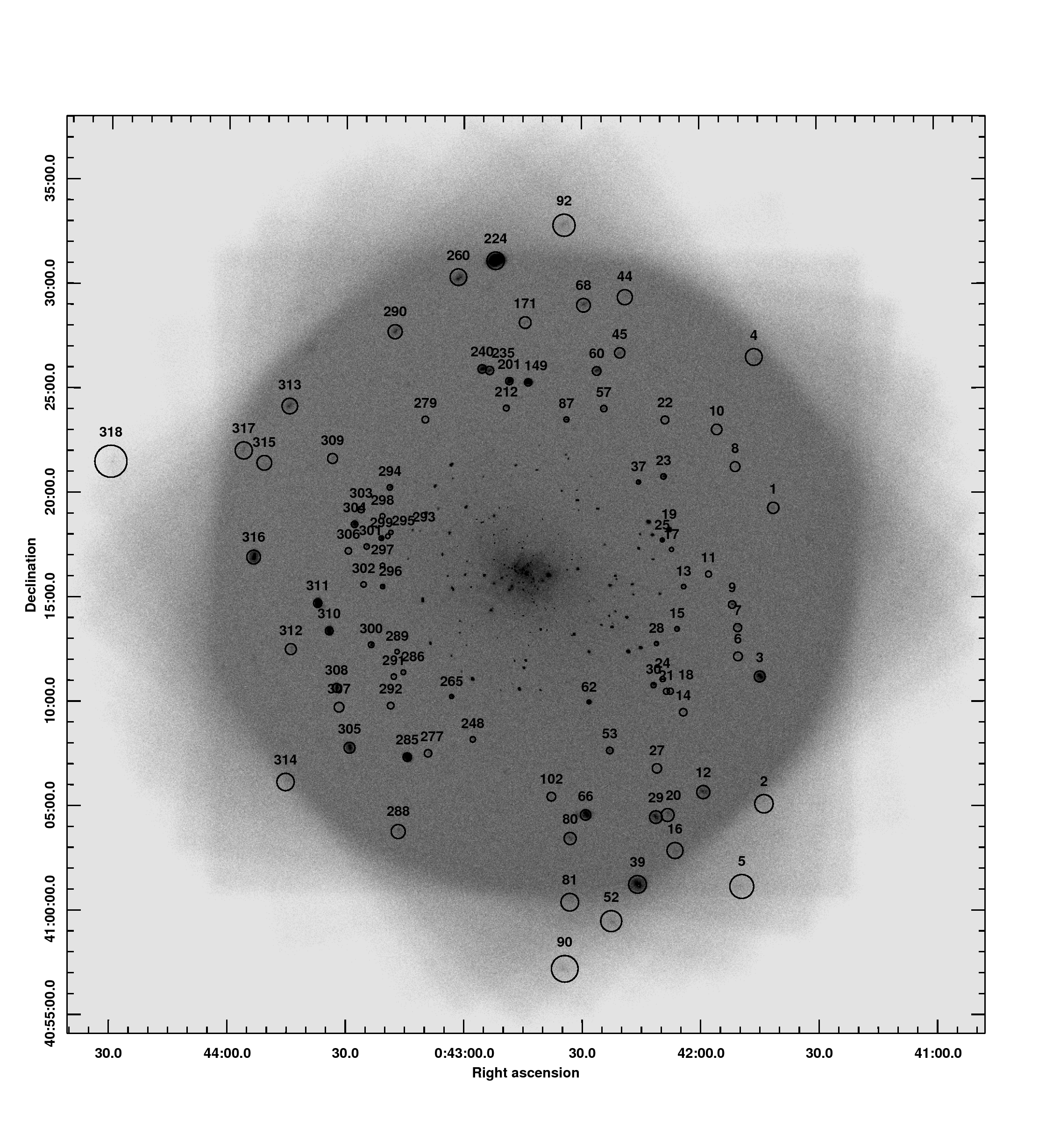}
  \caption{HRC-I merged low resolution image (2.08\arcsec \space pixels) with source positions and source No. from the catalogue. The image shows the overlap of the whole detector at different roll angles. The region size is the $\rm{90\%}$ encircled count fraction PSF.}
  \label{fig:low}
\end{figure*}

To investigate the X-ray variability of the point sources, we developed a Linux shell script, calling various CIAO tools, to obtain the source counts and upper limits for each source in each observation. We extracted source counts from the event files (with {\tt dmlist}), using circular extraction regions. As radii, we used the $\rm{90\%}$ encircled count fraction PSF radius at 1 keV, calculated with {\tt mkpsfmap} at the source position. Then we subtracted background counts calculated from the background maps accordingly (with {\tt dmstat}). Using the CIAO tool {\tt aprates} we then calculated the source statistics. We executed two {\tt aprates} runs with $\rm{68\%}$ and $\rm{99.7\%}$ confidence level for the errors and upper limits (corresponding to $\rm{1\sigma}$ and $\rm{3\sigma}$ confidence for a Gaussian distribution). Luminosities were calculated from the {\tt aprates} count rate output, as described in section \ref{sec:detection}.\\
For each source, we calculated the fraction (ndet) of the number of $\rm{>1\sigma}$ data points in the light curves over the number of observations, in which the source is in the detector's FOV. Additionally we introduced two measures of variability to the source catalogue \citep[see][]{2008A&A...480..599S, 2002ApJ...577..738K}.\\
One is the variability significance (svar)
\begin{equation}\label{eq:svar}
 \rm{svar = ({F_{max} - F_{min}})/{\sqrt{\sigma_{\rm{max}}^2 + \sigma_{\rm{min}}^2}}}
\end{equation}
where F is the flux and the $\rm{\sigma}$ are the errors associated with the flux. For the extreme values $\rm{F_{max}}$ and $\rm{F_{min}}$ (fmax and fmin in the primary source catalogue table) we used the flux values, where $\rm{F-\sigma}$ is maximal and $\rm{F+\sigma}$ is minimal respectively (adopted from Sturm et. al. 2012, submitted to A\&A). If a $\rm{3\sigma}$ upper limit, lower than $\rm{F_{min}}$ was available, we used it as the minimum flux instead, and in place of its uncertainty $\rm{\sigma_{min}}$ in equation (\ref{eq:svar}) we used the $\rm{1\sigma}$ upper limit.\\
The second variability parameter is the variability ratio (fvar), which stands for the ratio of maximum over minimum flux $\rm{F_{max}/F_{min}}$ of the source. To calculate these properties we used data points for which the significance of the respective source given by {\tt aprates} was above $\rm{1\sigma}$.\\
In addition to the overall variability significance svar between the maximum and minimum flux over 13 years, we calculated the maximum of the variability significance (ssvar), between neighbouring observations. The observations have different spacings in time but the maximum variability between them is still a good measure for variability on time scales from about ten days to one year.\\
Using the data in Table 2, we plotted the HRC-I light curves for all the sources over the last thirteen years \citep[visualized using matplotlib by][]{Hunter:2007}. The variability plots show brightness as count rates (label on right ordinate) and luminosity in M\,31 (label on left ordinate, assuming the power law spectrum from section \ref{sec:detection} and a distance of 780 kpc) with significance higher than $\rm{1\sigma}$, as a function of time in modified Julian date (MJD). The data points are filled circles (black) with $\rm{1\sigma}$ error-bars, showing the mean brightness during one observation. Whenever the source was outside the detector FOV of the relevant observation, there is no corresponding data point. If the source significance was below one, the $\rm{3\sigma}$ upper limit of the source flux is marked by a red triangle. The observations span about 12.5 years from 1999-11-30 to 2012-06-01. Sources No. 2, 4, 5, 8, 16, 22, 24, 44, 52, 90, 303, 315 and 318 have no detection $\rm{>1\sigma}$ in the individual observations and so we cannot show light curves for those sources.\\
There are 18 sources in the catalogue, for which the $\rm{90\%}$ encircled count fraction PSF overlaps with an adjacent source, thus causing some cross-talk (see ``XT`` flag in section \ref{sec:col}). As can be seen in Figs. \ref{fig:nucleus}, \ref{fig:high}, \ref{fig:mid} and \ref{fig:low} this overlap is mostly critical for the sources in the crowded core region of M\,31.\\
All light-curve plots are available in the online version of this publication in Fig. \ref{fig:lc_all}.
\begin{figure}
  \centering
  \resizebox{\hsize}{!}{\includegraphics[angle=0,clip=]{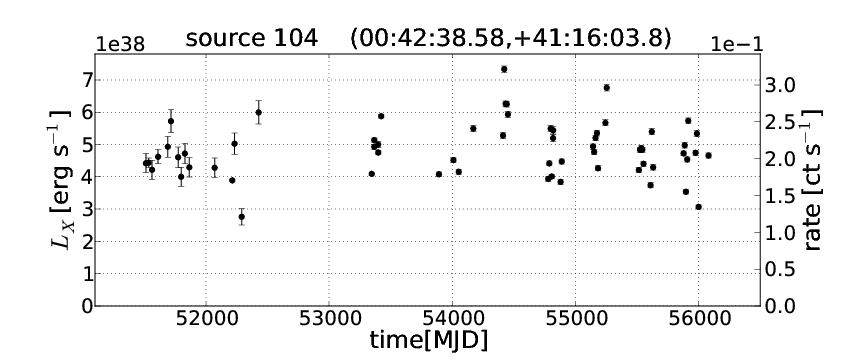}}
  \resizebox{\hsize}{!}{\includegraphics[angle=0,clip=]{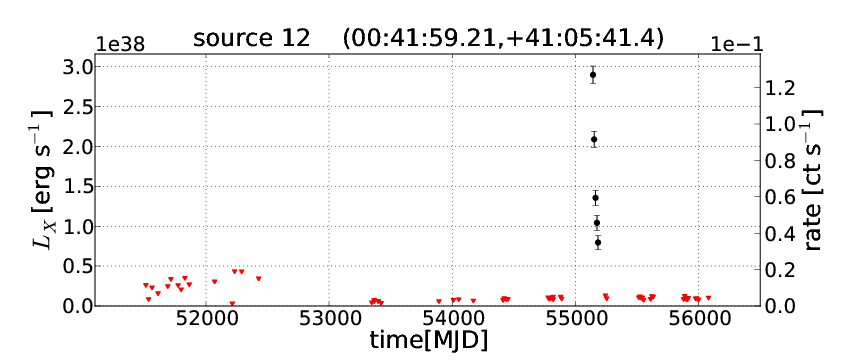}}
  \resizebox{\hsize}{!}{\includegraphics[angle=0,clip=]{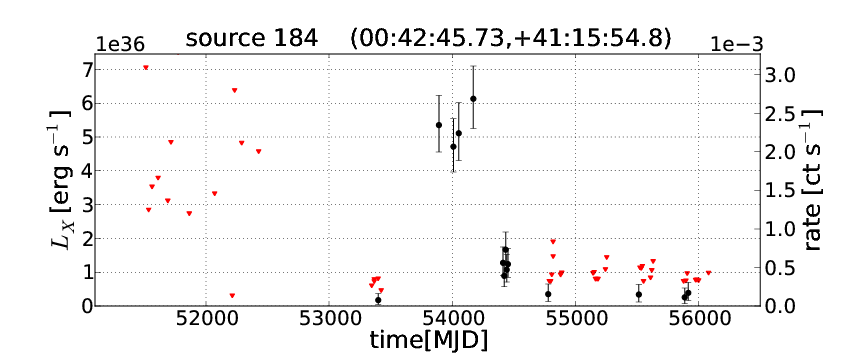}}
  \resizebox{\hsize}{!}{\includegraphics[angle=0,clip=]{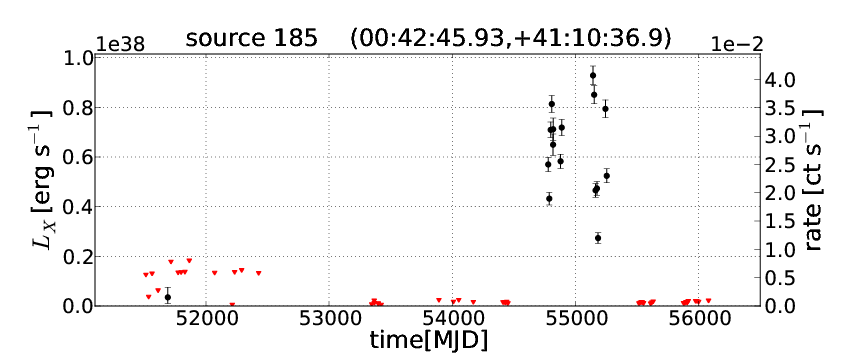}}
  \resizebox{\hsize}{!}{\includegraphics[angle=0,clip=]{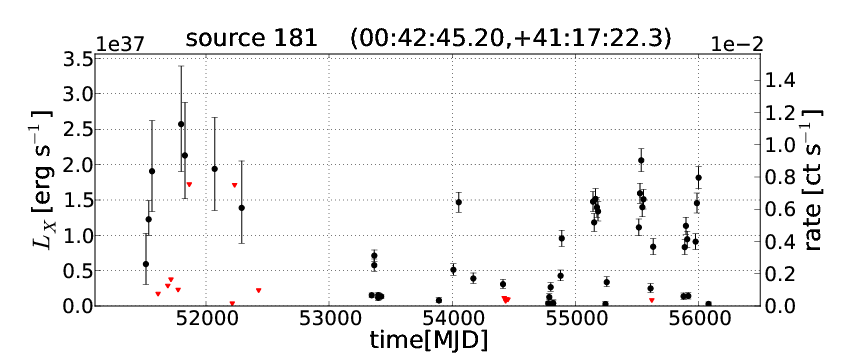}}
  \caption{Example of light curve plots showing X-ray luminosity (left axis, 0.2-10 keV) and photon count rate (right axis) over modified Julian date. Fluxes with $\rm{1\sigma}$ errors are given as black dots and $\rm{3\sigma}$ upper limits as red triangles.}
  \label{fig:lcclass}
\end{figure}

\subsection{Light-curve classes}
\label{sec:lc_class}
We derived two light-curve classes using the following selection criteria from the variability parameters defined in section \ref{sec:variability}:
\begin{itemize}
 \item Highly Variable (HV): $\rm{~ndet > 0.3 ~\& ~(ssvar > 4 ~or ~svar > 5)}$
 \item Outburst (OB): $\rm{~ndet < 0.3 ~\& ~fvar > 15 ~\& ~svar > 5}$
\end{itemize}
HV sources are detected in $\rm{>70\%}$ of the observations and show strong variability between neighbouring observations and/or high variability at longer time scales in the HRC-I monitoring (e.g. No. 104 in Fig. \ref{fig:lcclass}).\\
OB sources show a significant intensity increase by at least a factor of 15 and are not detected in at least $\rm{70\%}$ of the observations, when in the field of view. We further divide OBs into three subclasses:
\begin{itemize}
\item Short outbursts (SOB, e.g. No. 12 in Fig. \ref{fig:lcclass}), which decay to a quiescent level below the sensitivity limit in less than about 500 days.
\item Long outbursts (LOB, e.g. No. 184 in Fig. \ref{fig:lcclass}), which need more than 500 days to decay. 
\item Activity periods, which show strong variability, during an outburst in X-rays (AP, e.g. No. 185 in Fig. \ref{fig:lcclass}).
\end{itemize}
Some OBs may have been classified as transient X-ray sources in the literature, as they show one outburst and then disappear again. However, sources in M\,31 are only detected if they are brighter than about 1\ergs{36}. Therefore we cannot be sure, if a source is not still reasonably bright before and after OBs, since \citet{1998ASPC..137..506G} find quiescent states of Galactic XRBs to be much lower (between $\rm{\sim 10^{33} ~erg ~s^{-1}}$ and $\rm{\sim 10^{34} ~erg ~s^{-1}}$) than the sensitivity limit. Good examples for such behaviour are No. 181 (Fig. \ref{fig:lcclass}) or No. 152 and 198 in the catalogue, which show outbursts, but are still detected in between. Sources with very long APs can also be classified as HV instead of OB.\\
Fig. \ref{fig:fvar} shows a scatter plot of the variability factor (fvar) versus the maximum photon count rate for all sources with variability information in the catalogue (created using TOPCAT\footnote{http://www.starlink.ac.uk/topcat/}). The plot shows a branching of variable sources at higher luminosities. The OBs by definition have a higher fvar, the brighter they get, the HVs can have a low fvar at high luminosities (see also Fig. \ref{fig:hvxb_spread}). Due to the sensitivity limit of the observations, sources at lower luminosities generally have a smaller variability factor.

\subsection{Table description}
\label{sec:col}

The catalogue consists of two tables, containing the 318 sources.\\
The primary table with general source information (Table 1) and a secondary table with detailed source variability information (Table 2).\\
The primary table (Table 1) contains the source number (SRC\_ID, Col. 1), the RA and Dec coordinates (J2000) in degrees (RA, DEC, Col. 2, 3), three times the statistical $\rm{1\sigma}$ positional error (from {\tt wavdetect}, not including systematic errors due to centroid offsets far off-axis\footnote{http://cxc.harvard.edu/ciao/ahelp/wavdetect.html}) and the systematic catalogue error from offset correction and reference catalogue, see section \ref{sec:detection} (error, Col. 4), the net source counts and corresponding error (NET\_COUNTS, NET\_COUNTS\_ERR, Col. 5, 6), background counts in the source region (BKG\_COUNTS, Col. 7), the net count rate and corresponding error (NET\_RATE, NET\_RATE\_ERR, Col. 8, 9),  the background count rate and corresponding error (BKG\_RATE, BKG\_RATE\_ERR, Col. 10, 11), the exposure time of the most significant detection of the source (EXPTIME, Col. 12) and the highest source significance found in any observation (SRC\_SIGNIFICANCE, Col. 13). Those parameters are part of the {\tt wavdetect} output.\\
In addition we give the number of the highest resolution image the source is detected in (res, Col. 14, see section \ref{sec:detection}), the energy flux and its error (flux, flux\_err, Col. 15, 16), luminosity, assuming the source lies in M\,31 (see section \ref{sec:detection}) and error (lum, lum\_err, Col. 17, 18) and the distance to the nearest neighbouring source (dist\_NN, Col. 19).\\
From the light curves of each source we give the parameters svar, fvar, ndet, ssvar, fmax, fmin and errors (Col. 20 to 27) and the derived light curve classification (lc\_class, Col. 28), described in section \ref{sec:variability}. When a source is detected only in the merged images or only once in the detector FOV, we cannot give any light curve information and the variability values are set to 0.0.\\
We list the X-ray correlations of the sources with other X-ray catalogues, described in \ref{sec:corr} (XID, Col. 29). Due to the high X-ray source density in the central area of M\,31, there can be confusion of X-ray sources, especially in the case of catalogues obtained with lower resolution instruments, since they do not resolve all sources. The column contains a string of correlations, which provides the source No. in the catalogue, followed by the name of the reference catalogue, the source number or name and if available the type of the source, correlating with the catalogue source (e.g. [132, 'PFH2005', ' 310 XRB; ']).\\
Names of optical novae and globular clusters, associated with an X-ray source are listed separately (nova, glc, Col. 30, 31). For the globular clusters we also list the quality flag of the RBCv5 \citep[1: confirmed, 3: controversial object, 6: confirmed star, see][]{2007A&A...471..127G}, stating if the globular cluster has been confirmed. Source No. 85 in the catalogue correlates with a former GlC candidate and has since been confirmed to be a foreground star. Source No. 49 correlates with a controversial object from the RBCv5 and we classified it as GlC candidate.\\
We then give the source type (class, Col. 32), based on the identifications in the X-ray, radio and optical wavelengths, as described in section \ref{sec:corr} and the criteria summed up in Table \ref{tab:classes}. Candidates are designated by $<>$.\\
If necessary, we comment (comment, Col. 33) on special properties of a source. Comments refer to special light curve properties of a source (see section \ref{sec:lc_class}) or note the lack of a light curve (no lc) for some sources which are only found in the merged images, or too far off the optical axis (we did not include data points, where the extraction region overlaps with the detector edge, see section \ref{sec:variability}). For some sources we added a bad position flag (bad position) in the comments, since we found that the CIAO tool {\tt wavdetect} did not find the centroid position for some faint and far off-axis sources. If the identification with known X-ray sources or counterparts in radio or optical data, due to the bad position is not clear, we do not provide a classification for the sources based on these identifications (in Col. 32). If the $\rm{90\%}$ encircled count fraction extraction region of a source overlaps with another extraction region this is highlighted by ``XT`` followed by the overlapping source No. (e.g. XT[164]). For each statement in the ``comments'' column, which is not derived from our analysis, we give an abbreviation of the relevant reference.\\
The time variability table (Table 2) contains the observation numbers (ObsID, Col. 1), start date of the observation (date, 2), the modified Julian date at the start of the observation (MJD, Col. 3) and three columns for each source, giving its flux and upper and lower $\rm{68\%}$ confidence limits (Col. 4 to 957). If a source is not detected $\rm{>1\sigma}$, we give a $\rm{3\sigma}$ upper limit in the upper confidence limit column and zero for flux and lower confidence limit. If a source is out of the FOV the three flux columns are zero.\\
\begin{table*}
\caption[]{Prominent optical identifications up to about 1{\arcsec} separation from one of the X-ray sources.}
\begin{center}
\begin{tabular}{rllcrrrrrr}
\hline\hline\noalign{\smallskip}
\multicolumn{1}{l}{No. \tablefootmark{a}} &
\multicolumn{1}{l}{RA(J2000)} &
\multicolumn{1}{l}{Dec(J2000)} &
\multicolumn{1}{l}{XID \tablefootmark{b}} &
\multicolumn{1}{l}{OptID \tablefootmark{c}} &
\multicolumn{1}{l}{V} &
\multicolumn{1}{l}{B-V} &
\multicolumn{1}{l}{U-B} &
\multicolumn{1}{l}{V-R} &
\multicolumn{1}{l}{R-I} \\
\multicolumn{1}{l}{} &
\multicolumn{1}{l}{} &
\multicolumn{1}{l}{} &
\multicolumn{1}{l}{} &
\multicolumn{1}{l}{} &
\multicolumn{1}{l}{[mag]} &
\multicolumn{1}{l}{[mag]} &
\multicolumn{1}{l}{[mag]} &
\multicolumn{1}{l}{[mag]} &
\multicolumn{1}{l}{[mag]} \\
\noalign{\smallskip}\hline\noalign{\smallskip}
 26 &  00:42:10.30 &   +41:15:10.2 &        &  J004210.30+411510.6     &   19.31   &      1.80     &    1.20   &      0.69    &     1.12   \\
 37 &  00:42:15.54 &   +41:20:32.0 &        &  J004215.52+412031.5     &   20.52   &      0.42     &   -0.47   &      0.43    &     0.83   \\
 60 &  00:42:26.18 &   +41:25:51.1 &        &  J004226.17+412552.0     &   20.61   &      1.09     &   -0.80   &      0.23    &     0.44   \\
 75 &  00:42:32.07 &   +41:13:14.6 &   $\rm{<XRB>}$&  J004232.08+411315.2     &   20.75   &      0.92     &    0.74   &      0.48    &     0.57   \\
 85 &  00:42:33.48 &   +41:21:38.9 &   fgStar  &  J004233.56+412138.5     &   17.18   &      1.61     &    1.13   &      1.11    &            \\
 94 &  00:42:35.18 &   +41:14:21.4 &        &  EO                      &           &               &           &              &            \\
185 &  00:42:45.93 &   +41:10:36.9 &   $\rm{<XRB>}$&  J004245.84+411036.9     &   21.37   &      0.43     &   -0.73   &      0.41    &     0.51   \\
207 &  00:42:49.06 &   +41:19:46.1 &        &  J004249.07+411945.8     &   19.81   &      0.63     &   -0.02   &      0.84    &     0.60   \\
238 &  00:42:55.18 &   +41:18:36.3 &   $\rm{<XRB>}$&  EO                      &           &               &           &              &            \\
262 &  00:43:01.74 &   +41:17:27.3 &   $\rm{<fgStar>}$&  J004301.78+411726.5     &   18.05   &      1.16     &    0.80   &      0.72    &     0.75   \\
264 &  00:43:03.03 &   +41:20:41.2 &        &  EO                      &           &               &           &              &            \\
283 &  00:43:13.16 &   +41:18:13.3 &        &  J004313.18+411814.1     &   22.19   &      1.35     &           &      0.26    &     1.24   \\
284 &  00:43:14.34 &   +41:16:50.8 &        &  EO                      &           &               &           &              &            \\
289 &  00:43:17.00 &   +41:12:24.9 &   $\rm{<fgStar>}$&  J004317.04+411224.2     &   18.75   &      1.59     &    1.02   &      1.00    &     1.28   \\
291 &  00:43:17.83 &   +41:11:13.4 &   $\rm{<fgStar>}$&  J004317.91+411113.5     &   19.04   &      1.67     &    1.10   &      1.08    &     1.48   \\
292 &  00:43:18.62 &   +41:09:49.8 &    nova&  J004318.58+410949.1     &   21.34   &      0.18     &    0.01   &      0.41    &    -0.23   \\
294 &  00:43:18.91 &   +41:20:16.7 &   $\rm{<SSS>}$&  EO, J004318.89+412017.2 &   22.43   &      0.70     &   -1.45   &      0.49    &     0.94   \\
301 &  00:43:24.81 &   +41:17:26.5 &        &  J004324.81+411726.1     &   21.74   &     -0.00     &   -0.79   &      0.02    &    -0.72   \\
\noalign{\smallskip}\hline
\end{tabular}
\tablefoot{
\tablefoottext{a}{X-ray source number in the catalogue.}
\tablefoottext{b}{X-ray class of the source in the catalogue.}
\tablefoottext{c}{Source name of optical identification with a source from \citet{2006AJ....131.2478M}, or EO for emission line objects, found while inspecting the images.}
}
\end{center}
\label{tab:opt}
\end{table*}
Sources, which are not included in previous dedicated M\,31 catalogues are listed in Table \ref{tab:new}. For the new sources as for all other sources in the catalogue, we give a light curve classification (see section \ref{sec:lc_class}) and hardness information if possible. For some faint sources we can only give the positions since we do not have any significant light curve information.

\subsection{Optical counterparts}
\label{sec:oc}
The search for optical counterparts is important to further constrain the nature of an X-ray source. We checked finding charts derived from broad and narrow band images, provided by the Local Group Galaxy Survey project \citep[LGGS, see][]{2006AJ....131.2478M}. Where we found a prominent and not yet clearly identified optical source within the 3$\rm{\sigma}$ error circle and up to a separation of about 1{\arcsec} of the catalogue source, we list the source name from the \cite{2006AJ....131.2478M} catalogue, with visual magnitude and optical colours in Table \ref{tab:opt}. We regard an optical counterpart as prominent, if we visually identified it as the brightest and at the same time closest counterpart in the LGGS images. Table \ref{tab:opt} lists the closest LGGS counterpart to the corresponding X-ray position. Expected counterparts in this list could be galactic foreground stars, background galaxies or high-mass stars (as companion stars in high-mass X-ray binaries in M\,31, see \ref{sec:hmxrb}). We do not include in this list confirmed objects from earlier studies such as galaxies, foreground stars, globular clusters or supernova remnants.

\section{Discussion}
\label{sec:discussion}

In the central field of M\,31 there are many X-ray sources, which have previously been classified or identified into different classes according to their X-ray spectra or at least HRs, their time variability and optical and radio correlations (see X-ray catalogues in \ref{sec:catalogue}). 
Since HRC-I has only limited spectral resolution, we had to rely on previous catalogues or neighbouring \emph{Chandra} ACIS, \emph{Swift} XRT or \emph{XMM-Newton} EPIC observations, to estimate the hardness of a source (see section \ref{sec:HR}).\\
New in this catalogue, however, are the good source positions in the inner area of the field due to the high resolution of the HRC-I detector. The long integrated exposure improves the detection of faint sources. Table \ref{tab:classes} gives an overview of the criteria for different classifications in the catalogue, which are discussed in more detail throughout this section. Table \ref{tab:classes} also lists the statistics for how many sources are clearly identified in the catalogue and how many are classified using the selection criteria.\\
For most sources we were able to analyse time variability in detail. This enabled us to give a classification based on light curves, which has not been possible to this extent in previous catalogues. However, there are still sources without classification in the catalogue, because we do not have significant light curve information on them and thus can not obtain a light curve class (see section \ref{sec:variability}). \\
In general, the variability, which we infer from the light curves can be caused by intensity or spectral changes, resulting from anisotropic radiation of a rotating object, full or partial occultation of the X-ray source or a change in energy output of the emitting source.\\
In the catalogue we classify 195 sources as located in M\,31 ($\rm{61.3 ~\%}$) and 13 foreground and background sources ($\rm{4.1 ~\%}$). 110 sources remain unclassified ($\rm{34.6 ~\%}$, see Table \ref{tab:classes}). Details of the classifications are given below. We regard a source as \emph{identified} if the source has been clearly identified in previous studies and if this identification is compatible with the HRC-I analysis. Otherwise we regard a source as \emph{classified}, if it has only been a candidate in previous studies or if we get the classification only from the variability analysis (candidates indicated by $<>$ in the catalogue).\\

\begin{table*}
\caption[]{Source classes and criteria in the catalogue.}
\begin{center}
\begin{tabular}{llll}
\hline\hline\noalign{\smallskip}
\multicolumn{1}{l}{Class \tablefootmark{a}} &
\multicolumn{1}{l}{Criteria} &
\multicolumn{1}{l}{Identified} &
\multicolumn{1}{l}{Classified} \\
\noalign{\smallskip}\hline\noalign{\smallskip}
XRB  & hard X-ray source, HV, OB & 3 & 101 \\
GlC  & correlation with optically confirmed GlC, hard spectrum & 32 & 1 \\
SSS  & X-rays mostly below 1 keV & 0 & 6  \\
nova & correlation with optical nova & 45 &    \\
SNR  & softer X-ray source and/or extent, correlation with radio or optical SNR & 6 & 1 \\
AGN  & hard X-rays, radio identification & 1 & 1  \\
fgStar  & softer X-rays, optical identification & 7 & 4  \\
unclassified & & 110 & \\
\noalign{\smallskip}\hline
\end{tabular}
\tablefoot{
\tablefoottext{a}{Classes are adopted from \citet{2011A&A...534A..55S,2005A&A...434..483P}. XRB: X-ray binary, GlC: globular cluster, SSS: super-soft source, nova: optical nova, SNR: supernova remnant, AGN: active galactic nucleus, fgStar: foreground star.}
}
\end{center}
\label{tab:classes}
\end{table*}

\subsection{X-ray binaries (XRB)}
\label{sec:xrb}

According to theoretical models, X-ray binaries (XRB) are thought to consist of a compact object (black hole, neutron star or white dwarf) and a donor star of various types, from which matter is accreted onto the compact object.\\
From the phenomenological point of view, we divide X-ray binaries into globular cluster (GlC) sources (optically identified), XRBs in the field with a hard X-ray spectrum (probably a black hole or a neutron star as the compact object) and XRBs with a super-soft (SSS, nova) X-ray spectrum (probably a white dwarf as the compact object).\\
While for hard sources a significant part of the photons are detected at energies $\rm{> 1 ~keV}$, for SSS sources the majority of photons are detected below 1 keV.

\subsubsection{Globular cluster sources (GlC)}
\label{sec:gc}

\begin{figure}
  \centering
  \resizebox{\hsize}{!}{\includegraphics[angle=0,clip=]{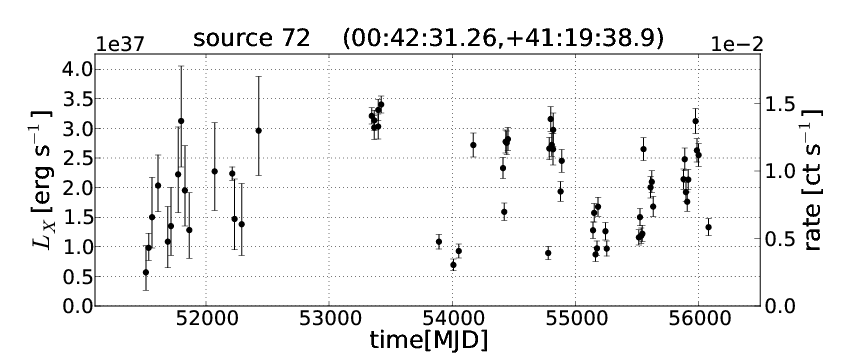}}
  \caption{Light curve plot of the GlC X-ray source No. 72, showing X-ray luminosity (left axis, 0.2-10\,keV) and photon count rate (right axis) over modified Julian date.}
  \label{fig:72}
\end{figure}

We find 32 of the X-ray sources in the catalogue to be associated with confirmed M\,31 GlCs (excluding novae). From the variability criteria (see section \ref{sec:lc_class}), we classify 16 of them as highly variable (HV, e.g. No. 72, shown in Fig. \ref{fig:72}) and 2 as OB. 14 GlC sources have no variability class, because they are too faint in X-rays to detect significant variability. In addition there are two GlC sources associated with optical novae (No. 81 and 150 in section \ref{sec:sss}). 
The errors of both the X-ray positions and the globular cluster catalogue (section \ref{sec:catalogue}) are small and the density of globular clusters in the central region of M\,31 is too low to lead to a significant contribution of chance coincidences. This allows us to state with high confidence, that the X-ray sources, we find correlated with GlCs are actually inside an M\,31 globular cluster.\\
\citet{1975ApJ...199L.143C} and \citet{1975MNRAS.172P..15F} found, that bright X-ray sources in GlCs in the Milky Way are LMXBs and that the density of LMXBs per stellar mass in GlCs is higher, than in the field. They explain this fact by two different dynamical mechanisms of binary formation (three body scattering and tidal capture). Based on these findings, we assume, that all the GlC X-ray sources are LMXB candidates. Since all the GlC sources with significant detections in their light curves show high variability between neighbouring observations and are thus classified as HV or OB sources, we classify all sources in the field showing the variability class HV or OB as XRB candidates. As shown in Table \ref{tab:classes} and the following sections, other source types can either be excluded due to time variability behaviour, spectral information or identifications at other wavelengths.\\
The classification scheme gets further support by the fact, that similar time variability in the 0.2 to 10 keV X-ray energies has been observed in Galactic X-ray sources, which have been confirmed to be XRBs \citep[e.g.][]{2010ApJ...719..201H}. The variability in the HRC-I light curves might be enhanced by spectral changes of a source.\\
As mentioned above, the GlC sources which have not been classified by the light curve variability are most probably also LMXBs. Therefore, also in the field, several LMXBs will not be classified by the light curve criterion.\\
There are no other source types, besides the SSS emission of optical novae in M\,31, that correlate with the OB and HV sources and their distribution becomes more dense toward the centre of M\,31, following the density of stellar populations in the galaxy. Thus we are very confident, that the OB and HV sources are actually XRB candidates located in M\,31, as we would expect background or foreground objects to be randomly distributed across the FOV.

\subsubsection{XRB variability}
\label{sec:gc}

\begin{figure*}
  \centering
  \resizebox{\hsize}{!}{\includegraphics[angle=0,clip=]{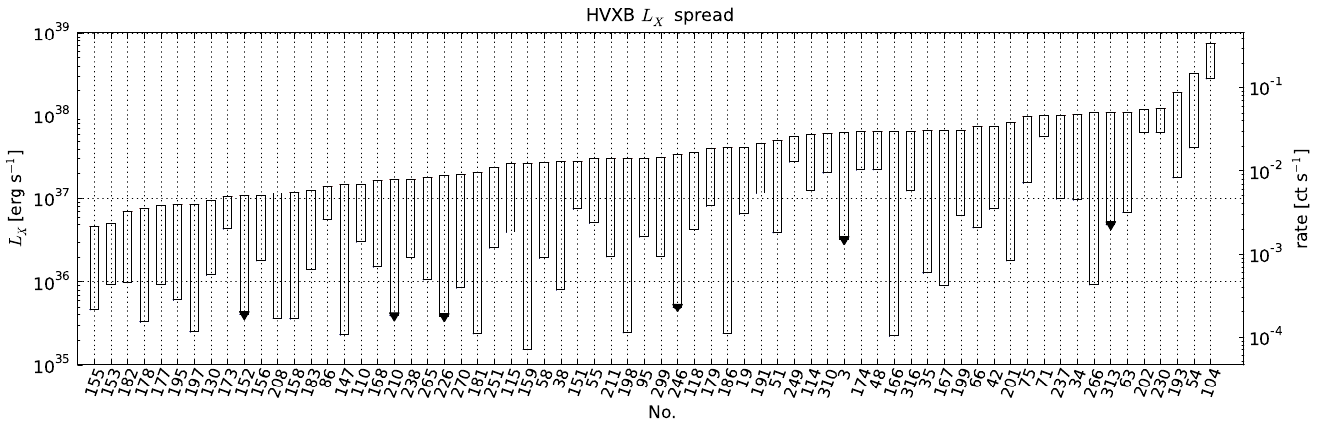}}
  \caption{Variability plot showing the spread of luminosity (left y-axis) and count rate (right y-axis) for all HV XRB sources (HVXB) in the field of M\,31. The x-axis is labelled with the catalogue source number (No.). If the minimum count rate is derived from an upper limit of the light curve, the lower count rate is marked by a downward arrow.}
  \label{fig:hvxb_spread}
\end{figure*}

\begin{figure*}
  \sidecaption
  \includegraphics[width=12cm]{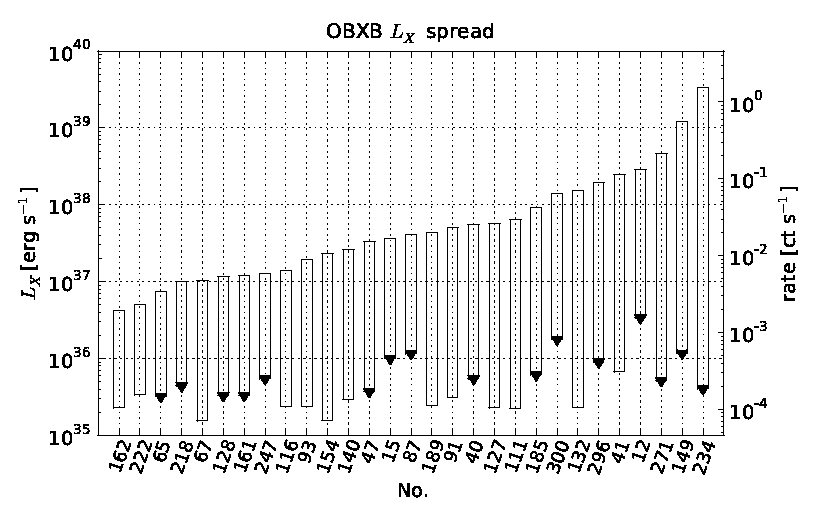}
  \caption{Variability plot showing the spread of luminosity (left y-axis) and count rate (right y-axis) for all OB XRB sources (OBXB) in the field of M\,31. The x-axis is labelled with the catalogue source number (No.). If the minimum count rate is derived from an upper limit of the light curve, the lower count rate is marked by a downward arrow.}
  \label{fig:obxb_spread}
\end{figure*}

Fig. \ref{fig:hvxb_spread} and Fig. \ref{fig:obxb_spread} show the spread of X-ray luminosity $\rm{L_X}$, we observe for HV and OB XRB (HVXB, OBXB) sources in the field of M\,31. The OBXBs all show a high variability factor (as seen in Fig. \ref{fig:fvar}). Several sources are detected in a very deep observation at a flux below the most stringent upper limits. Most HVXBs do not fall below the sensitivity limit during the observed period of thirteen years, but show a more confined $\rm{L_X}$ range. Some of them have a particularly low $\rm{L_X}$ spread, which could hint at an underlying source class within the HVXBs.\\
\\
{\bf Phenomenological discussion of some XRB candidates, with hard X-ray spectrum:}\\
\\
No. 3 shows an AP, starting in 2004 \citep[see][]{2006ApJ...645..277T}. It was first detected by \citet{2007ATel.1171....1G}.\\
No. 47 shows a SOB starting on 2009-11-07, followed by an AP lasting until the end of the HRC-I monitoring in 2012. In an ACIS-I observation (ObsID 11275 from 2009-11-11), the source shows a hard spectrum (see Table \ref{tab:HR}).\\
No. 49 is an X-ray burster \citep[see][]{2005A&A...430L..45P}, showing a highly variable light curve. The source is associated with a GlC candidate (see RBCv5 in section \ref{sec:corr}).\\
No. 54 shows two high luminosity states, each lasting for several years. During the low state (luminosity around 5\ergs{37}), the variability is low and during the high state (luminosity around 2\ergs{38}), variability is high. From 2008 until 2010-02-15 (ObsID 9825) the source steadily brightens until it reaches the high state. During this high state, the source shows variability similar to AP sources. This system is discussed as being a black-hole X-ray binary, showing spectral state changes in \citet{2011ApJ...743..185B}.\\
No. 57 shows two APs in 2004 \citep{ 2008A&A...480..599S} and 2007 \citep{2008ATel.1805....1H}. The source is not bright enough to classify it as OB (see criteria in section \ref{sec:lc_class}). We still classify the source as XRB candidate, based on the literature.\\
No. 58 is not confirmed as a fgStar candidate as in \citet{2011A&A...534A..55S}, as there is no star visible in the \citet{2006AJ....131.2478M} images and we find a highly variable light curve. Therefore we classify the source as an XRB candidate.\\
No. 65 shows a SOB, lasting a few months and starting on 2008-09-01 (in ACIS-I ObsID 9523) with a hard X-ray spectrum (see Table \ref{tab:HR}). In an observation on 2008-07-15 (ObsID 9522) it was not detected. It is further detected with ACIS-I on 2008-10-13 (ObsID 9524) and not detected after 2008-11-27 (ObsID 9521), which constrains the duration of the outburst.\\
No. 67 shows an AP starting 2009-05-29 \citep[see][]{2009ATel.1978....1G, 2009ATel.2074....1K}. The source was active untill 2011-03-10 in HRC (ObsID 13180).\\
No. 82 shows a SOB in the globular cluster B112, starting 2012-06-01 \citep[][]{2012ATel.4164....1H}. In following monitoring observations with the \emph{Swift} XRT, the source showed a rapid decay within a few weeks. It is one out of only two OB sources we found in globular clusters (see section \ref{sec:gc}).\\
No. 84 is a very faint source in the catalogue. \citet{2005ApJ...631..832W} detect the source in an outburst in their ACIS-I monitoring program and discuss it as a recurrent transient, since it was already detected by the ROSAT HRI instrument \citep[see][]{1993ApJ...410..615P}.\\
No. 104 is the most luminous HV source in the catalogue. It has been discussed as a Z-track XRB by \citet{2003A&A...411..553B}.\\
No. 116 shows a SOB, which is only detected with HRC-I, but not discovered with ACIS. We extract 34 counts in the soft HRC-I band (PI channel 50 to 130), 69 counts in the medium band (131 to 200) and 25 counts in the hard band (201 to 280). This count distribution hints at a hard X-ray spectrum \citep[see][]{2007A&A...465..375P}.\\
No. 127 shows a SOB in an HRC-I observation starting 2006-11-13. ACIS-I ObsID 07064 (2006-12-04) indicates a hard X-ray spectrum (see Table \ref{tab:HR}).\\
No. 128 shows an AP lasting from 2006-12-04 (ACIS-I ObsID 7064), until 2010-02-04 (ACIS-I ObsID 11278) with a hard X-ray spectrum (see Table \ref{tab:HR}).\\
No. 132 shows a LOB, which we cannot clearly constrain, since the source is already bright in the first HRC-I observation. It was discovered by \citet{1999IAUC.7291....1M} in October 1999. \citet{2001ApJ...563L.119T} argued, that the source shows an outburst of at least 400 days, with a plateau phase of about one year, which should indicate a black-hole binary.\\
No. 149 is the second ultra-luminous X-ray (ULX) source, detected in M\,31 \citep[see][]{2012ATel.3890....1H, 2012ATel.3921....1H}. It was also observed with \emph{Swift} XRT \citep[see][]{2013MNRAS.428.2480E} and in radio observations by \citet{2013Natur.493..187M}.\\
No. 153 and 158 are clearly resolved as two sources in the catalogue \citep[see also][]{2002ApJ...577..738K}. In \citet{2012ApJ...756...32B} the source is treated as one source. We give two light curves, which show some cross talk.\\
No. 161 shows a SOB and is first detected in ACIS-I ObsID 12973, starting 2011-08-25. In the following ACIS-I observation (ObsID 12974) from 2011-09-28 the source can be classified as hard (see Table \ref{tab:HR}).\\
No. 181 is reported as an X-ray transient by \citet{2001IAUC.7659....1S}. In the HRC-I monitoring light curve, with lower sensitivity limit, we detect the source as HV.\\
No. 185 shows an AP from 2008-05-31 (ACIS-I ObsID 9529) till 2010-03-05 (ACIS-I ObsID 11279). In HRC-I observations we detect the source as HV. The source has a hard spectrum, derived from \emph{XMM-Newton} observations \citep{2008ATel.1647....1P}.\\
No. 201 shows a long-term variability, with two LOBs at the beginning and end of HRC-I observations. After the second outburst, the source stays in a high-luminosity state, which slowly declines over the course of three years and is still at 3\ergs{37} at the end of the HRC-I monitoring.\\
No. 222 shows a SOB in just one of the HRC-I observations (ObsID 7284 on 2006-09-30). ACIS-I ObsID 7064 on 2006-12-04 indicates a hard spectrum (see Table \ref{tab:HR}).\\
No. 234 is the first ULX source found in M\,31 \citep[see][]{2009ATel.2356....1H, 2012A&A...538A..49K, 2012MNRAS.420.2969M, 2012ApJ...756...32B, 2012A&A...542A.120N}.\\
No. 237 is a HV source. In Nov. 2010 it dropped in luminosity from $\rm{\sim 10^{38} ~erg ~s^{-1}}$ and after that showed high variability. This indicates a spectral state change.\\
No. 240 shows rather slow variability. It is not classified as HV as the count rate has large errors, when the source is brightest.\\
No. 247 has a SOB starting 2010-10-19 \citep{2012ApJ...756...32B} in ACIS-I ObsID 12160, where we find a hard spectrum (see Table \ref{tab:HR}).\\
No. 259 shows a faint AP from 2006-06-05, until 2010-12-12, which is not significant enough for the outburst criteria. ACIS-I observation ObsID 7064 on 2006-12-04, indicates a hard spectrum (see Table \ref{tab:HR}).\\
No. 275 is an X-ray burster in a globular cluster system, which is discussed by \citet{2005A&A...430L..45P}.\\
No. 276 shows dips with a 107 min X-ray period \citep{2004A&A...419.1045M} and was thus classified as an X-ray binary. In the HRC-I light curve the source does not show high enough variability to be classified as HV.\\
No. 278 is at the position of a confirmed globular cluster at the edge of a SNR in the optical images \citep[see][]{2006AJ....131.2478M}, which might cause some contamination of the light curve of the GlC source. However, the SNR has to be very faint in X-rays, since we do not see an extended source at the position in the merged X-ray images.\\
No. 279 shows a AP starting 2004-05-23, discussed in \citet{2005ApJ...632.1086W}. The last detection in the HRC-I light curve is on 2005-02-21.\\
No. 285 shows a very peculiar light curve with two rather well separated luminosity states and has previously been found to have a 2.78 h period by \citet{2002ApJ...581L..27T}. \citet{2006MNRAS.366..287B} argue, that there is another component to the variability, caused by precession of the accretion disc of the compact object.\\
No. 296 shows a SOB, starting 2008-08-21, discovered with \emph{Swift} XRT \citep{2008ATel.1674....1P, 2008ATel.1693....1G, 2008ATel.1806....1H, 2008ATel.1859....1G}.\\
No. 299 is a HV source. One of its brightness peaks has been reported by \citet{2009ATel.2294....1N}.\\
No. 300 shows one SOB and is only detected with HRC-I. To estimate the hardness of the spectrum, we extract 629 counts in the soft HRC-I band (PI channel 50 to 130), 1440 counts in the medium band (131 to 200) and 479 counts in the hard band (201 to 280). This distribution indicates a hard X-ray spectrum \citep{2007A&A...465..375P}.\\
\\
{\bf Recurrent outburst hard X-ray sources:}\\
\\
No. 15 shows three APs in the light curve, the first of which is discussed in \citet{2001A&A...378..800O} and \citet{2006ApJ...643..356W}. The source is discussed as a recurrent transient in \citet{2011A&A...534A..55S}, since the source was already observed with the ROSAT observatory \citep[see][]{1993ApJ...410..615P}. We detect the latest AP, with the source still active at the end of the HRC-I monitoring on 2012-06-01.\\
No. 40 shows many recurrent SOBs about every 1.2 years \citep[see][]{2007ATel.1307....1H, 2010ATel.2730....1P, 2007ATel.1328....1G}.\\
No. 111 shows recurrent SOBs. One is discussed in \citet{2001ATel...79....1G}. The source was not active in ACIS-I 9521 (2008-11-27) and shows one outburst in HRC-I ObsID 10684 (2008-12-18). We detected a second SOB starting in ACIS-I ObsID 12164 (2011-02-16), which we also observe one day later in HRC-I ObsID 13178. If there is a quasi-periodic recurrence time, the HRC-I data would suggest an outburst period of $\sim$2 years.\\
No. 140 shows recurrent SOBs and a hard spectrum in ACIS-I \citep{2002ApJ...577..738K} and \emph{XMM-Newton} EPIC observations, separated by $\sim$6 years.\\
No. 147 shows either short recurrent OBs about every year, which are missed in some of the HRC-I monitoring observations or it shows two APs.\\
No. 162 shows a SOB in HRC-I ObsID 7283 (2006-06-05). ACIS-I observation (ObsID 7138 from 2006-06-09) closely after the HRC-I detection (ObsID 7283 from 2006-06-05) indicates a hard spectrum (see Table \ref{tab:HR}). The earliest detection of the outburst was visible in an ACIS-I observation from 2006-05-26 with a total of 9 counts. In earlier observations there was an outburst only seen with ACIS-I in an observation starting on 1999-10-13 (ObsID 303), where we extract 26 counts. Due to this outburst the source is part of the \citet{2002ApJ...577..738K} catalogue and thus is classified as a source with recurrent outbursts.\\
No. 189 shows recurrent SOBs in the HRC-I light curve. The source brightens from 2009-11-07 till 2010-02-26. ACIS-I observation ObsID 11279 on 2010-03-05 indicates a hard spectrum (see Table \ref{tab:HR}). The source is active till ACIS-I ObsID 11839 from 2010-06-23.\\
No. 246 shows multiple SOBs, for which we catch three in a very bright state in the HRC-I observations. There are many additional observations with the \emph{Chandra} ACIS detector (see Table \ref{tab:246obs}), filling in some of the HRC-I observation gaps. The source seems to have bright outbursts about every year, lasting up to about three months.\\
\begin{table}
\caption[]{Additional \emph{Chandra} detections for recurrent outburst source No. 246.}
\begin{center}
\begin{tabular}{lrr}
\hline\hline\noalign{\smallskip}
\multicolumn{1}{l}{Instrument} &
\multicolumn{1}{l}{ObsID\tablefootmark{a}} &
\multicolumn{1}{l}{Date\tablefootmark{b}} \\
\noalign{\smallskip}\hline\noalign{\smallskip}
 ACIS-I &   303 & 1999-10-13 \\
 ACIS-I &   312 & 2000-08-27 \\
 ACIS-S &  1581 & 2000-12-13 \\
 ACIS-S &  1854 & 2001-01-13 \\
 ACIS-S &  1575 & 2001-10-05 \\
 ACIS-I &  7136 & 2006-01-06 \\
 ACIS-I &  7064 & 2006-12-04 \\
 ACIS-I &  8183 & 2007-01-14 \\
 ACIS-I &  8191 & 2007-06-18 \\
 ACIS-I &  8192 & 2007-07-05 \\
 ACIS-I &  9523 & 2008-09-01 \\
 ACIS-I & 10715 & 2009-09-18 \\
 ACIS-I & 10716 & 2009-09-25 \\
 ACIS-I & 11838 & 2010-05-27 \\
 ACIS-I & 12161 & 2010-11-16 \\
 ACIS-I & 12162 & 2010-12-12 \\
 ACIS-I & 12973 & 2011-08-25 \\
 ACIS-S & 14197 & 2011-09-01 \\
 ACIS-S & 14198 & 2011-09-06 \\
\noalign{\smallskip}\hline
\end{tabular}
\tablefoot{
\tablefoottext{a}{Observation identification number.}
\tablefoottext{b}{Start date of the observation.}
}
\end{center}
\label{tab:246obs}
\end{table}
No. 290 shows OBs, lasting up to several months, about every year during the time span covered by HRC-I observations. Table \ref{tab:290obs} lists additional observations from other instruments, where the source is visible in the X-ray images, which fill in some of the time, covered by HRC-I light curves. The \emph{XMM-Newton} EPIC HR of the source indicates a hard X-ray spectrum \citep[see 2XMM catalogue by][]{2009A&A...493..339W}.\\
\begin{table}
\caption[]{Additional detections for recurrent outburst source No. 290.}
\begin{center}
\begin{tabular}{lrr}
\hline\hline\noalign{\smallskip}
\multicolumn{1}{l}{Instrument} &
\multicolumn{1}{l}{ObsID\tablefootmark{a}} &
\multicolumn{1}{l}{Date\tablefootmark{b}} \\
\noalign{\smallskip}\hline\noalign{\smallskip}
 ACIS-S &  1582 & 2001-02-18 \\
 ACIS-S &  2900 & 2002-11-29 \\
 XMM-EPIC & 0405320701 & 2006-12-31 \\
 XMM-EPIC & 0405320801 & 2007-01-16 \\
 XMM-EPIC & 0405320901 & 2007-02-05 \\
 ACIS-I &  8184 & 2007-02-14 \\
 ACIS-I &  8185 & 2007-03-10 \\
 XMM-EPIC & 0505720301 & 2008-01-08 \\
 XMM-EPIC & 0505720401 & 2008-01-18 \\
 XMM-EPIC & 0505720501 & 2008-01-27 \\
 XMM-EPIC & 0505720601 & 2008-02-07 \\
 ACIS-I & 10553 & 2009-03-11 \\
\noalign{\smallskip}\hline
\end{tabular}
\tablefoot{
\tablefoottext{a}{Observation identification number.}
\tablefoottext{b}{Start date of the observation.}
}
\end{center}
\label{tab:290obs}
\end{table}
No. 313 shows recurrent OBs, first discovered with the Einstein and ROSAT X-ray observatories. In \emph{XMM-Newton} EPIC observations it showed a harder X-ray spectrum \citep[see 2XMM catalogue by][]{2009A&A...493..339W}. In the HRC-I light curves we detect four OBs. It shows both LOBs and SOBs, with the SOBs being about nine months apart. Table \ref{tab:313obs} gives additional observations, where the source is detected.\\
\begin{table}
\caption[]{Additional observations for recurrent outburst source No. 313.}
\begin{center}
\begin{tabular}{lrr}
\hline\hline\noalign{\smallskip}
\multicolumn{1}{l}{Instrument} &
\multicolumn{1}{l}{ObsID\tablefootmark{a}} &
\multicolumn{1}{l}{Date\tablefootmark{b}} \\
\noalign{\smallskip}\hline\noalign{\smallskip}
 ACIS-S &   309 & 2000-06-01 \\
 ACIS-S &  1581 & 2000-12-13 \\
 ACIS-I &  4680 & 2003-12-27 \\
 ACIS-I &  4681 & 2004-01-31 \\
 XMM-EPIC & 0405320501 & 2006-07-02 \\
 XMM-EPIC & 0405320601 & 2006-08-09 \\
 XMM-EPIC & 0405320701 & 2006-12-31 \\
 ACIS-I &  8183 & 2007-01-14 \\
 XMM-EPIC & 0405320801 & 2007-01-16 \\
 XMM-EPIC & 0405320901 & 2007-02-05 \\
 ACIS-I &  8184 & 2007-02-14 \\
 ACIS-I &  8185 & 2007-03-10 \\
 ACIS-I &  7068 & 2007-06-02 \\
 XMM-EPIC & 0505720501 & 2008-01-27 \\
 XMM-EPIC & 0505720601 & 2008-02-07 \\
 XMM-EPIC & 0650560201 & 2010-12-26 \\
 XMM-EPIC & 0650560301 & 2011-01-04 \\
 XMM-EPIC & 0650560501 & 2011-01-25 \\
 XMM-EPIC & 0650560601 & 2011-02-03 \\
 ACIS-I & 12164 & 2011-02-16 \\
 ACIS-S & 14197 & 2011-09-01 \\
 ACIS-S & 14198 & 2011-09-06 \\
 XMM-EPIC & 0674210201 & 2011-12-28 \\
 XMM-EPIC & 0674210301 & 2012-01-07 \\
 XMM-EPIC & 0674210401 & 2012-01-15 \\
 XMM-EPIC & 0674210501 & 2012-01-21 \\
 XMM-EPIC & 0674210601 & 2012-01-31 \\
\noalign{\smallskip}\hline
\end{tabular}
\tablefoot{
\tablefoottext{a}{Observation identification number.}
\tablefoottext{b}{Start date of the observation.}
}
\end{center}
\label{tab:313obs}
\end{table}

\subsubsection{GlC population}
\label{sec:glcpop}

\begin{figure}
  \centering
  \resizebox{\hsize}{!}{\includegraphics[angle=0,clip=]{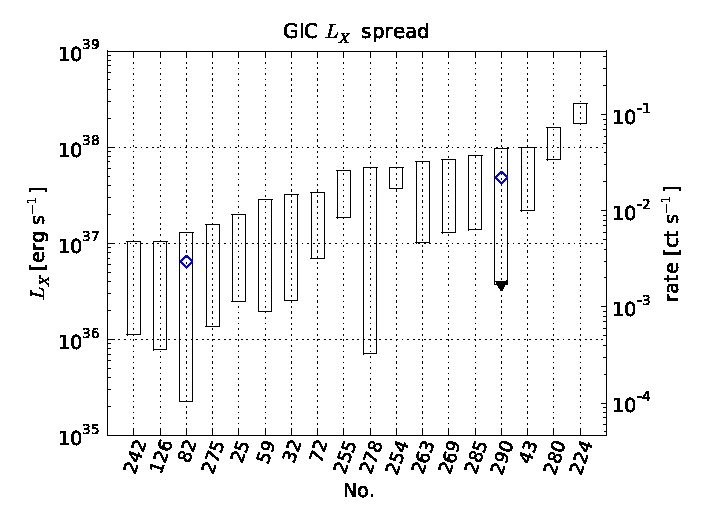}}
  \caption{Variability plot showing the spread of luminosity (left y-axis) and count rate (right y-axis) for all OB or HV GlC sources. The x-axis shows the source No. in the catalogue. If the minimum count rate is derived from an upper limit of the light curve, the lower count rate is marked by a downward arrow. The blue diamonds mark the sources which are classified as OB sources in the catalogue.}
  \label{fig:glc_spread}
\end{figure}

We found that after applying the variability criteria for XRB selection, the ratio of hard XRBs classified as HV (HVXB) to hard XRBs classified as OB (OBXB) is 2.3/1 in the field of M\,31. For the hard XRBs found in GlCs we obtained a ratio of 7.5/1 (HVXB/OBXB). For this statistical analysis we only used sources inside the $\rm{\sim70 \%}$ effective exposure radius (see Fig. \ref{fig:exp}), to avoid effects by the inhomogeneous coverage outside this circle. Fig. \ref{fig:glc_spread} visualises, that most GlC sources are HVXB sources with rather low variability factor. This can also be seen in Fig. \ref{fig:fvar}, where the GlC sources concentrate in the HV source branch. Since we have fewer sources (and thus higher statistical uncertainty) in the GlCs, we estimated the probability that this difference could occur by chance. We used Binomial statistics and assumed that the true ratio is 2.3/1 (from the field sources, where we have a sample of 97 sources, 29 OBXB and 68 HVXB). The probability to obtain a ratio of 7.5/1 after 17 trials (2 OB and 15 HV sources in GlCs) is only $\rm{8\%}$. This may indicate, that the abundance fraction of outburst sources in GlCs might be lower than in the field. The reason for this possible difference could be differing formation processes of LMXBs in the GlC and field environments. However, there is also the possibility of additional OBs in GlCs, masked by interfering HV sources, which could not be resolved if they were located in the same GlC. A higher OB fraction in the field might also be caused by high-mass X-ray binaries \citep[HMXBs, e.g. Be-XRBs with Type II outburst, see][]{2011Ap&SS.332....1R} in the sample, which were not optically identified. It has been discussed by \citet{1975ApJ...199L.143C}, \citet{1975MNRAS.172P..15F} and recently by \citet{2007MNRAS.380.1685V} that XRBs in GlCs are likely to form through tidal capture in the dense star population of a globular cluster, which they also use to explain an over-abundance of XRBs in GlCs.\\
Overall, the $\rm{\sim70 \%}$ effective exposure radius of the catalogue contains 154 confirmed GlCs from the RBCv5 (see section \ref{sec:corr}). These are associated with 27 X-ray sources, suggesting that at least $\rm{17\%}$ of GlCs in M\,31 should host an XRB. This is a very conservative lower limit, since the observations of M\,31 are not deep enough to detect all XRB sources.

\subsubsection{Low mass X-ray binaries (LMXB)}
\label{sec:lmxrb}
Following the variability analysis, the majority of X-ray sources in M\,31 seem to be LMXBs (see section \ref{sec:gc}). If one of those variable sources shows a hard X-ray spectrum, we classify it as an XRB candidate ($\rm{<XRB>}$). The HRC-I observations mainly cover the bulge of M\,31, which hosts a predominantly old stellar population. Therefore, we would only expect very few HMXB candidates in the catalogue \citep[for an estimate see][]{2007A&A...468...49V}.\\
Whether the compact object is a neutron star or a black hole can to some extent be decided from X-ray spectra or some variability properties \citep[soft X-ray sources with transient outbursts are most likely black hole systems, see][]{2000IAUS..195...37T}. The variability of XRBs in the light curves could in part be due to the transition from one spectral state into another, where hardness ratios in the 0.2-10 keV X-ray band would change. Such state changes in X-ray binaries have been discussed by \citet{1989A&A...225...79H}. For similar results from the RXTE satellite see \citet{2010ApJ...719..201H}.\\
The Eddington luminosity (proportional to the mass of the compact object in XRBs) describes the limit at which the outward photon pressure becomes so high, that no further material can be gravitationally accreted. So from the luminosities at which different spectral states of a source exist, a rough estimate for the mass of the compact object can be derived \citep[see discussion of No. 54 in][]{2011ApJ...743..185B}. If the mass is between about 1.4 solar masses ($\rm{M_{\odot}}$) and 3.0 $\rm{M_{\odot}}$, we would expect a neutron star. If the mass exceeds 3.0 $\rm{M_{\odot}}$ the compact object should be a black hole \citep[see][]{2000IAUS..195...37T}.\\
Our sample contains 4 sources exceeding the Eddington limit for a 3.0 $\rm{M_{\odot}}$ object. Two of those are the ULX sources, which have already been discussed as BH XRBs in previous studies \citep[e.g.][]{2012A&A...538A..49K, 2013Natur.493..187M}. From HRC-I data alone a clear distinction between BH and NS systems was not possible since there was not sufficient simultaneous spectral information.\\
Short bursts (at time scales of $\sim$ one minute) or pulsations in the short-term light curves (during an observation) of a source are strong evidence for a neutron star binary. Bursts \citep[as have been found for No. 49 and No. 275 by][]{2005A&A...430L..45P} are thought to arise from short nuclear burning flashes on the surface of a neutron star. For a recent review on neutron star binaries see \citet{2010AdSpR..45..949B}.\\
\citet{1998ASPC..137..506G} discuss the separation of some galactic neutron star and black hole binaries, using the variability range of OBs. The sensitivity limit of the HRC-I observations for sources in M\,31 does not allow to measure high enough ratios between maximum and minimum X-ray luminosity to apply the criteria, proposed by \citet{1998ASPC..137..506G}, to the sample of sources.

\subsubsection{High mass X-ray binaries (HMXB)}
\label{sec:hmxrb}
HMXBs are expected in star forming regions, which are mainly located in the disk of M\,31. So far there are no confirmed HMXBs in M\,31. We would classify a source as HMXB candidate if it shows XRB-like time variability in the light curves (see criteria in section \ref{sec:gc}), has a hard X-ray spectrum and shows an optical counterpart with colours consistent with a high-mass star in the \citet{2006AJ....131.2478M} images.\\
Source No. 301 is bright in X-rays, has a prominent optical counterpart (LGGS and Hubble space telescope (HST) images) and shows no radio emission in the reference catalogues. The colours and magnitude of the optical counterpart (see Table \ref{tab:opt}) and the hardness ratios of the X-ray emission would be consistent with a HMXB in M\,31 (see Sturm et al. 2012, submitted to A\&A, for a comparison with HMXBs in the Small Magellanic Cloud). However the errors of the X-ray hardness and the optical colours are too large to exclude an AGN as the source of the X-rays. We do not detect $\rm{H\alpha}$ emission at the source position and have no spectrum of the optical counterpart. The HRC-I X-ray light curve is too faint to show significant variability.\\

\subsubsection{Super Soft Sources (SSS) and Novae (nova)}
\label{sec:sss}
\begin{figure}
  \centering
  \resizebox{\hsize}{!}{\includegraphics[angle=0,clip=]{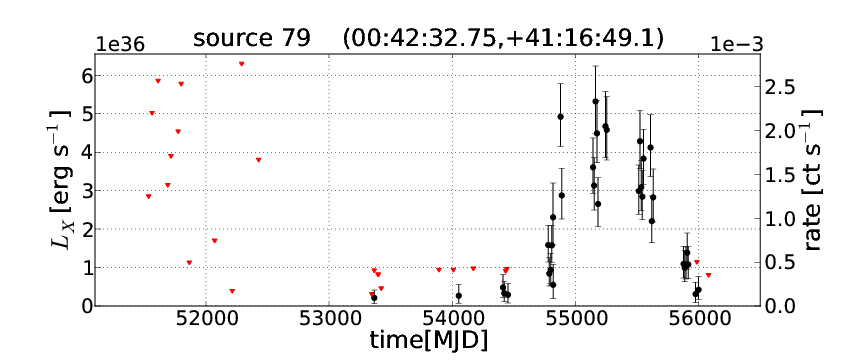}}
  \resizebox{\hsize}{!}{\includegraphics[angle=0,clip=]{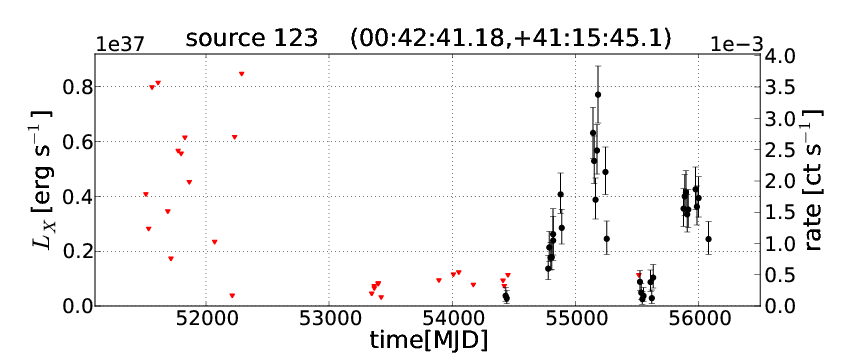}}
  \caption{Light curve plots of LOB X-ray sources associated with optical novae, showing X-ray luminosity (left axis, 0.2-10\,keV) and photon count rate (right axis) over modified Julian date.}
  \label{fig:79_123}
\end{figure}
\begin{figure}
  \centering
  \resizebox{\hsize}{!}{\includegraphics[angle=0,clip=]{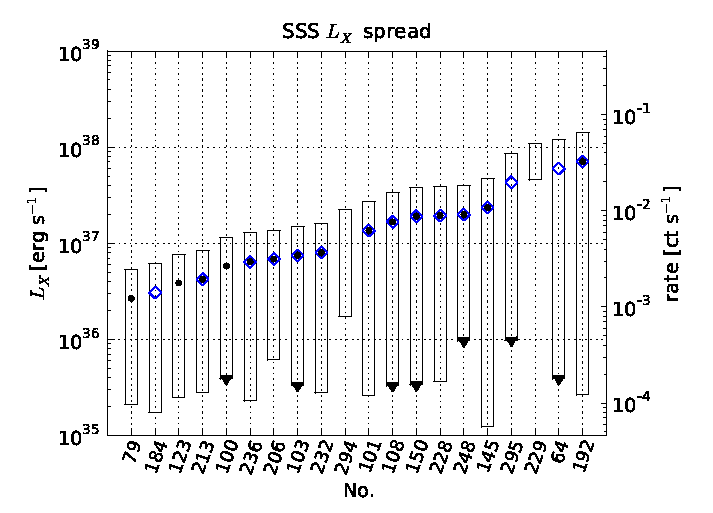}}
  \caption{Variability plot showing the spread of luminosity (left y-axis) and count rate (right y-axis) for all OB or HV SSS or nova sources. The x-axis is labelled with the catalogue source number (No.). If the minimum count rate is derived from an upper limit of the light curve, the lower count rate is marked by a downward arrow. The blue diamonds mark the sources which are classified as OB sources in the catalogue. The black dots mark the sources, which are associated to optical novae.}
  \label{fig:sss_spread}
\end{figure}

The class of SSS in M\,31 is thought to be associated with cataclysmic variables (CV), which are binary systems with a white dwarf as the accreting object. Those systems are also responsible for nova explosions \citep[see][]{2007A&A...465..375P}. Most of the sources with super-soft hardness ratios (photon counts mostly below 1 keV) are associated with nova explosions, which are discussed in detail in \citet{2005A&A...442..879P}, \citet{2007A&A...465..375P}, \citet{2010A&A...523A..89H} and \citet{2011A&A...533A..52H}. A nova explosion is thought to be caused by a thermonuclear runaway in the hydrogen layer, which is accumulated on the surface of the white dwarf via accretion from the outer envelope of a companion star. Mainly depending on the mass of the white dwarf and the accretion rate of the system there can also be a state where the accumulated matter on the white dwarf surface undergoes steady hydrogen burning without a nova explosion \citep[see][]{2007ApJ...663.1269N}. We detect six SSSs in the catalogue, which are not associated with known optical novae (No. 64, 184, 229, 294, 295, 314).\\
The catalogue contains 45 X-ray sources associated with optical novae, 12 of which have not previously been published and will be discussed in detail by Henze et al. 2013 (in preparation). One type of novae shows very short outbursts in the light curves (12 classified as OB sources), while others show a very steady X-ray outburst over several years. Due to these long outbursts sources No. 79, 100 and 123 are classified as HV sources, as they show the longest outbursts we observe for novae (e.g. Fig. \ref{fig:79_123}, \ref{fig:sss_spread}). No. 123 even shows a re-brightening in the 2011/2012 monitoring observations (see Fig. \ref{fig:79_123}).\\
Fig. \ref{fig:sss_spread} gives an overview of SSSs and novae classified as significantly time variable in the catalogue. It can be seen, that most novae are classified as OB sources. There are three OB SSS without optical nova association (No. 184, 295, 64), which might be X-rays from a nova missed in optical observations. No. 294 and 229 are HV SSSs and show a particularly low variability factor, which as described above could be caused by steady hydrogen burning on the surface of a white dwarf.\\
\\
{\bf Discussion of individual SSS:}\\
\\
No. 64 is a SSS showing an SOB, which is not associated with a reported optical nova \citep[see][]{2010A&A...523A..89H}. However, the nova outburst might have been missed in optical observations.\\
No. 81 and No. 150 are classical novae in a M\,31 globular clusters \citep[see][]{2009A&A...500..769H, 2013A&A...549A.120H}.\\
No. 184 is not detected in a 29.07 ks ACIS-I observation (ObsID 7064 on 2006-12-04), where it should be bright according to the HRC-I light curve (see No. 184 in Fig. \ref{fig:lcclass}), which shows a plateau during this time. This indicates a super-soft X-ray spectrum, since the \emph{Chandra} ACIS detector has a lower sensitivity than HRC-I at lower energies \citep[see][]{2007A&A...465..375P}. For a 50 eV black-body spectrum, the conversion factor from HRC-I to ACIS-I count rates would be 0.081 according to WebPIMMS. Thus converting the HRC-I (ObsID 7285) count rate of $\rm{1.95 \cdot 10^{-3} \enspace counts~s^{-1}}$, we would still expect 4.5 photon counts during the ACIS-I observation. This means, that either the spectrum is even softer, than we estimate above, or the source was unexpectedly fainter during the time of the ACIS-I observation and then re-brightened. The HRC-I light curve is typical for a SSS phase of an optical nova several months after outburst, but no known optical nova correlates with its position.\\
No. 205 is detected at the position of nova M31N 2002-08b \citep[optical light curve see][]{2012A&A...537A..43L}. In two $\rm{H\alpha}$ images from the \citet{2006AJ....131.2478M} survey on 2002-09-12, there is bright $\rm{H\alpha}$ emission visible at the source position, which is most likely associated with the nova. This is the first detection of X-rays from M31N 2002-08b (see Table \ref{tab:new}).\\
No. 229 is a super-soft source candidate, showing a 217s pulse period \citep[see][]{2008ApJ...676.1218T}.\\
No. 294 is a HV and classified as SSS by \citet{2005A&A...434..483P}, but does not show a nova outburst and is detected throughout the thirteen years of monitoring. It also has a optical counterpart in an $\rm{H\alpha}$ (see Table \ref{tab:opt}) image of the LGGS.\\
No. 295 is a super-soft source, which could be the counterpart of a nova explosion missed in the optical. It has a 865s period, which is discussed in \citet{2001A&A...378..800O}. Its spectrum is similar to galactic super-soft sources \citep[see][]{2001ApJ...563L.119T}.\\
No. 314 shows a faint SOB \citep[see][]{2007ATel.1296....1H}. The source re-brightened in a \emph{Swift} follow-up observation and has a super-soft spectrum according to \citet{2007ATel.1334....1K}. This may be emission from a missed optical nova.\\

\subsection{Supernova remnants (SNR)}
\label{sec:snr}

After a supernova explosion, a shock front propagates through the interstellar medium (ISM), heating it to temperatures where thermal X-rays are emitted. There can also be strong radio emission from synchrotron radiation and optical emission lines are seen, where the ISM is ionized. Depending on the age of the SNR, the extent can be large enough to be resolved as an extended source of X-rays, radio or optical light even at the distance of M\,31. The catalogue contains six sources, thought to be associated with supernova remnants from the \citet{2012A&A...544A.144S}  catalogue (some of their sources, overlapping with the FOV are not contained in the catalogue due to lower sensitivity of the HRC-I detector (compared to \emph{XMM-Newton} EPIC) toward the edges of the FOV). Aside from those, we classified one new SNR candidate (No. 272), which shows X-ray extent, radio emission and hardness ratios \citep[obtained from][]{2011A&A...534A..55S} consistent with thermal X-ray emission. No. 304 is a confirmed SNR and is discussed as being associated with a low-mass X-ray binary by \citet{2005ApJ...634..365W}. 
\begin{figure*}
  \begin{minipage}{0.24\linewidth}
    \centering
    \includegraphics[width=\linewidth]{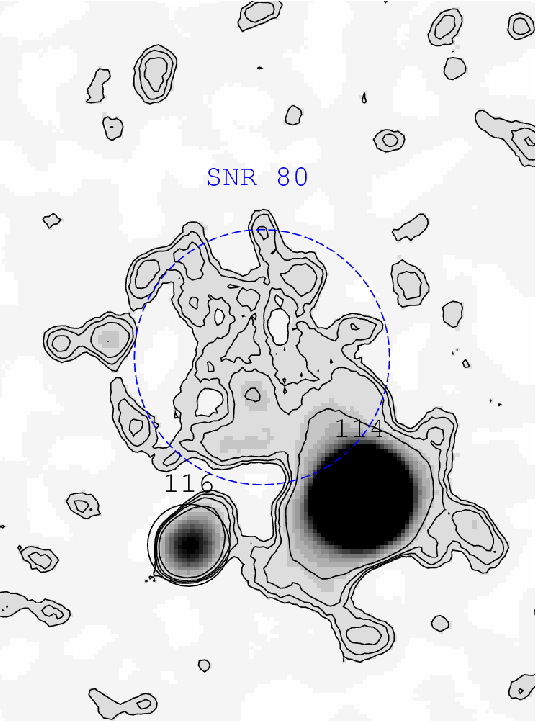}
  \end{minipage}
  \begin{minipage}{0.24\linewidth}
    \centering
    \includegraphics[width=\linewidth]{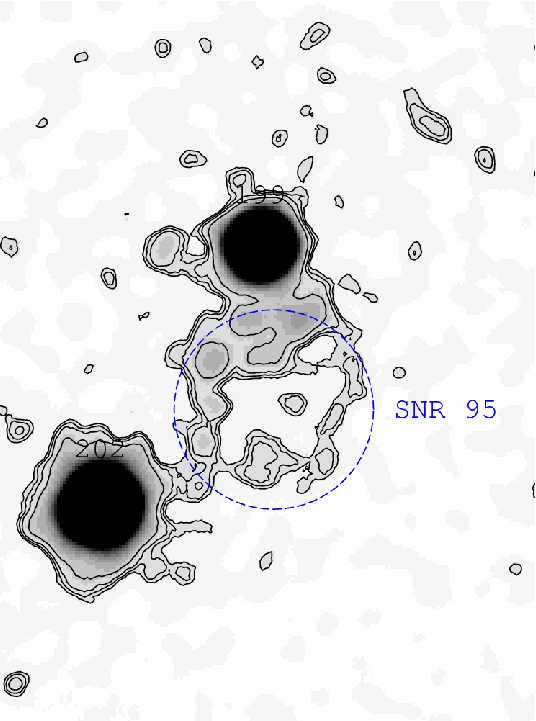}
  \end{minipage}
  \begin{minipage}{0.24\linewidth}
    \centering
    \includegraphics[width=\linewidth]{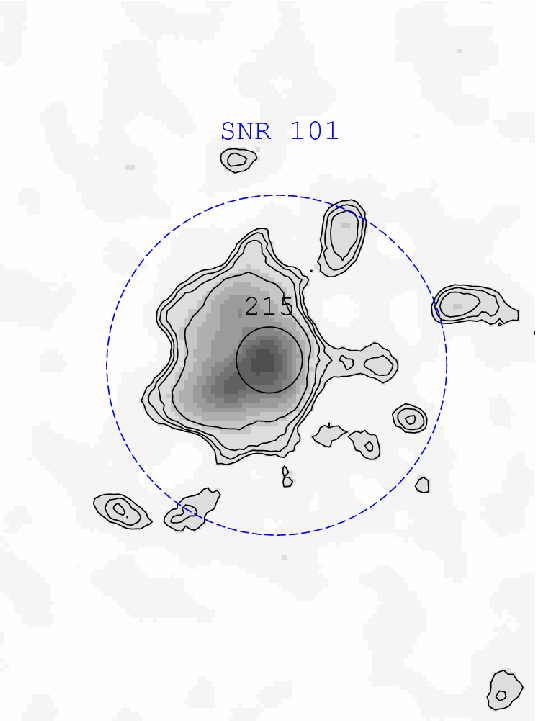}
  \end{minipage}
  \begin{minipage}{0.24\linewidth}
    \centering
    \includegraphics[width=\linewidth]{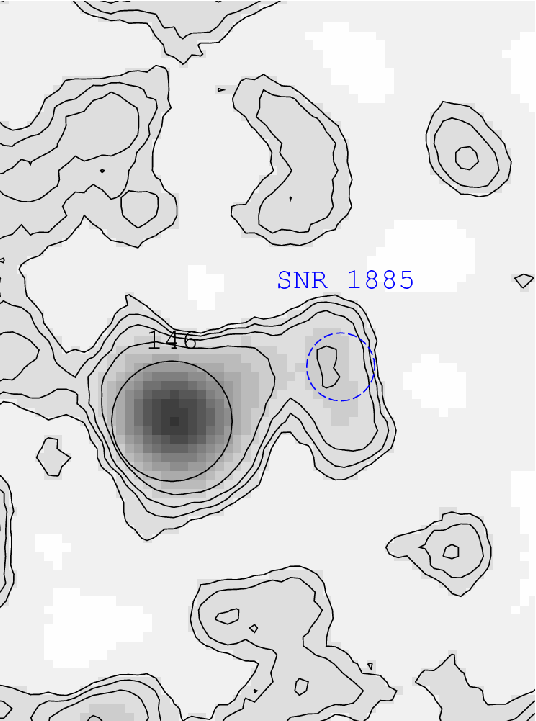}
  \end{minipage}
  \caption{Contour plots of the logarithmically scaled merged HRC-I images with dashed circular regions at the radio positions with approximate radio radii \citep{1999ApJ...514..195F, 2001AIPC..565..433S} of the known supernova remnants within 90{\arcsec} of the center of M\,31. The images are smoothed with a Gaussian function of 7 pixel FWHM and contour levels are at 0.53, 0.56, 0.6 and 0.7 counts per pixel. North is up and East is left. The zoom is slightly different (for region size and SNR names see Table \ref{tab:snr}). The X-ray catalogue source regions are circles with $\sim 90\%$ encircled count fraction radius.}
   \label{fig:snr}
\end{figure*}
By visually inspecting the merged high resolution image and smoothing with a 7 pixel FWHM Gaussian function, we found extended X-ray sources at the positions of the three radio SNRs discussed in \citet{2001AIPC..565..433S} (see Fig. \ref{fig:snr}). Of these remnants only SNR 101 has a counterpart in the source catalogue (No. 215). SNR 101 and SNR 95 have already been detected in X-rays and are discussed in \citet{2003ApJ...590L..21K}. However our deeper image of SNR 95 is much more convincing. For SNR 80 \citep[{[B90]80 in}][]{1990ApJS...72..761B} we report the first observation in X-rays (Fig. \ref{fig:snr} leftmost image), close to another bright X-ray source. In the 1.02 Ms merged high resolution image we find 446 source counts in a visually determined region (see Table \ref{tab:snr}), which corresponds to a significance of about $\rm{3.2\sigma}$.\\
\\
In the right-most image of Fig. \ref{fig:snr} we see a very dim X-ray emission at the position of SN 1885A \citep{1999ApJ...514..195F}, which is not included in the catalogue. Source properties were extracted from a circular region with a radius of 0.7{\arcsec} \citep[see estimated extent in][]{2001AIPC..565..433S} from the merged high resolution image at the position of SN 1885A. We obtain 28 counts and a significance of $\rm{2.6\sigma}$, which is not sufficient to state a clear detection. To the south-east of SN 1885A, lies a transient X-ray source, which we found not to be associated with the supernova in accordance to \citet{2002ApJ...578..114K}, since the positional errors plus the maximum extent of the source are clearly separated. Table \ref{tab:snr} shows the circular regions we estimated for each SNR and the counts and significance we obtained for each of these regions using the CIAO tool {\tt aprates} to compute source statistics.

\begin{table*}
\caption[]{Source statistics for SNRs within 90{\arcsec} of the center of M\,31, extracted from the 1.02 Ms merged high resolution image.}
\begin{center}
\begin{tabular}{lrrrrr}
\hline\hline\noalign{\smallskip}
\multicolumn{1}{l}{SNR ID \tablefootmark{a}} &
\multicolumn{1}{l}{RA(J2000) [h:m:s]} &
\multicolumn{1}{l}{Dec(J2000) [d:m:s]} &
\multicolumn{1}{l}{Counts \tablefootmark{b}} &
\multicolumn{1}{l}{Significance \tablefootmark{b}} &
\multicolumn{1}{l}{Diameter [{\arcsec}] \tablefootmark{c}} \\
\noalign{\smallskip}\hline\noalign{\smallskip}
SNR 80      & 00:42:40.37 & +41:15:52.2 & 446 & 3.2 & 7.8 \\
SNR 95      & 00:42:47.81 & +41:15:26.3 & 731 & 4.5 & 8.7 \\
SNR 101     & 00:42:50.44 & +41:15:56.4 & 856 & 6.7 & 7.9 \\
SNR 1885A \tablefootmark{d} & 00:42:42.89 & +41:16:05.0 & 28 & 2.6 & 1.4 \\
\noalign{\smallskip}\hline
\end{tabular}
\tablefoot{
\tablefoottext{a}{Taken from \citet{2001AIPC..565..433S}.}
\tablefoottext{b}{Calculated with CIAO tool {\tt aprates}.}
\tablefoottext{c}{Visually estimated from smoothed merge image.}
\tablefoottext{d}{Position taken from \citet{1999ApJ...514..195F}.}
}
\end{center}
\label{tab:snr}
\end{table*}

\subsection{Foreground stars (fgStar) and background galaxies (AGN, Gal, GCl)}
\label{sec:fgbg}
\begin{figure}
  \centering
  \resizebox{\hsize}{!}{\includegraphics[angle=0,clip=]{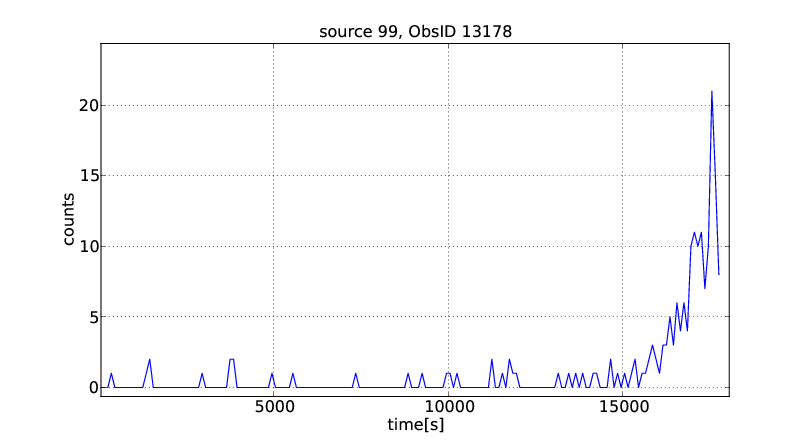}}
  \caption{Light curve of No. 99 during HRC-I observation 13178, showing an X-ray flare at the end of the observation. The plot displays counts over time in seconds from the start of the observation, with a binning of 100s.}
  \label{fig:99flare}
\end{figure}

There are sources in the catalogue, which are not located in M\,31. There are X-ray-bright foreground stars in our own galaxy, which we identified by correlating the source positions with known foreground stars (fgStar) in other X-ray catalogues (see section \ref{sec:catalogue}) and visually inspecting images from the LGGS for clearly visible foreground stars. The advantage of the catalogue for these identifications is, that they get more precise for most sources, due to the good positional accuracy of \emph{Chandra} HRC-I, and the long exposure available to us. The catalogue contains four confirmed fgStars and seven fgStar candidates, which are statistically distributed across the FOV as expected. For No. 85, 262, 289 and 291 we give the colours of the optical counterparts in Table \ref{tab:opt}.\\
The luminosities of fgStars are erroneously overestimated in the catalogue, since they are much closer to us than M\,31 and we used the distance to M\,31 for the calculation of all luminosities.\\
Active galactic nuclei (AGN) in the background should show no X-ray extent in the images, have a hard spectrum and show moderate variability \citep[see][]{2012ApJ...757...40B}. They are super-massive black holes at the centre of a galaxy, accreting matter. The real X-ray luminosity is underestimated in the catalogue, since those sources are much further away from us than M\,31. AGN can also show strong radio emission. The radio emission in AGN is expected to arise from synchrotron radiation connected to acceleration processes in the jet of the super massive black hole \citep[see][]{2010ApJ...720.1066C}. The catalogue only contains one confirmed AGN and one candidate, which we adopted from previous X-ray catalogues (section \ref{sec:catalogue}). They show very faint light curves and thus we cannot draw any conclusions about variability.\\
We did not find any identifications of X-ray sources with normal galaxies (Gal) or galaxy clusters (GCl) in the background nor with CVs or other known sources in the foreground of the M\,31 central field.\\
\\
{\bf Discussion of individual foreground and background objects:}\\
\\
No. 44 and 45 show radio emission \citep[see B90:][]{1990ApJS...72..761B}, and a harder spectrum \citep[][]{2011A&A...534A..55S}, which would suggest they are AGN candidates.\\
No. 85 correlates with the position of a confirmed foreground star from the RBCv5 \citep[see][]{2007A&A...471..127G} and the light curve shows one significant outburst, which is caused by a short flare in the light curve of HRC-I ObsID 10882.\\
No. 99 is a fgStar candidate with time variability due to flaring activity in HRC-I ObsID 13178 (see observation light curve in Fig. \ref{fig:99flare}).\\
No. 207 shows an optical counterpart (Table \ref{tab:opt}) and radio emission (source [B90]96). From the two extended optical sources, the radio emission and the soft X-ray spectrum at the position, the source could be a galaxy merger in the background, where the X-ray emission is caused by gas of the ISM, heated during the collision.\\

\subsection{M\,31 core}
\label{sec:core}

For the largely overlapping central sources of M\,31 we propose a new distribution of discrete X-ray sources in the catalogue (see Fig. \ref{fig:nucleus}).\\
No. 164 and 165 are at the position of the double core of M\,31 \citep[see][]{2010ApJ...710..755G}. They show variability very similar to the XRB light curves and are classified as HV sources in the catalogue. In Hubble Space Telescope images of the M\,31 core (M\,31*), a double structure, corresponding to the two X-ray source positions of No. 164 and 165 is visible as the most luminous region around the dynamical centre of the galaxy. This circum-nuclear environment is discussed as a region of very high star density around a central black hole by \citet{2005ApJ...631..280B} and \citet{2009MNRAS.397..148L}. \citet{2011ApJ...728L..10L} recently discussed No. 164 as the X-ray source of the central super-massive black hole (M\,31*) and give a detailed \emph{Chandra} light curve with an outburst in X-rays. The maximum of this outburst has only been observed with \emph{Chandra} ACIS and is thus not visible in the HRC-I light curve of No. 164.\\
No. 167 is a very soft X-ray source to the south of the core of M\,31 \citep[see][]{2000ApJ...537L..23G}.\\
No. 170 and 172 are close to 167 and show cross talk from this source, so that we cannot classify them separately.

\section{Summary and conclusions}
\label{sec:conclusion}

We present the deepest X-ray catalogue of the M\,31 central field available to date, which contains 318 discrete X-ray sources with detailed variability information (spanning 13 years). From the HRC-I data analysis, we are able to provide a crude classification of X-ray binary (XRB) candidates based on their long-term time variability. The catalogue contains light curves for 97 XRB candidates in the field of M\,31 and 18 XRB candidates in globular clusters (GlC).\\
This classification of XRBs from time variability is the major step forward of the HRC-I analysis compared to previous studies. Many of the XRB candidates, had only been classified as $\rm{<hard>}$ X-ray sources in one of the most recent X-ray catalogues of M\,31 by \citet{2011A&A...534A..55S}.\\
$\rm{36 ~\%}$ of all sources in the catalogue are classified as XRB candidates with hard X-ray spectrum (X-rays from optical novae make up another $\rm{14 ~\%}$ of the catalogue sources). Nine XRB candidate sources show recurrent outbursts. The analysis may suggest, that outburst sources are less frequent in GlCs than in the field of M\,31. We detected one SNR in the central region of M\,31, which is seen in X-rays for the first time and discovered one previously unknown SNR candidate. We present some promising optical counterparts to the X-ray sources, which could be spectrally identified with future optical follow-up observations.\\
We observed 61 new sources, which have not yet been published in a dedicated M\,31 catalogue. 17 of those sources have never been detected before, 12 sources are new X-ray sources associated with optical novae, 19 sources have only been detected in the \emph{Chandra} source catalogue \citep[see][]{2010ApJS..189...37E} and 13 sources have only been discussed in separate papers or Astronomer's Telegrams, but are not included in previous catalogues.\\
The catalogue contains a large sample of XRBs, with variability behaviour and can thus provide important input for a better understanding of XRBs and the X-ray source population of the M\,31 central field in general.

\begin{acknowledgements}

We thank the anonymous referee for constructive comments that helped to improve the clarity of the paper. 
This research has made use of data obtained from the \emph{Chandra} Data Archive and the \emph{Chandra} Source Catalog, and software provided by the \emph{Chandra} X-ray Center (CXC) in the application packages CIAO, ChIPS, and Sherpa. 
Based on observations obtained with \emph{XMM-Newton}, an ESA science mission with instruments and contributions directly funded by ESA Member States and NASA. 
The XMM-Newton project is supported by the Bundesministerium für Wirtschaft und Technologie/Deutsches Zentrum für Luft- und Raumfahrt (BMWI/DLR FKZ 50 OX 0001) and the Max Planck Society. 
This research has made use of NASA's Astrophysics Data System. 
This research has made use of the VizieR catalogue access tool, CDS, Strasbourg, France. 
This research has made use of SAOImage DS9, developed by Smithsonian Astrophysical Observatory. 
This publication makes use of data products from the Two Micron All Sky Survey, which is a joint project of the University of Massachusetts and the Infrared Processing and Analysis Center/California Institute of Technology, funded by the National Aeronautics and Space Administration and the National Science Foundation. 
Based on observations made with the NASA/ESA Hubble Space Telescope, obtained from the data archive at the Space Telescope Science Institute. STScI is operated by the Association of Universities for Research in Astronomy, Inc. under NASA contract NAS 5-26555. 
The Second Palomar Observatory Sky Survey (POSS-II) was made by the California Institute of Technology with funds from the National Science Foundation, the National Geographic Society, the Sloan Foundation, the Samuel Oschin Foundation, and the Eastman Kodak Corporation. The Oschin Schmidt Telescope is operated by the California Institute of Technology and Palomar Observatory. 
We acknowledge the use of NASA's \emph{SkyView} facility (http://skyview.gsfc.nasa.gov) located at NASA Goddard Space Flight Center. 
We acknowledge the use of public data from the \emph{Swift} data archive. 
This research has made use of data and/or software provided by the High Energy Astrophysics Science Archive Research Center (HEASARC), which is a service of the Astrophysics Science Division at NASA/GSFC and the High Energy Astrophysics Division of the Smithsonian Astrophysical Observatory. 
This research has made use of the SIMBAD database, operated at CDS, Strasbourg, France. 
M. Henze acknowledges support from the BMWI/DLR grant FKZ 50 OR 1010 and from an ESA fellowship. 
R. Sturm acknowledges support from the BMWI/DLR grant FKZ 50 OR 0907. 

\end{acknowledgements}

\bibliographystyle{aa}
\bibliography{auto,general}

\begin{table*}
\caption[]{New X-ray sources not previously published in dedicated M\,31 catalogues (section \ref{sec:corr}).}
\begin{center}
\begin{tabular}{rllrrrlrl}
\hline\hline\noalign{\smallskip}
\multicolumn{1}{l}{No.} &
\multicolumn{1}{l}{RA [h:m:s]} &
\multicolumn{1}{l}{Dec [d:m:s]} &
\multicolumn{1}{l}{Sig. \tablefootmark{a}} &
\multicolumn{1}{l}{$L_X$[erg $\rm{s^{-1}}$] \tablefootmark{b}} &
\multicolumn{1}{l}{svar} &
\multicolumn{1}{l}{CSC name} &
\multicolumn{1}{l}{Ref. \tablefootmark{e}} &
\multicolumn{1}{l}{Comment} \\
\noalign{\smallskip}\hline\noalign{\smallskip}  
   11 &  00:41:57.74 &  +41:16:07.3 &      4.99 &    6.98E+36 &       1.72 &                         &   &\\                               
   12 &  00:41:59.21 &  +41:05:41.4 &     48.56 &    3.56E+38 &      26.09 &                         &   1&SOB \\
   21 &  00:42:08.53 &  +41:10:30.8 &     10.26 &    1.23E+37 &       4.96 &     J004208.3+411029    &   &\\                    
   29 &  00:42:11.25 &  +41:04:29.8 &     13.86 &    6.20E+37 &       4.98 &     J004211.3+410430    &   2&\\
   41 &  00:42:17.29 &  +41:15:37.6 &    315.73 &    2.92E+38 &      42.67 &     J004217.3+411537    &   3&SOB \\
   47 &  00:42:21.29 &  +41:15:52.6 &     48.01 &    3.94E+37 &      15.22 &                         &   &AP \\                              
   65 &  00:42:29.08 &  +41:15:47.6 &     20.93 &    8.84E+36 &       7.29 &     J004229.1+411547    &   &SOB \\                                                  
   67 &  00:42:29.10 &  +41:13:48.7 &     18.90 &    1.09E+37 &       7.93 &                         &   4&AP \\                               
   74 &  00:42:31.70 &  +41:14:54.2 &      8.83 &    3.10E+36 &       4.37 &     J004231.7+411454    &   &\\                                                  
   82 &  00:42:33.26 &  +41:17:42.7 &     30.70 &    1.42E+37 &       9.28 &                         &   5&SOB, GlC \\
   83 &  00:42:33.26 &  +41:15:55.6 &      5.21 &    1.27E+35 &       1.15 &                         &   &\\                               
   88 &  00:42:34.02 &  +41:10:57.2 &      5.48 &    5.02E+36 &       1.93 &                         &   &M31N 2009-05b\tablefootmark{d} \\                   
   94 &  00:42:35.18 &  +41:14:21.4 &      7.20 &    2.40E+36 &       3.66 &     J004235.1+411421    &   &\\                                                  
   96 &  00:42:36.08 &  +41:20:22.5 &      6.11 &    1.82E+36 &       2.66 &     J004236.1+412022    &   &\\                                                  
   97 &  00:42:36.20 &  +41:18:01.9 &      8.62 &    3.17E+35 &       3.11 &                         &   &M31N 2009-08e\tablefootmark{d} \\                   
   98 &  00:42:36.57 &  +41:15:54.8 &      7.00 &    9.66E+35 &       1.96 &     J004236.5+411554    &   &\\                                                  
  103 &  00:42:38.33 &  +41:16:30.9 &     49.83 &    1.67E+37 &      10.44 &                         &   &SOB, M31N 2011-11e\tablefootmark{d} \\                
  105 &  00:42:38.67 &  +41:17:23.5 &      5.01 &    1.32E+36 &       2.80 &     J004238.6+411723    &   &\\                                                  
  107 &  00:42:38.78 &  +41:15:57.6 &      6.14 &    1.28E+35 &       1.13 &                         &   &\\                               
  108 &  00:42:39.02 &  +41:13:25.7 &     60.36 &    3.35E+37 &      14.74 &                         &   &SOB, M31N 2011-01b\tablefootmark{d} \\                
  113 &  00:42:39.80 &  +41:14:54.5 &      7.17 &    2.31E+36 &       3.33 &                         &   &\\                               
  116 &  00:42:40.46 &  +41:15:46.2 &     46.68 &    1.49E+37 &      10.05 &     J004240.4+411545    &   &SOB \\                                                  
  120 &  00:42:40.88 &  +41:16:09.8 &      5.04 &    8.82E+34 &       2.08 &     J004240.8+411609    &   &\\                                                  
  121 &  00:42:40.91 &  +41:16:29.0 &      5.77 &    1.11E+35 &       3.43 &                         &   &\\                               
  125 &  00:42:41.21 &  +41:17:01.4 &      6.60 &    2.06E+36 &       3.58 &                         &   &M31N 2009-08c\tablefootmark{d} \\                   
  127 &  00:42:41.46 &  +41:16:17.9 &    169.32 &    6.48E+37 &      20.38 &     J004241.4+411617    &   &SOB \\                                                  
  128 &  00:42:41.61 &  +41:14:37.2 &     45.11 &    1.53E+36 &       9.18 &     J004241.6+411437    &   &AP \\                                                  
  135 &  00:42:42.37 &  +41:15:45.1 &     12.25 &    2.63E+35 &       3.11 &     J004242.3+411544    &   &\\                                                  
  139 &  00:42:42.69 &  +41:14:46.6 &      8.05 &    1.16E+36 &       2.31 &     J004242.6+411446    &   &\\                                                  
  143 &  00:42:42.76 &  +41:16:14.8 &      9.60 &    2.15E+35 &       2.99 &                         &   &M31N 2010-01d\tablefootmark{d} \\                   
  145 &  00:42:43.00 &  +41:15:10.3 &    140.38 &    5.02E+37 &      18.07 &                         &   &SOB, M31N 2011-02b\tablefootmark{d} \\                
  149 &  00:42:43.68 &  +41:25:18.5 &    402.97 &    1.33E+39 &      87.30 &                         &   6&SOB, ULX2 \\
  150 &  00:42:43.71 &  +41:12:43.0 &     76.05 &    4.49E+37 &      16.68 &                         &   &SOB, M31N 2010-10f\tablefootmark{d}, GlC \\
  154 &  00:42:43.80 &  +41:16:12.7 &     71.80 &    2.31E+36 &      13.41 &     J004243.8+411612    &   &SOB \\                                                  
  159 &  00:42:43.93 &  +41:16:10.8 &    178.98 &    6.16E+36 &      14.41 &     J004243.9+411610    &   7&AP\\
  161 &  00:42:43.99 &  +41:16:37.1 &     39.04 &    1.19E+37 &       9.49 &                         &   &SOB \\                                                
  176 &  00:42:45.04 &  +41:15:21.1 &     22.86 &    5.43E+35 &       5.41 &                         &   &M31N 2009-05a\tablefootmark{d} \\      
  180 &  00:42:45.10 &  +41:15:42.8 &     43.98 &    1.22E+36 &       6.22 &                         &   &\\                               
  184 &  00:42:45.73 &  +41:15:54.8 &     24.20 &    5.53E+35 &       6.62 &     J004245.7+411554    &   &LOB \\                                                  
  185 &  00:42:45.93 &  +41:10:36.9 &    131.43 &    1.88E+37 &      24.50 &     J004245.9+411036    &   8&AP\\
  189 &  00:42:46.42 &  +41:16:10.2 &    118.98 &    4.78E+37 &      17.65 &     J004246.4+411610    &   &SOB \\                                                  
  203 &  00:42:48.63 &  +41:17:59.5 &      7.52 &    2.67E+36 &       3.59 &                         &   &\\                                      
  205 &  00:42:48.70 &  +41:16:26.6 &     11.19 &    2.27E+35 &       1.76 &                         &   &M31N 2002-08b\tablefootmark{d} \\ 
  209 &  00:42:49.16 &  +41:16:23.9 &      9.06 &    2.66E+35 &       1.79 &                         &   &\\                               
  210 &  00:42:49.21 &  +41:16:01.4 &    200.04 &    8.23E+36 &      10.89 &     J004249.2+411601    &   7&AP\\
  216 &  00:42:50.55 &  +41:16:37.2 &      5.64 &    9.82E+34 &       1.90 &                         &   &\\                                                   
  220 &  00:42:51.62 &  +41:15:12.6 &      5.47 &    1.13E+35 &       1.29 &                         &   &\\                               
  222 &  00:42:51.79 &  +41:17:27.3 &     15.63 &    5.73E+36 &       5.65 &     J004251.8+411726    &   &SOB \\                                                  
  234 &  00:42:53.17 &  +41:14:23.1 &   4291.34 &    3.57E+39 &     153.35 &                         &   9&SOB, ULX1\\
  241 &  00:42:55.48 &  +41:16:08.0 &      9.55 &    1.55E+36 &       2.56 &     J004255.4+411608    &   &\\                                                  
  243 &  00:42:55.75 &  +41:17:52.8 &      4.99 &    1.87E+36 &       2.99 &                         &   &M31N 2011-10d\tablefootmark{d} \\                   
  244 &  00:42:55.99 &  +41:17:21.5 &      9.05 &    3.50E+36 &       4.39 &                         &   &\\                                             
  247 &  00:42:56.97 &  +41:20:05.1 &     18.79 &    1.36E+37 &       8.82 &                         &   10&\\
  248 &  00:42:57.74 &  +41:08:13.1 &     22.50 &    4.95E+37 &      13.22 &                         &   11&SOB, M31N 2010-10e\tablefootmark{d}\\
  257 &  00:43:00.76 &  +41:13:48.8 &      5.26 &    2.66E+35 &       1.01 &                         &   &\\                               
  259 &  00:43:01.15 &  +41:13:17.7 &     13.57 &    1.09E+36 &       4.06 &                         &   &\\                               
  296 &  00:43:20.71 &  +41:15:31.7 &    119.56 &    2.01E+38 &      36.46 &     J004320.7+411531    &   12&SOB\\                                                  
  297 &  00:43:20.76 &  +41:16:32.6 &      4.15 &    4.26E+36 &       0.74 &                         &   &\\                               
  300 &  00:43:23.60 &  +41:12:44.7 &    151.95 &    1.56E+38 &      43.59 &     J004323.6+411244    &   &SOB \\                                                  
  302 &  00:43:25.57 &  +41:15:37.7 &      7.04 &    1.12E+37 &       3.25 &                         &   &GlC \\                               
  314 &  00:43:45.29 &  +41:06:10.0 &     14.67 &    8.30E+37 &       3.59 &     J004345.4+410611    &   13&GlC\\
\noalign{\smallskip}\hline
\end{tabular}
\tablefoot{
\tablefoottext{a}{Highest {\tt wavdetect} significance.}
\tablefoottext{b}{X-ray luminosity (0.2-10 keV) of most significant detection.}
\tablefoottext{d}{Novae will be analysed in Henze et al. 2013 (in preparation).}
\tablefoottext{e}{Ref.: (1)~\cite{2009ATel.2262....1H}; (2) \cite{2007ATel.1171....1G}; (3) \cite{2006ATel..881....1H}; (4) \cite{2009ATel.1978....1G}; (5) \cite{2012ATel.4164....1H}; (6) \cite{2012ATel.3890....1H}; (7) \cite{2008ATel.1698....1G}; (8) \cite{2008ATel.1647....1P}; (9) \cite{2009A&A...500..769H}; (10) \cite{2012ApJ...756...32B}; (11) \cite{2010ATel.3038....1P}; (12) \cite{2008ATel.1674....1P}; (13) \cite{2007ATel.1296....1H}}.}
\end{center}
\label{tab:new}
\end{table*}

\Online
\begin{appendix}

\section{Observations}
\label{sec:obs}
\begin{table*}
\caption[]{\emph{Chandra} HRC-I observations used in this study.}
\begin{center}
\begin{tabular}{rrrrrrr}
\hline\hline\noalign{\smallskip}
\multicolumn{1}{l}{ObsID} &
\multicolumn{1}{l}{Exposure[s] \tablefootmark{a}} &
\multicolumn{1}{l}{RA(J2000) [deg]} &
\multicolumn{1}{l}{Dec(J2000) [deg]} &
\multicolumn{1}{l}{Roll angle [deg]} &
\multicolumn{1}{l}{MJD \tablefootmark{b}} &
\multicolumn{1}{l}{UT \tablefootmark{c}} \\
\noalign{\smallskip}\hline\noalign{\smallskip}
00267    & 1268  & 10.68195  & 41.26369  & 264.83    & 51512.77  & 1999-11-30.77 \\
00268    & 5177  & 10.68383  & 41.26435  & 282.14    & 51535.10  & 1999-12-23.10 \\
00269    & 1204  & 10.68627  & 41.26340  & 298.54    & 51562.91  & 2000-01-19.91 \\
00271    & 2463  & 10.68926  & 41.26492  & 331.70    & 51611.76  & 2000-03-08.76 \\
00272    & 1202  & 10.68767  & 41.27255  & 82.55    & 51690.70  & 2000-05-26.70 \\
00273    & 1190  & 10.68734  & 41.27320  & 101.67    & 51716.20  & 2000-06-21.20 \\
00275    & 1188  & 10.68291  & 41.27314  & 134.00    & 51774.43  & 2000-08-18.43 \\
00276    & 1182  & 10.68046  & 41.27222  & 152.42    & 51798.86  & 2000-09-11.86 \\
00277    & 1187  & 10.67952  & 41.26924  & 192.16    & 51829.18  & 2000-10-12.18 \\
00278    & 1182  & 10.68094  & 41.26537  & 250.96    & 51865.19  & 2000-11-17.19 \\
01570    & 1183  & 10.68592  & 41.27310  & 94.69    & 52070.93  & 2001-06-10.93 \\
01912    & 46731  & 10.67112  & 41.26732  & 226.83   & 52213.99  & 2001-10-31.99 \\
02904    & 1181  & 10.68094  & 41.26564  & 253.90   & 52232.88  & 2001-11-19.88 \\
02905    & 1094  & 10.68620  & 41.26497  & 296.74   & 52290.20  & 2002-01-16.20 \\
02906    & 1187  & 10.68861  & 41.27227  & 88.37    & 52427.90  & 2002-06-02.90 \\
05925    & 46288  & 10.68295  & 41.26512  & 270.75    & 53345.76  & 2004-12-06.76 \\
06177    & 20037  & 10.68427  & 41.26494  & 285.27    & 53366.31  & 2004-12-27.31 \\
05926    & 28265  & 10.68433  & 41.26482  & 285.63    & 53366.84  & 2004-12-27.84 \\
06202    & 18045  & 10.68601  & 41.26498  & 303.50    & 53398.06  & 2005-01-28.06 \\
05927    & 26999  & 10.68605  & 41.26490  & 303.50    & 53398.83  & 2005-01-28.83 \\
05928    & 44855  & 10.68742  & 41.26528  & 319.07    & 53422.68  & 2005-02-21.68 \\
07283    & 19941  & 10.68668  & 41.27304  & 90.26   & 53891.29  & 2006-06-05.29 \\
07284    & 20002  & 10.67974  & 41.27133  & 173.77   & 54008.89  & 2006-09-30.89 \\
07285    & 18516  & 10.68043  & 41.26645  & 245.26   & 54052.30  & 2006-11-13.30 \\
07286    & 18707  & 10.68829  & 41.26613  & 333.63   & 54170.61  & 2007-03-11.61 \\
08526    & 19944  & 10.67992  & 41.26685  & 236.71   & 54411.64  & 2007-11-07.64 \\
08527    & 19980  & 10.68098  & 41.26611  & 250.73   & 54421.76  & 2007-11-17.76 \\
08528    & 19975  & 10.68183  & 41.26549  & 262.91   & 54432.79  & 2007-11-28.79 \\
08529    & 18922  & 10.68262  & 41.26536  & 270.77   & 54441.57  & 2007-12-07.57 \\
08530    & 19866  & 10.68330  & 41.26507  & 278.33   & 54451.49  & 2007-12-17.49 \\
09825    & 20212  & 10.67993  & 41.26690  & 234.91   & 54778.34  & 2008-11-08.34 \\
09826    & 19918  & 10.68095  & 41.26603  & 250.89   & 54787.14  & 2008-11-17.14 \\
09827    & 19962  & 10.68197  & 41.26556  & 263.10   & 54798.24  & 2008-11-28.24 \\
09828    & 19967  & 10.68257  & 41.26534  & 271.22   & 54807.41  & 2008-12-07.41 \\
09829    & 10075  & 10.68351  & 41.26532  & 279.41   & 54818.02  & 2008-12-18.02 \\
10838    & 10041  & 10.68331  & 41.26521  & 279.41   & 54818.49  & 2008-12-18.49 \\
10683    & 19897  & 10.68687  & 41.26546  & 315.66   & 54878.90  & 2009-02-16.90 \\
10684    & 18697  & 10.68756  & 41.26565  & 322.39   & 54888.17  & 2009-02-26.17 \\
10882    & 18846  & 10.68010  & 41.26661  & 239.91   & 55142.23  & 2009-11-07.23 \\
10883    & 18326  & 10.68083  & 41.26613  & 249.41   & 55151.24  & 2009-11-16.24 \\
10884    & 18358  & 10.68186  & 41.26558  & 262.25   & 55162.63  & 2009-11-27.63 \\
10885    & 18267  & 10.68271  & 41.26531  & 272.30   & 55173.94  & 2009-12-08.94 \\
10886    & 18343  & 10.68341  & 41.26513  & 278.92   & 55182.90  & 2009-12-17.90 \\
11808    & 17114  & 10.68681  & 41.26536  & 314.83   & 55242.86  & 2010-02-15.86 \\
11809    & 18415  & 10.68747  & 41.26558  & 322.27   & 55253.27  & 2010-02-26.27 \\
12110    & 19954  & 10.68059  & 41.26639  & 246.42   & 55514.17  & 2010-11-14.17 \\
12111    & 19866  & 10.68144  & 41.26587  & 257.32   & 55523.18  & 2010-11-23.18 \\
12112    & 19944  & 10.68230  & 41.26531  & 267.64   & 55533.66  & 2010-12-03.66 \\
12113    & 18974  & 10.68299  & 41.26516  & 274.87   & 55542.56  & 2010-12-12.56 \\
12114    & 19977  & 10.68367  & 41.26520  & 281.67   & 55552.18  & 2010-12-22.18 \\
13178    & 17453  & 10.68688  & 41.26540  & 315.50   & 55609.15  & 2011-02-17.15 \\
13179    & 17482  & 10.68753  & 41.26564  & 322.79   & 55619.25  & 2011-02-27.25 \\
13180    & 17023  & 10.68819  & 41.26608  & 332.15   & 55630.12  & 2011-03-10.12 \\
13227    & 19987  & 10.68029  & 41.26644  & 243.25   & 55877.10  & 2011-11-12.10 \\
13228    & 19003  & 10.68122  & 41.26587  & 254.80   & 55886.22  & 2011-11-21.22 \\
13229    & 19567  & 10.68213  & 41.26552  & 264.98   & 55895.98  & 2011-11-30.98 \\
13230    & 18932  & 10.68288  & 41.26521  & 273.93   & 55906.56  & 2011-12-11.56 \\
13231    & 19487  & 10.68355  & 41.26516  & 280.22   & 55915.33  & 2011-12-20.33 \\
13278    & 18973  & 10.68683  & 41.26548  & 315.75   & 55974.76  & 2012-02-17.76 \\
13279    & 18825  & 10.68765  & 41.26564  & 323.40   & 55985.26  & 2012-02-28.26 \\
13280    & 19285  & 10.68859  & 41.26629  & 335.95   & 55999.21  & 2012-03-13.21 \\
13281    & 18907  & 10.68685  & 41.27307  & 87.90   & 56079.90  & 2012-06-01.90 \\

\noalign{\smallskip}\hline
\end{tabular}
\tablefoot{
\tablefoottext{a}{Exposure after good time interval correction.}
\tablefoottext{b}{Modified Julian Date at the start of the observation.}
\tablefoottext{c}{Universal Time at the start of the observation.}
}
\end{center}
\label{tab:cat}
\end{table*}

\end{appendix}

\begin{appendix}

\section{HRC-I light curves}
\label{sec:lc_all}

\begin{figure*}
\begin{minipage}{0.5\linewidth}
\includegraphics[width=\linewidth]{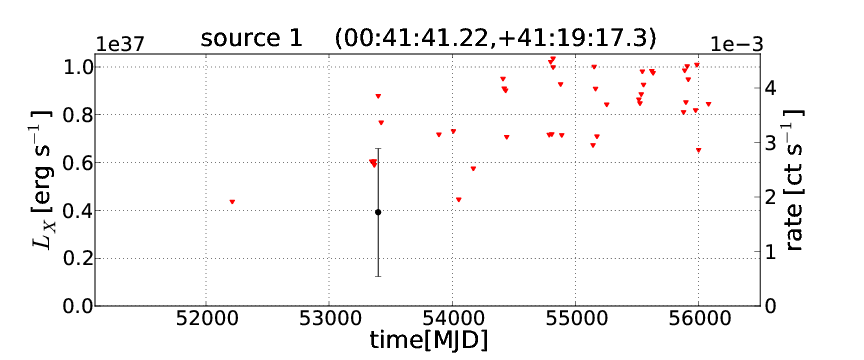}
\includegraphics[width=\linewidth]{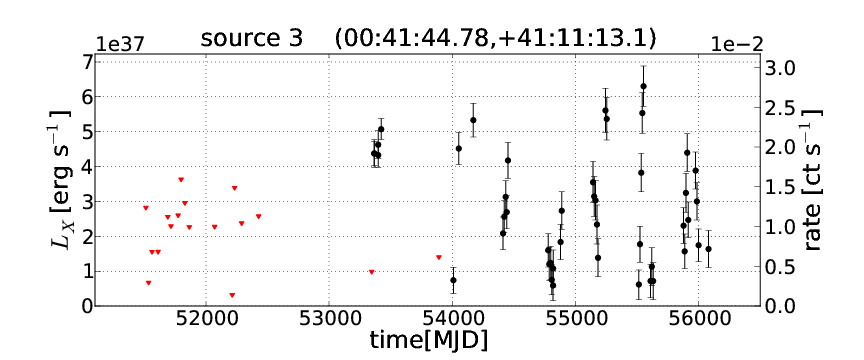}
\includegraphics[width=\linewidth]{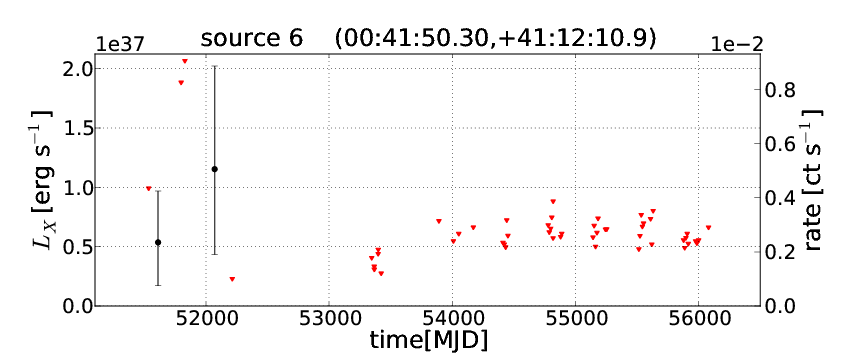}
\includegraphics[width=\linewidth]{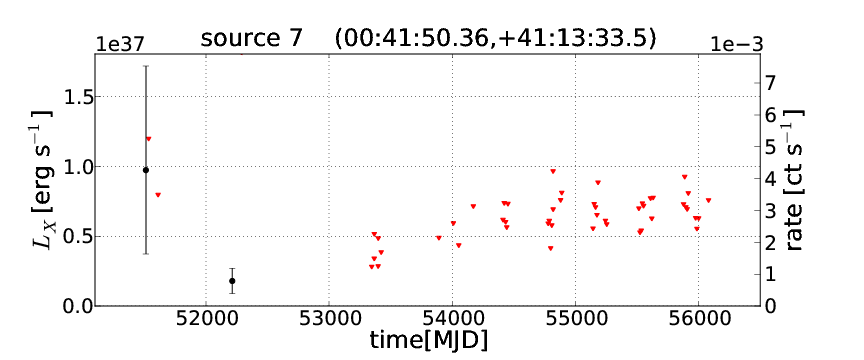}
\includegraphics[width=\linewidth]{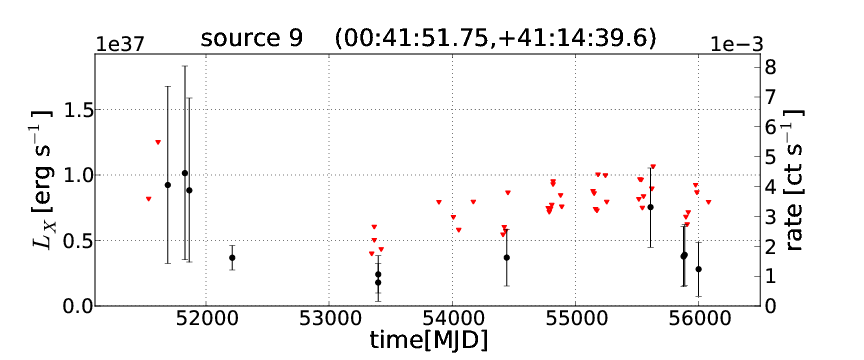}
\includegraphics[width=\linewidth]{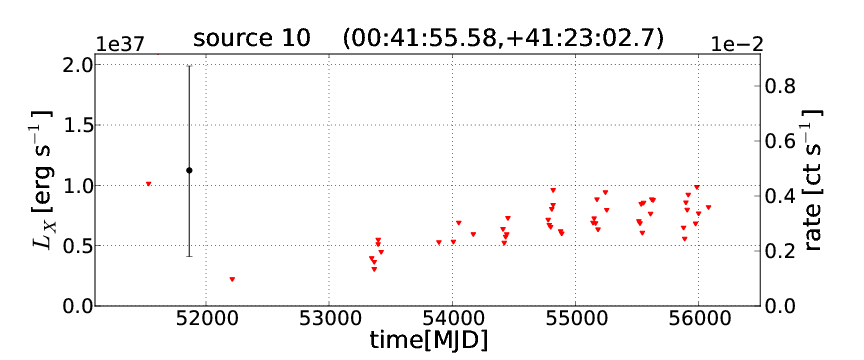}
\end{minipage}
\begin{minipage}{0.5\linewidth}
\includegraphics[width=\linewidth]{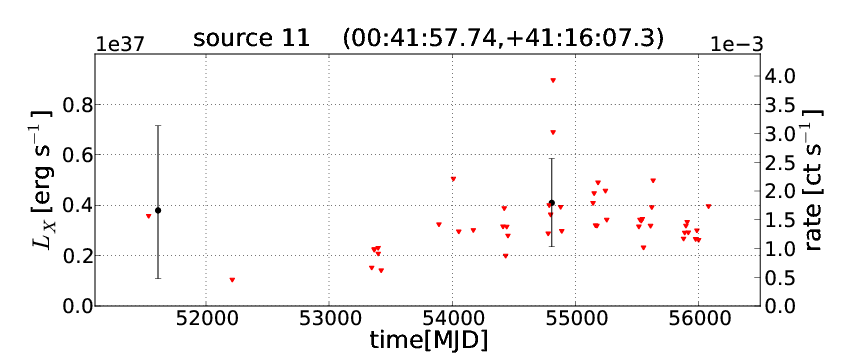}
\includegraphics[width=\linewidth]{plots_new/12_light.png}
\includegraphics[width=\linewidth]{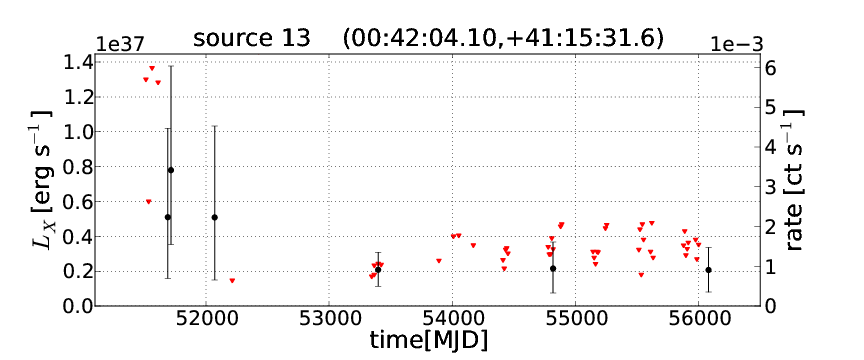}
\includegraphics[width=\linewidth]{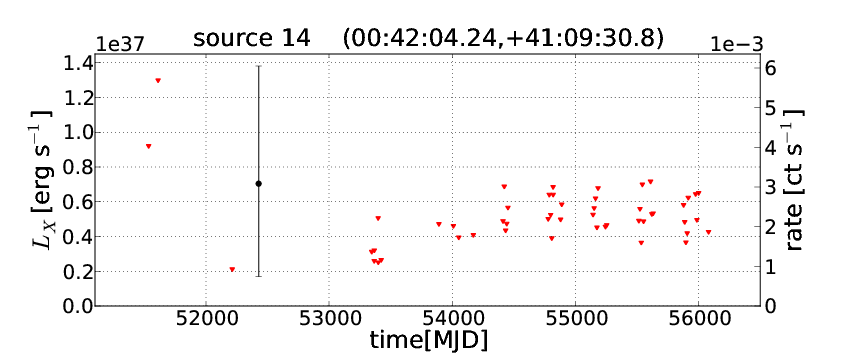}
\includegraphics[width=\linewidth]{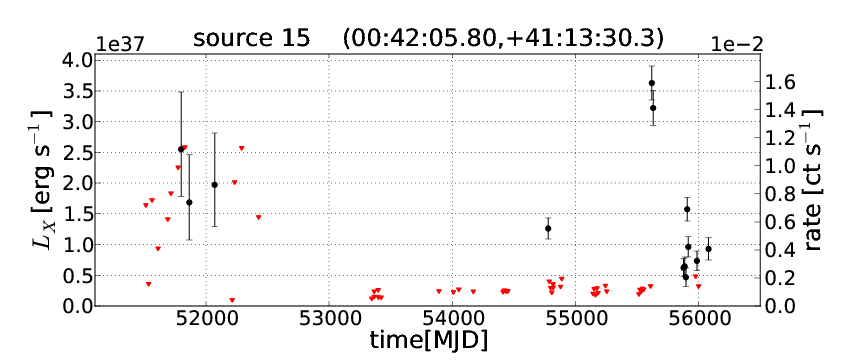}
\includegraphics[width=\linewidth]{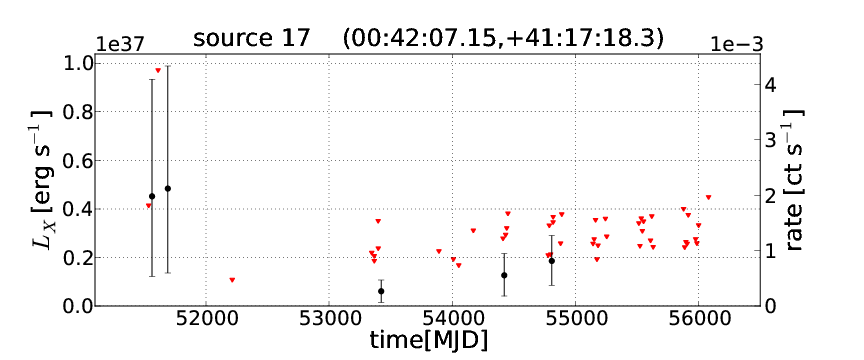}
\end{minipage}
\caption{Light curve plots showing X-ray luminosity (left axis, 0.2-10\,keV) and photon count rate (right axis) over modified Julian date for all catalogue sources. Fluxes with $\rm{1\sigma}$ errors are given as black dots and $\rm{3\sigma}$ upper limits as red triangles.}
\label{fig:lc_all}
\end{figure*}

\addtocounter{figure}{-1} 

\begin{figure*}
\begin{minipage}{0.5\linewidth}
\includegraphics[width=\linewidth]{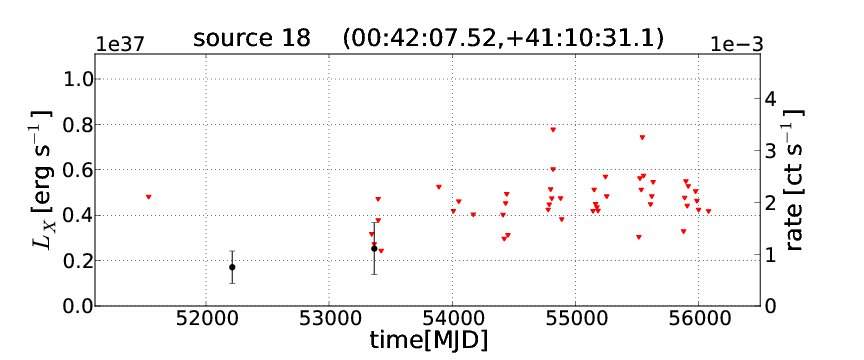}
\includegraphics[width=\linewidth]{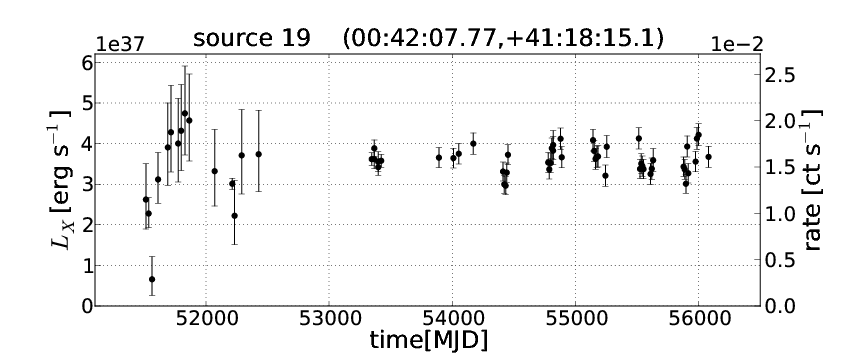}
\includegraphics[width=\linewidth]{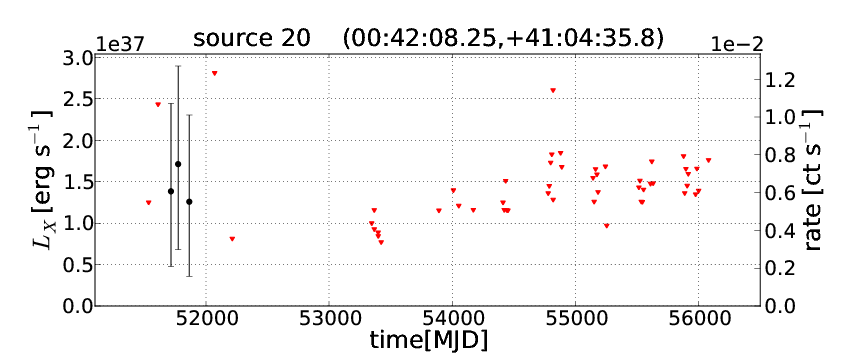}
\includegraphics[width=\linewidth]{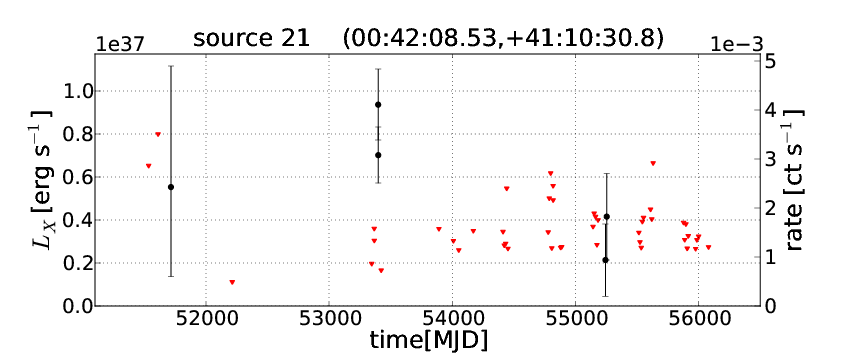}
\includegraphics[width=\linewidth]{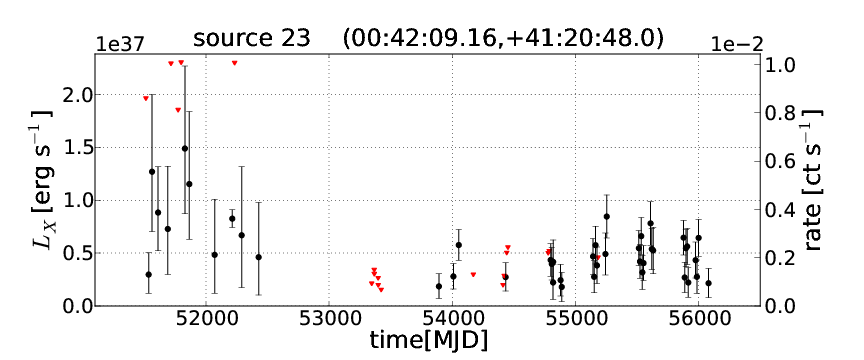}
\includegraphics[width=\linewidth]{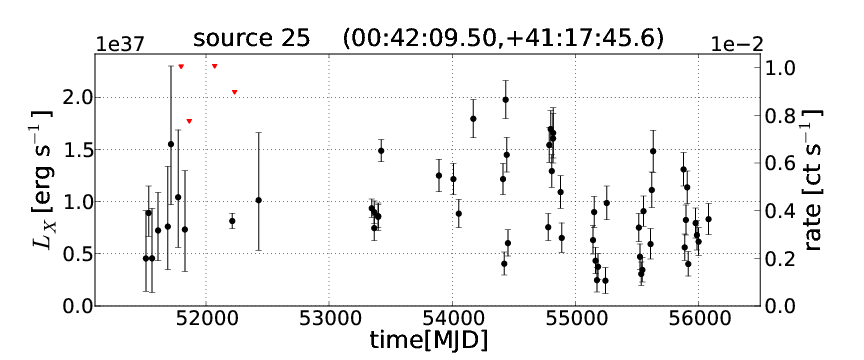}
\end{minipage}
\begin{minipage}{0.5\linewidth}
\includegraphics[width=\linewidth]{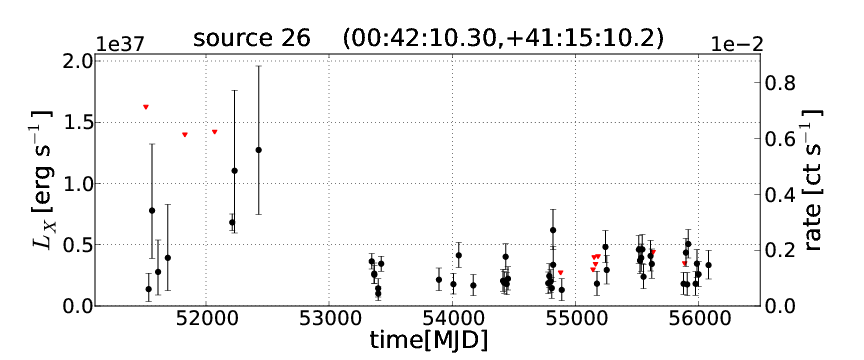}
\includegraphics[width=\linewidth]{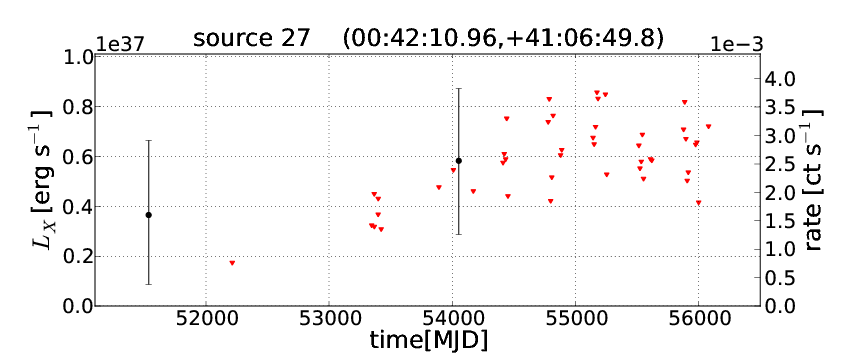}
\includegraphics[width=\linewidth]{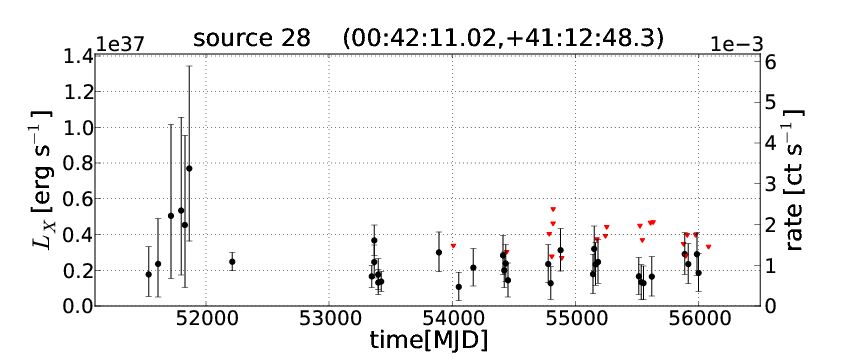}
\includegraphics[width=\linewidth]{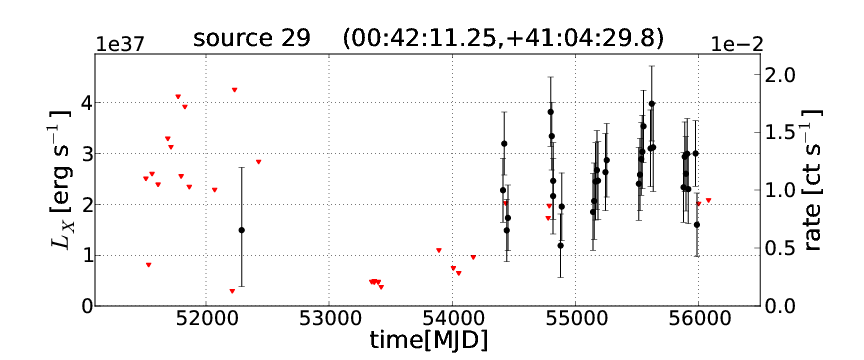}
\includegraphics[width=\linewidth]{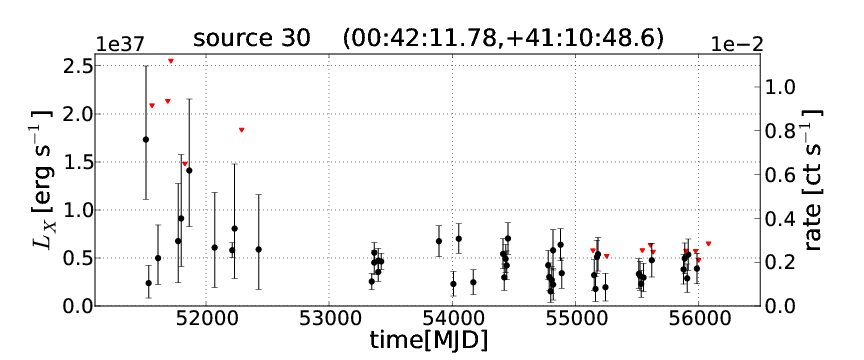}
\includegraphics[width=\linewidth]{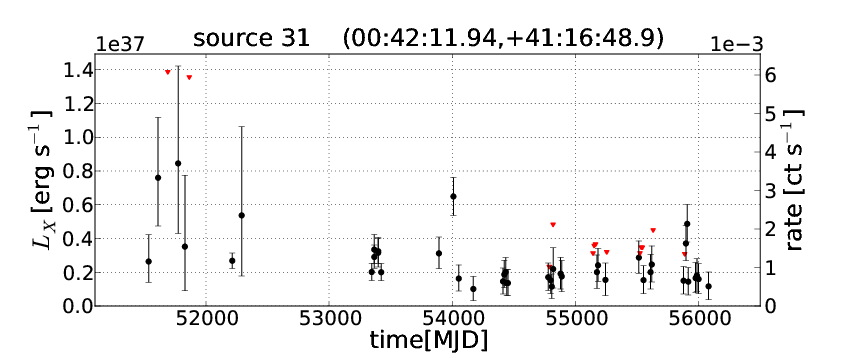}
\end{minipage}
\caption{continued.}
\label{fig:lc_all}
\end{figure*}

\addtocounter{figure}{-1} 

\begin{figure*}
\begin{minipage}{0.5\linewidth}
\includegraphics[width=\linewidth]{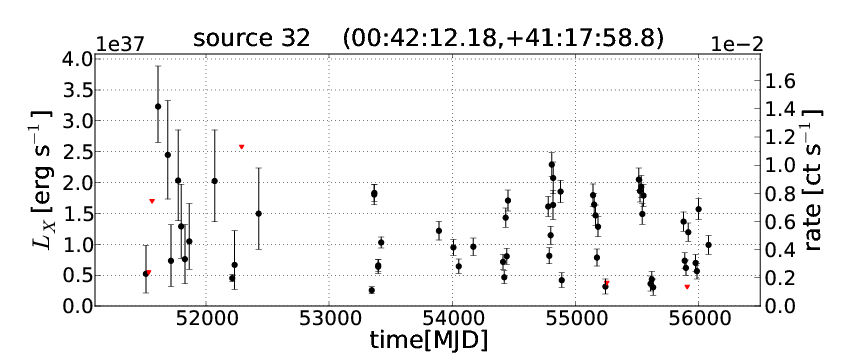}
\includegraphics[width=\linewidth]{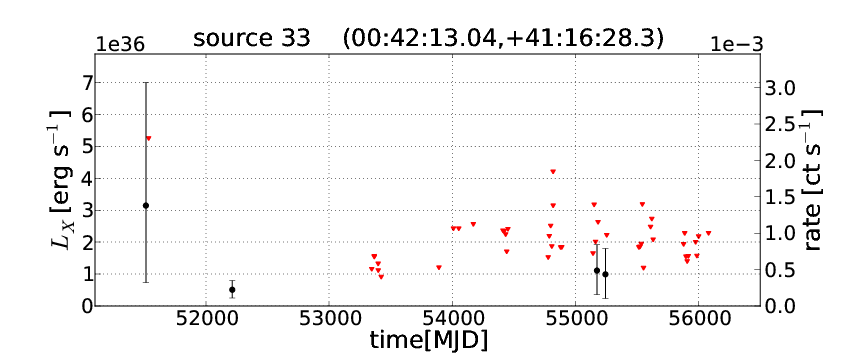}
\includegraphics[width=\linewidth]{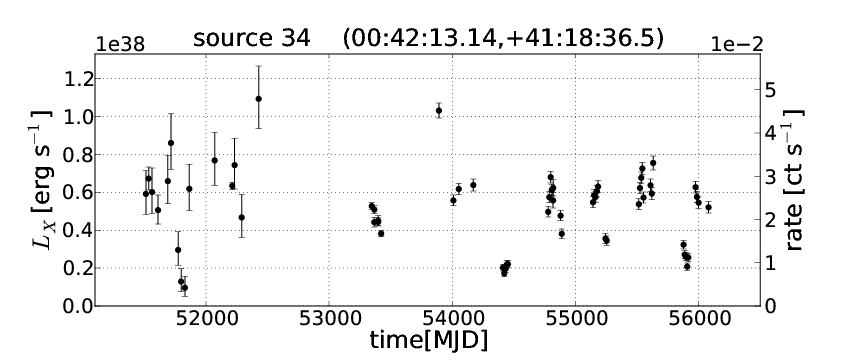}
\includegraphics[width=\linewidth]{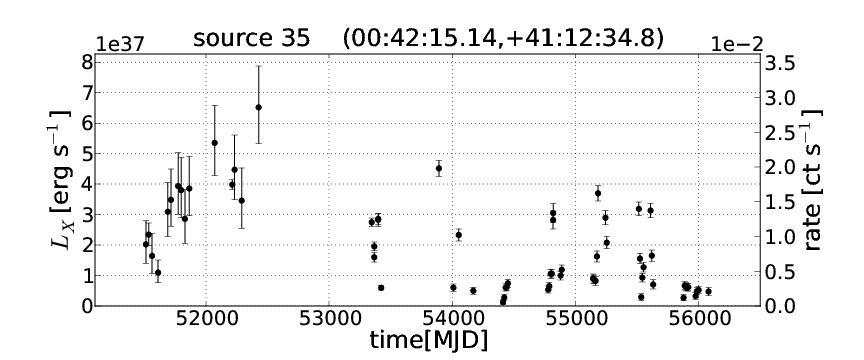}
\includegraphics[width=\linewidth]{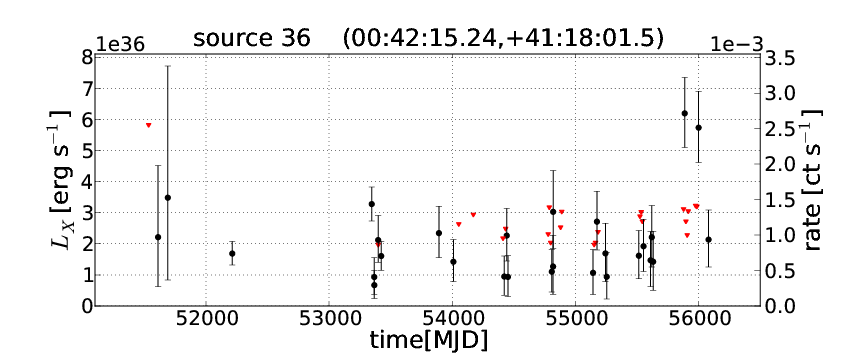}
\includegraphics[width=\linewidth]{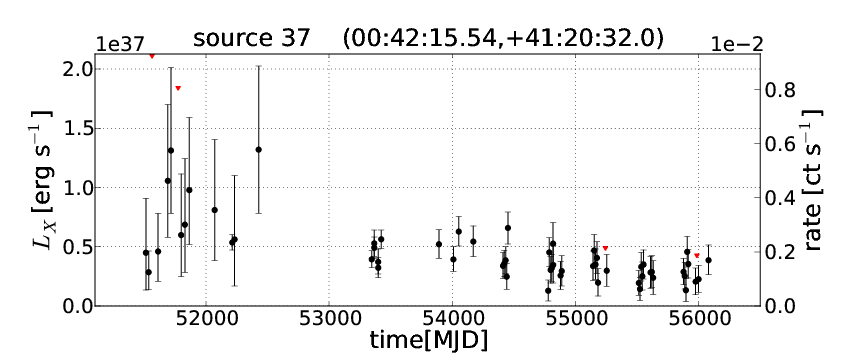}
\end{minipage}
\begin{minipage}{0.5\linewidth}
\includegraphics[width=\linewidth]{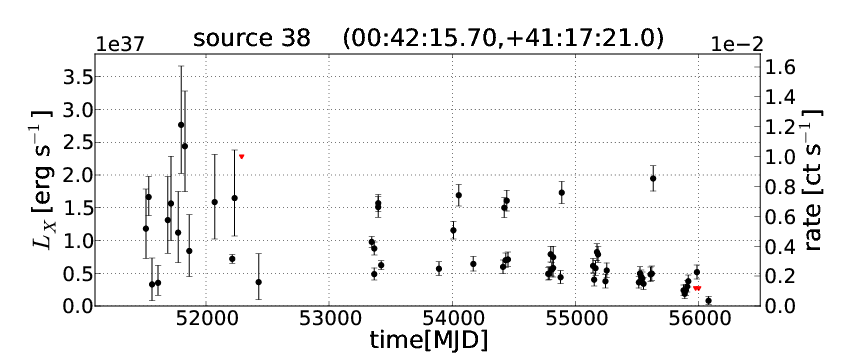}
\includegraphics[width=\linewidth]{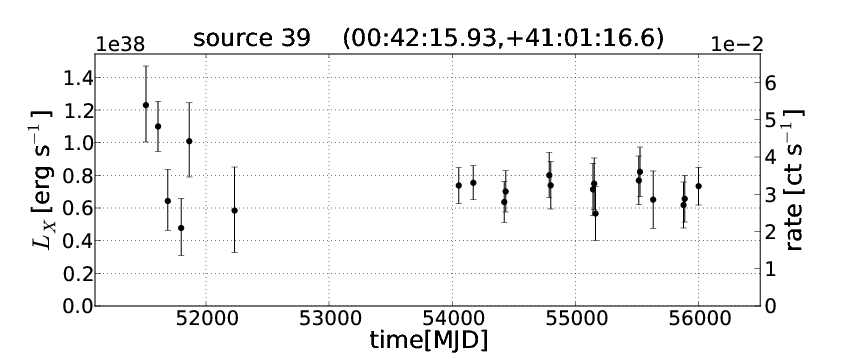}
\includegraphics[width=\linewidth]{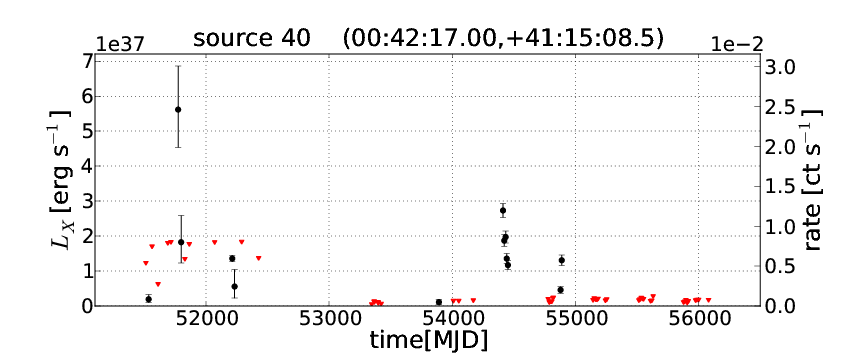}
\includegraphics[width=\linewidth]{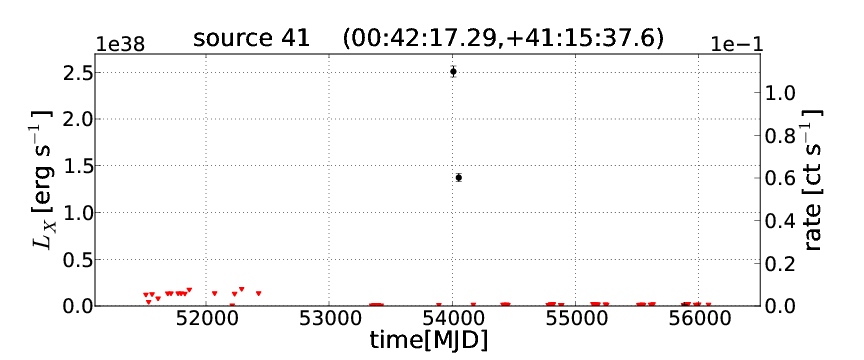}
\includegraphics[width=\linewidth]{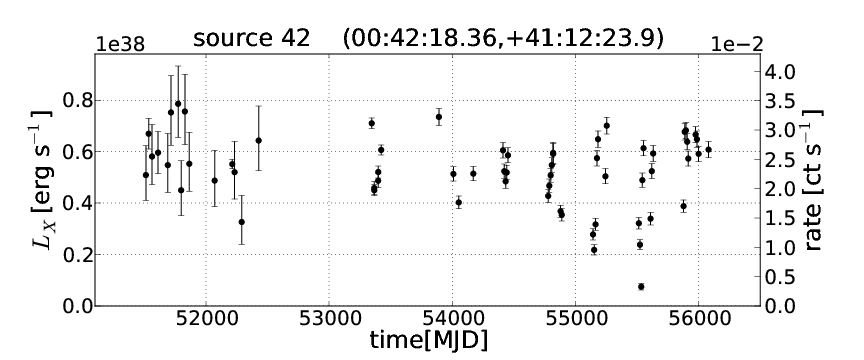}
\includegraphics[width=\linewidth]{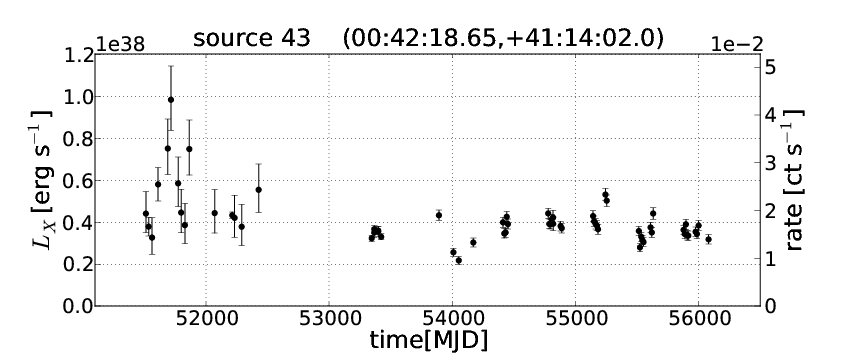}
\end{minipage}
\caption{continued.}
\label{fig:lc_all}
\end{figure*}

\addtocounter{figure}{-1} 

\begin{figure*}
\begin{minipage}{0.5\linewidth}
\includegraphics[width=\linewidth]{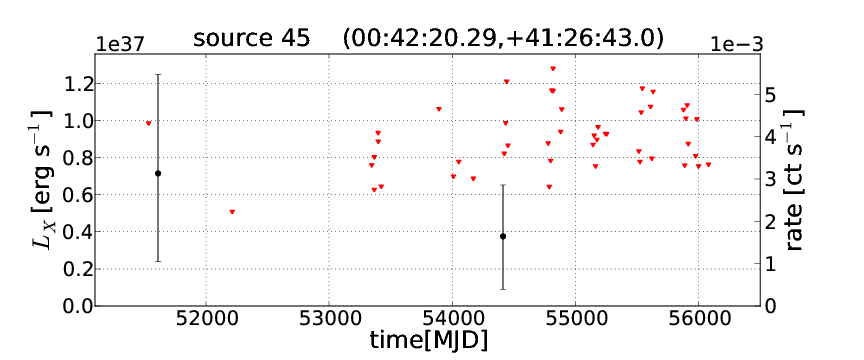}
\includegraphics[width=\linewidth]{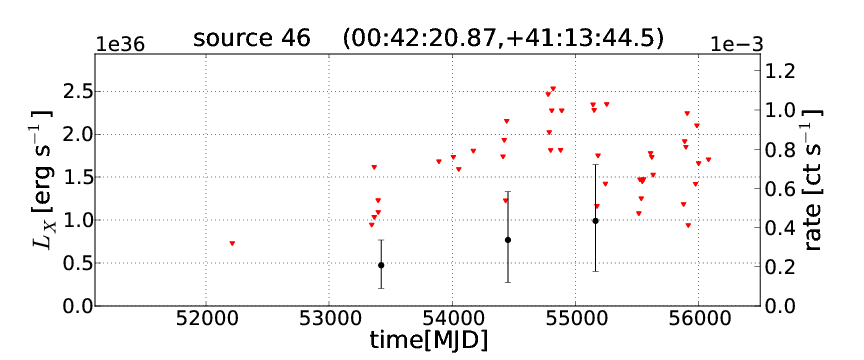}
\includegraphics[width=\linewidth]{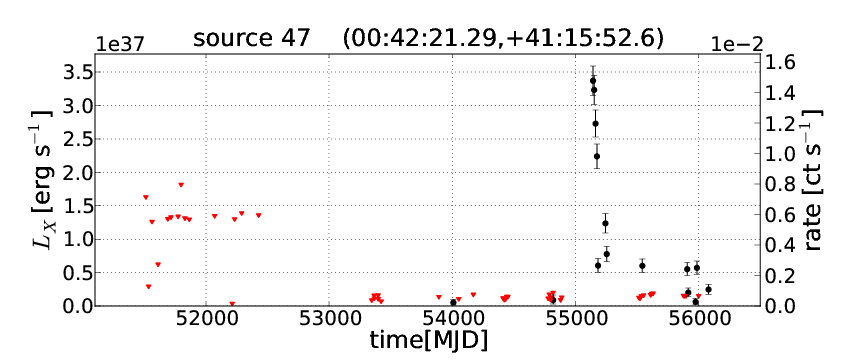}
\includegraphics[width=\linewidth]{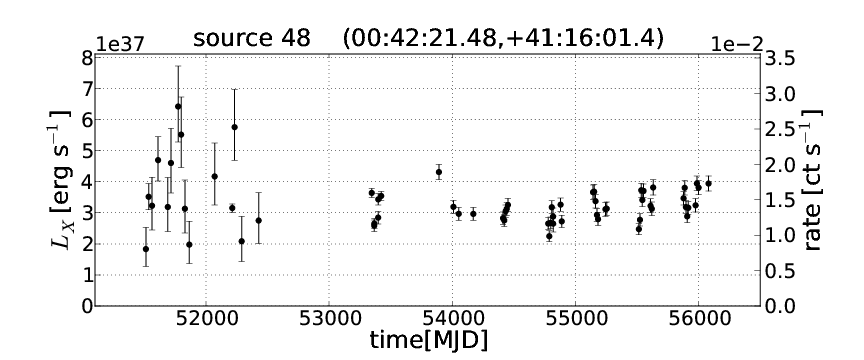}
\includegraphics[width=\linewidth]{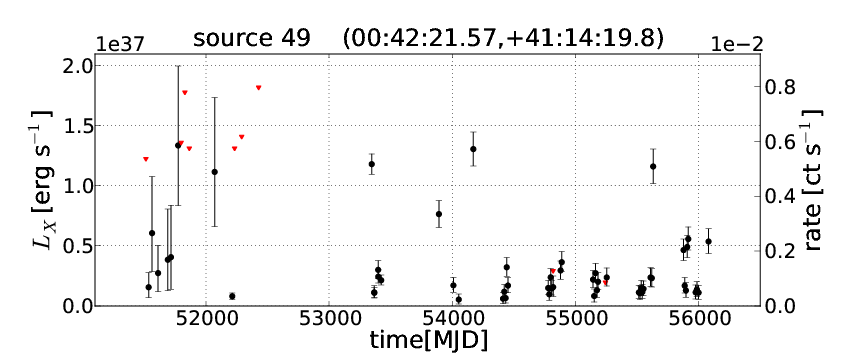}
\includegraphics[width=\linewidth]{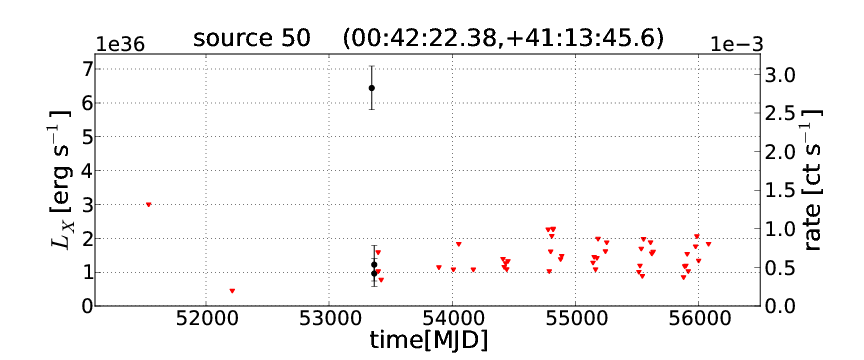}
\end{minipage}
\begin{minipage}{0.5\linewidth}
\includegraphics[width=\linewidth]{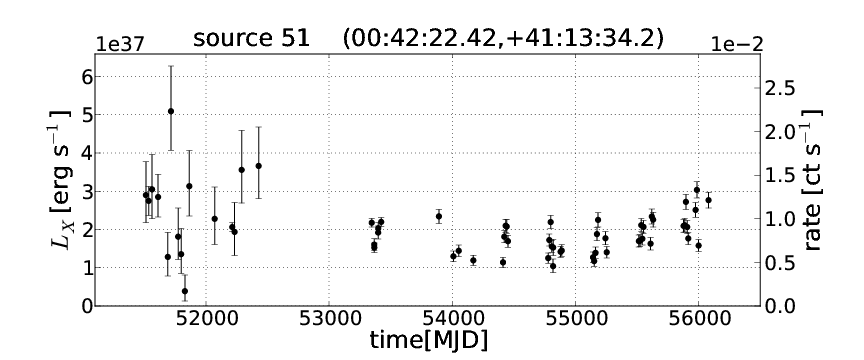}
\includegraphics[width=\linewidth]{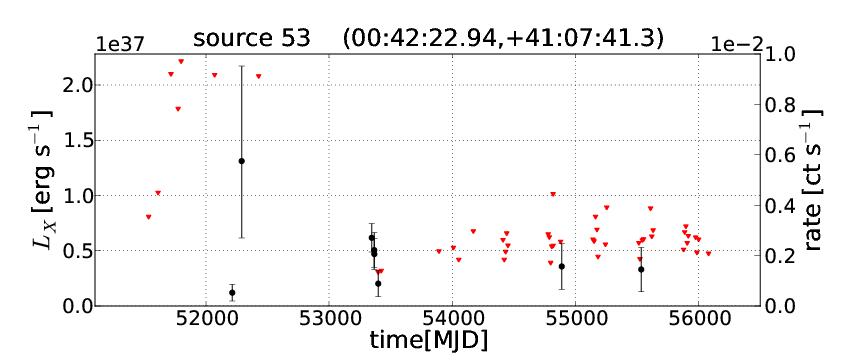}
\includegraphics[width=\linewidth]{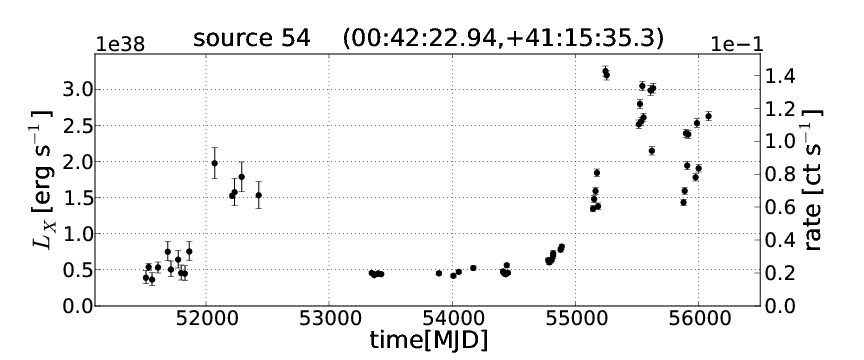}
\includegraphics[width=\linewidth]{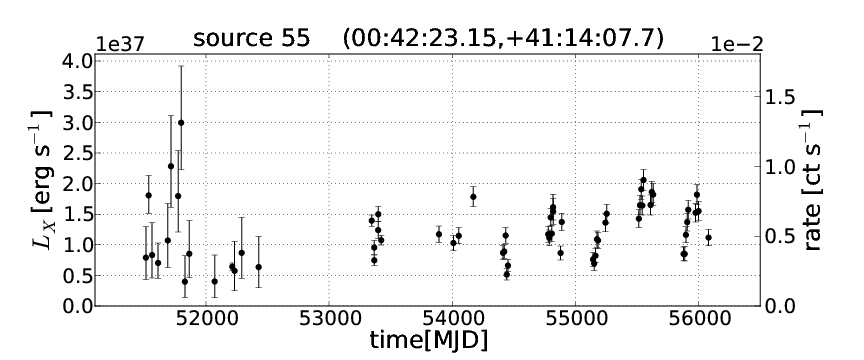}
\includegraphics[width=\linewidth]{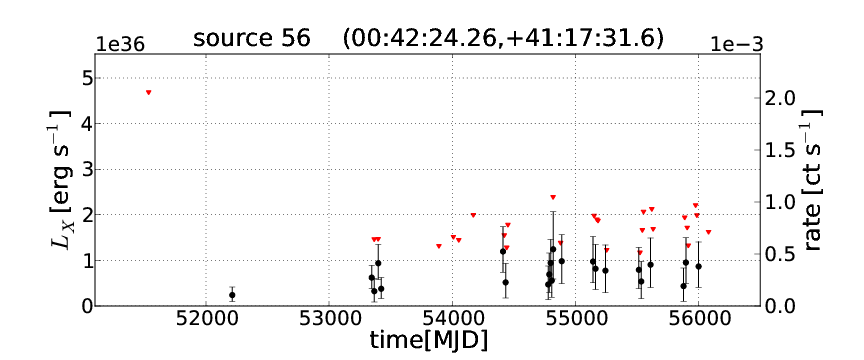}
\includegraphics[width=\linewidth]{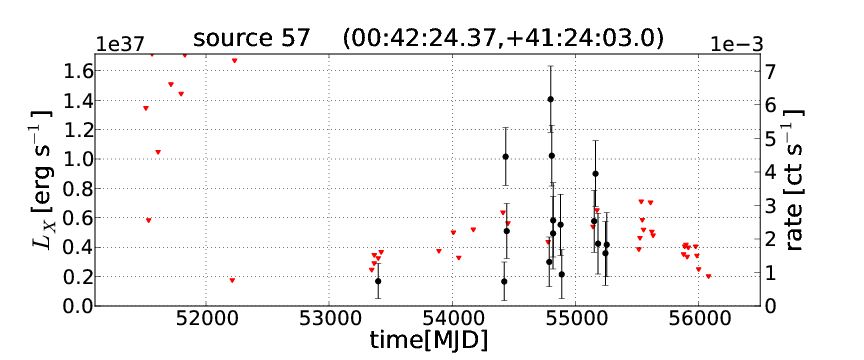}
\end{minipage}
\caption{continued.}
\label{fig:lc_all}
\end{figure*}

\addtocounter{figure}{-1} 

\begin{figure*}
\begin{minipage}{0.5\linewidth}
\includegraphics[width=\linewidth]{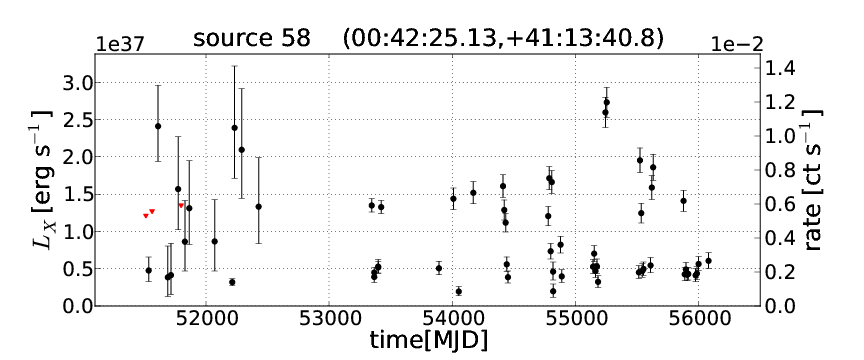}
\includegraphics[width=\linewidth]{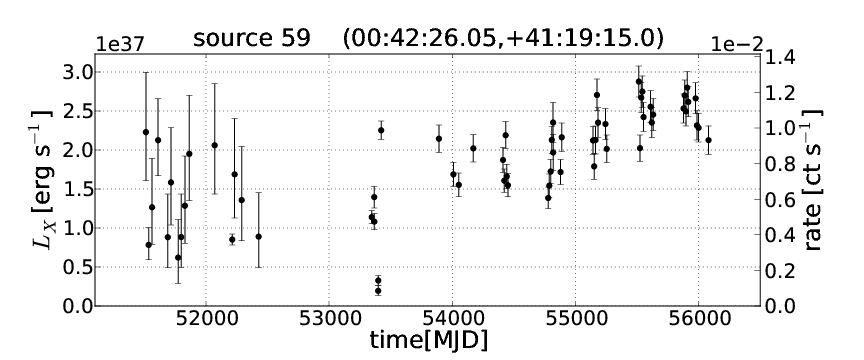}
\includegraphics[width=\linewidth]{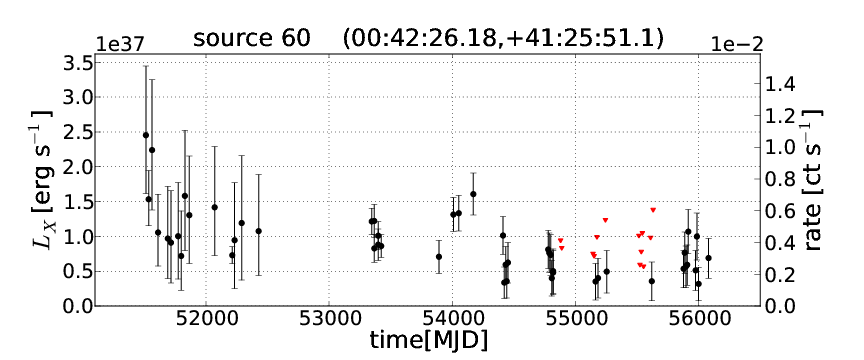}
\includegraphics[width=\linewidth]{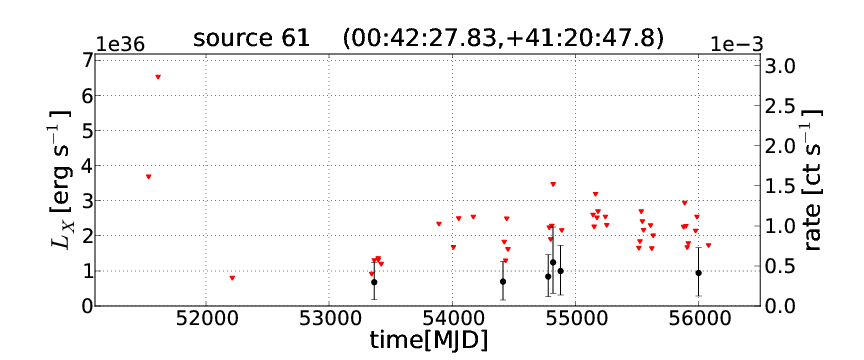}
\includegraphics[width=\linewidth]{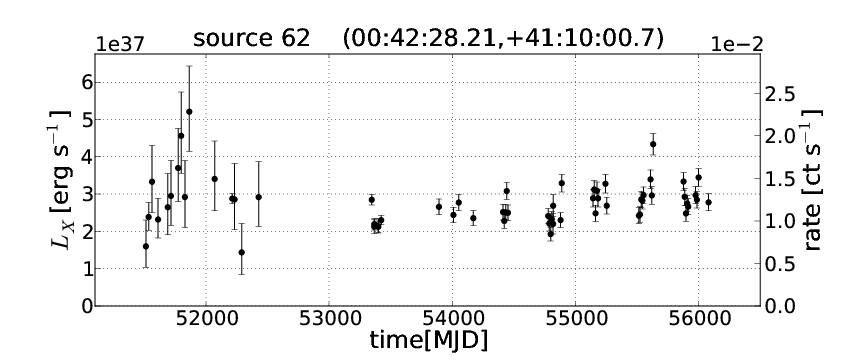}
\includegraphics[width=\linewidth]{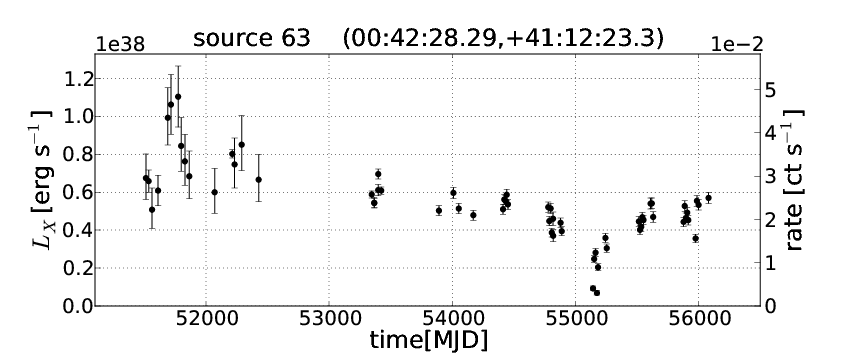}
\end{minipage}
\begin{minipage}{0.5\linewidth}
\includegraphics[width=\linewidth]{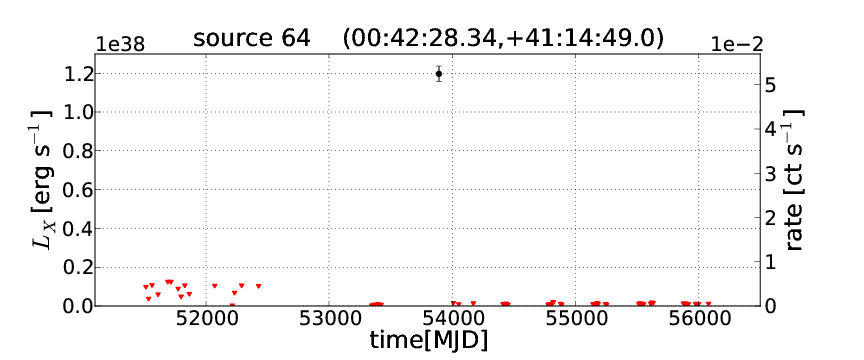}
\includegraphics[width=\linewidth]{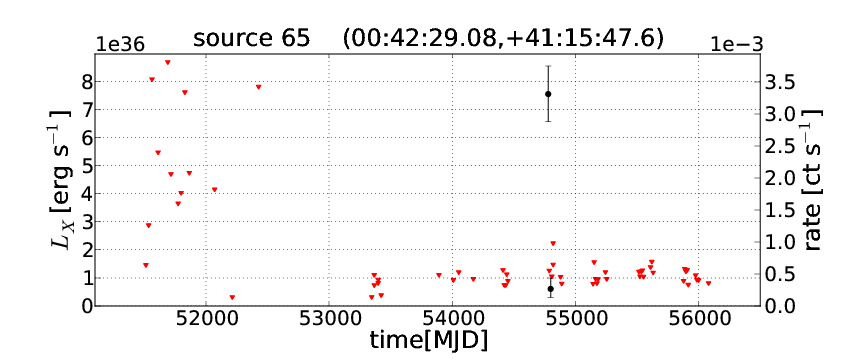}
\includegraphics[width=\linewidth]{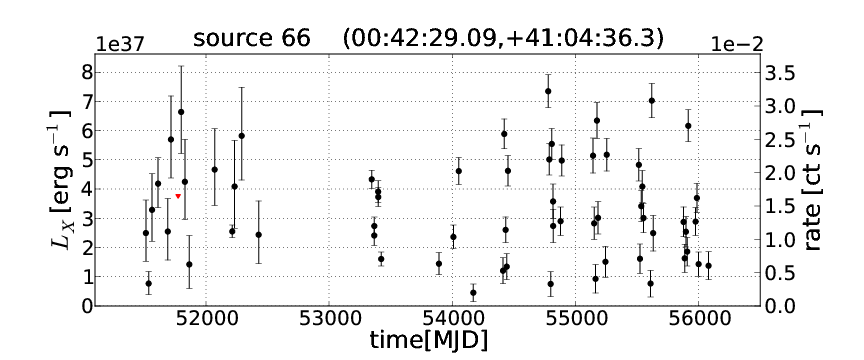}
\includegraphics[width=\linewidth]{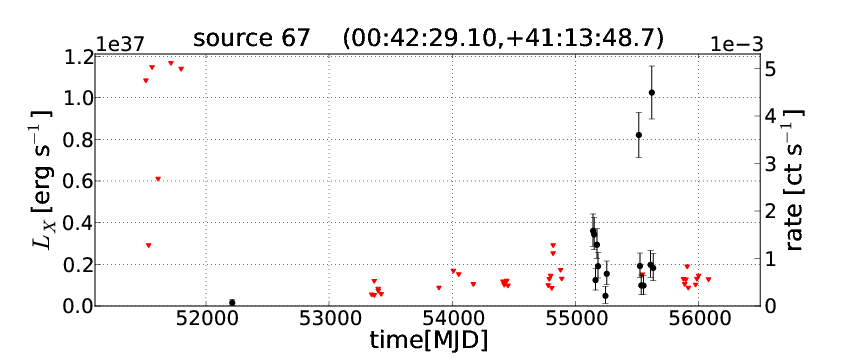}
\includegraphics[width=\linewidth]{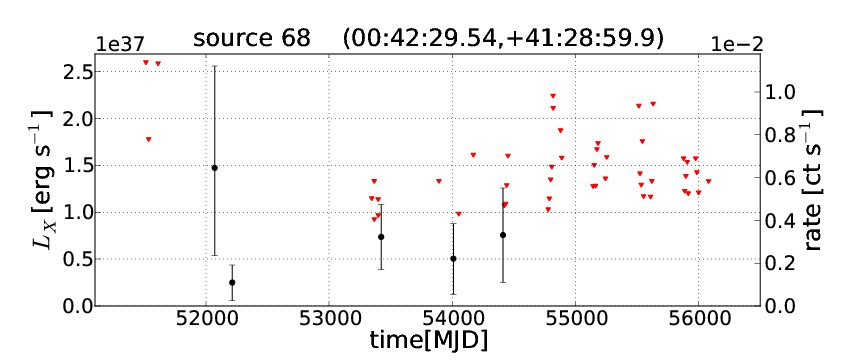}
\includegraphics[width=\linewidth]{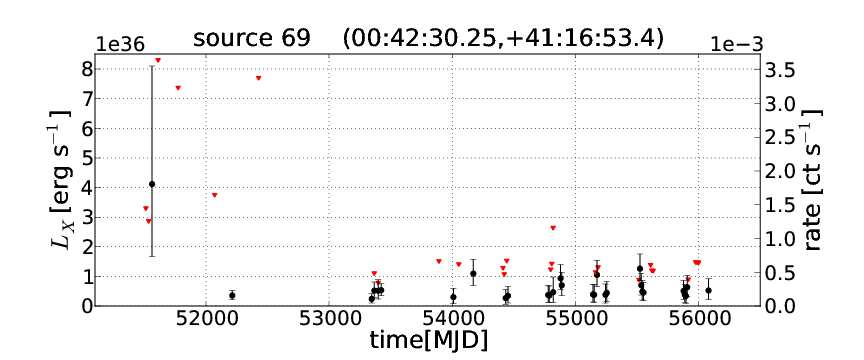}
\end{minipage}
\caption{continued.}
\label{fig:lc_all}
\end{figure*}

\addtocounter{figure}{-1} 

\begin{figure*}
\begin{minipage}{0.5\linewidth}
\includegraphics[width=\linewidth]{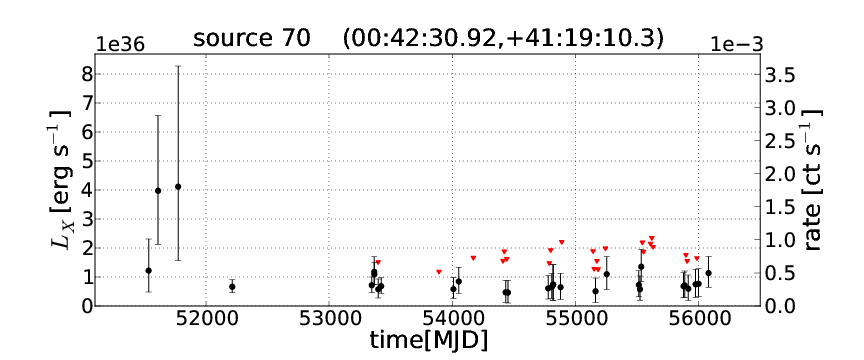}
\includegraphics[width=\linewidth]{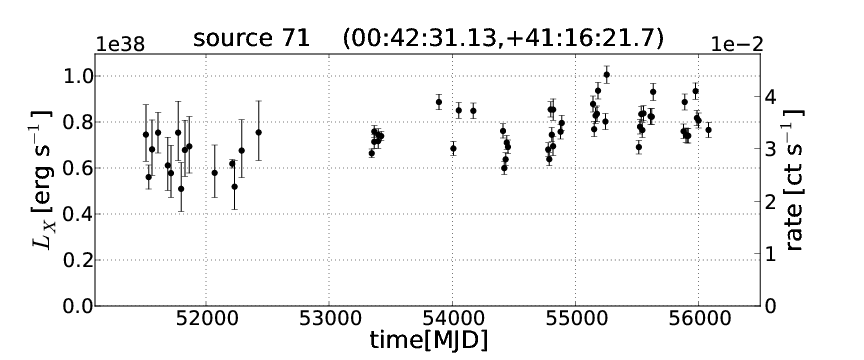}
\includegraphics[width=\linewidth]{plots_new/72_light.png}
\includegraphics[width=\linewidth]{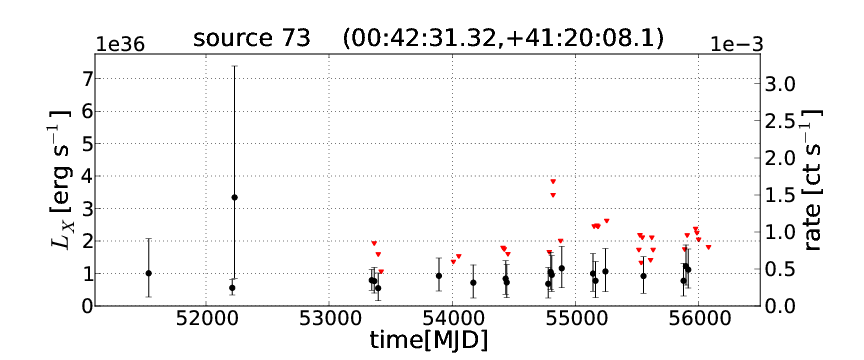}
\includegraphics[width=\linewidth]{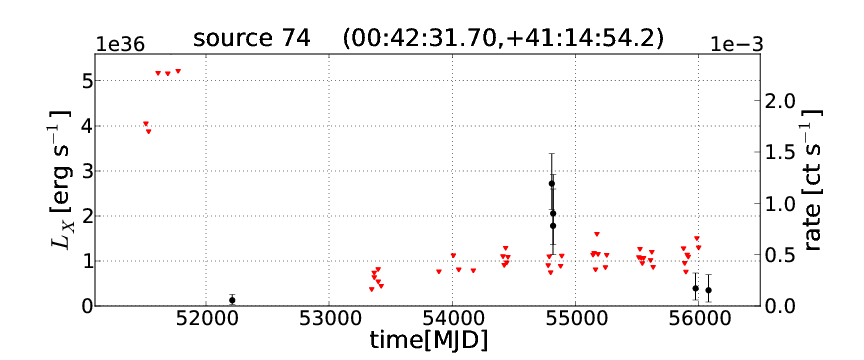}
\includegraphics[width=\linewidth]{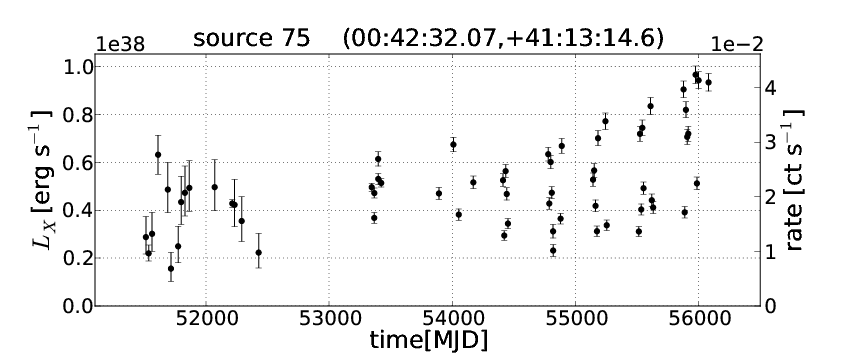}
\end{minipage}
\begin{minipage}{0.5\linewidth}
\includegraphics[width=\linewidth]{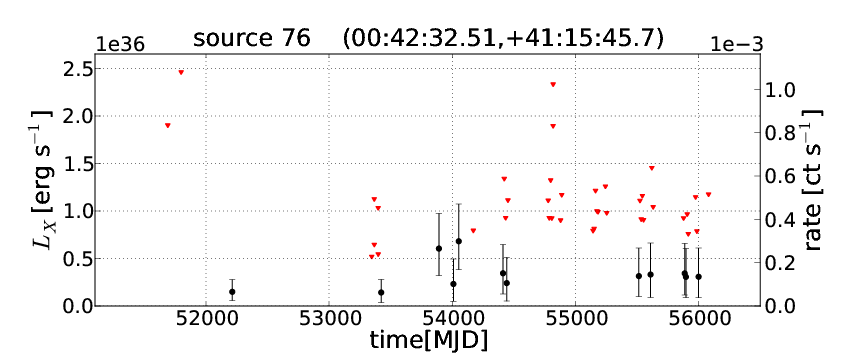}
\includegraphics[width=\linewidth]{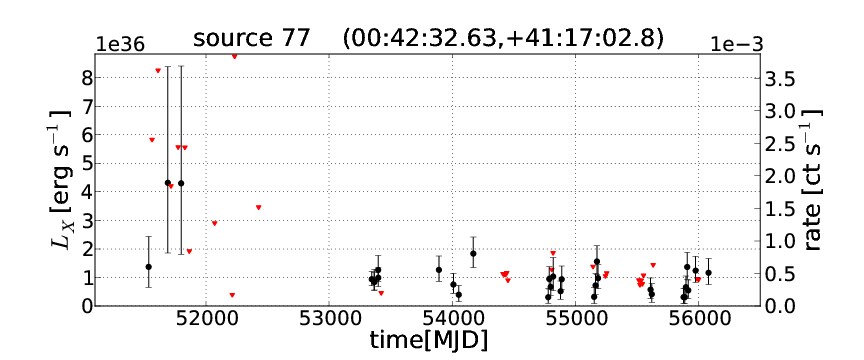}
\includegraphics[width=\linewidth]{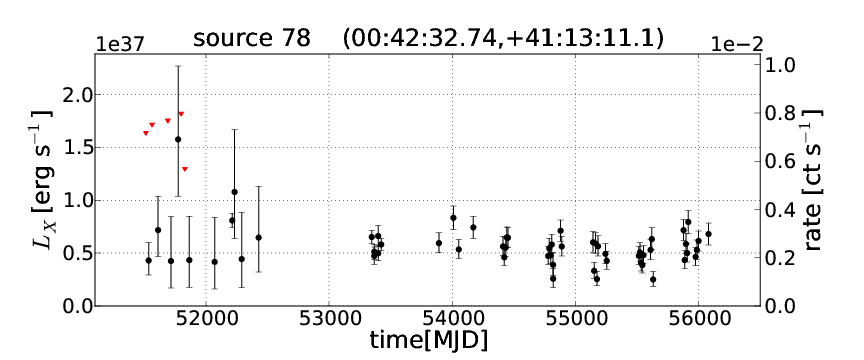}
\includegraphics[width=\linewidth]{plots_new/79_light.png}
\includegraphics[width=\linewidth]{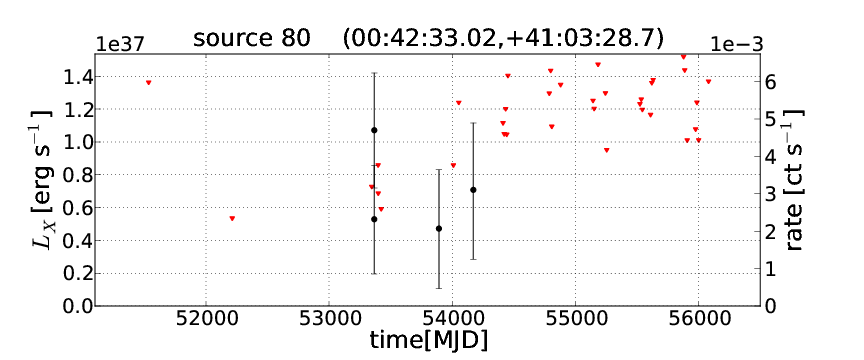}
\includegraphics[width=\linewidth]{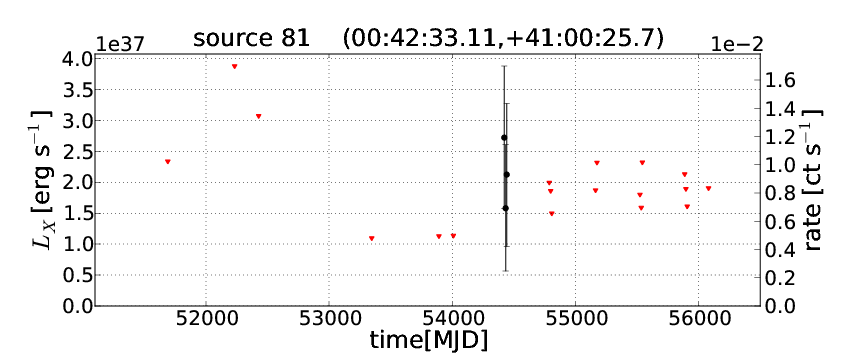}
\end{minipage}
\caption{continued.}
\label{fig:lc_all}
\end{figure*}

\addtocounter{figure}{-1} 

\begin{figure*}
\begin{minipage}{0.5\linewidth}
\includegraphics[width=\linewidth]{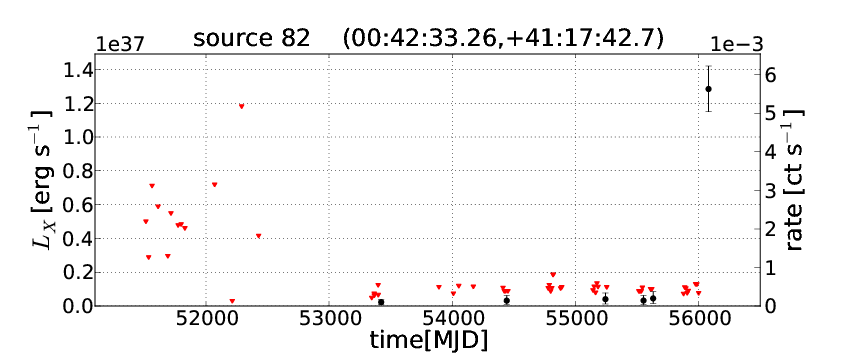}
\includegraphics[width=\linewidth]{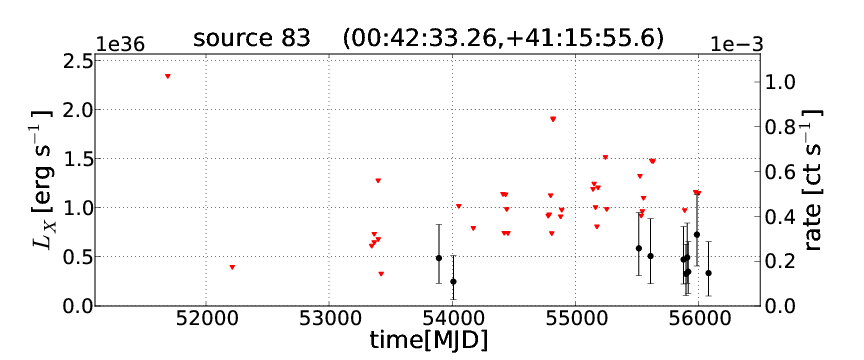}
\includegraphics[width=\linewidth]{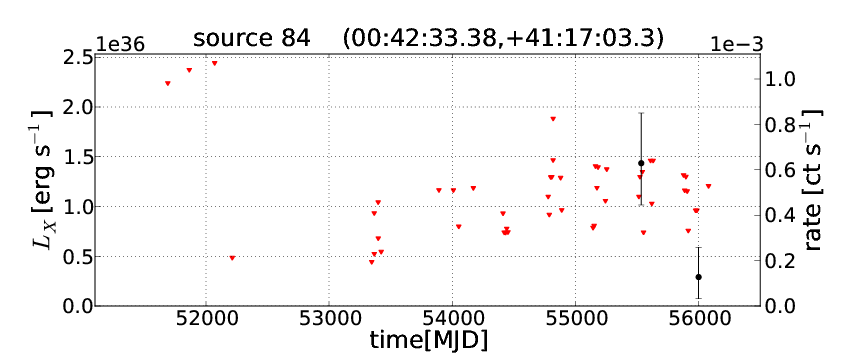}
\includegraphics[width=\linewidth]{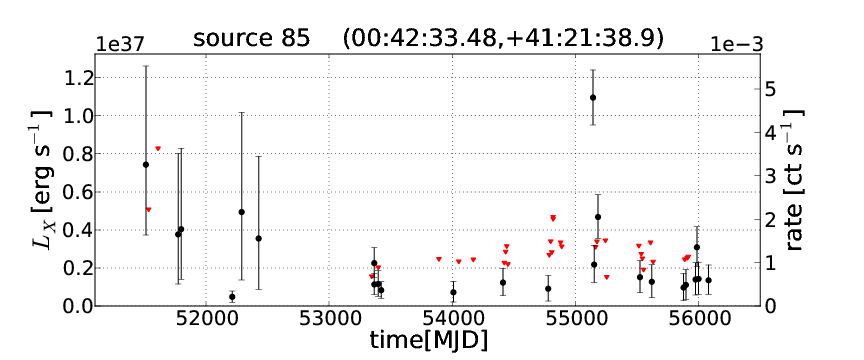}
\includegraphics[width=\linewidth]{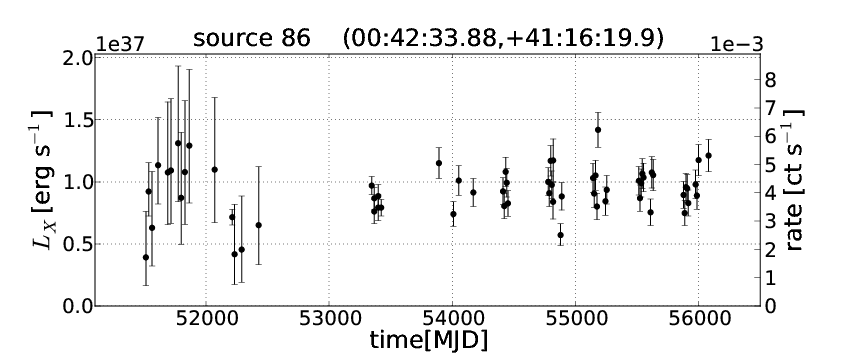}
\includegraphics[width=\linewidth]{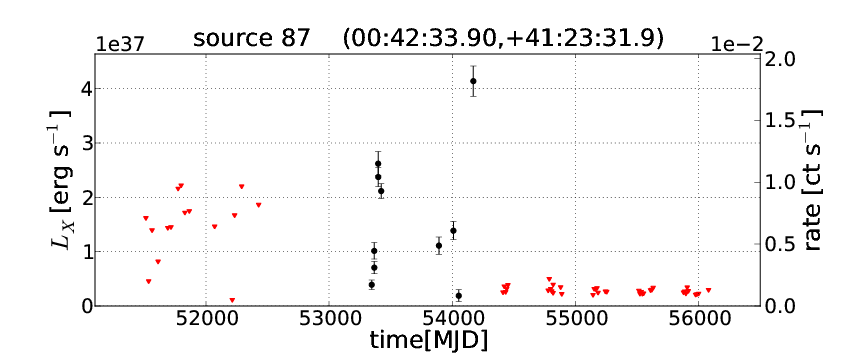}
\end{minipage}
\begin{minipage}{0.5\linewidth}
\includegraphics[width=\linewidth]{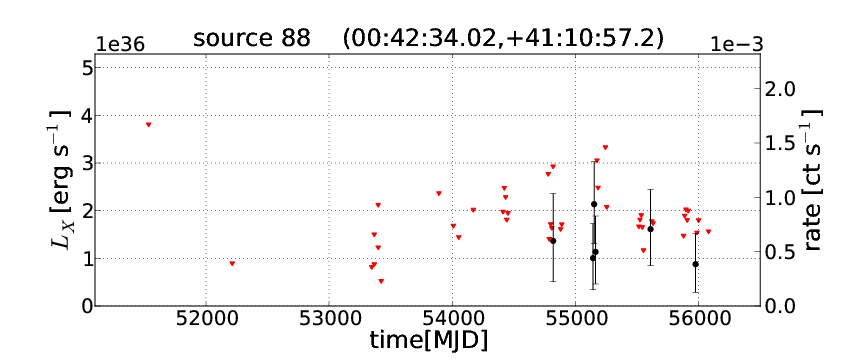}
\includegraphics[width=\linewidth]{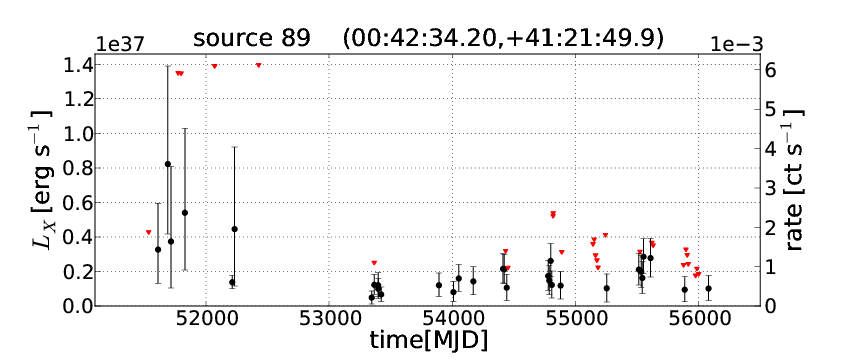}
\includegraphics[width=\linewidth]{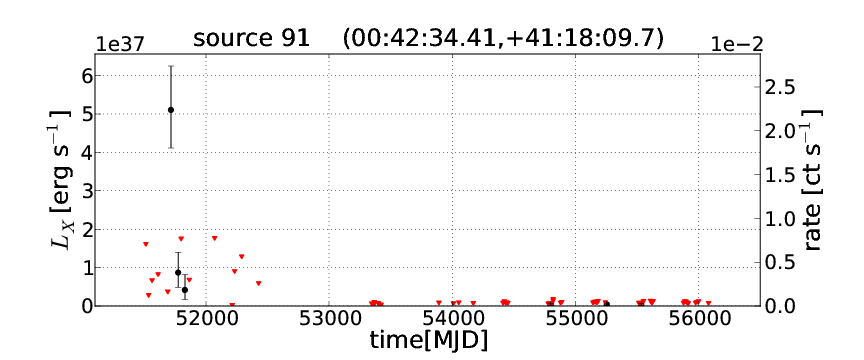}
\includegraphics[width=\linewidth]{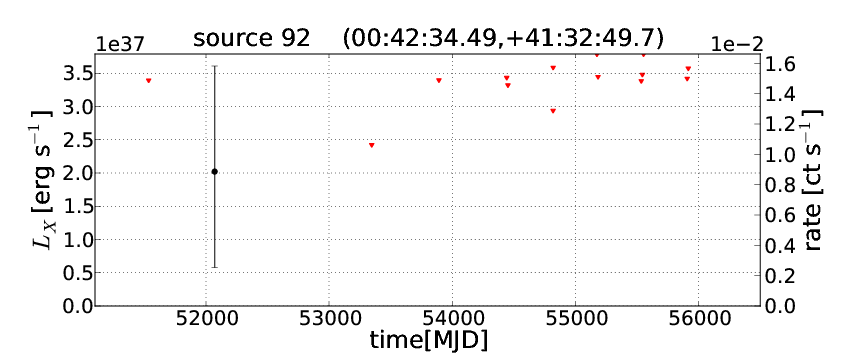}
\includegraphics[width=\linewidth]{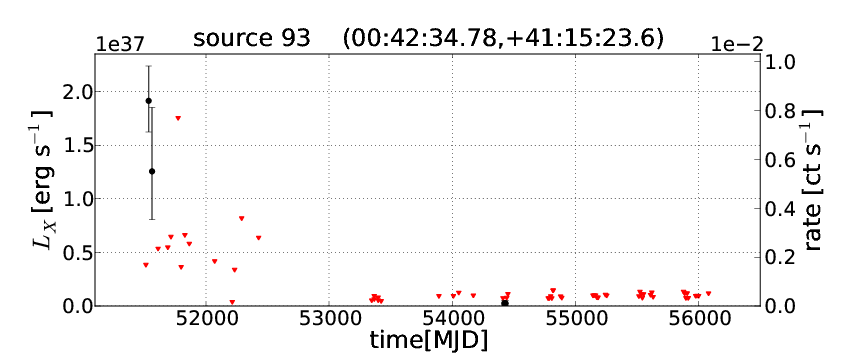}
\includegraphics[width=\linewidth]{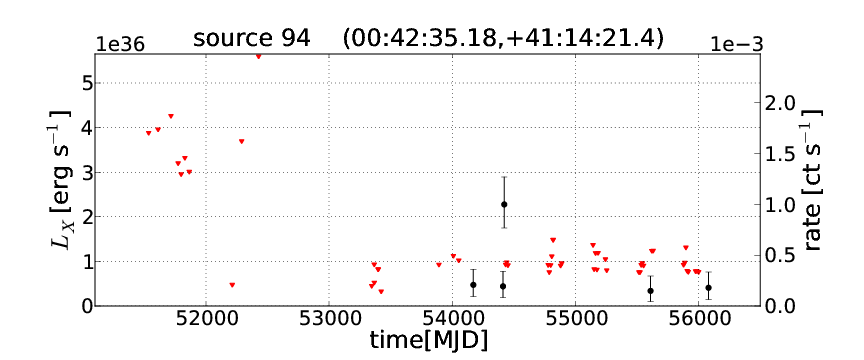}
\end{minipage}
\caption{continued.}
\label{fig:lc_all}
\end{figure*}

\addtocounter{figure}{-1} 

\FloatBarrier

\begin{figure*}
\begin{minipage}{0.5\linewidth}
\includegraphics[width=\linewidth]{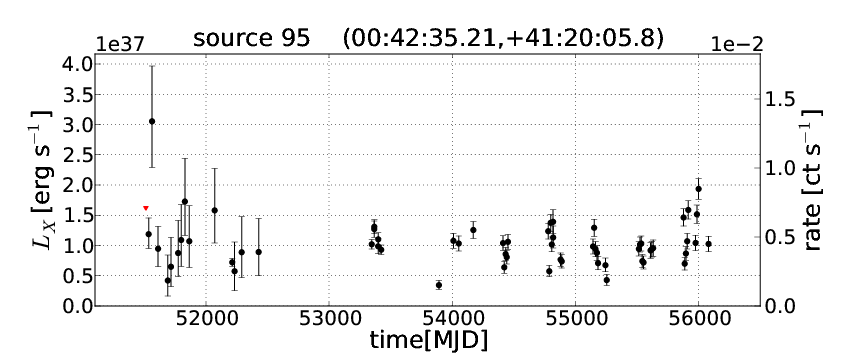}
\includegraphics[width=\linewidth]{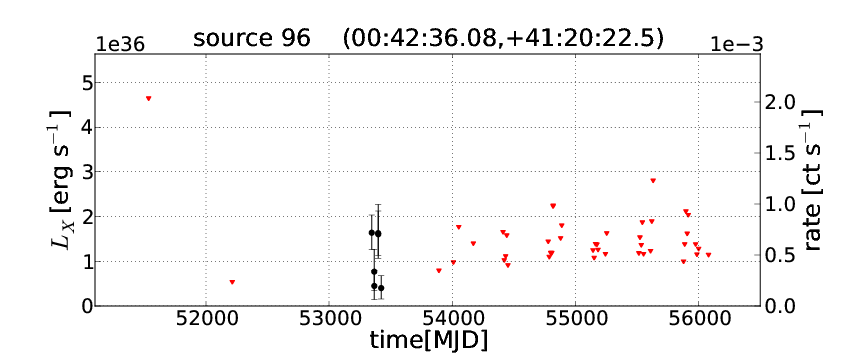}
\includegraphics[width=\linewidth]{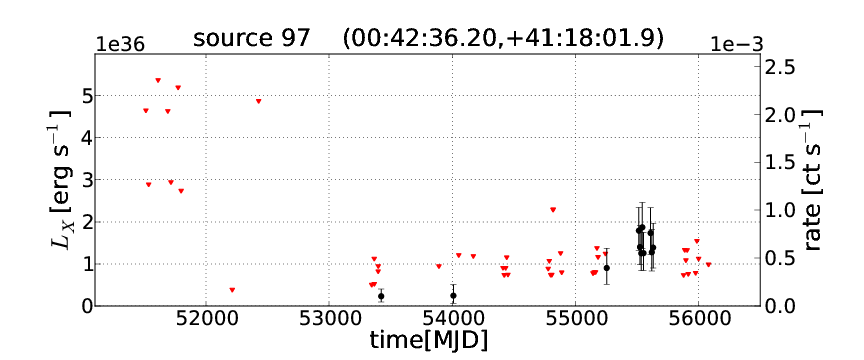}
\includegraphics[width=\linewidth]{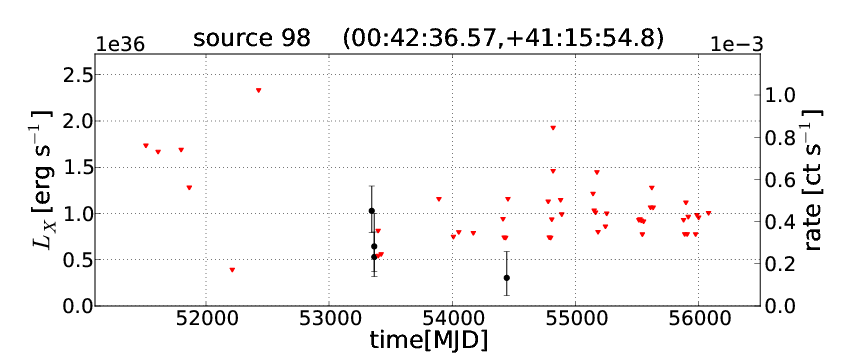}
\includegraphics[width=\linewidth]{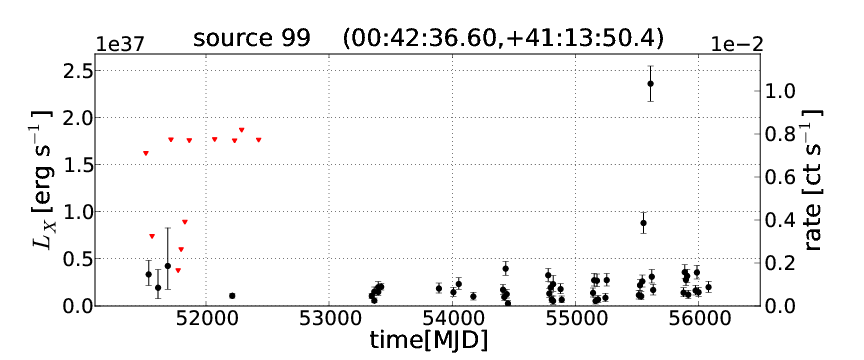}
\includegraphics[width=\linewidth]{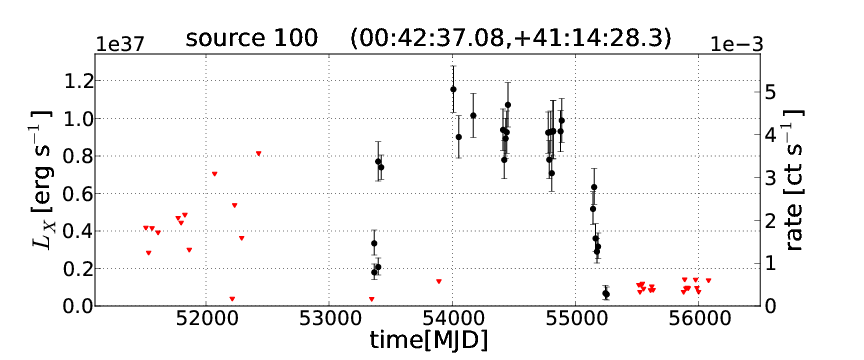}
\end{minipage}
\begin{minipage}{0.5\linewidth}
\includegraphics[width=\linewidth]{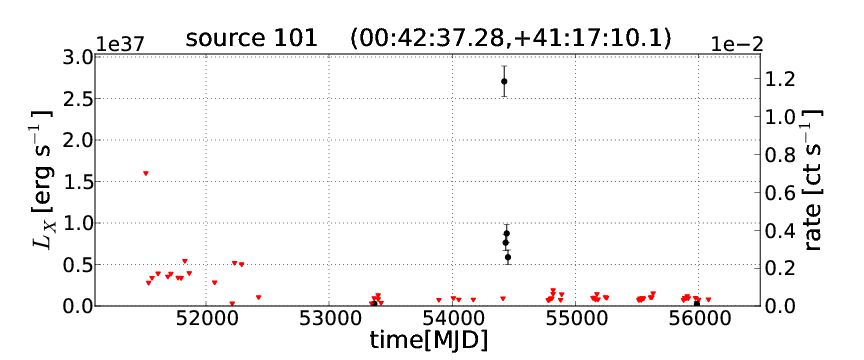}
\includegraphics[width=\linewidth]{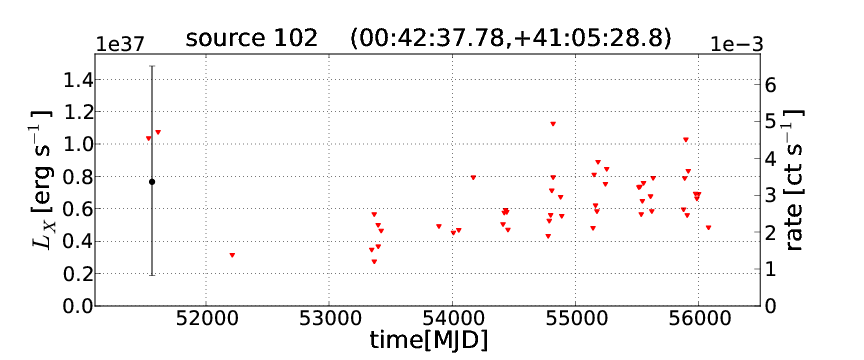}
\includegraphics[width=\linewidth]{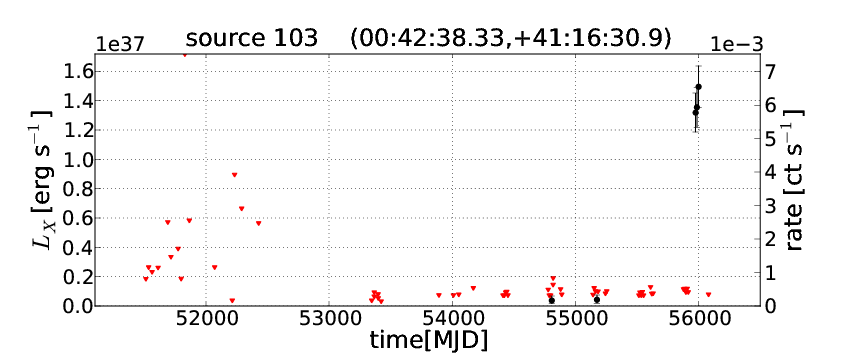}
\includegraphics[width=\linewidth]{plots_new/104_light.png}
\includegraphics[width=\linewidth]{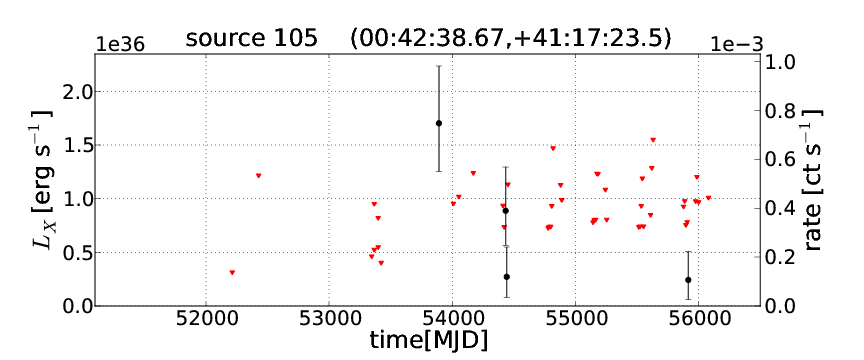}
\includegraphics[width=\linewidth]{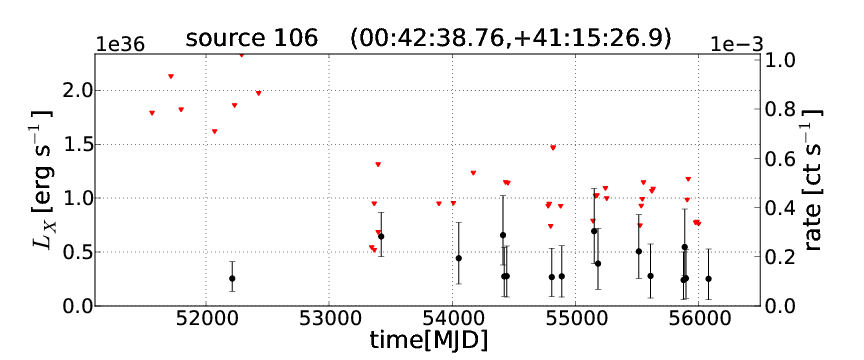}
\end{minipage}
\caption{continued.}
\label{fig:lc_all}
\end{figure*}

\addtocounter{figure}{-1} 

\begin{figure*}
\begin{minipage}{0.5\linewidth}
\includegraphics[width=\linewidth]{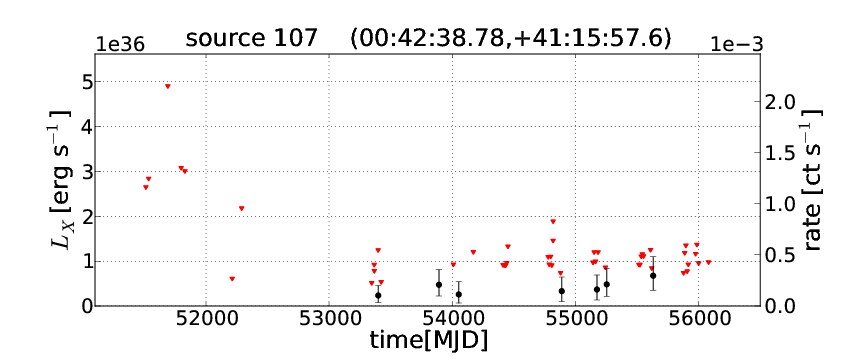}
\includegraphics[width=\linewidth]{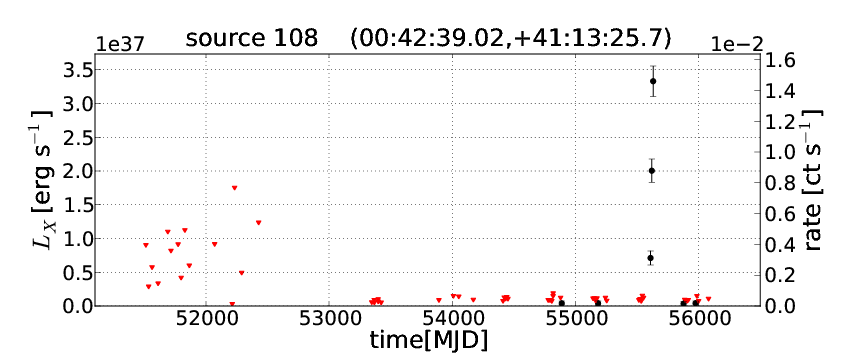}
\includegraphics[width=\linewidth]{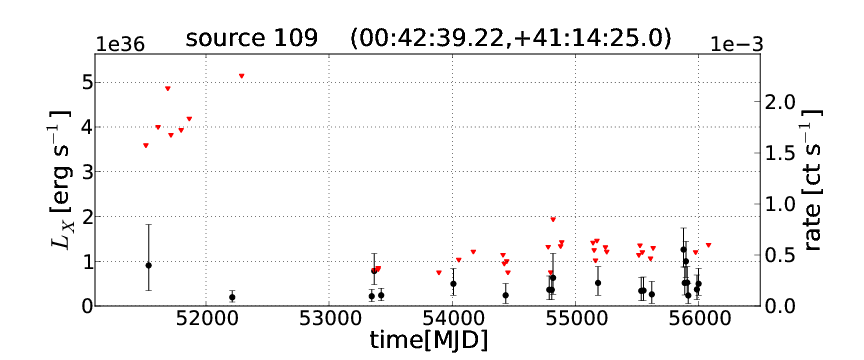}
\includegraphics[width=\linewidth]{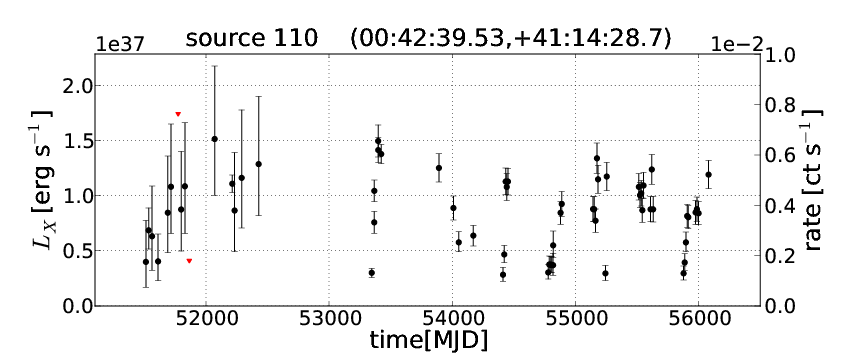}
\includegraphics[width=\linewidth]{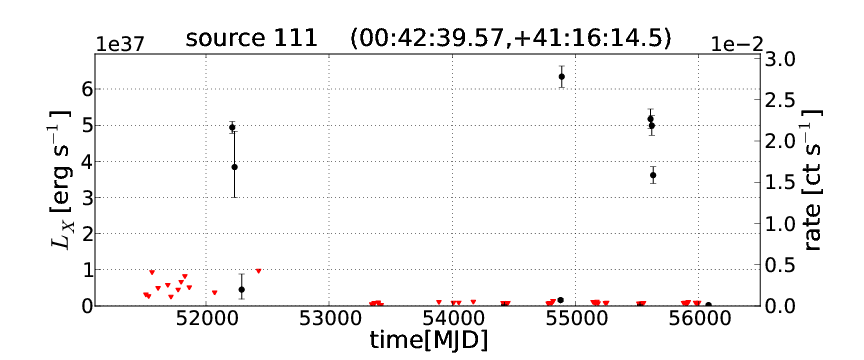}
\includegraphics[width=\linewidth]{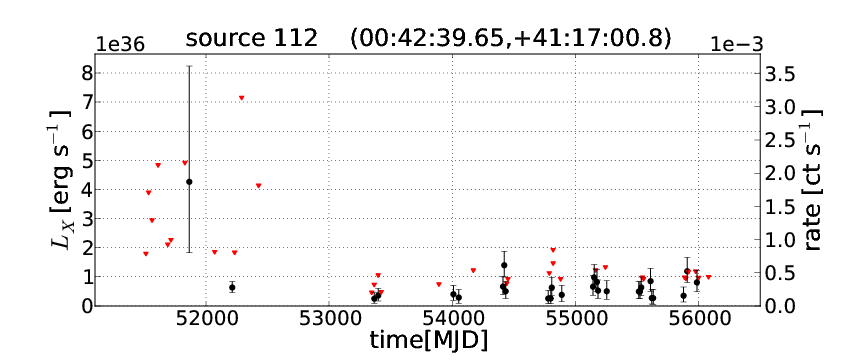}
\end{minipage}
\begin{minipage}{0.5\linewidth}
\includegraphics[width=\linewidth]{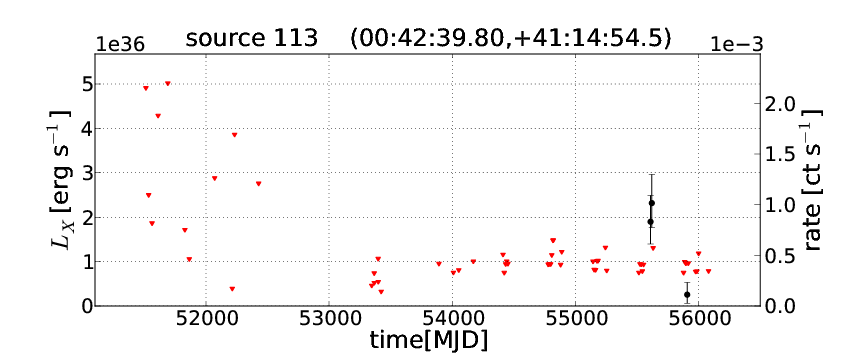}
\includegraphics[width=\linewidth]{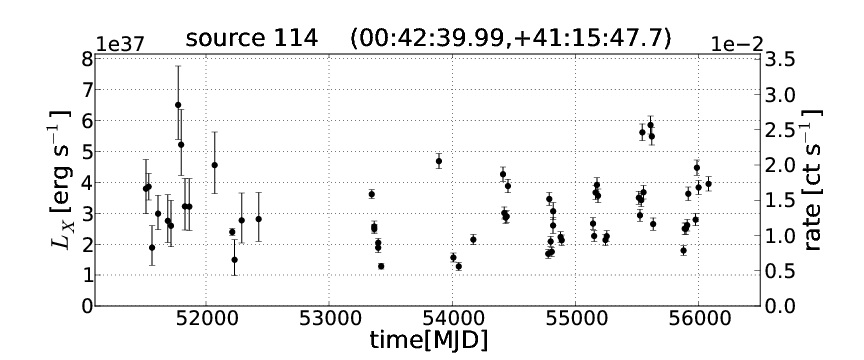}
\includegraphics[width=\linewidth]{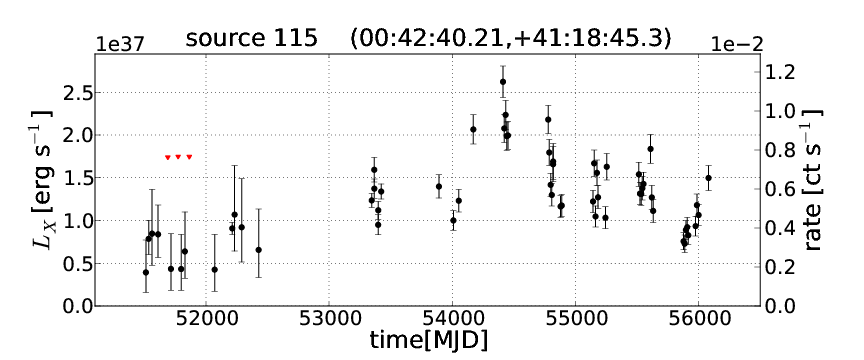}
\includegraphics[width=\linewidth]{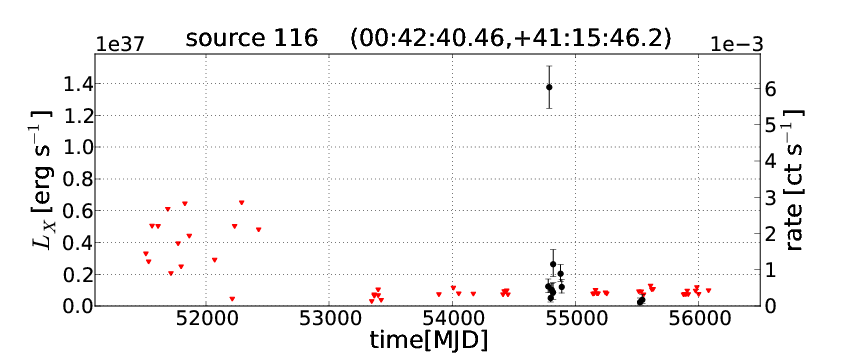}
\includegraphics[width=\linewidth]{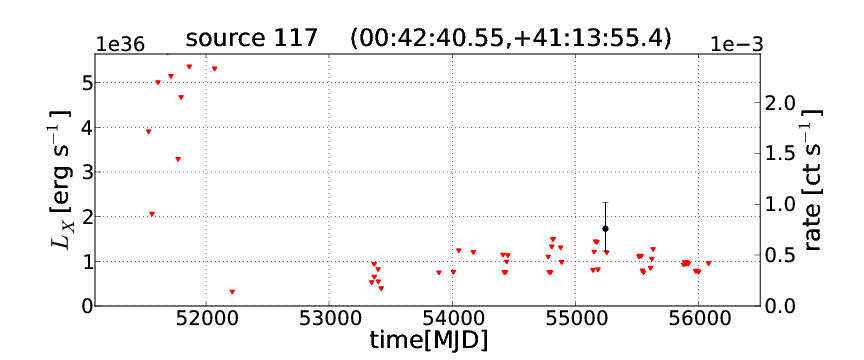}
\includegraphics[width=\linewidth]{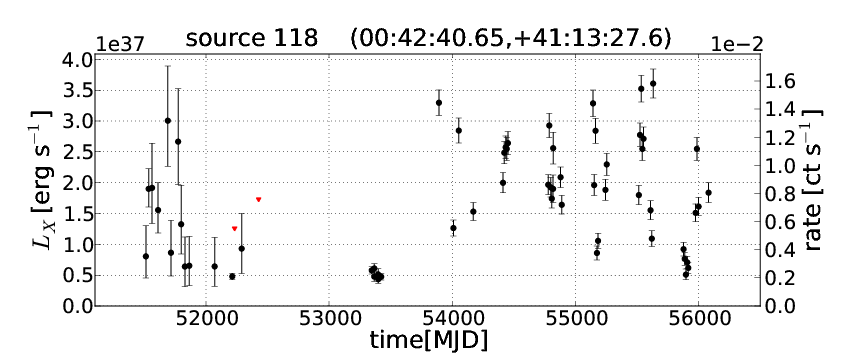}
\end{minipage}
\caption{continued.}
\label{fig:lc_all}
\end{figure*}

\addtocounter{figure}{-1} 

\begin{figure*}
\begin{minipage}{0.5\linewidth}
\includegraphics[width=\linewidth]{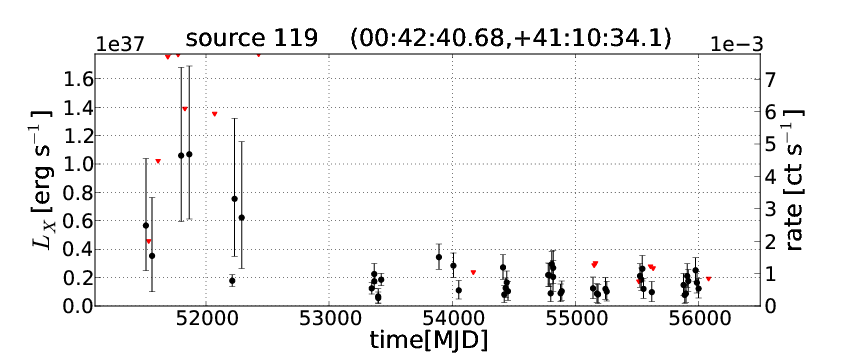}
\includegraphics[width=\linewidth]{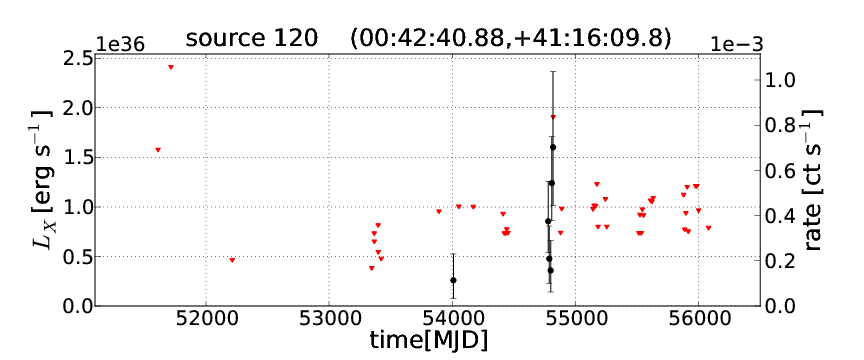}
\includegraphics[width=\linewidth]{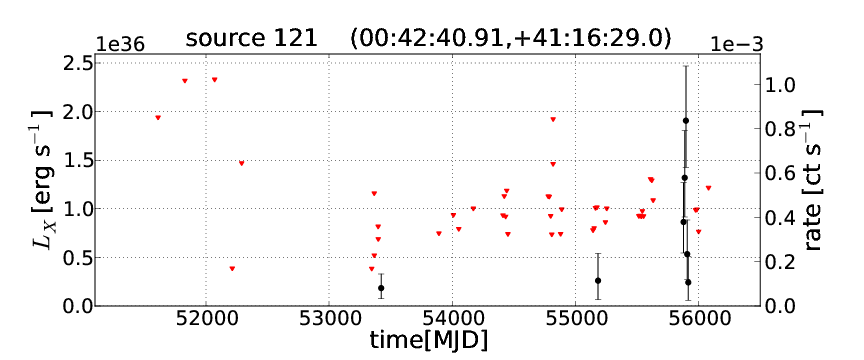}
\includegraphics[width=\linewidth]{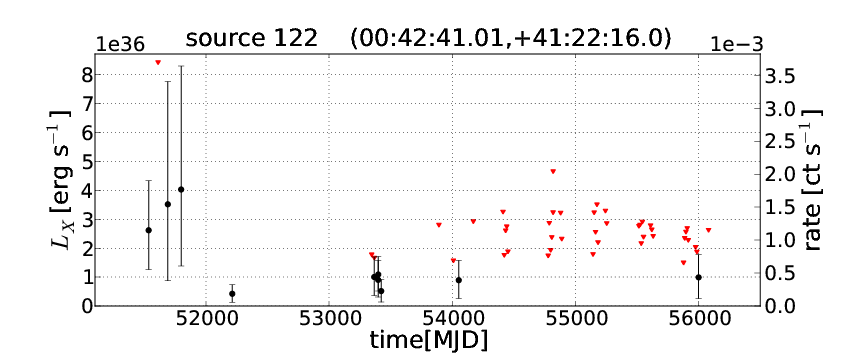}
\includegraphics[width=\linewidth]{plots_new/123_light.png}
\includegraphics[width=\linewidth]{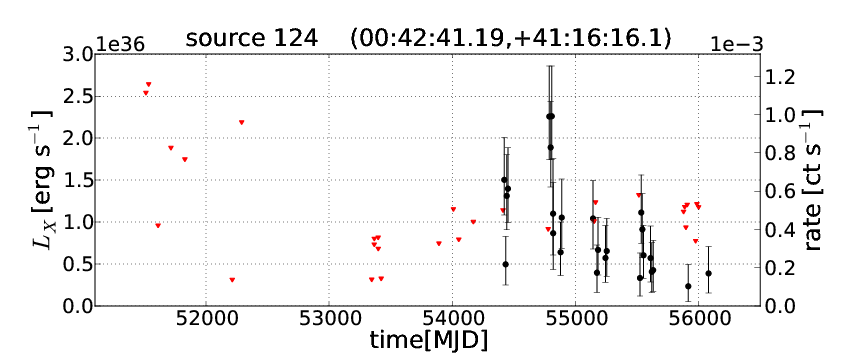}
\end{minipage}
\begin{minipage}{0.5\linewidth}
\includegraphics[width=\linewidth]{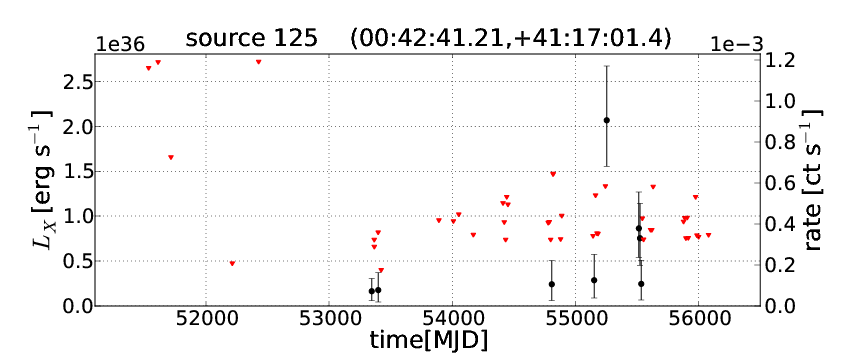}
\includegraphics[width=\linewidth]{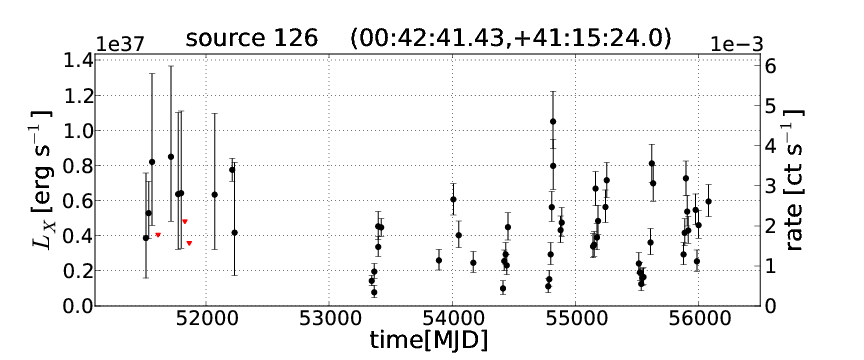}
\includegraphics[width=\linewidth]{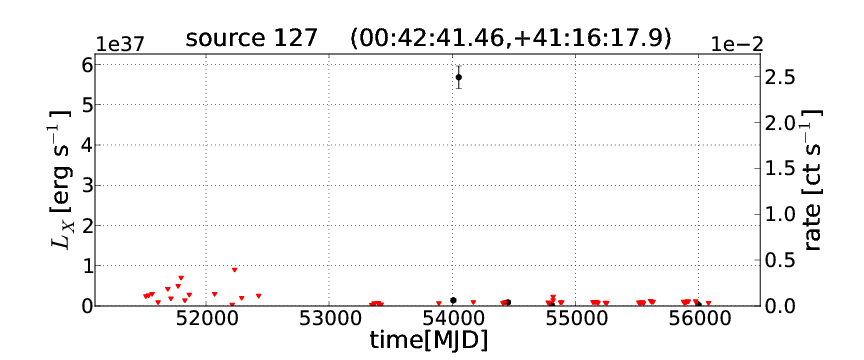}
\includegraphics[width=\linewidth]{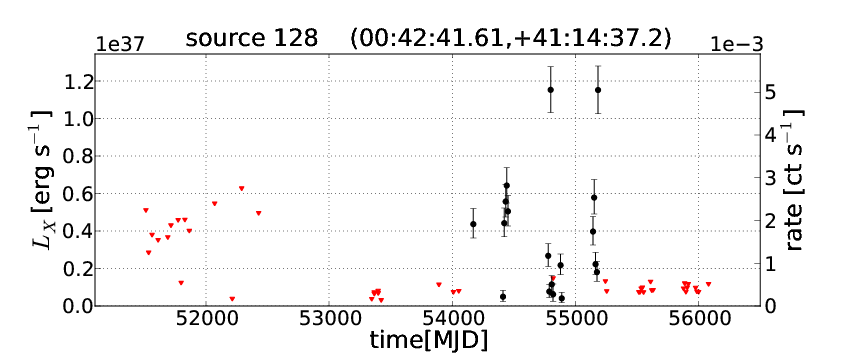}
\includegraphics[width=\linewidth]{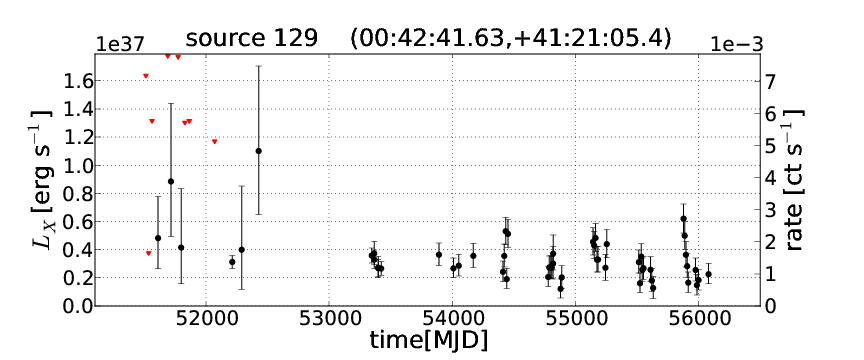}
\includegraphics[width=\linewidth]{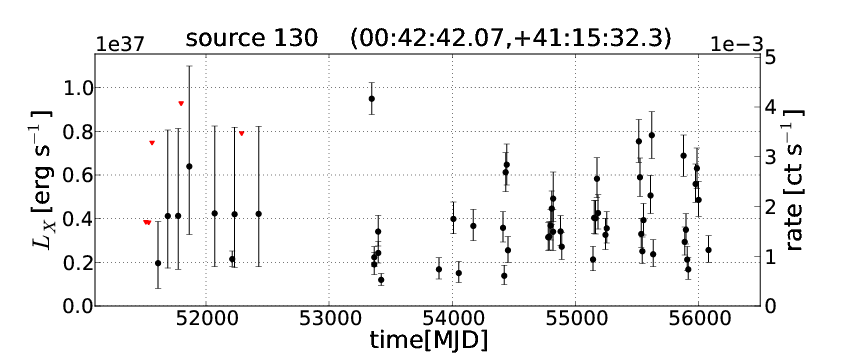}
\end{minipage}
\caption{continued.}
\label{fig:lc_all}
\end{figure*}

\addtocounter{figure}{-1} 

\begin{figure*}
\begin{minipage}{0.5\linewidth}
\includegraphics[width=\linewidth]{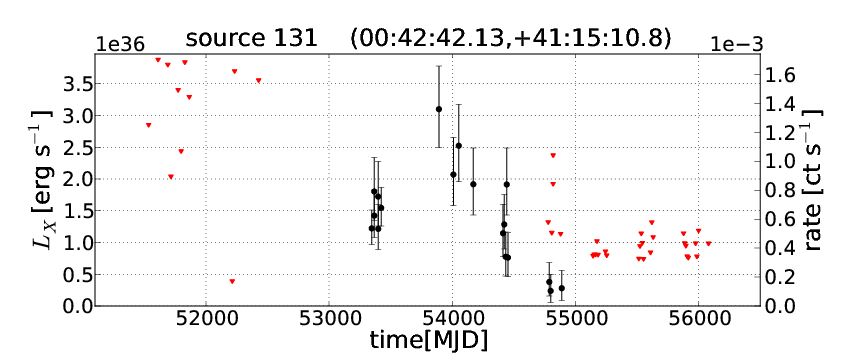}
\includegraphics[width=\linewidth]{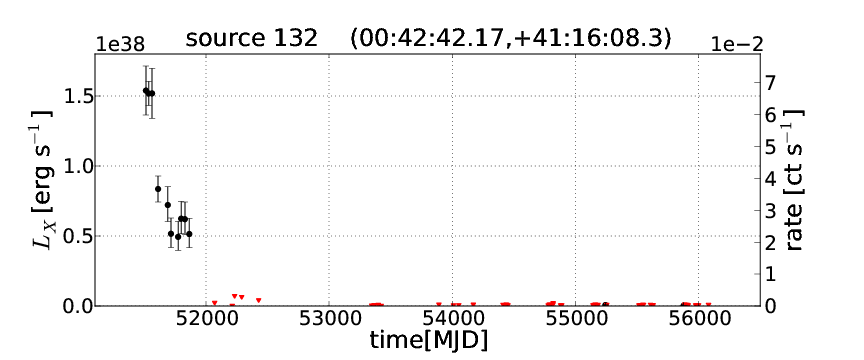}
\includegraphics[width=\linewidth]{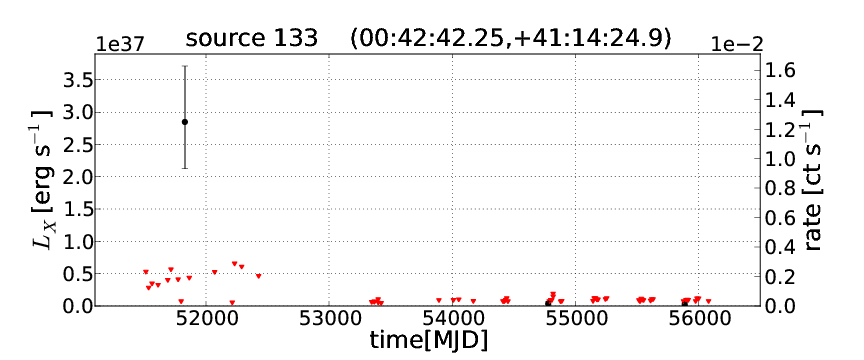}
\includegraphics[width=\linewidth]{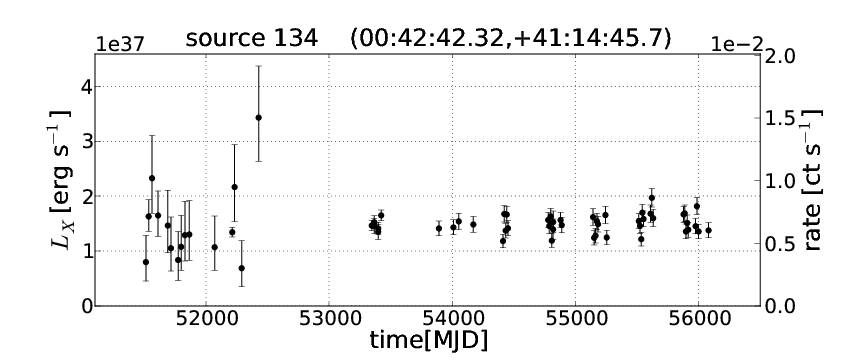}
\includegraphics[width=\linewidth]{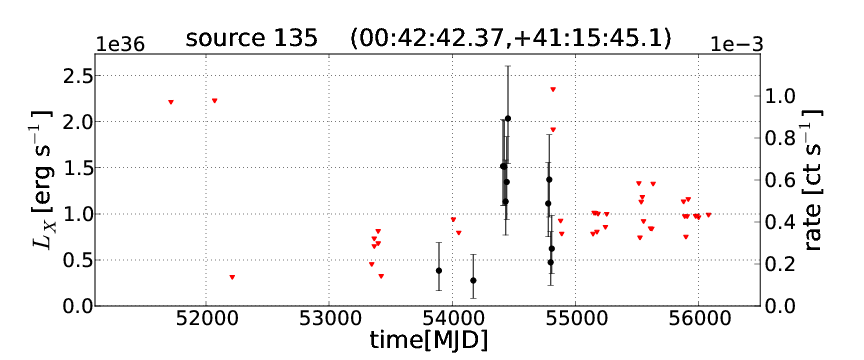}
\includegraphics[width=\linewidth]{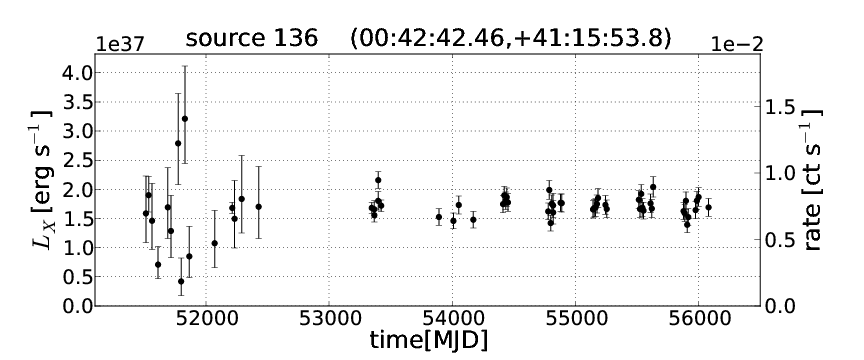}
\end{minipage}
\begin{minipage}{0.5\linewidth}
\includegraphics[width=\linewidth]{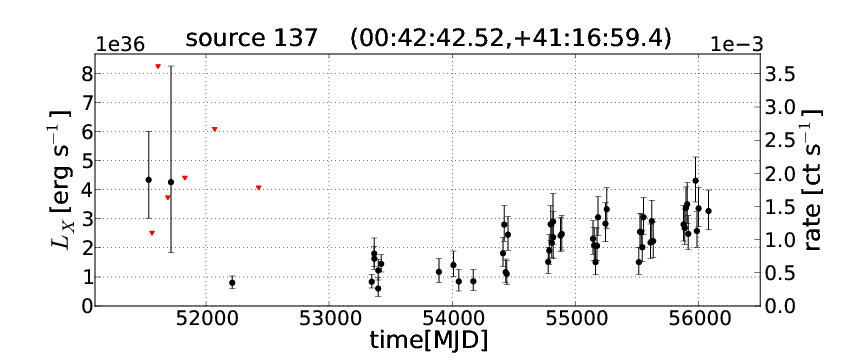}
\includegraphics[width=\linewidth]{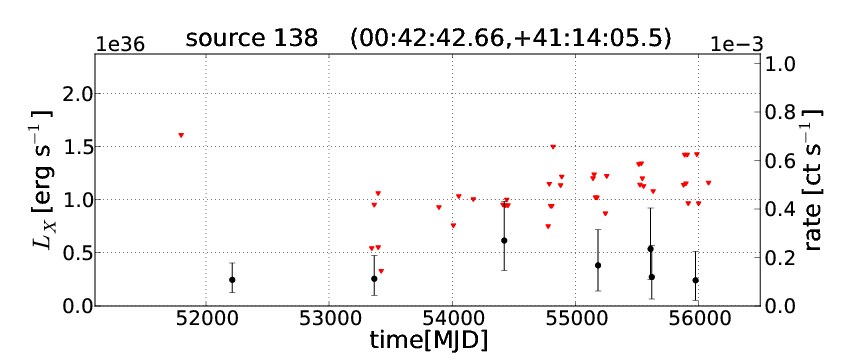}
\includegraphics[width=\linewidth]{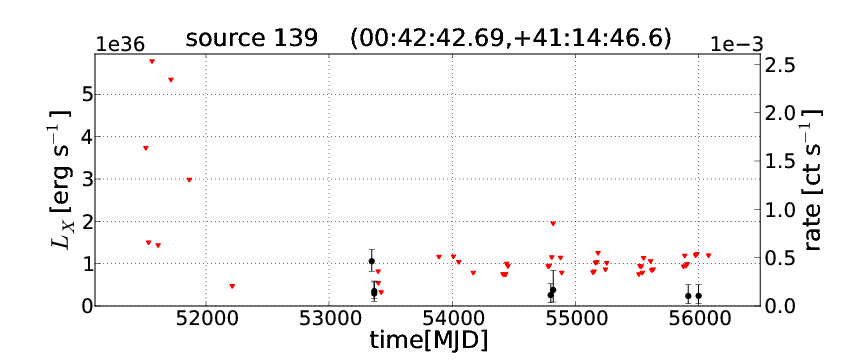}
\includegraphics[width=\linewidth]{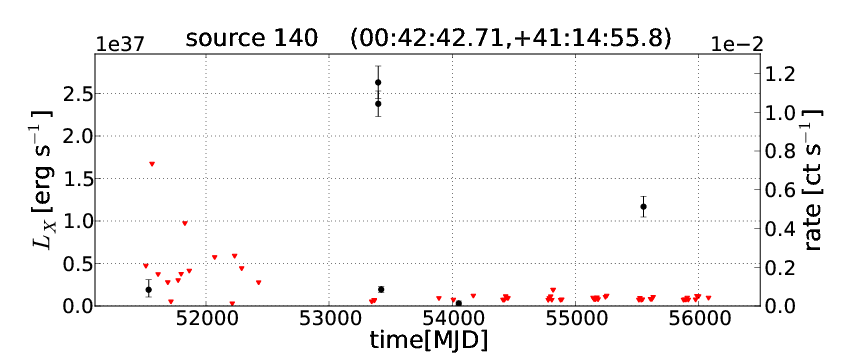}
\includegraphics[width=\linewidth]{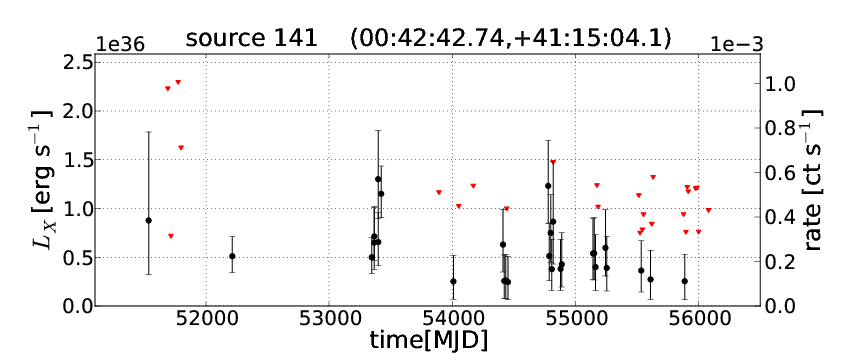}
\includegraphics[width=\linewidth]{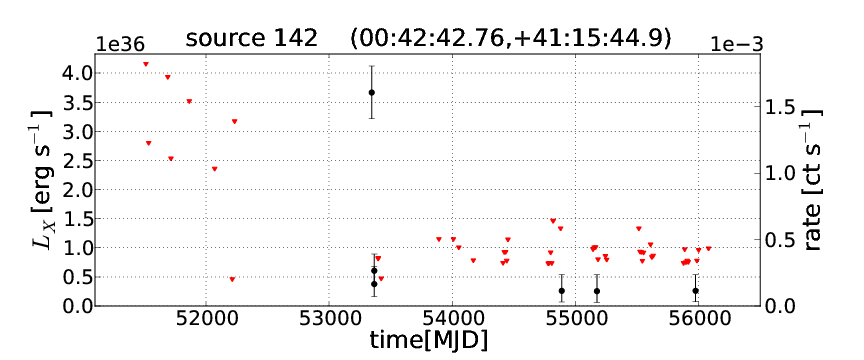}
\end{minipage}
\caption{continued.}
\label{fig:lc_all}
\end{figure*}

\addtocounter{figure}{-1} 

\begin{figure*}
\begin{minipage}{0.5\linewidth}
\includegraphics[width=\linewidth]{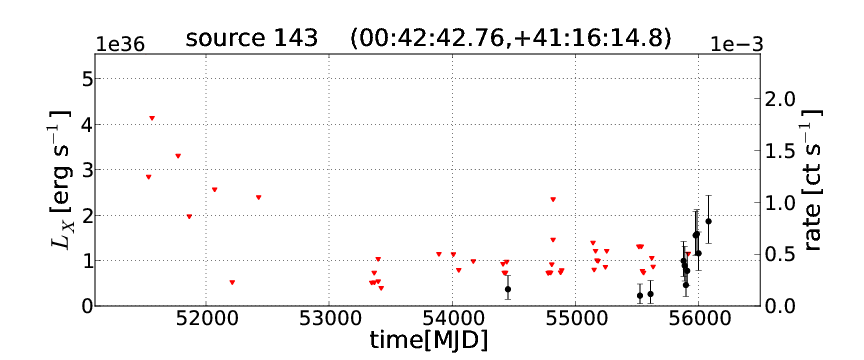}
\includegraphics[width=\linewidth]{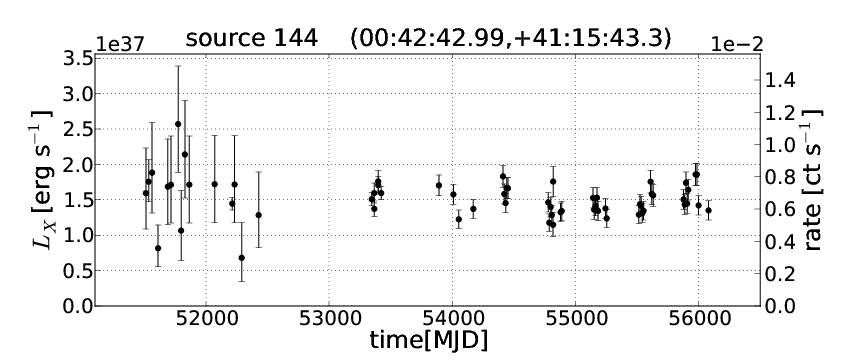}
\includegraphics[width=\linewidth]{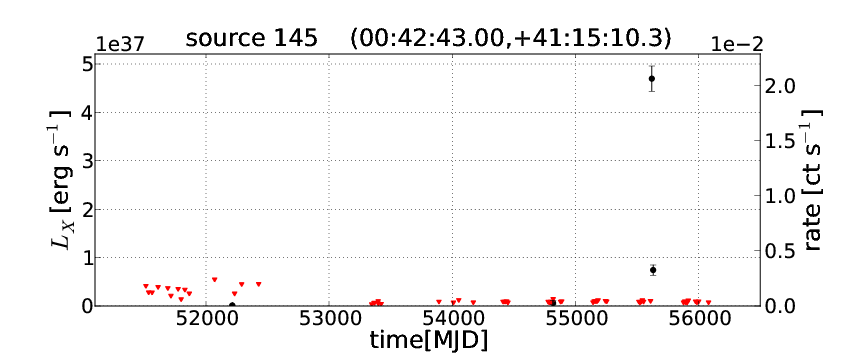}
\includegraphics[width=\linewidth]{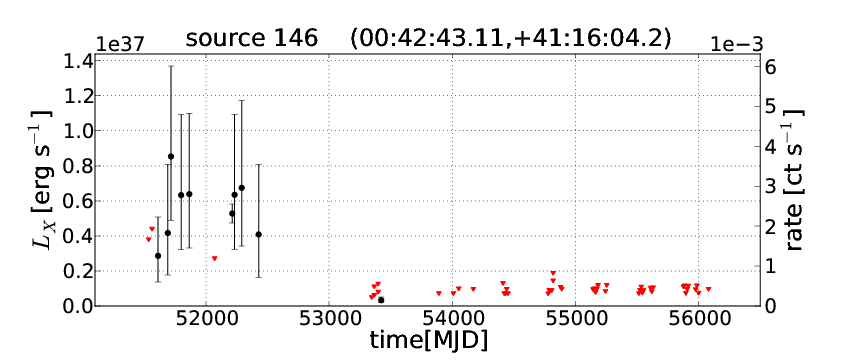}
\includegraphics[width=\linewidth]{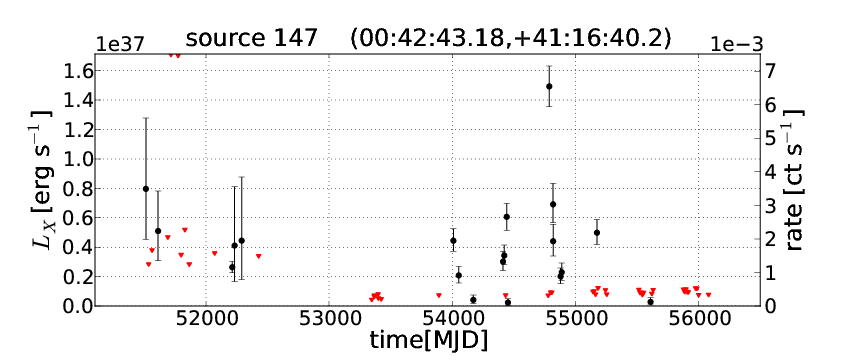}
\includegraphics[width=\linewidth]{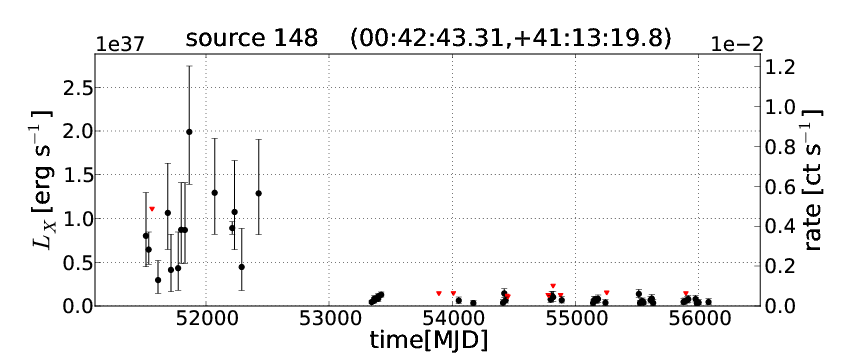}
\end{minipage}
\begin{minipage}{0.5\linewidth}
\includegraphics[width=\linewidth]{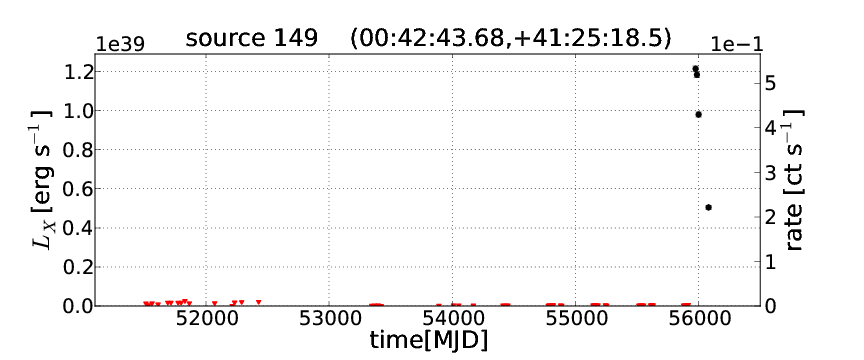}
\includegraphics[width=\linewidth]{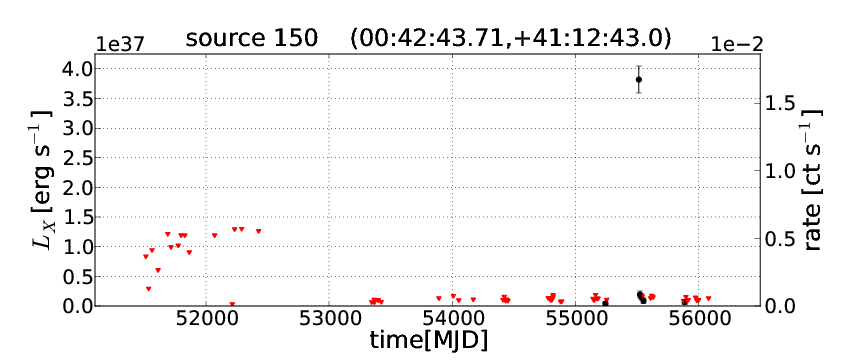}
\includegraphics[width=\linewidth]{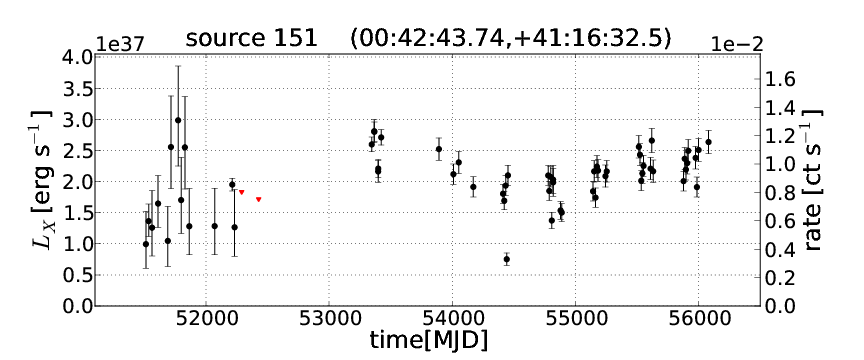}
\includegraphics[width=\linewidth]{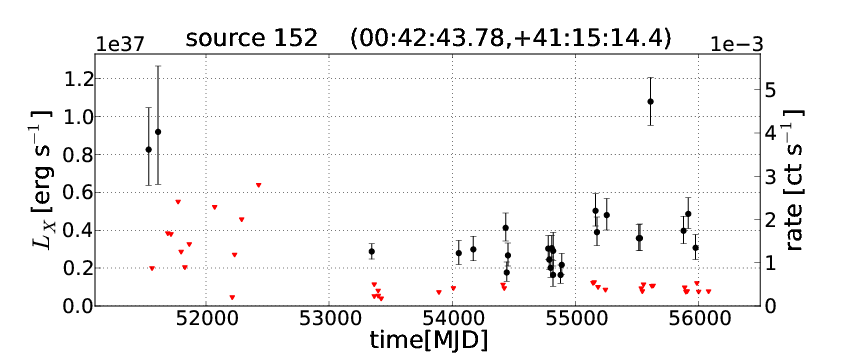}
\includegraphics[width=\linewidth]{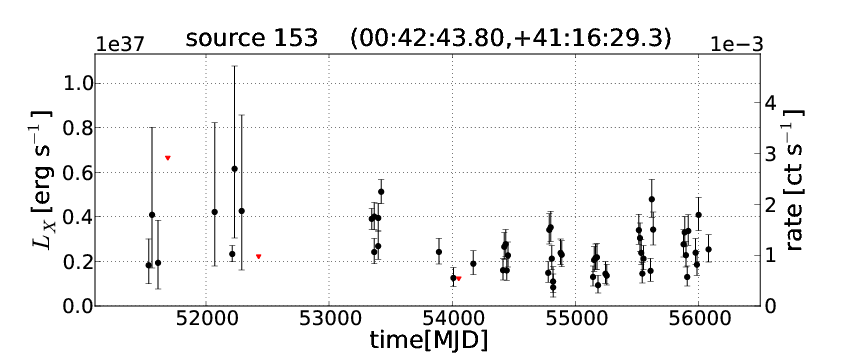}
\includegraphics[width=\linewidth]{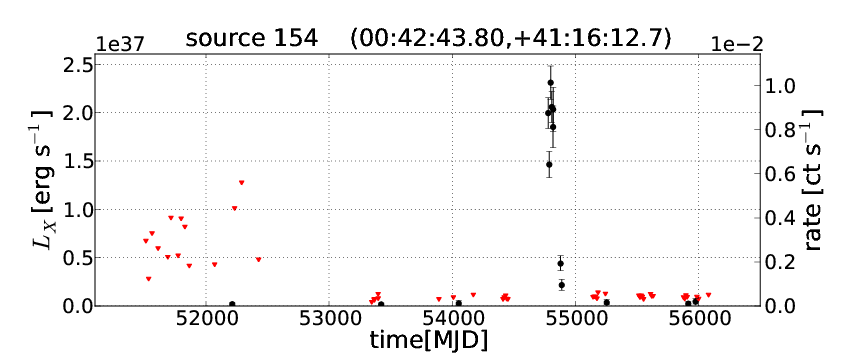}
\end{minipage}
\caption{continued.}
\label{fig:lc_all}
\end{figure*}

\addtocounter{figure}{-1} 

\begin{figure*}
\begin{minipage}{0.5\linewidth}
\includegraphics[width=\linewidth]{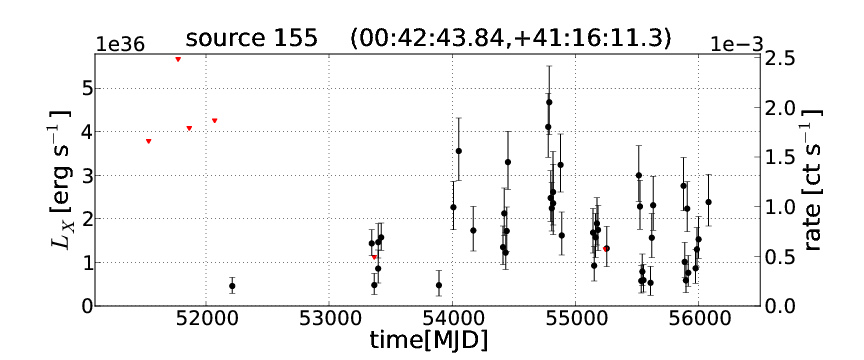}
\includegraphics[width=\linewidth]{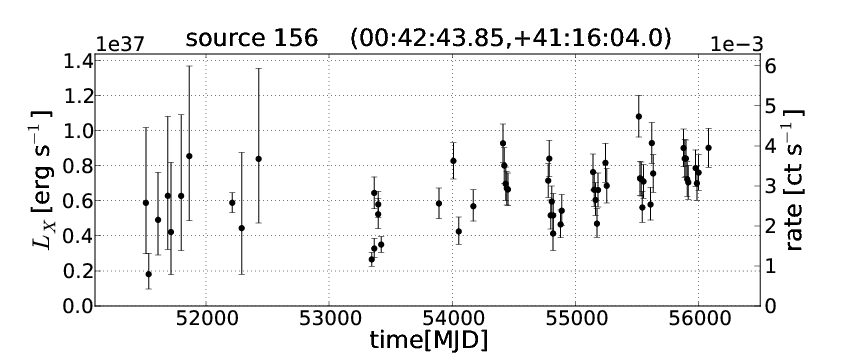}
\includegraphics[width=\linewidth]{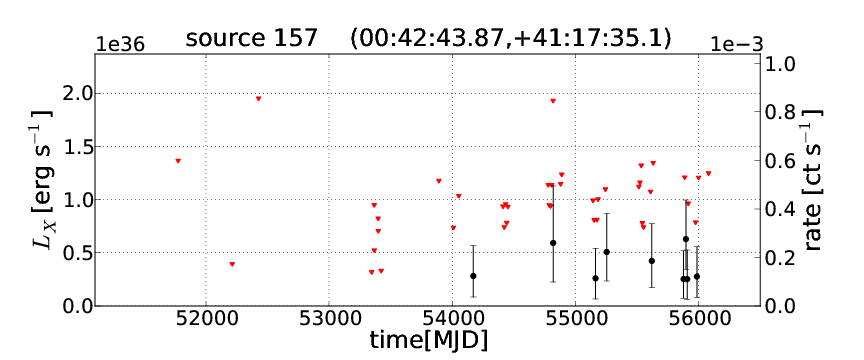}
\includegraphics[width=\linewidth]{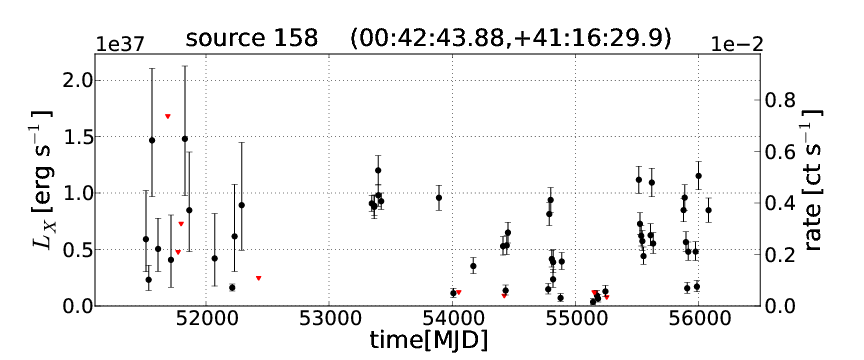}
\includegraphics[width=\linewidth]{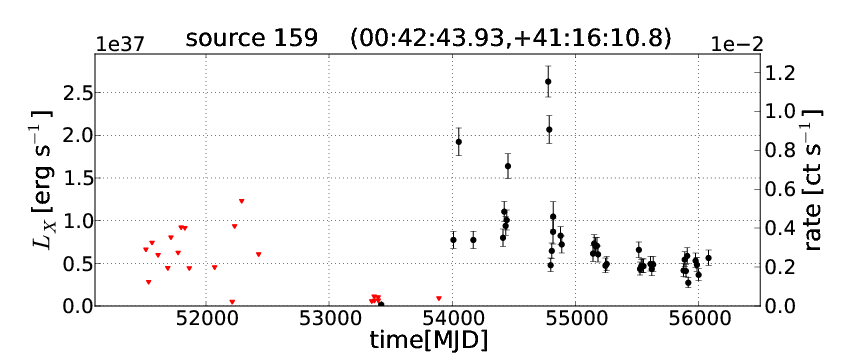}
\includegraphics[width=\linewidth]{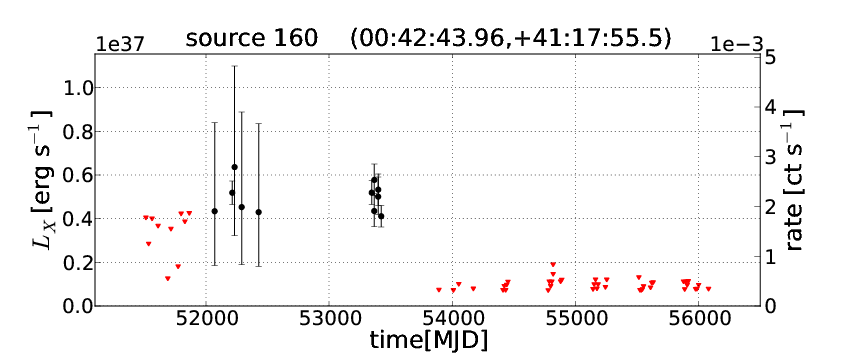}
\end{minipage}
\begin{minipage}{0.5\linewidth}
\includegraphics[width=\linewidth]{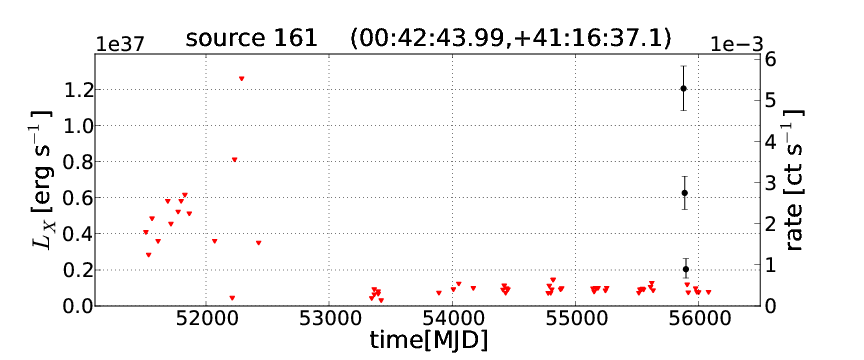}
\includegraphics[width=\linewidth]{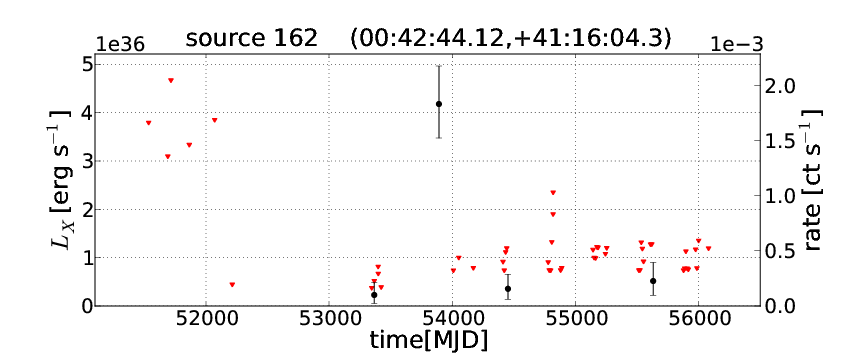}
\includegraphics[width=\linewidth]{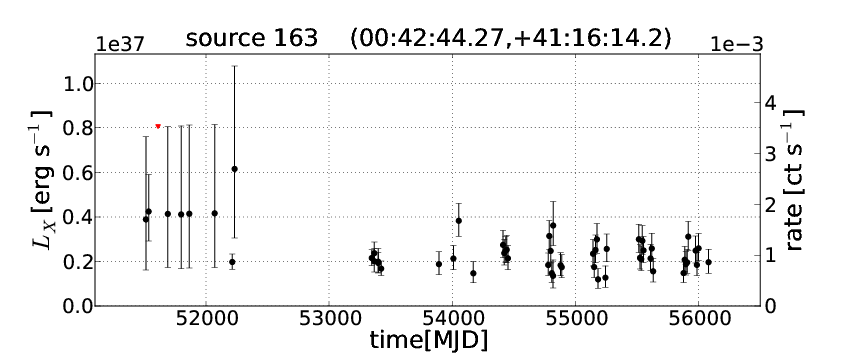}
\includegraphics[width=\linewidth]{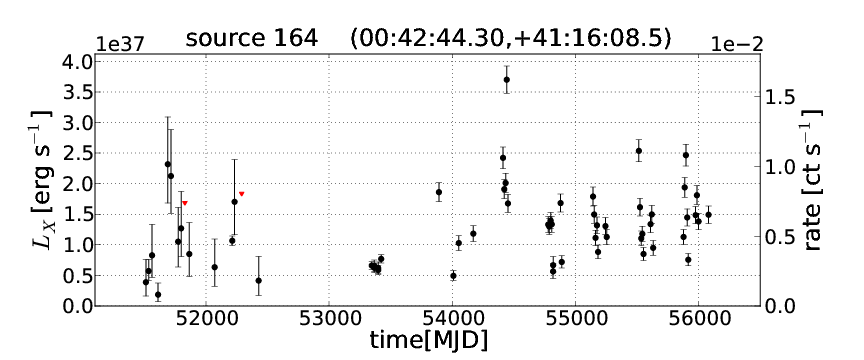}
\includegraphics[width=\linewidth]{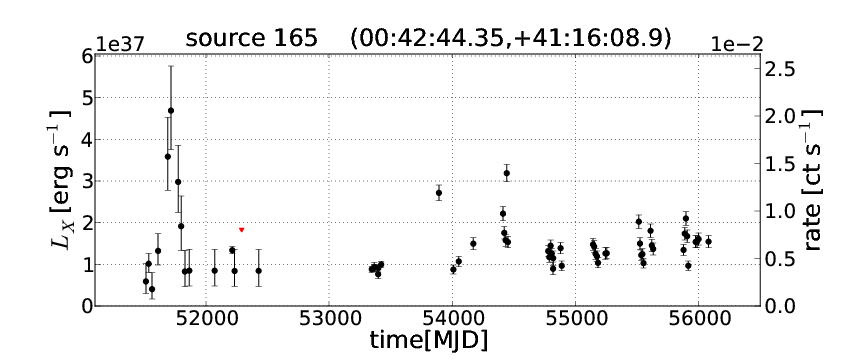}
\includegraphics[width=\linewidth]{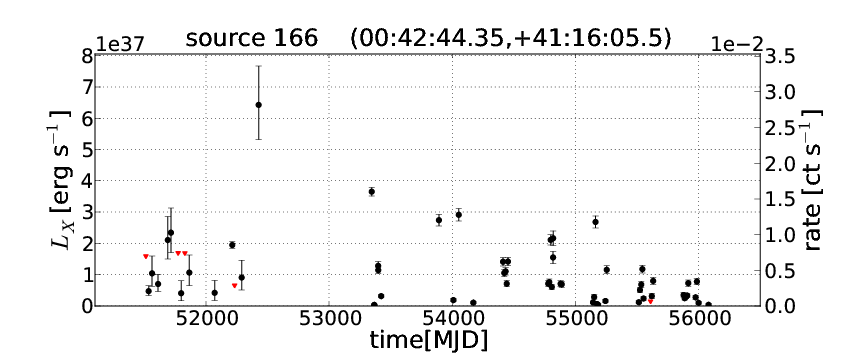}
\end{minipage}
\caption{continued.}
\label{fig:lc_all}
\end{figure*}

\addtocounter{figure}{-1} 

\begin{figure*}
\begin{minipage}{0.5\linewidth}
\includegraphics[width=\linewidth]{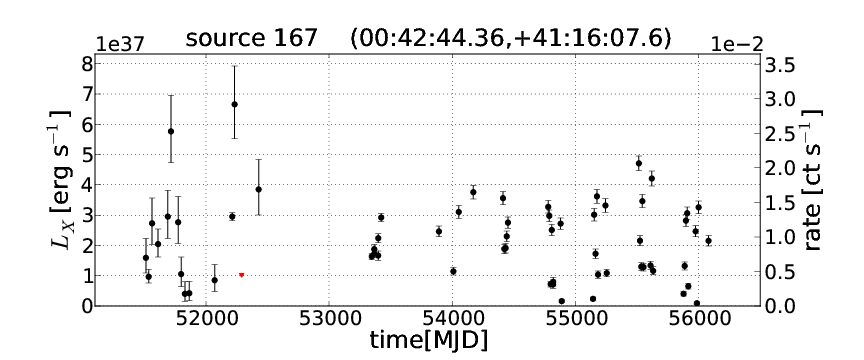}
\includegraphics[width=\linewidth]{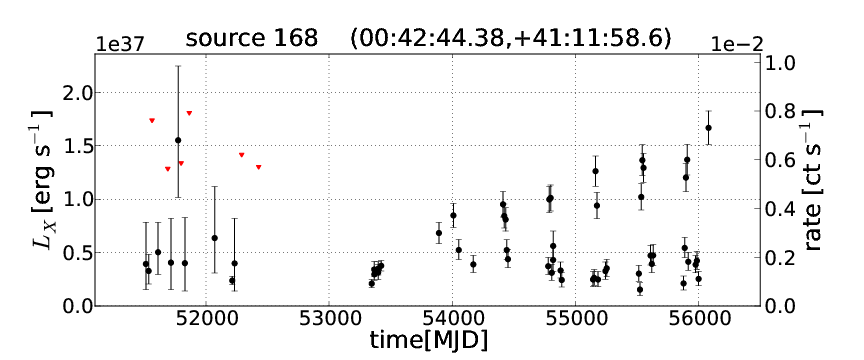}
\includegraphics[width=\linewidth]{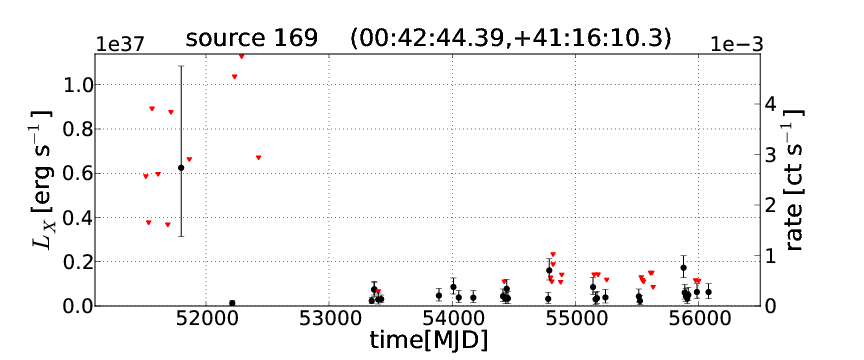}
\includegraphics[width=\linewidth]{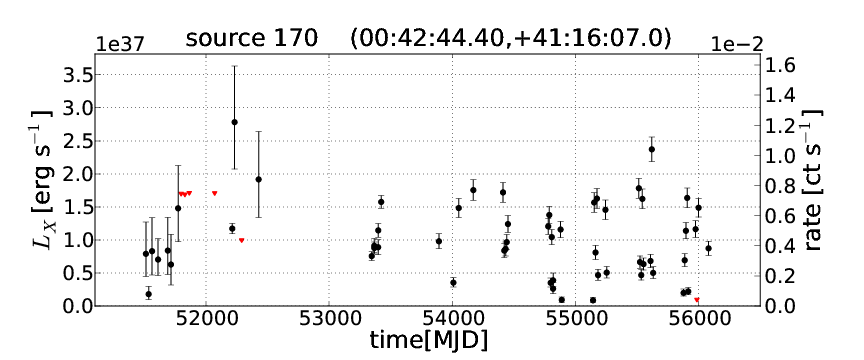}
\includegraphics[width=\linewidth]{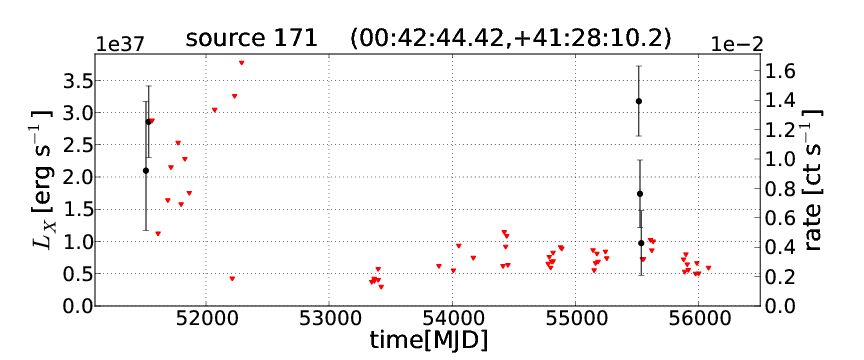}
\includegraphics[width=\linewidth]{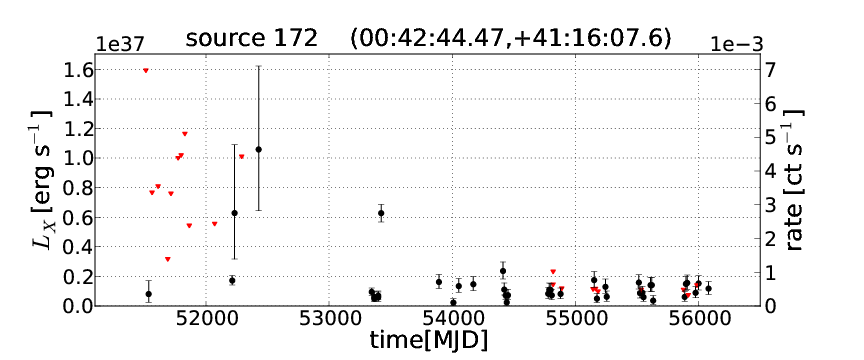}
\end{minipage}
\begin{minipage}{0.5\linewidth}
\includegraphics[width=\linewidth]{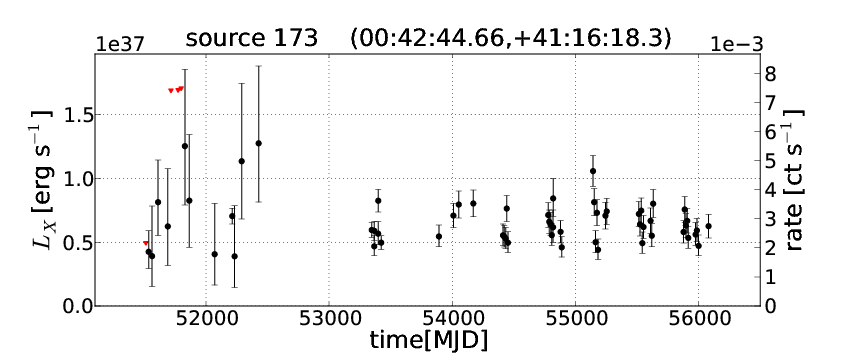}
\includegraphics[width=\linewidth]{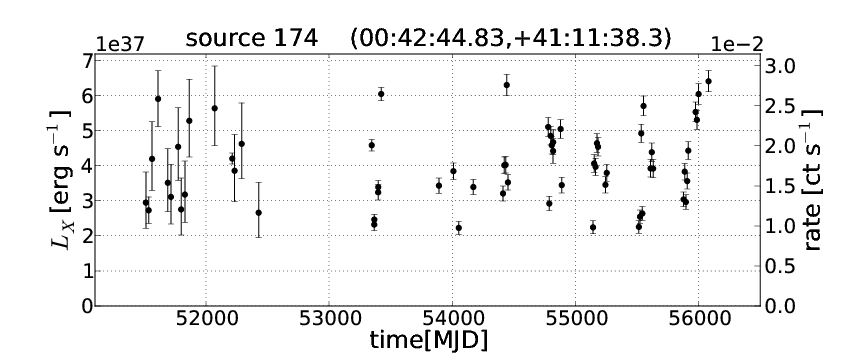}
\includegraphics[width=\linewidth]{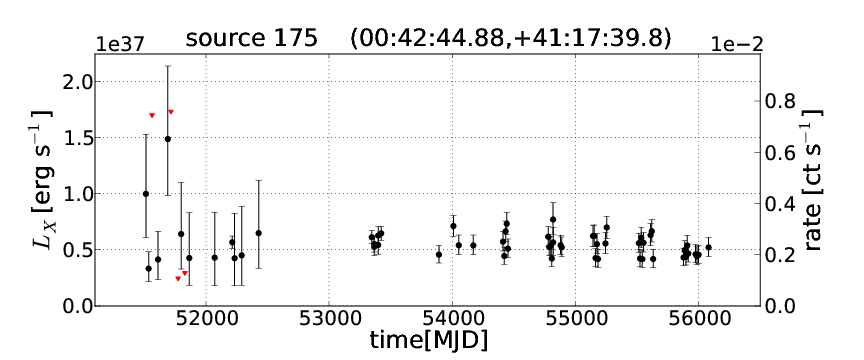}
\includegraphics[width=\linewidth]{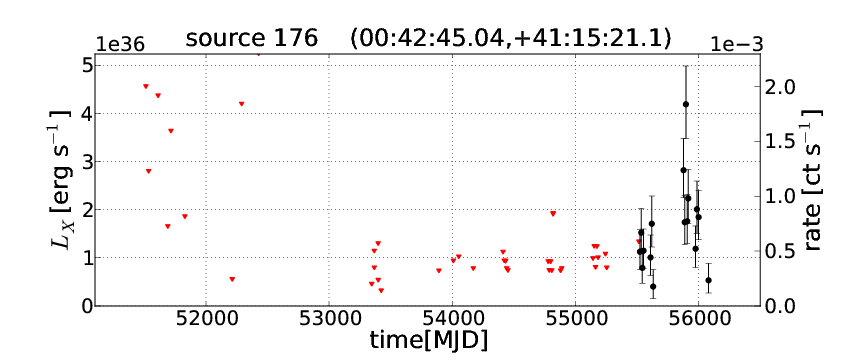}
\includegraphics[width=\linewidth]{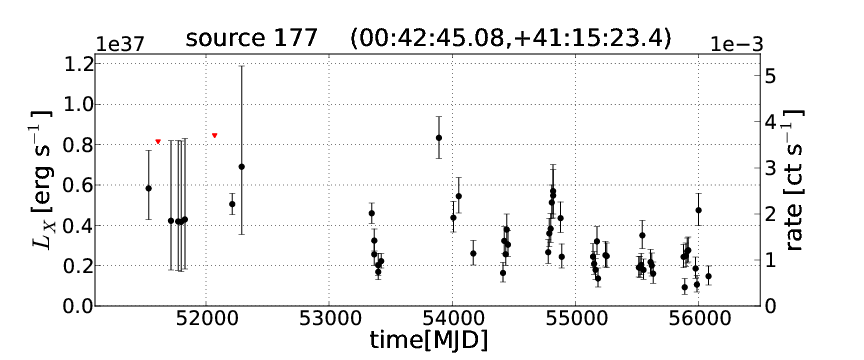}
\includegraphics[width=\linewidth]{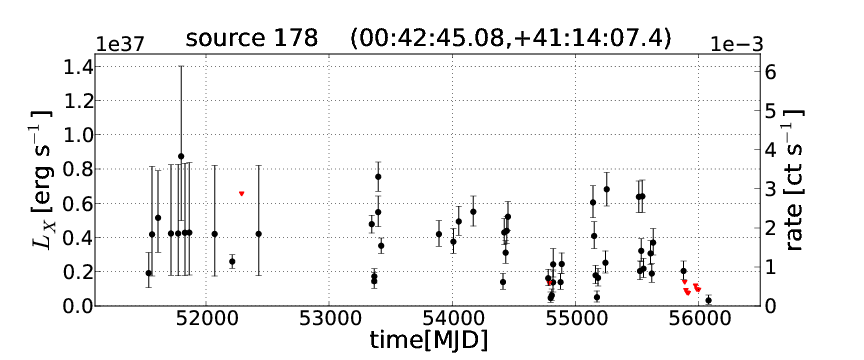}
\end{minipage}
\caption{continued.}
\label{fig:lc_all}
\end{figure*}

\addtocounter{figure}{-1}

\begin{figure*}
\begin{minipage}{0.5\linewidth}
\includegraphics[width=\linewidth]{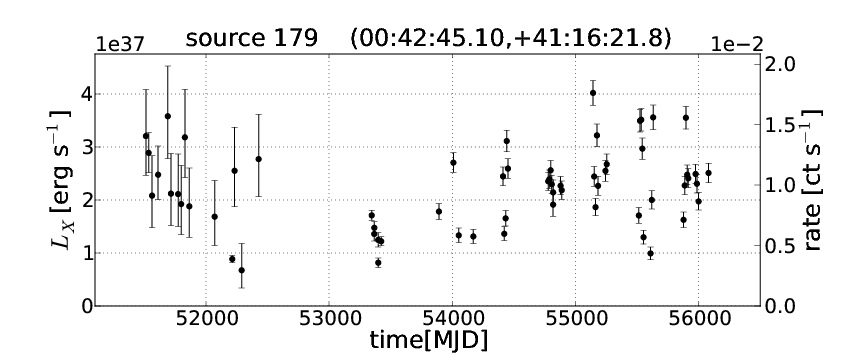}
\includegraphics[width=\linewidth]{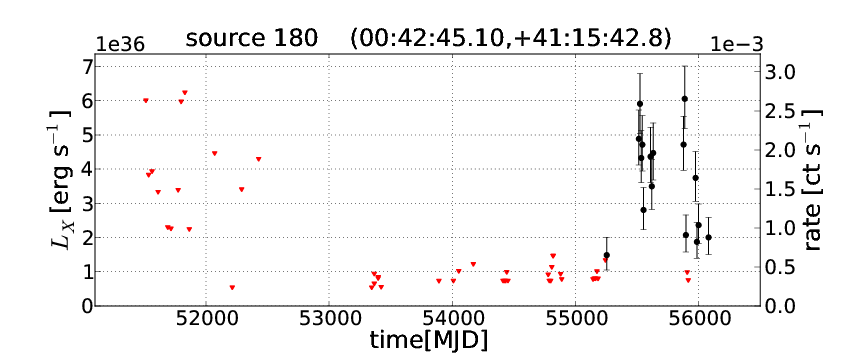}
\includegraphics[width=\linewidth]{plots_new/181_light.png}
\includegraphics[width=\linewidth]{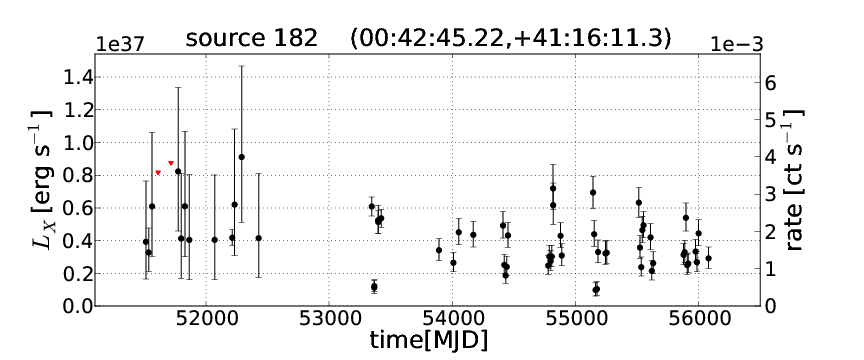}
\includegraphics[width=\linewidth]{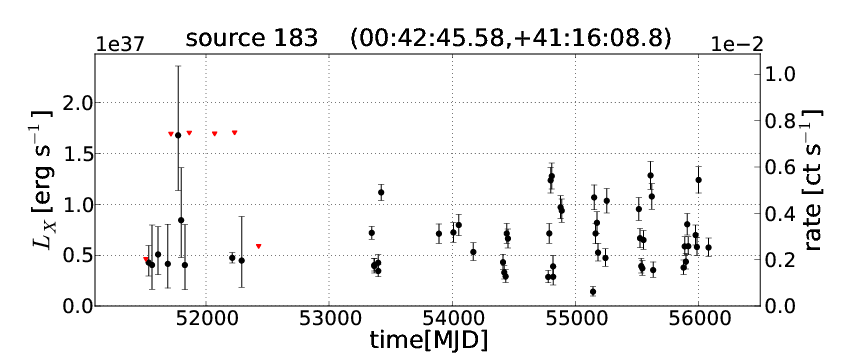}
\includegraphics[width=\linewidth]{plots_new/184_light.png}
\end{minipage}
\begin{minipage}{0.5\linewidth}
\includegraphics[width=\linewidth]{plots_new/185_light.png}
\includegraphics[width=\linewidth]{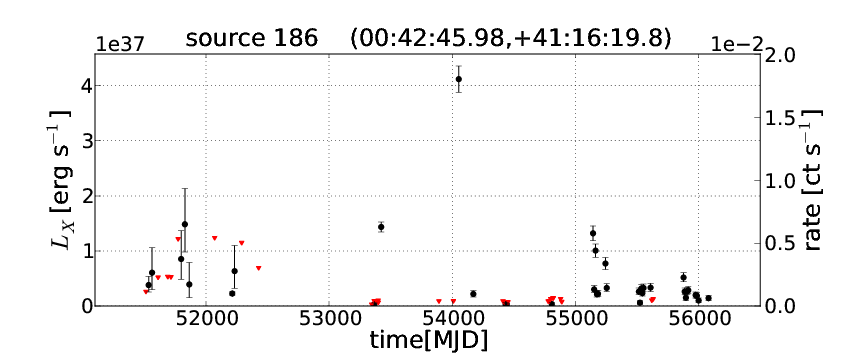}
\includegraphics[width=\linewidth]{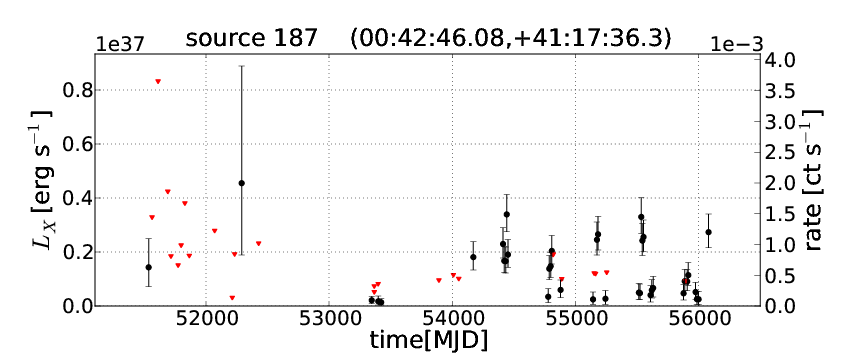}
\includegraphics[width=\linewidth]{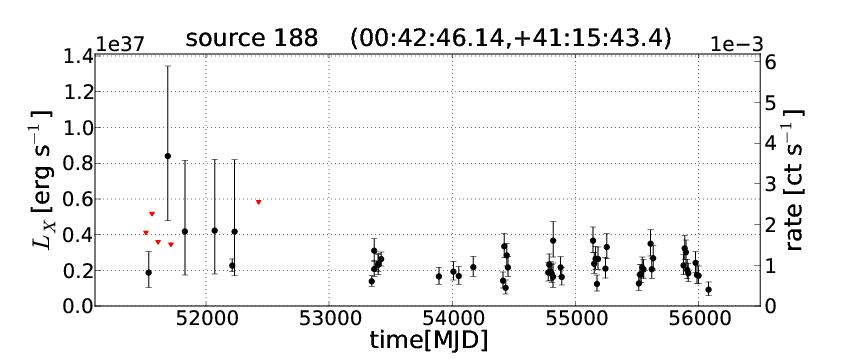}
\includegraphics[width=\linewidth]{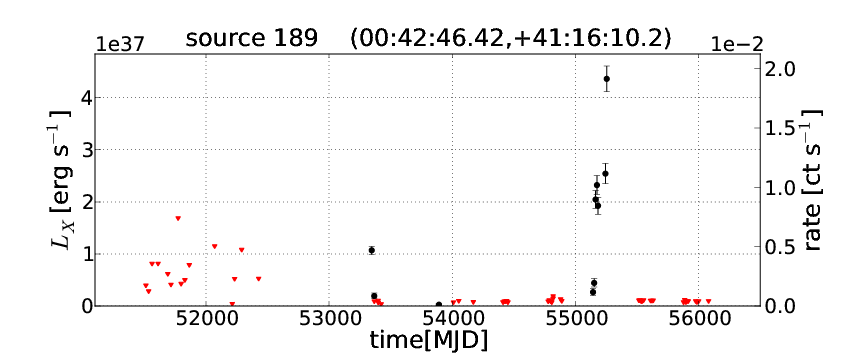}
\includegraphics[width=\linewidth]{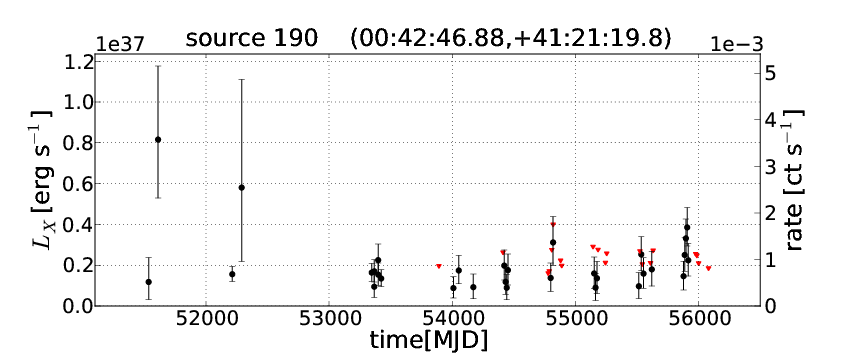}
\end{minipage}
\caption{continued.}
\label{fig:lc_all}
\end{figure*}

\addtocounter{figure}{-1} 

\begin{figure*}
\begin{minipage}{0.5\linewidth}
\includegraphics[width=\linewidth]{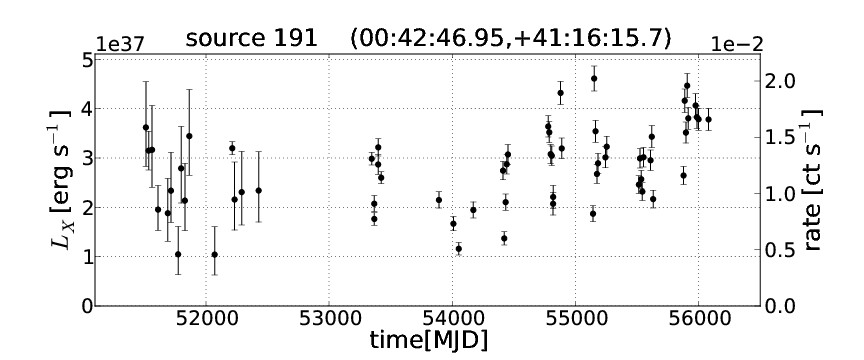}
\includegraphics[width=\linewidth]{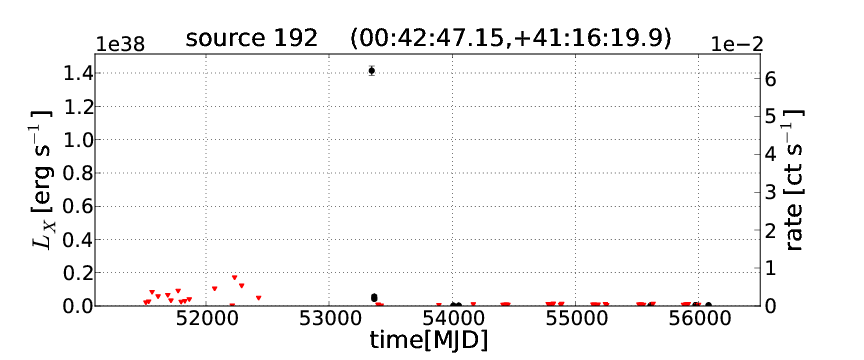}
\includegraphics[width=\linewidth]{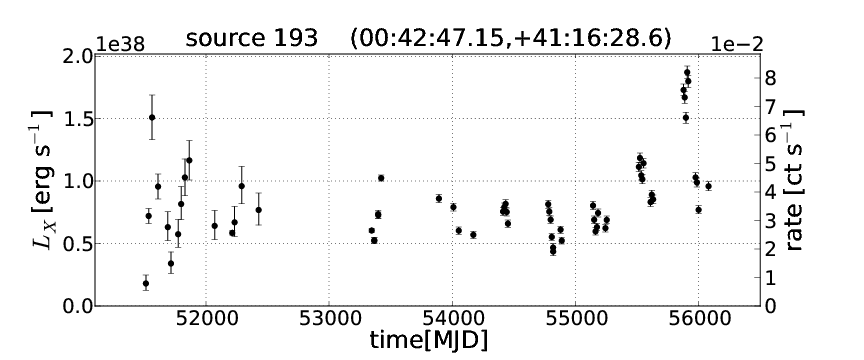}
\includegraphics[width=\linewidth]{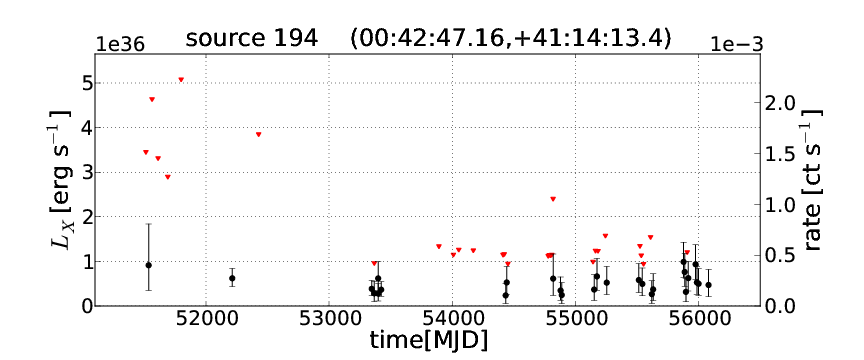}
\includegraphics[width=\linewidth]{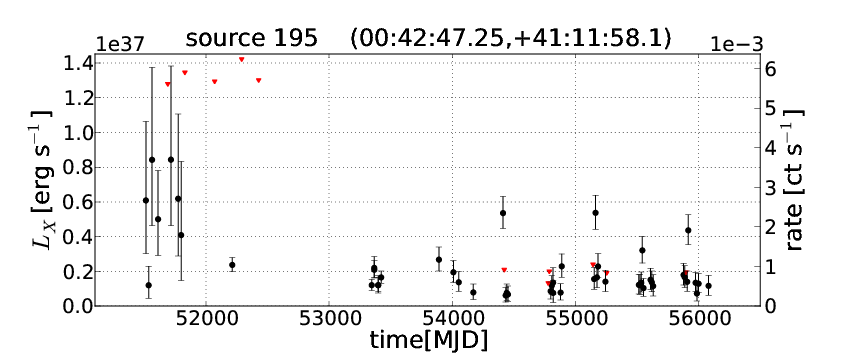}
\includegraphics[width=\linewidth]{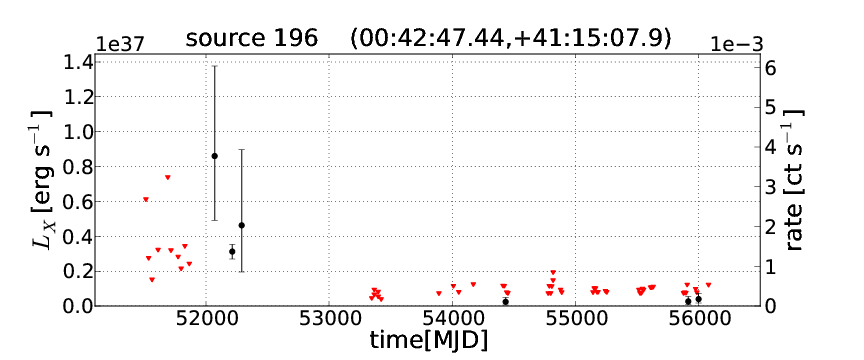}
\end{minipage}
\begin{minipage}{0.5\linewidth}
\includegraphics[width=\linewidth]{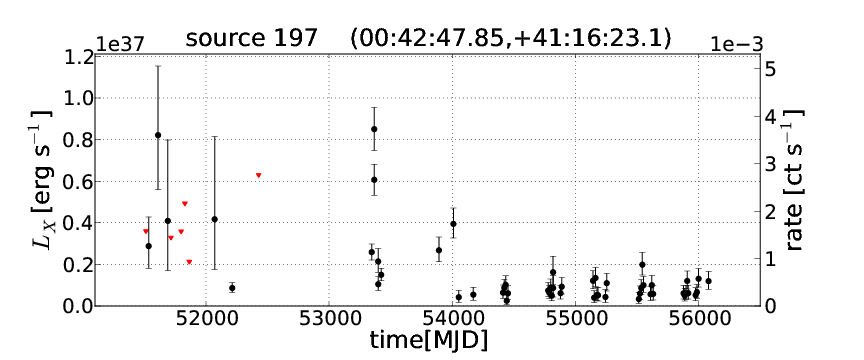}
\includegraphics[width=\linewidth]{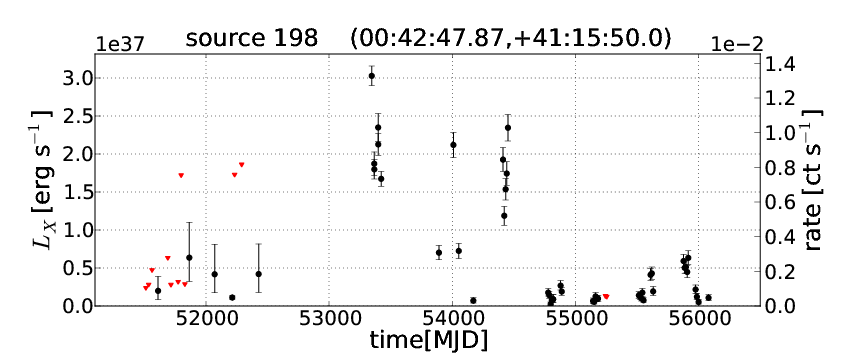}
\includegraphics[width=\linewidth]{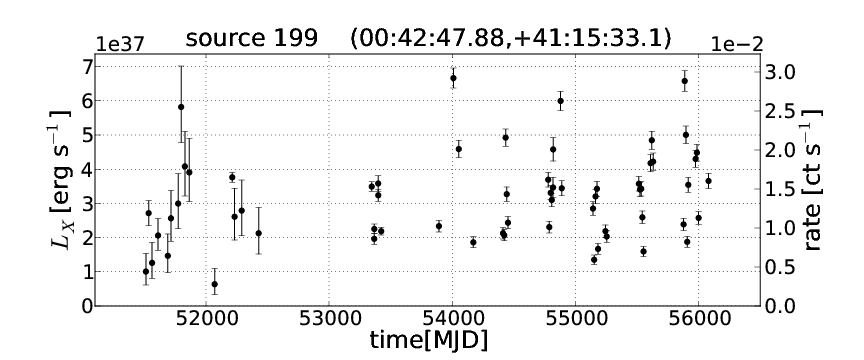}
\includegraphics[width=\linewidth]{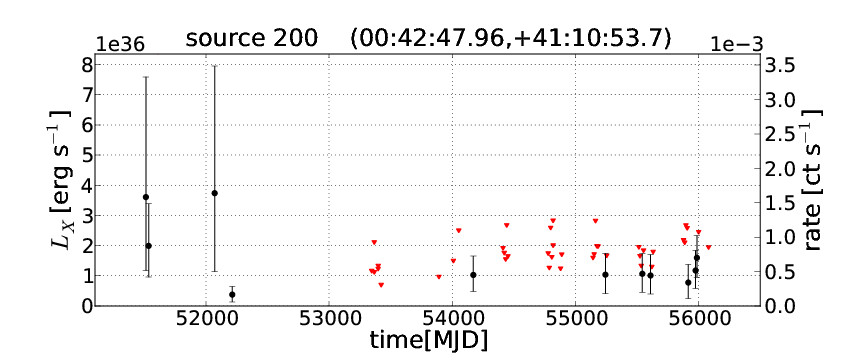}
\includegraphics[width=\linewidth]{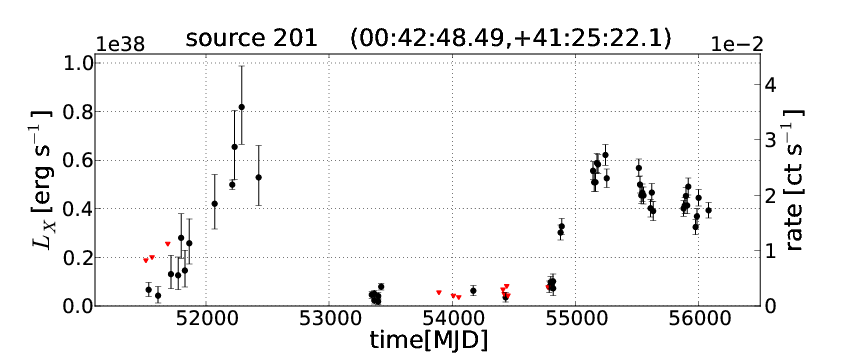}
\includegraphics[width=\linewidth]{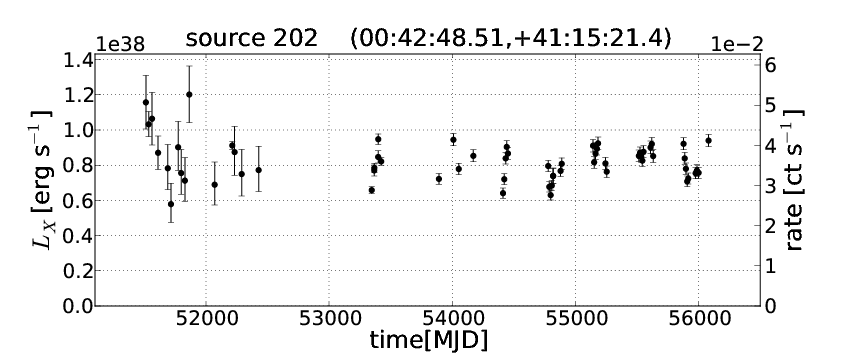}
\end{minipage}
\caption{continued.}
\label{fig:lc_all}
\end{figure*}

\addtocounter{figure}{-1} 

\begin{figure*}
\begin{minipage}{0.5\linewidth}
\includegraphics[width=\linewidth]{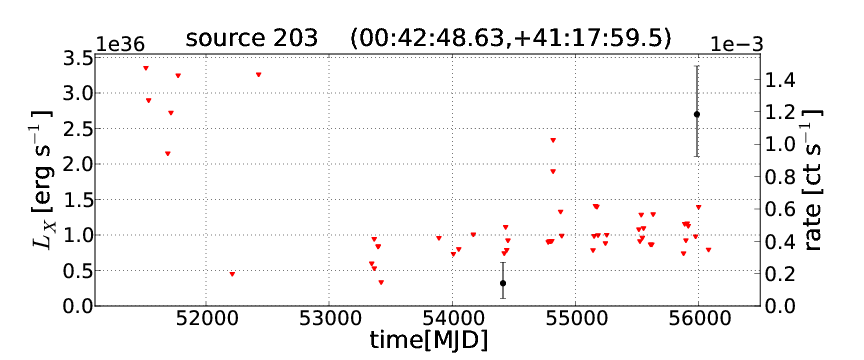}
\includegraphics[width=\linewidth]{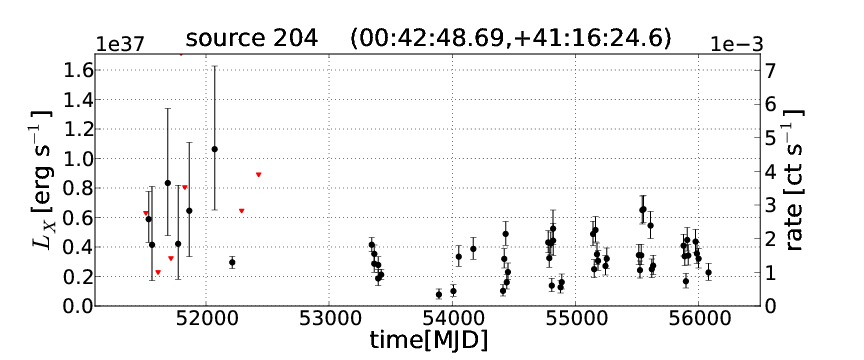}
\includegraphics[width=\linewidth]{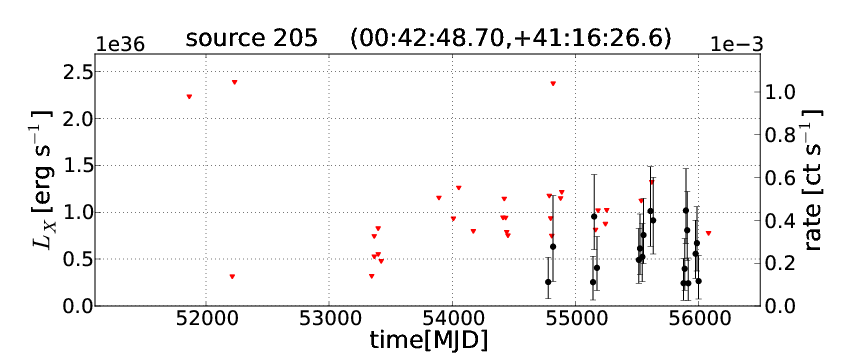}
\includegraphics[width=\linewidth]{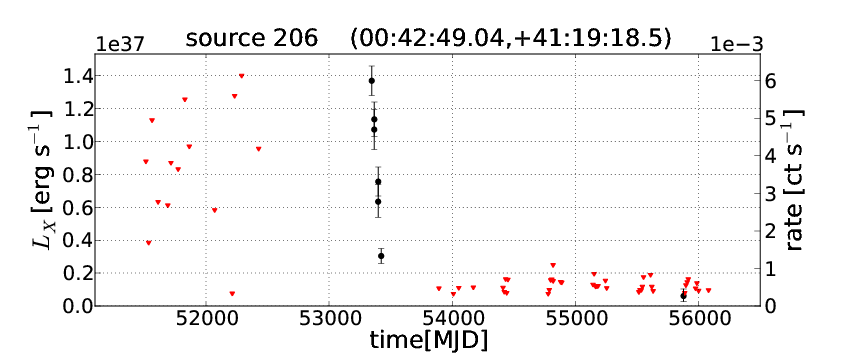}
\includegraphics[width=\linewidth]{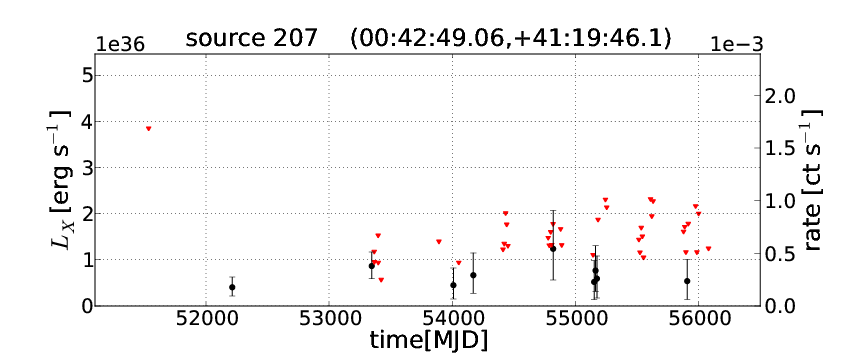}
\includegraphics[width=\linewidth]{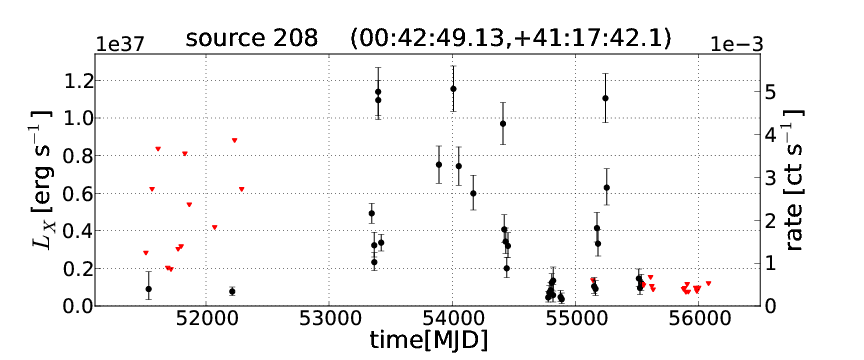}
\end{minipage}
\begin{minipage}{0.5\linewidth}
\includegraphics[width=\linewidth]{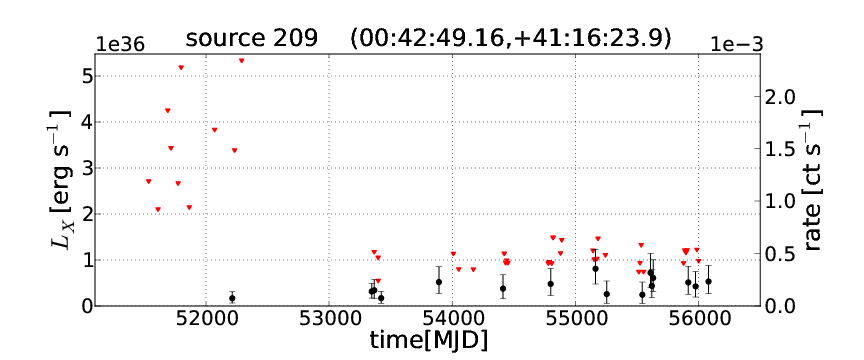}
\includegraphics[width=\linewidth]{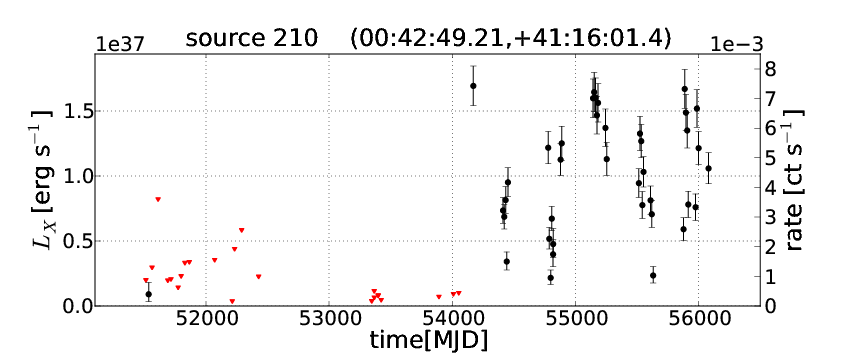}
\includegraphics[width=\linewidth]{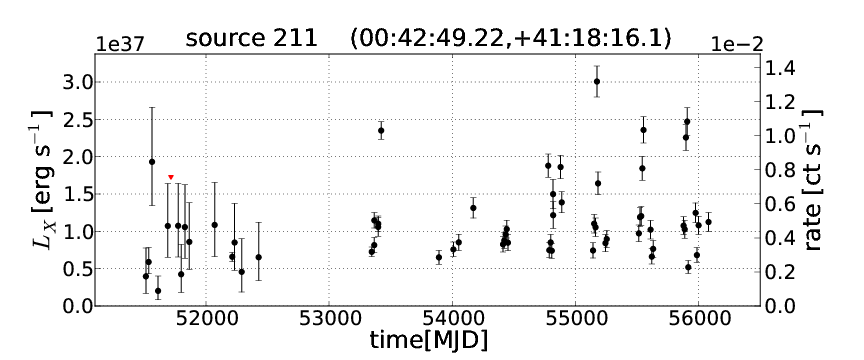}
\includegraphics[width=\linewidth]{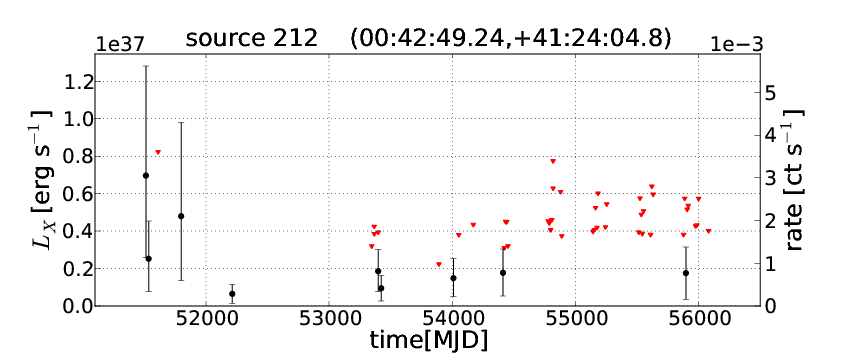}
\includegraphics[width=\linewidth]{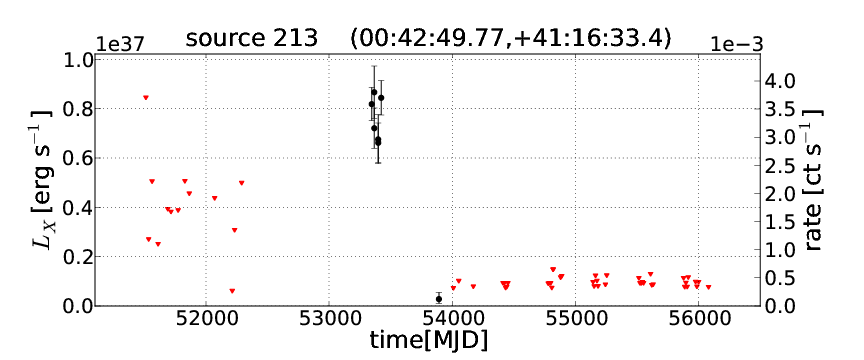}
\includegraphics[width=\linewidth]{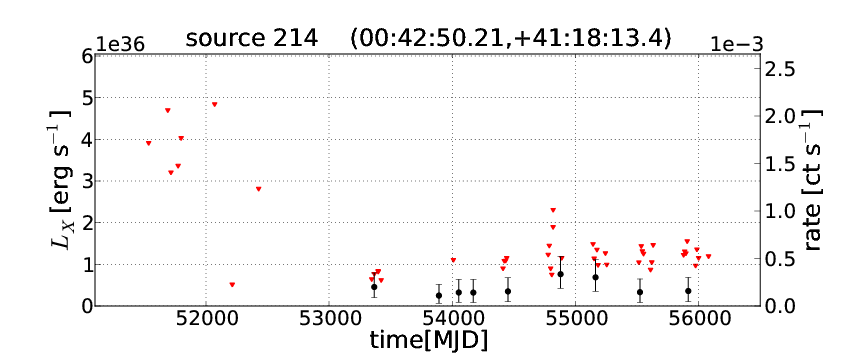}
\end{minipage}
\caption{continued.}
\label{fig:lc_all}
\end{figure*}

\addtocounter{figure}{-1} 

\begin{figure*}
\begin{minipage}{0.5\linewidth}
\includegraphics[width=\linewidth]{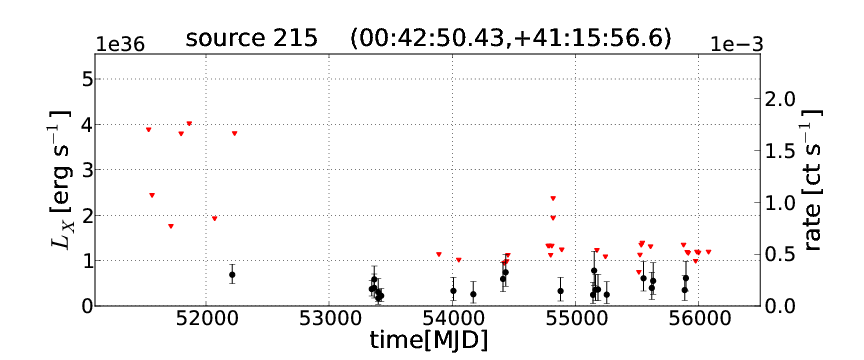}
\includegraphics[width=\linewidth]{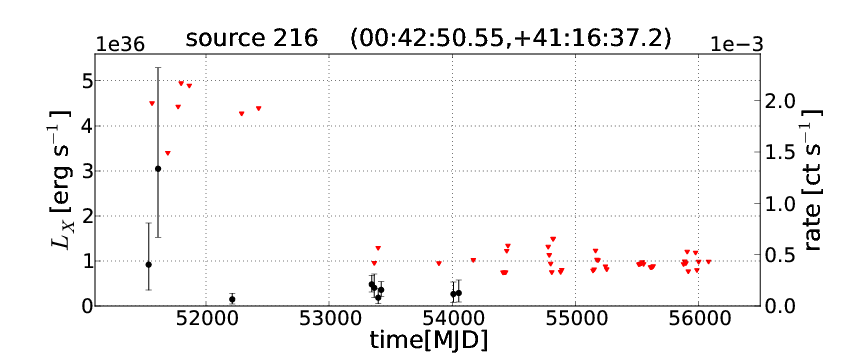}
\includegraphics[width=\linewidth]{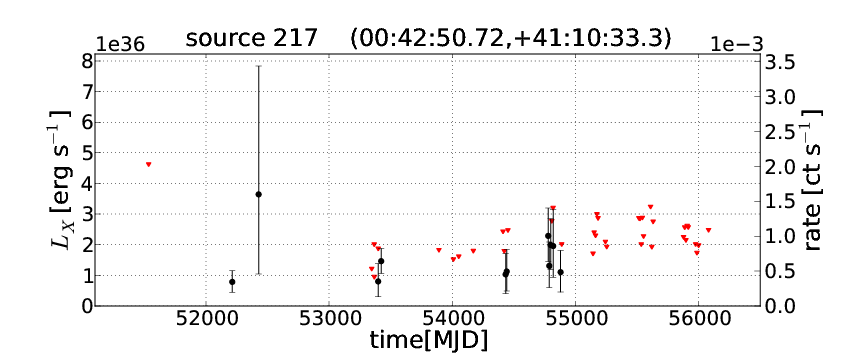}
\includegraphics[width=\linewidth]{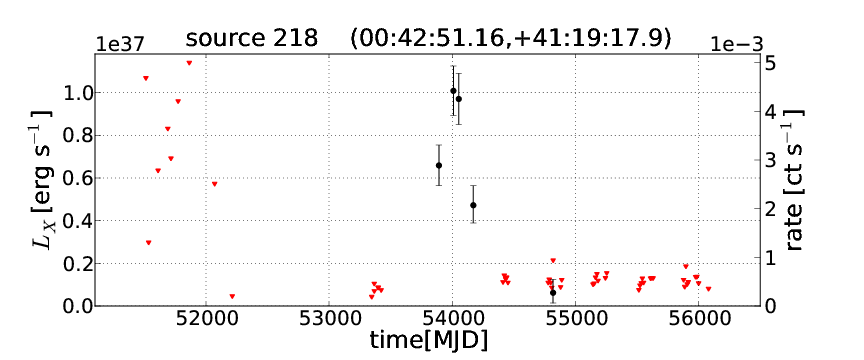}
\includegraphics[width=\linewidth]{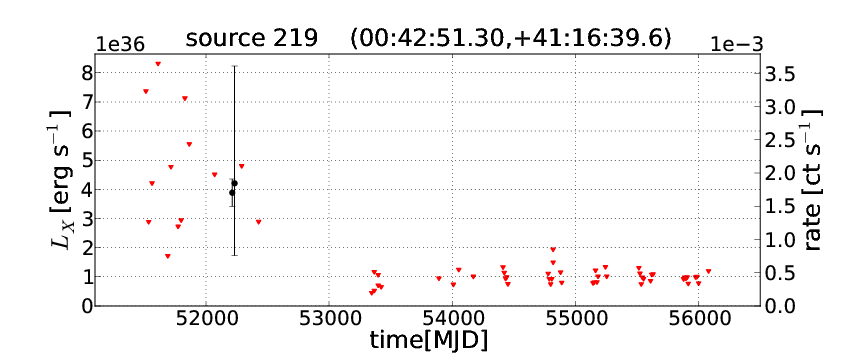}
\includegraphics[width=\linewidth]{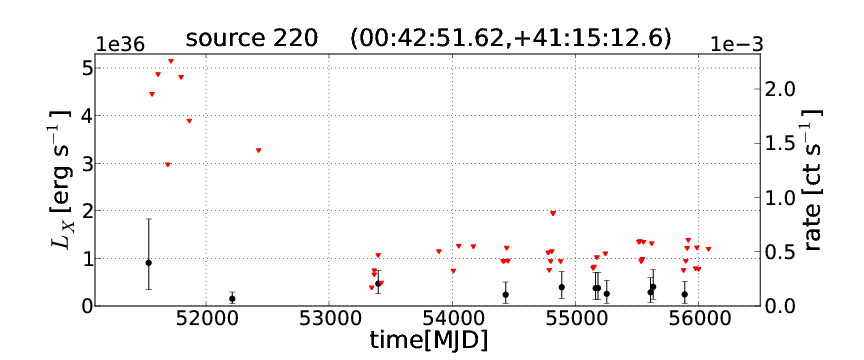}
\end{minipage}
\begin{minipage}{0.5\linewidth}
\includegraphics[width=\linewidth]{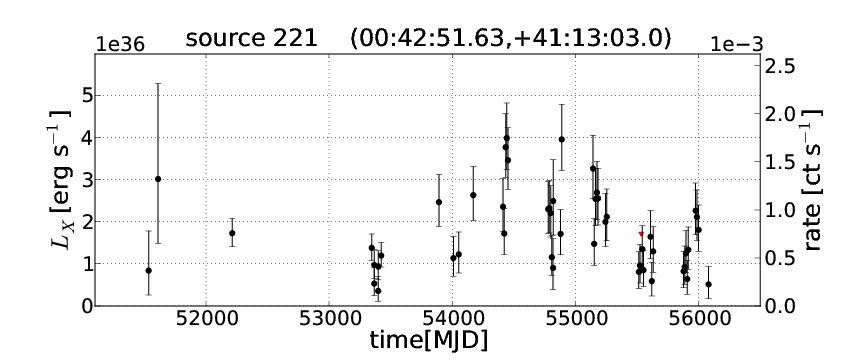}
\includegraphics[width=\linewidth]{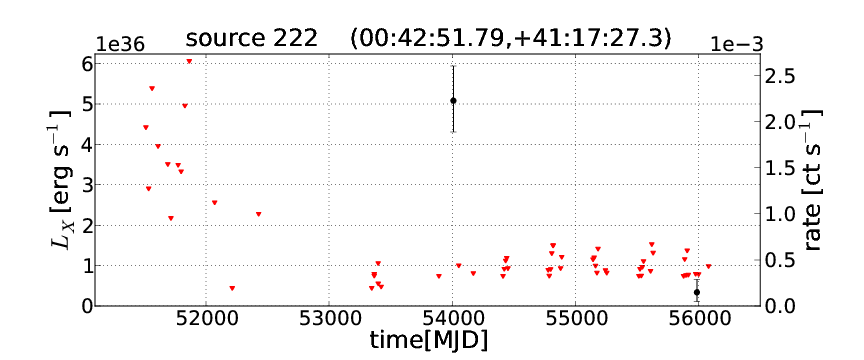}
\includegraphics[width=\linewidth]{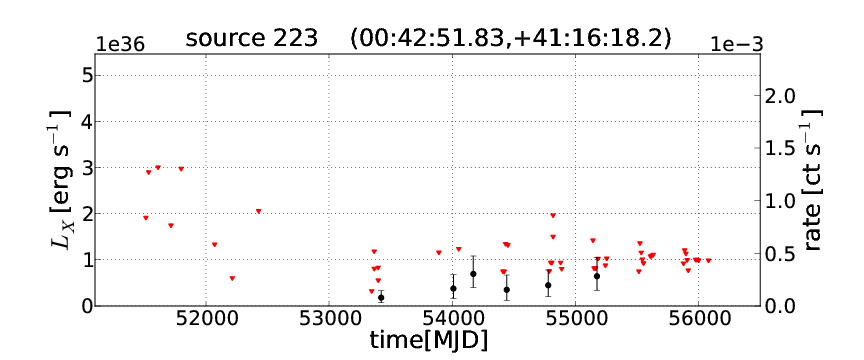}
\includegraphics[width=\linewidth]{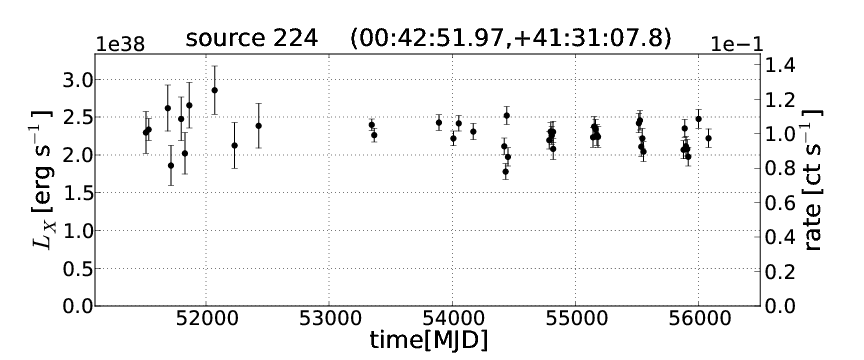}
\includegraphics[width=\linewidth]{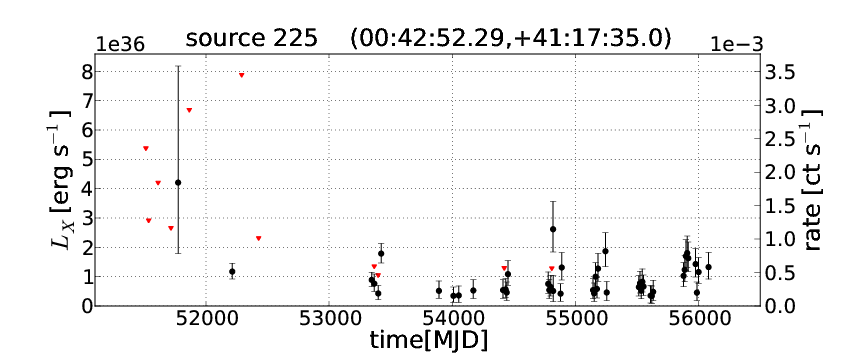}
\includegraphics[width=\linewidth]{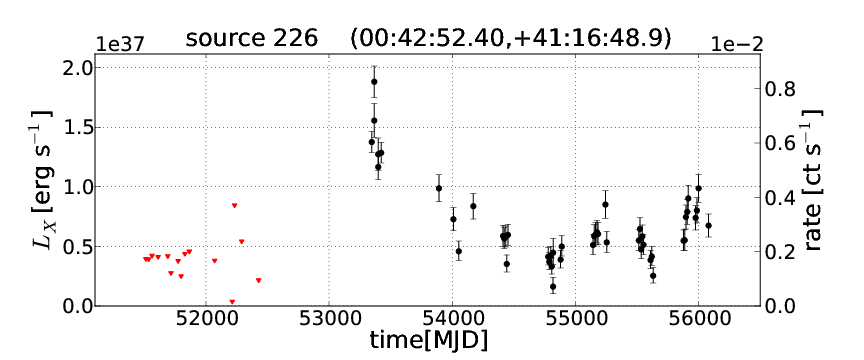}
\end{minipage}
\caption{continued.}
\label{fig:lc_all}
\end{figure*}

\addtocounter{figure}{-1} 

\FloatBarrier

\begin{figure*}
\begin{minipage}{0.5\linewidth}
\includegraphics[width=\linewidth]{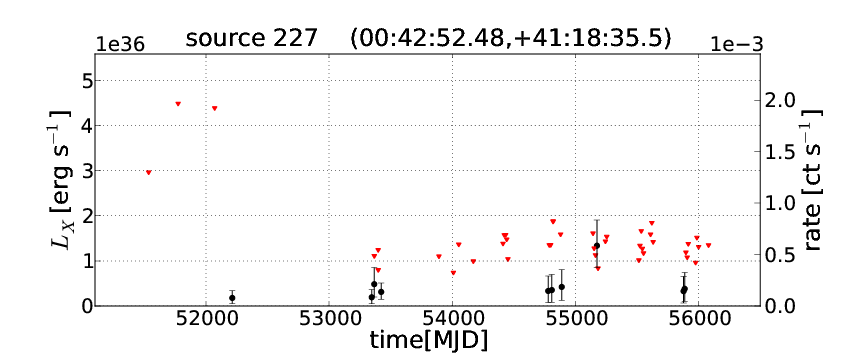}
\includegraphics[width=\linewidth]{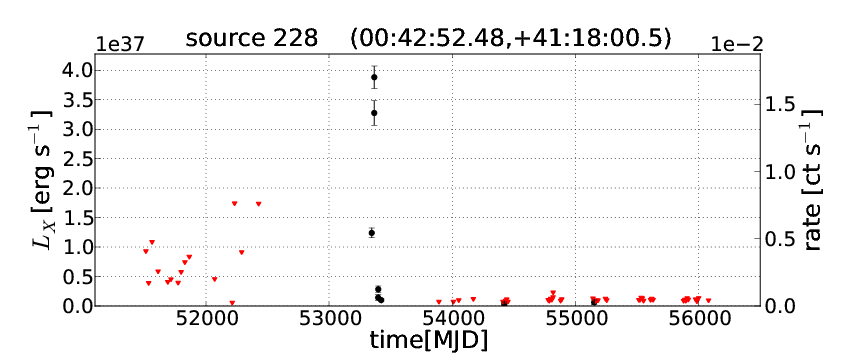}
\includegraphics[width=\linewidth]{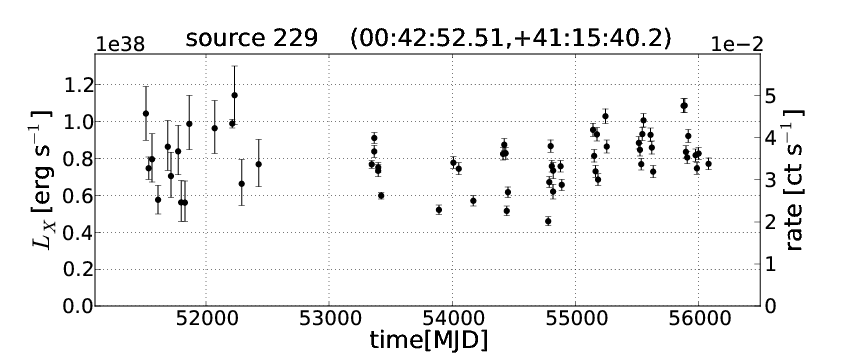}
\includegraphics[width=\linewidth]{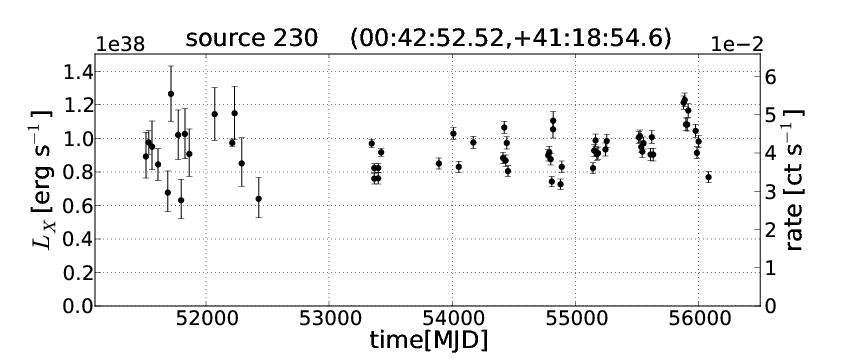}
\includegraphics[width=\linewidth]{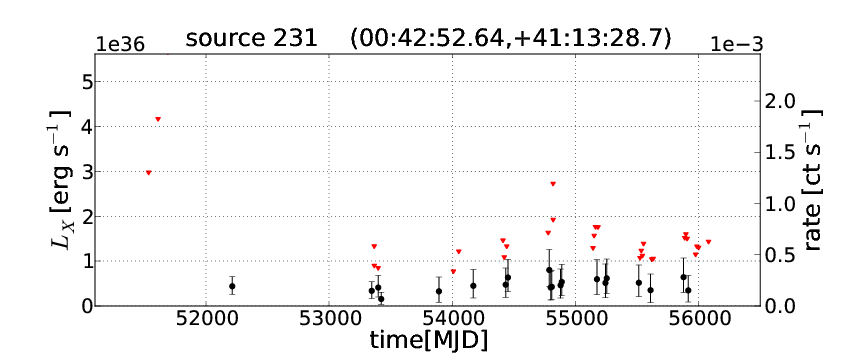}
\includegraphics[width=\linewidth]{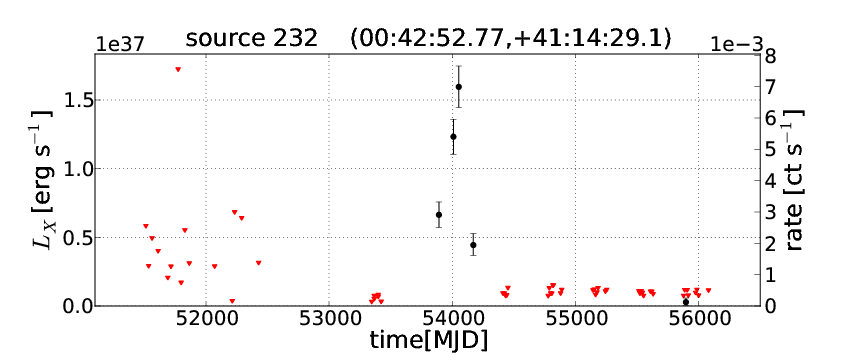}
\end{minipage}
\begin{minipage}{0.5\linewidth}
\includegraphics[width=\linewidth]{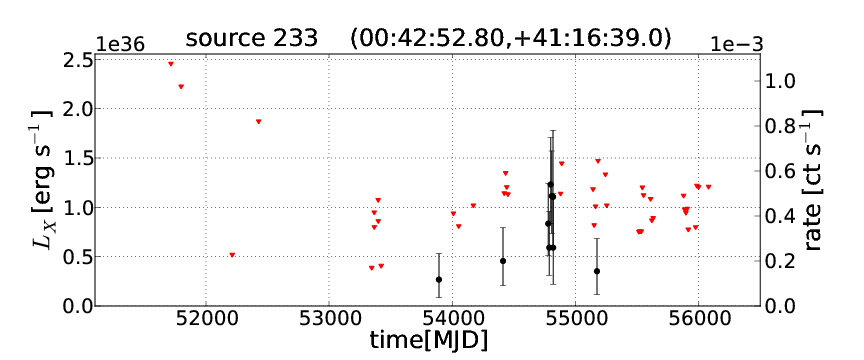}
\includegraphics[width=\linewidth]{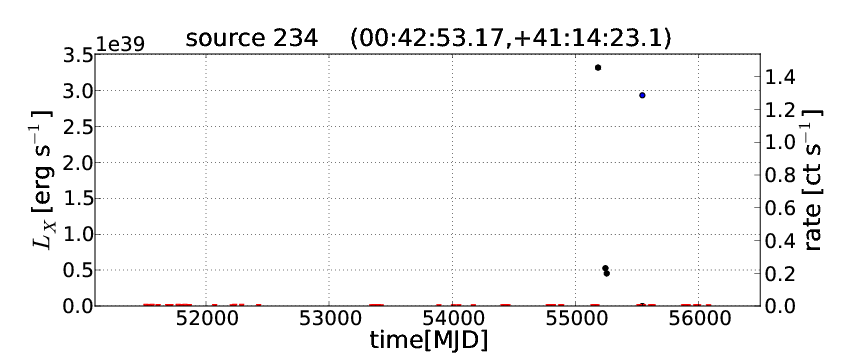}
\includegraphics[width=\linewidth]{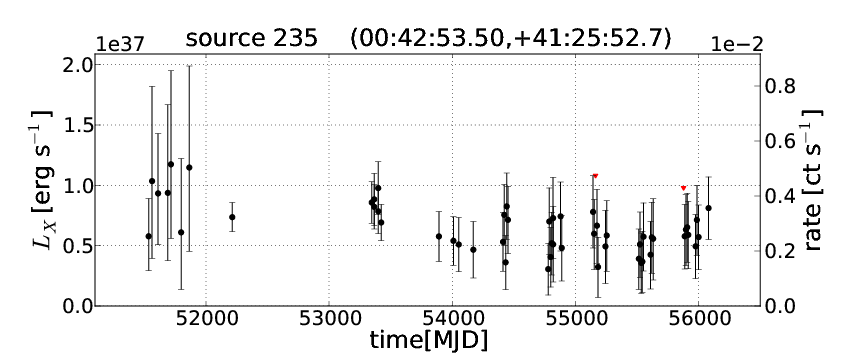}
\includegraphics[width=\linewidth]{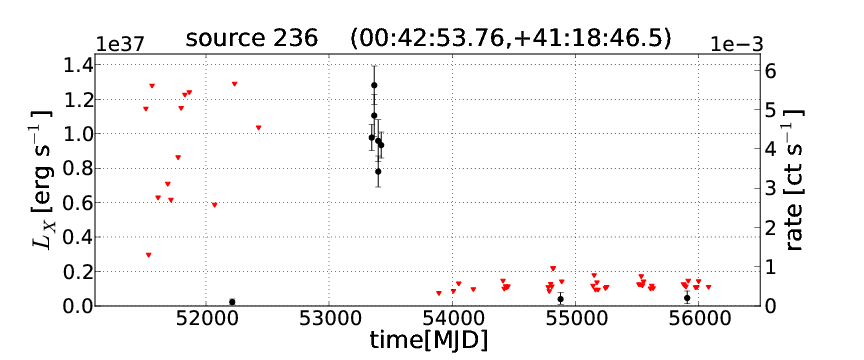}
\includegraphics[width=\linewidth]{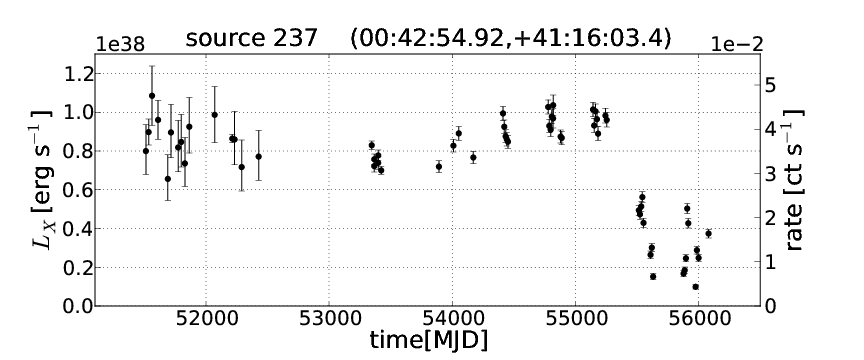}
\includegraphics[width=\linewidth]{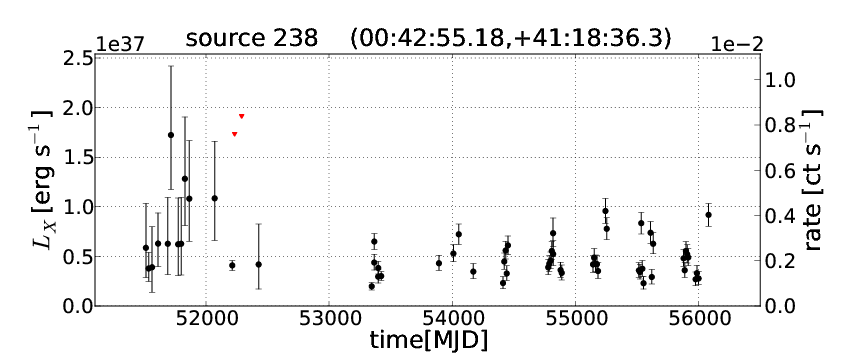}
\end{minipage}
\caption{continued.}
\label{fig:lc_all}
\end{figure*}

\addtocounter{figure}{-1} 

\begin{figure*}
\begin{minipage}{0.5\linewidth}
\includegraphics[width=\linewidth]{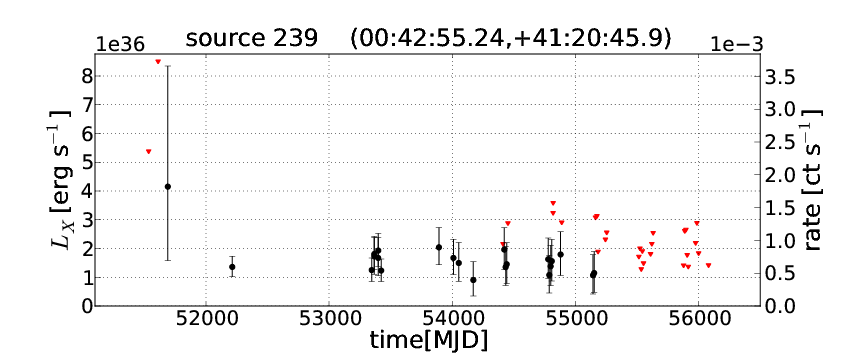}
\includegraphics[width=\linewidth]{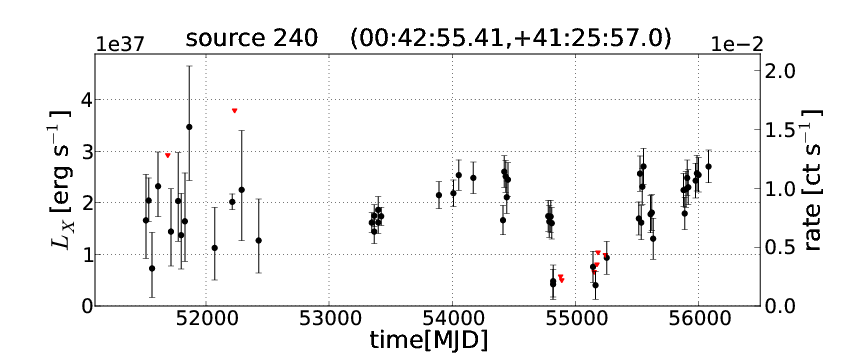}
\includegraphics[width=\linewidth]{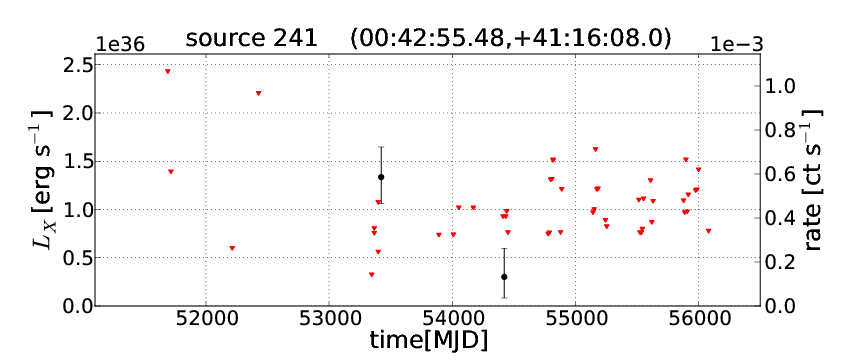}
\includegraphics[width=\linewidth]{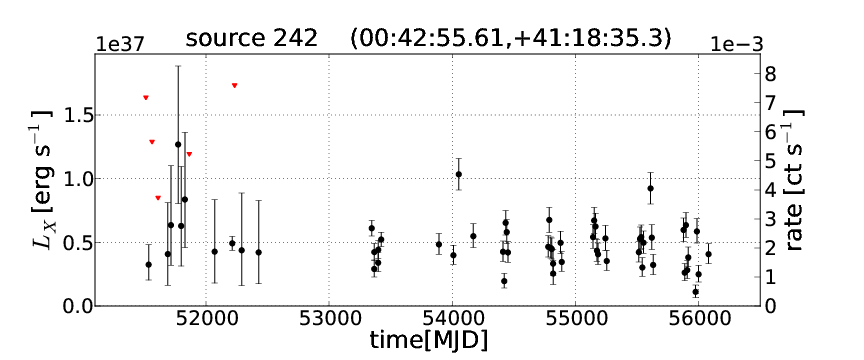}
\includegraphics[width=\linewidth]{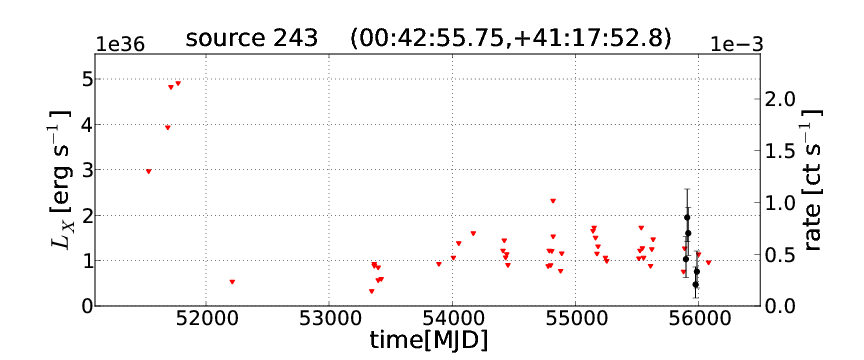}
\includegraphics[width=\linewidth]{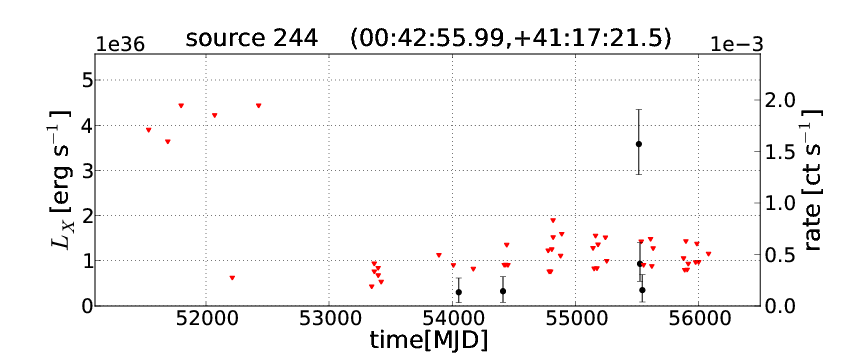}
\end{minipage}
\begin{minipage}{0.5\linewidth}
\includegraphics[width=\linewidth]{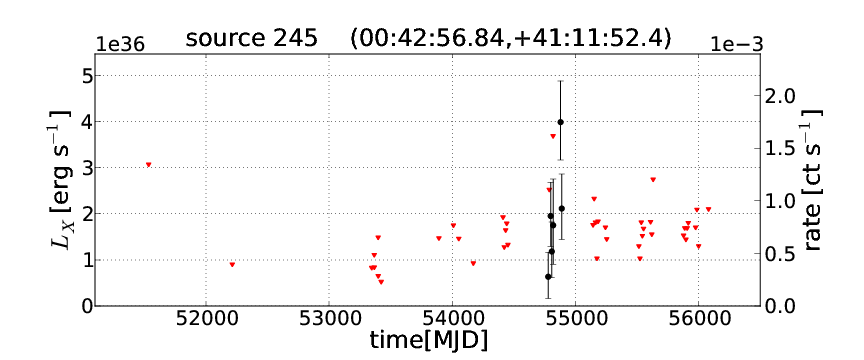}
\includegraphics[width=\linewidth]{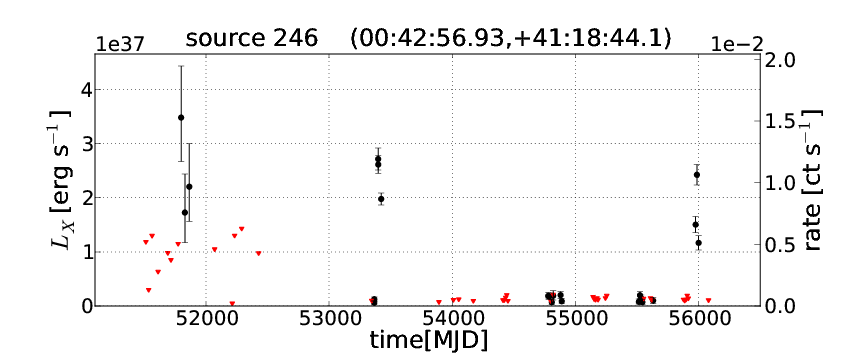}
\includegraphics[width=\linewidth]{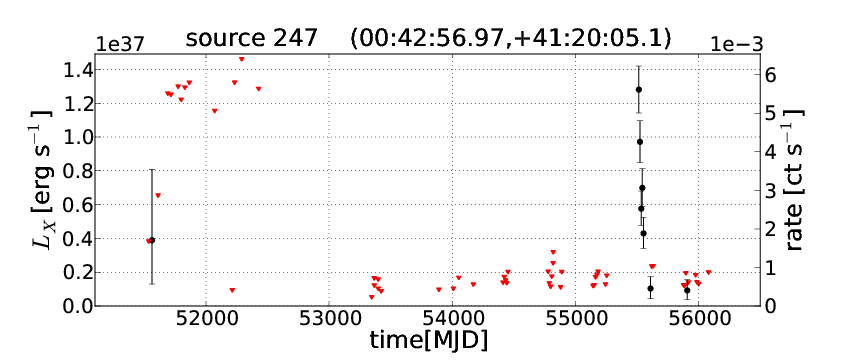}
\includegraphics[width=\linewidth]{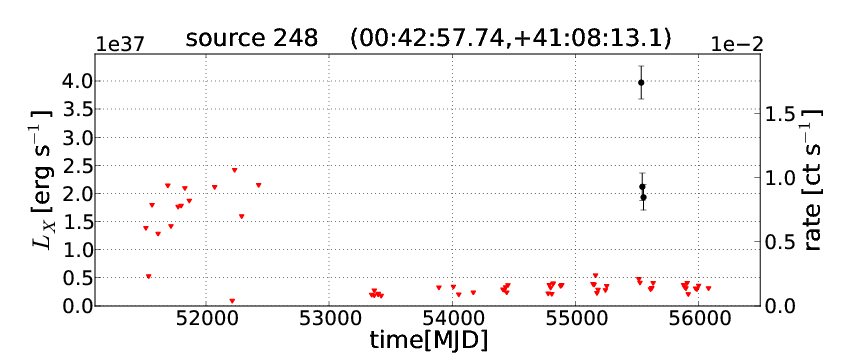}
\includegraphics[width=\linewidth]{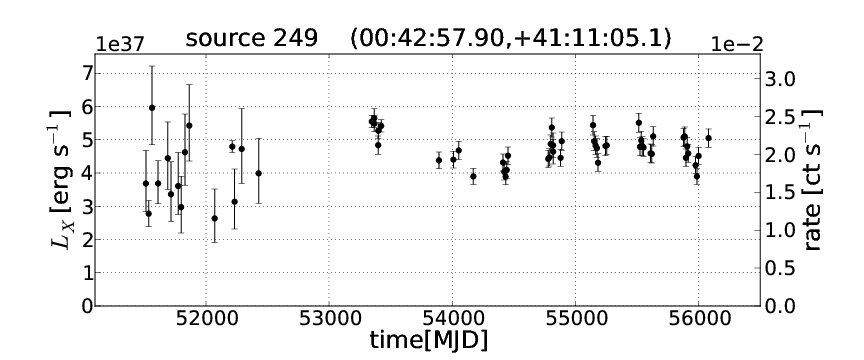}
\includegraphics[width=\linewidth]{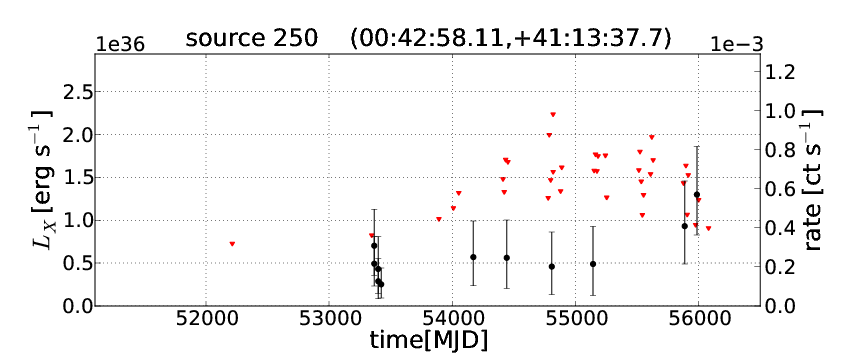}
\end{minipage}
\caption{continued.}
\label{fig:lc_all}
\end{figure*}

\addtocounter{figure}{-1} 

\begin{figure*}
\begin{minipage}{0.5\linewidth}
\includegraphics[width=\linewidth]{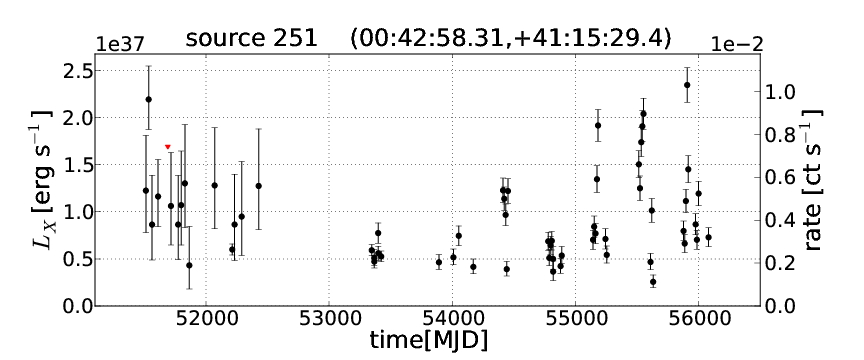}
\includegraphics[width=\linewidth]{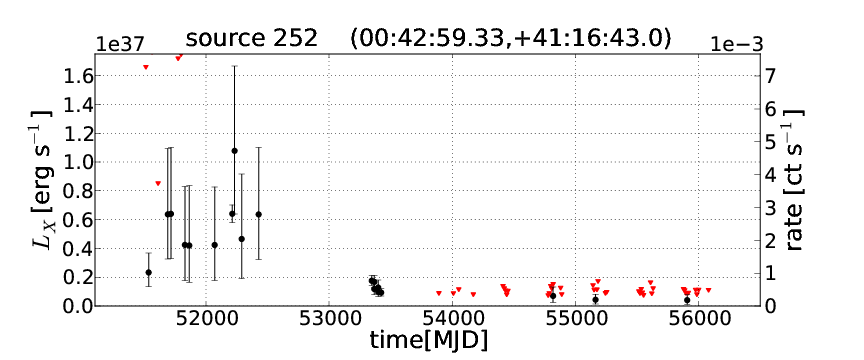}
\includegraphics[width=\linewidth]{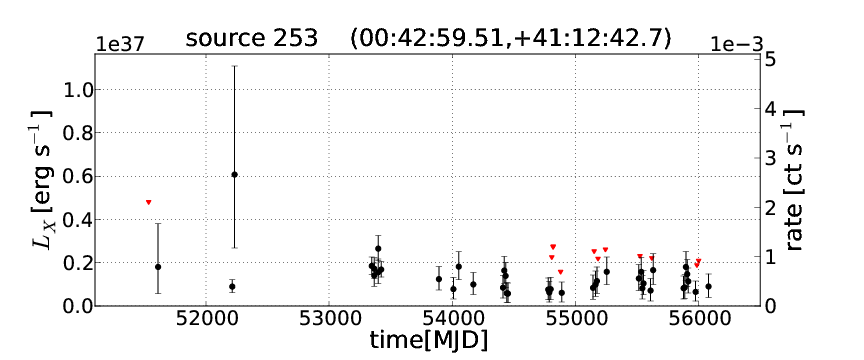}
\includegraphics[width=\linewidth]{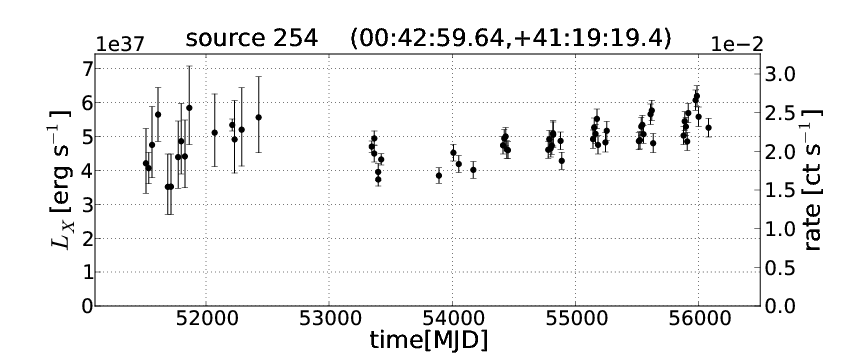}
\includegraphics[width=\linewidth]{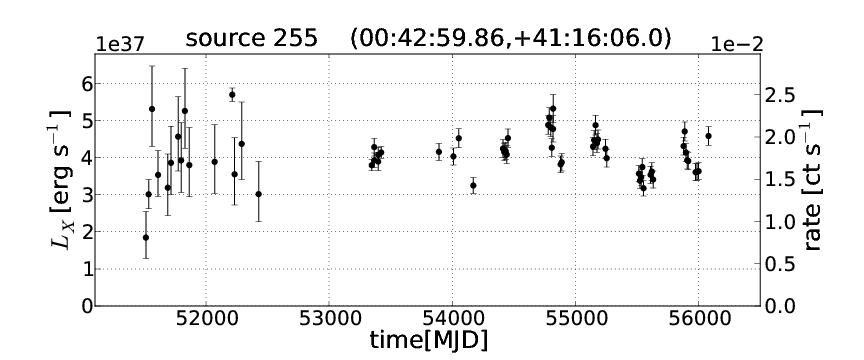}
\includegraphics[width=\linewidth]{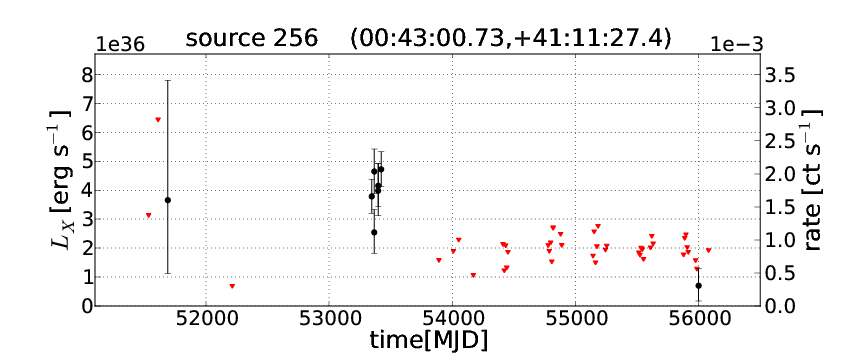}
\end{minipage}
\begin{minipage}{0.5\linewidth}
\includegraphics[width=\linewidth]{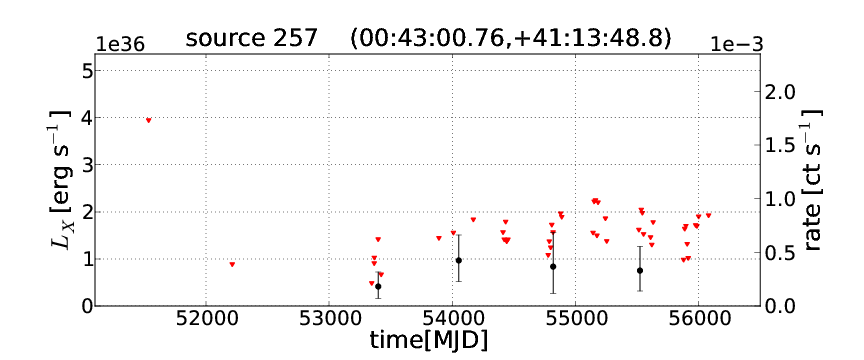}
\includegraphics[width=\linewidth]{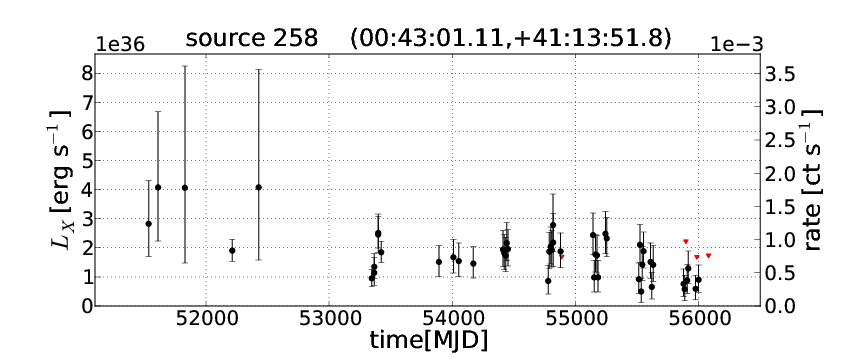}
\includegraphics[width=\linewidth]{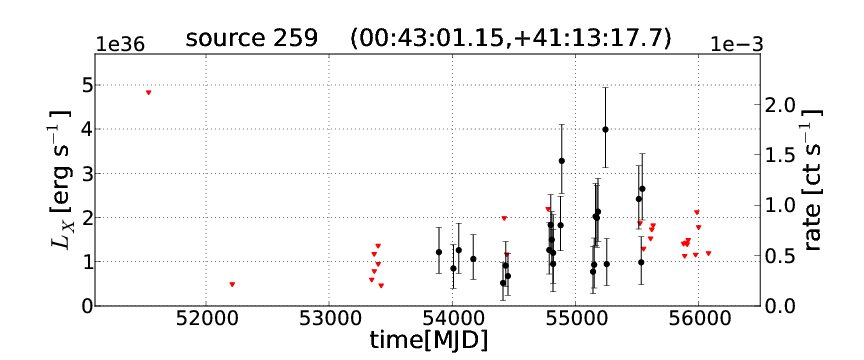}
\includegraphics[width=\linewidth]{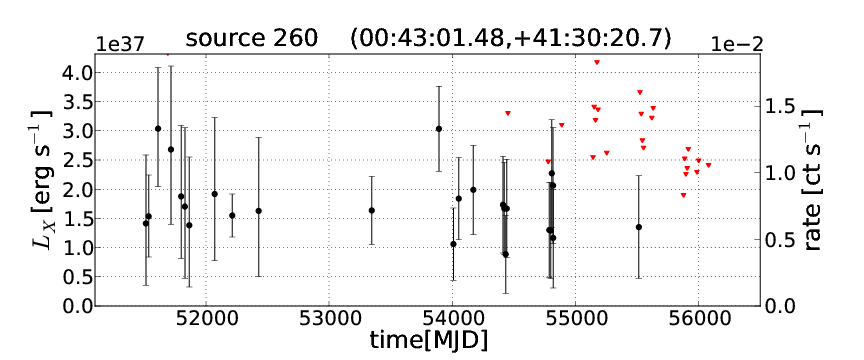}
\includegraphics[width=\linewidth]{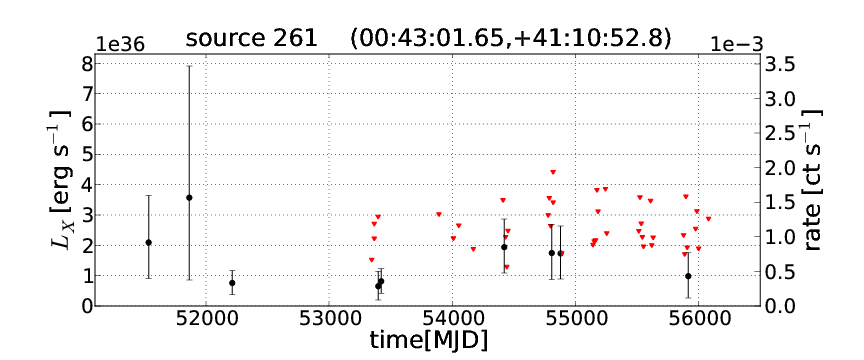}
\includegraphics[width=\linewidth]{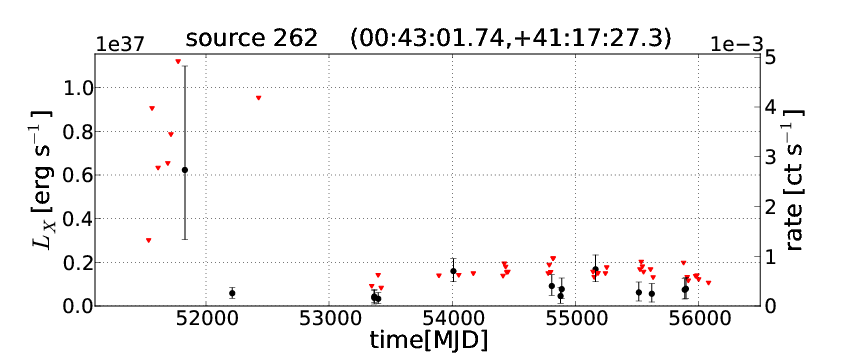}
\end{minipage}
\caption{continued.}
\label{fig:lc_all}
\end{figure*}

\addtocounter{figure}{-1} 

\begin{figure*}
\begin{minipage}{0.5\linewidth}
\includegraphics[width=\linewidth]{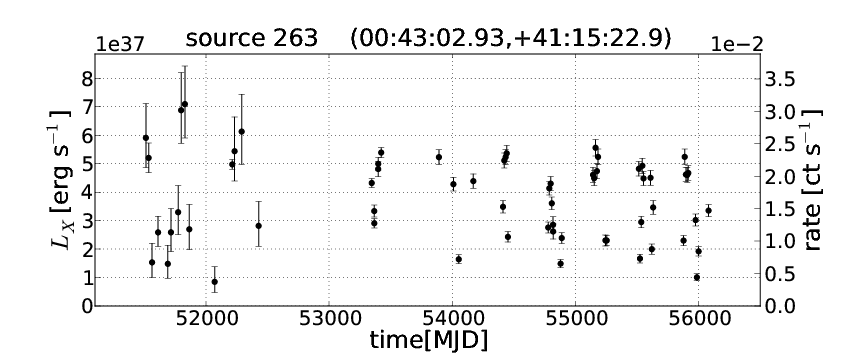}
\includegraphics[width=\linewidth]{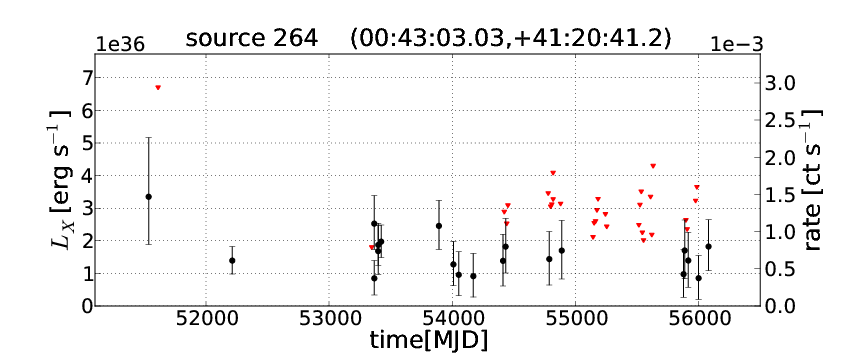}
\includegraphics[width=\linewidth]{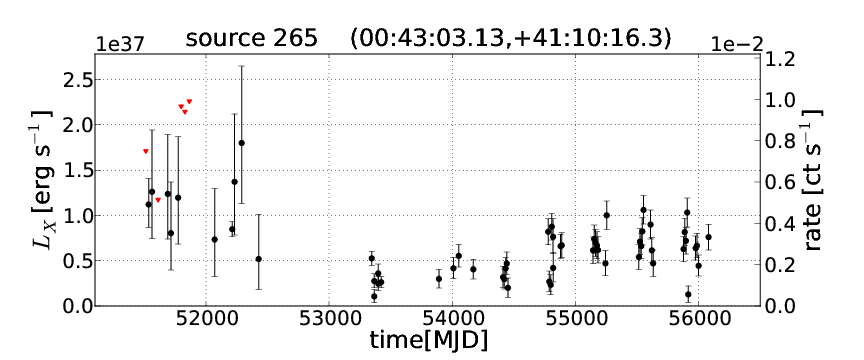}
\includegraphics[width=\linewidth]{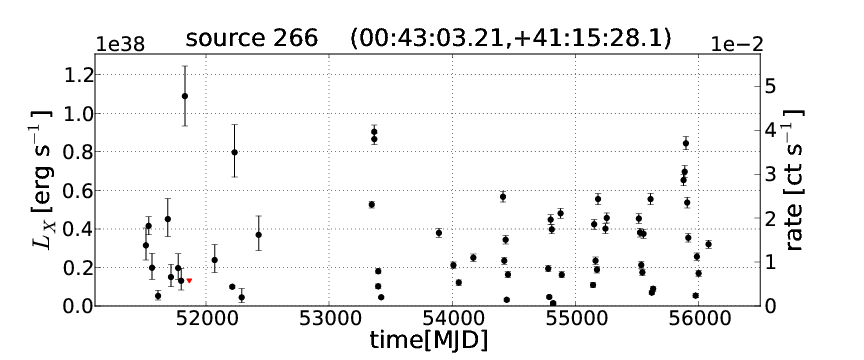}
\includegraphics[width=\linewidth]{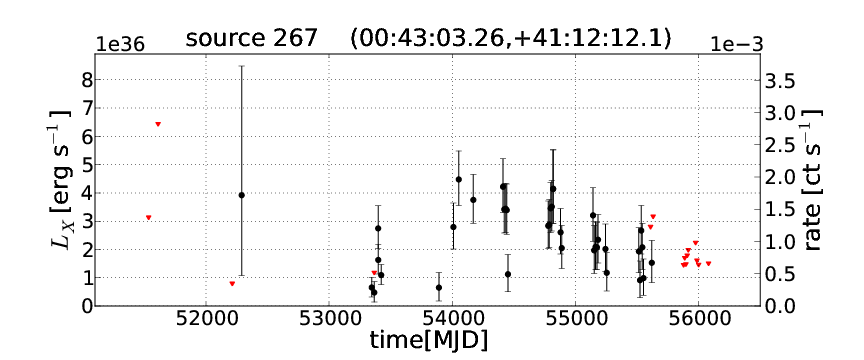}
\includegraphics[width=\linewidth]{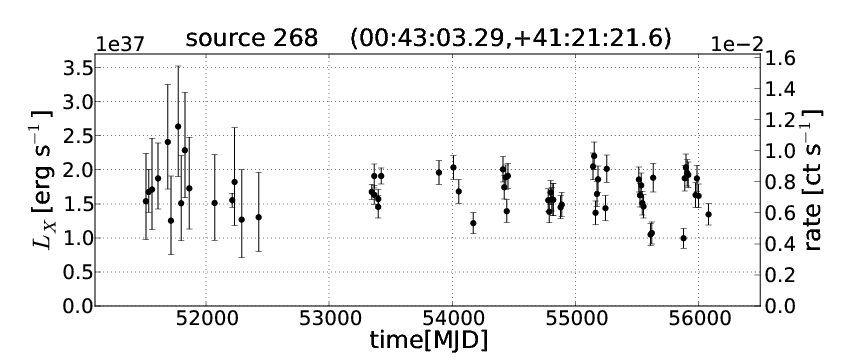}
\end{minipage}
\begin{minipage}{0.5\linewidth}
\includegraphics[width=\linewidth]{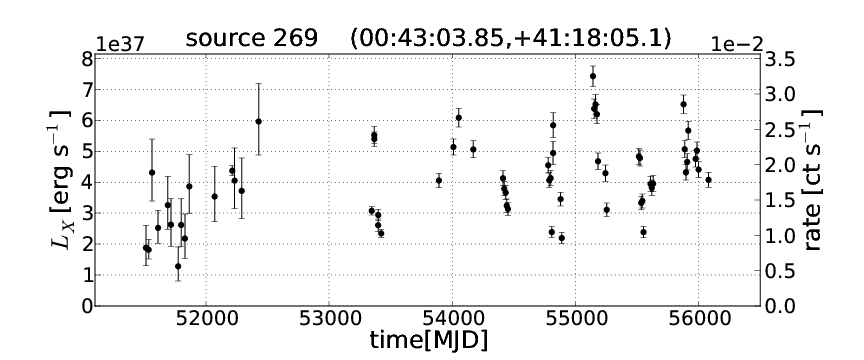}
\includegraphics[width=\linewidth]{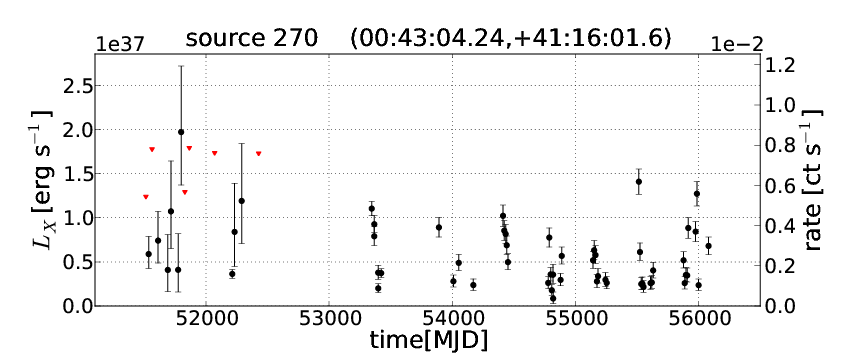}
\includegraphics[width=\linewidth]{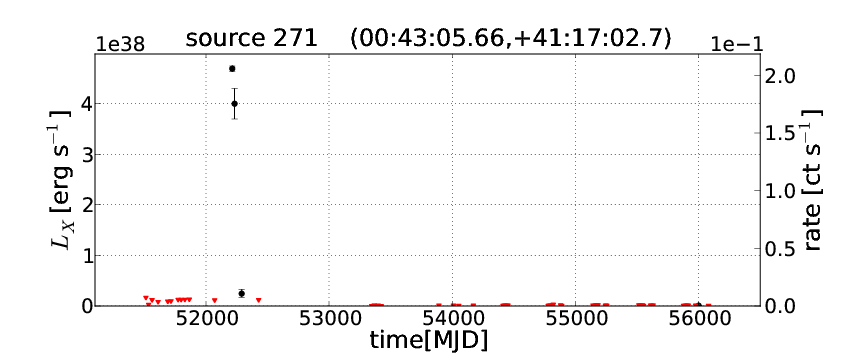}
\includegraphics[width=\linewidth]{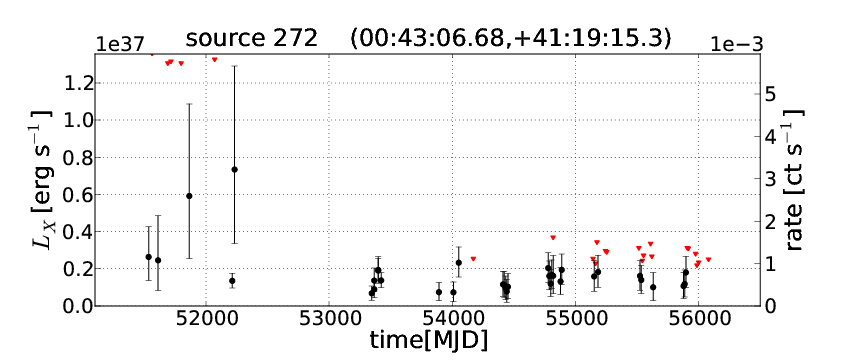}
\includegraphics[width=\linewidth]{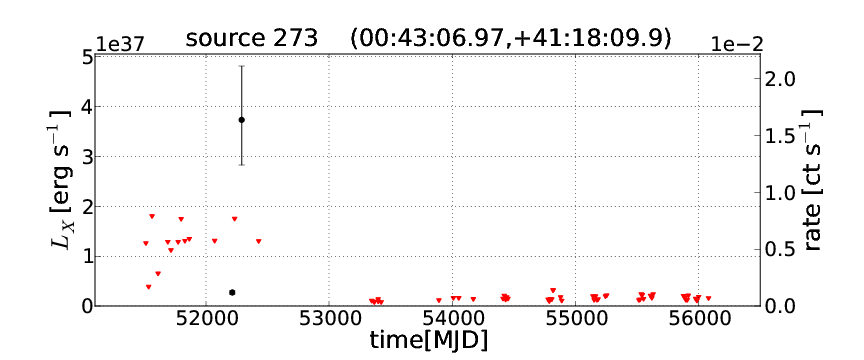}
\includegraphics[width=\linewidth]{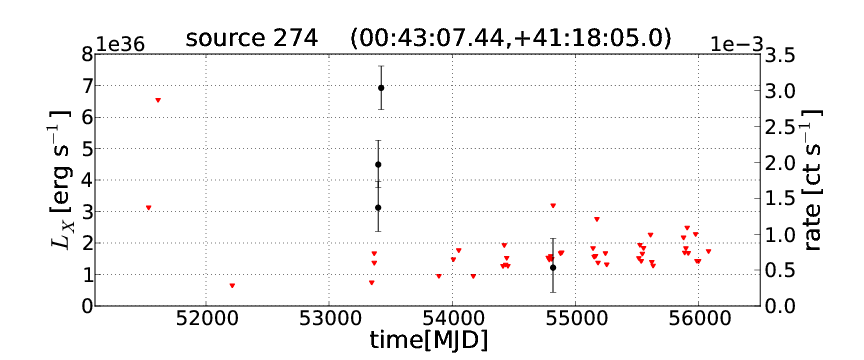}
\end{minipage}
\caption{continued.}
\label{fig:lc_all}
\end{figure*}

\addtocounter{figure}{-1} 

\begin{figure*}
\begin{minipage}{0.5\linewidth}
\includegraphics[width=\linewidth]{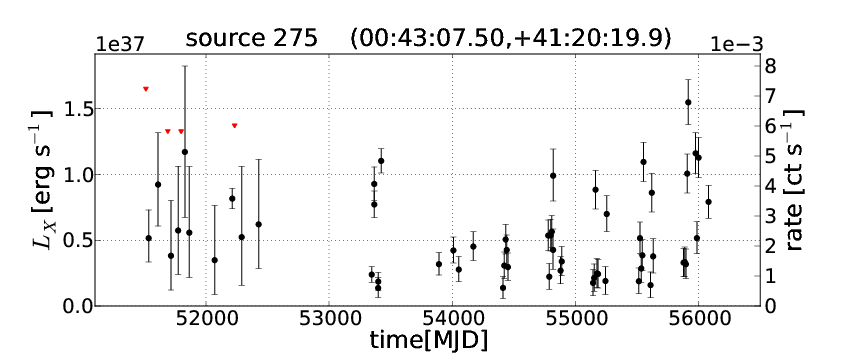}
\includegraphics[width=\linewidth]{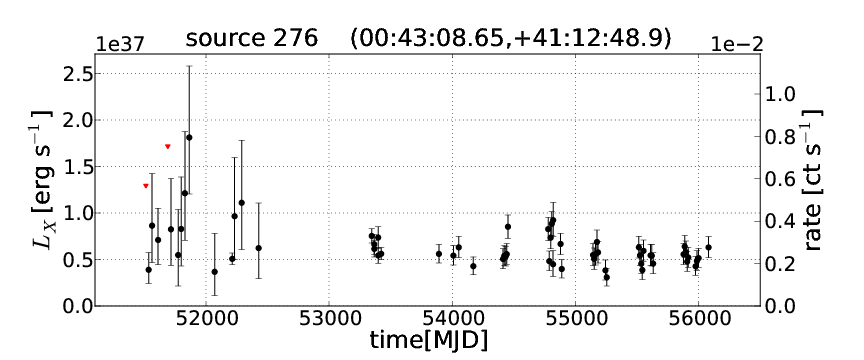}
\includegraphics[width=\linewidth]{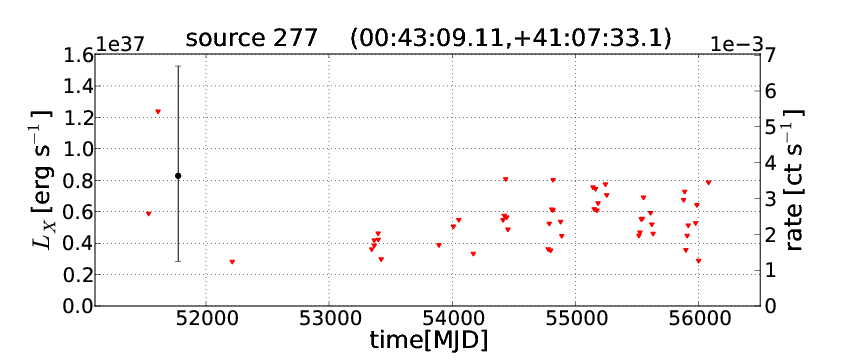}
\includegraphics[width=\linewidth]{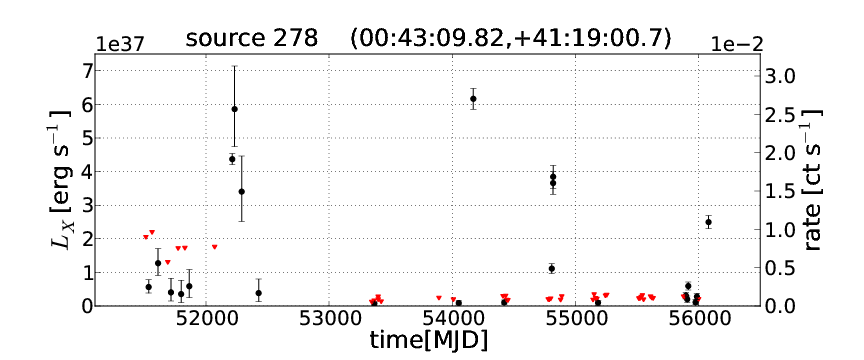}
\includegraphics[width=\linewidth]{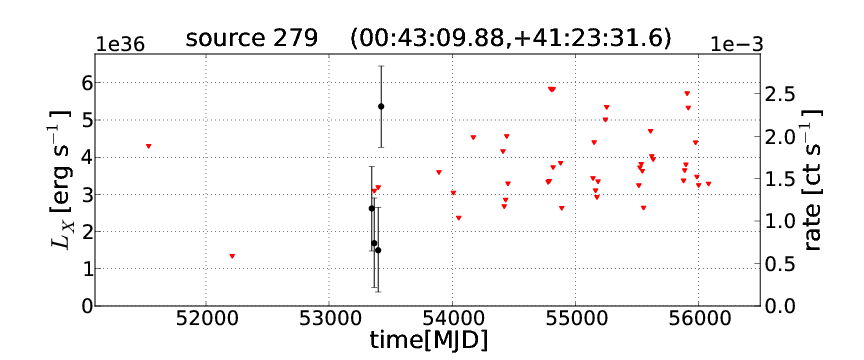}
\includegraphics[width=\linewidth]{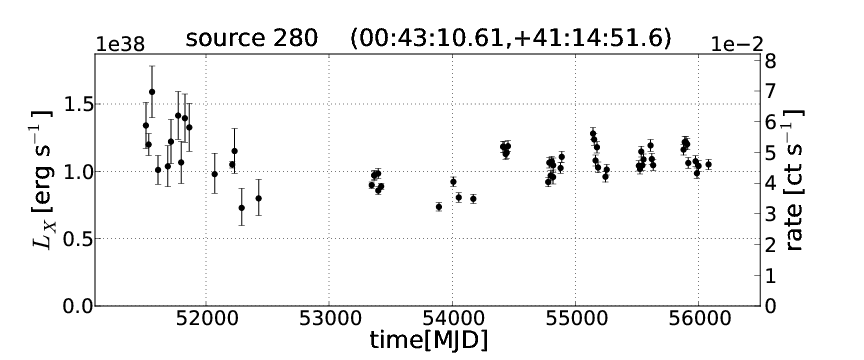}
\end{minipage}
\begin{minipage}{0.5\linewidth}
\includegraphics[width=\linewidth]{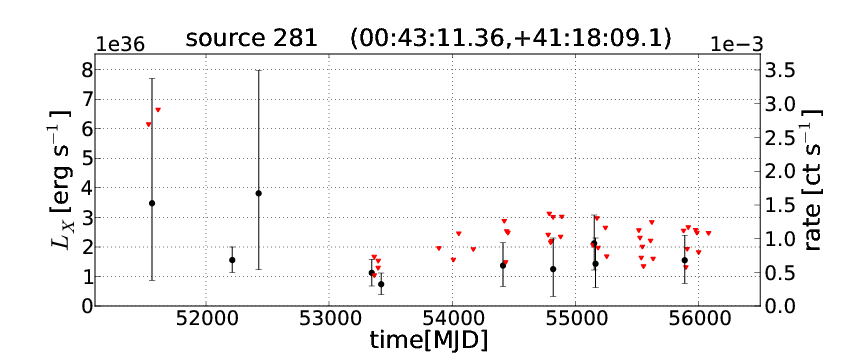}
\includegraphics[width=\linewidth]{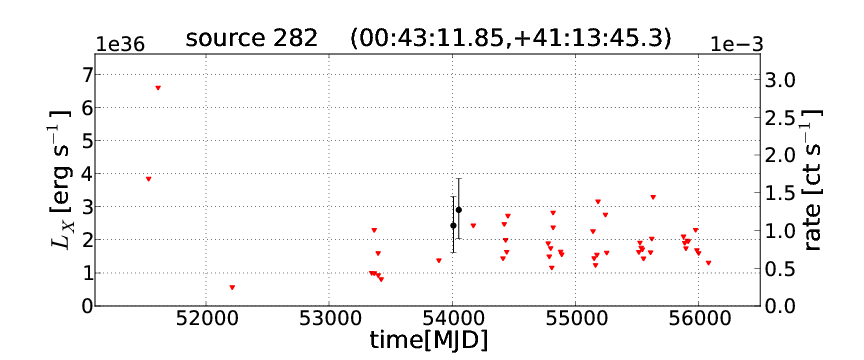}
\includegraphics[width=\linewidth]{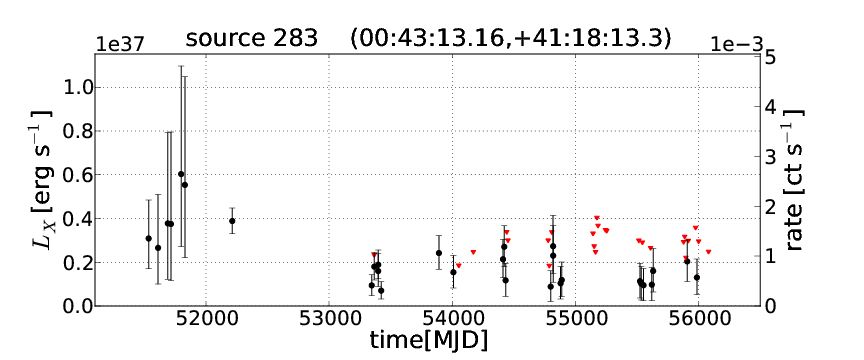}
\includegraphics[width=\linewidth]{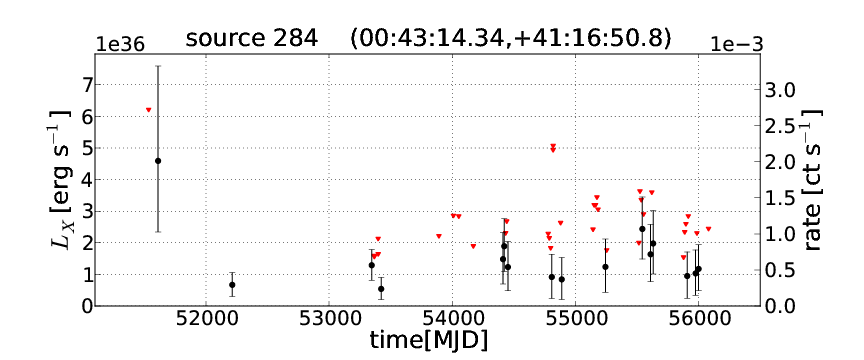}
\includegraphics[width=\linewidth]{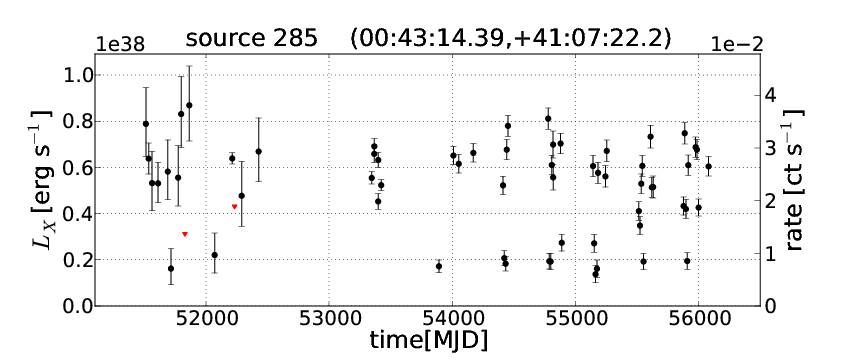}
\includegraphics[width=\linewidth]{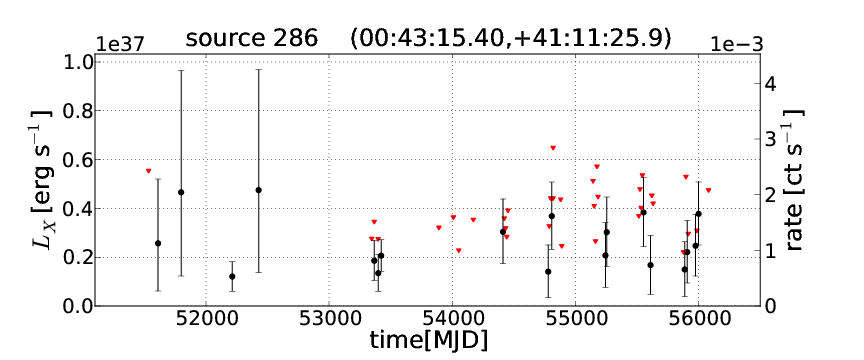}
\end{minipage}
\caption{continued.}
\label{fig:lc_all}
\end{figure*}

\addtocounter{figure}{-1} 

\begin{figure*}
\begin{minipage}{0.5\linewidth}
\includegraphics[width=\linewidth]{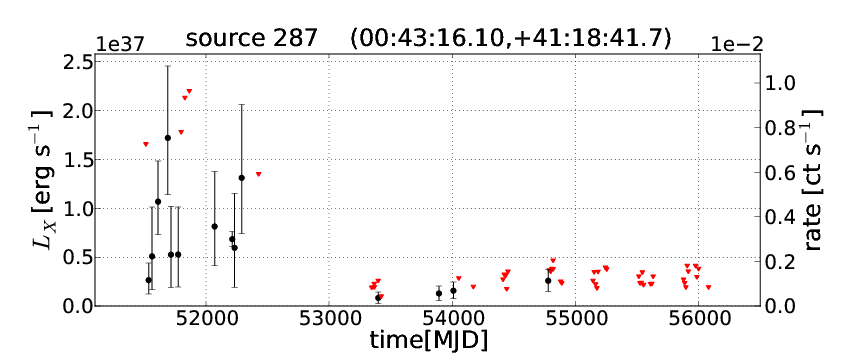}
\includegraphics[width=\linewidth]{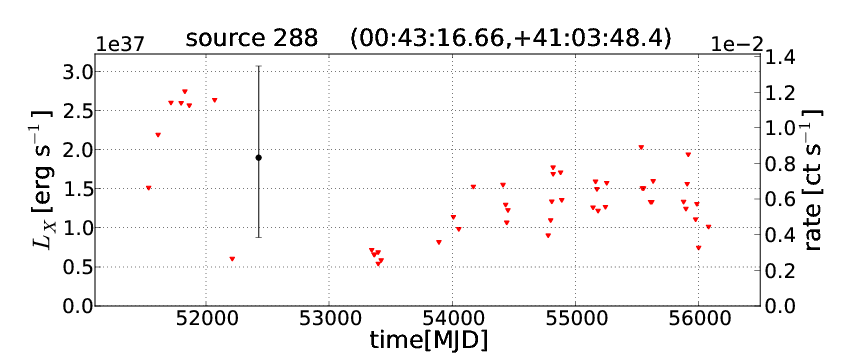}
\includegraphics[width=\linewidth]{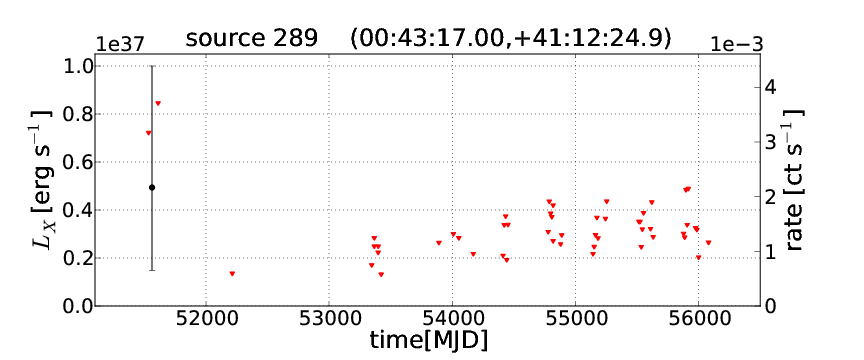}
\includegraphics[width=\linewidth]{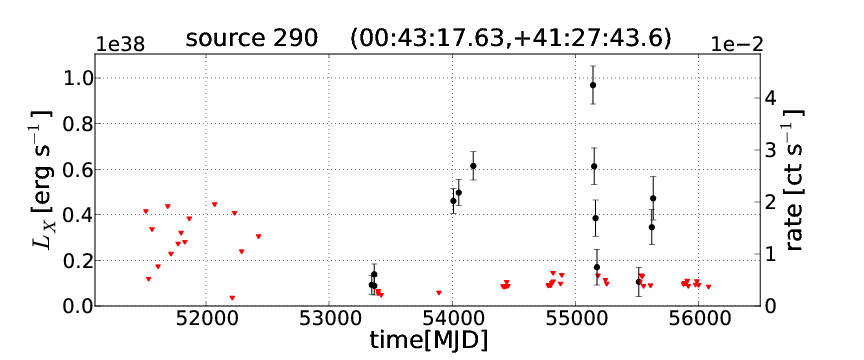}
\includegraphics[width=\linewidth]{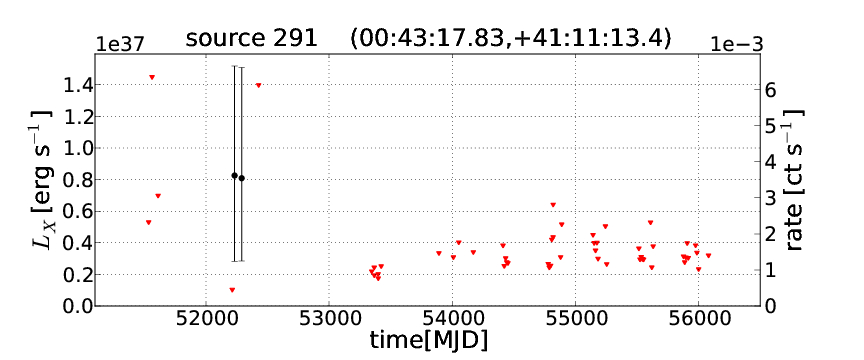}
\includegraphics[width=\linewidth]{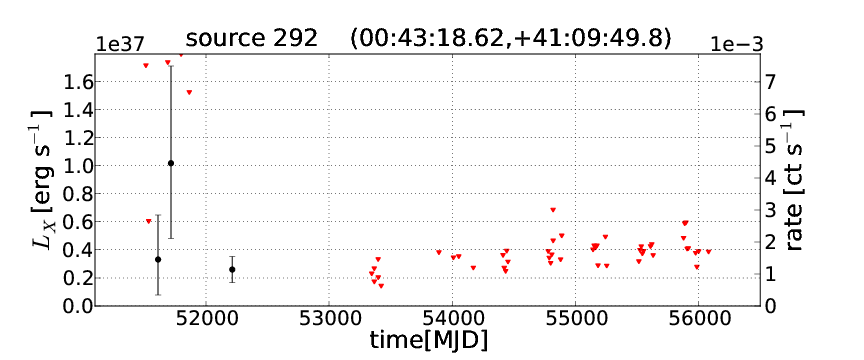}
\end{minipage}
\begin{minipage}{0.5\linewidth}
\includegraphics[width=\linewidth]{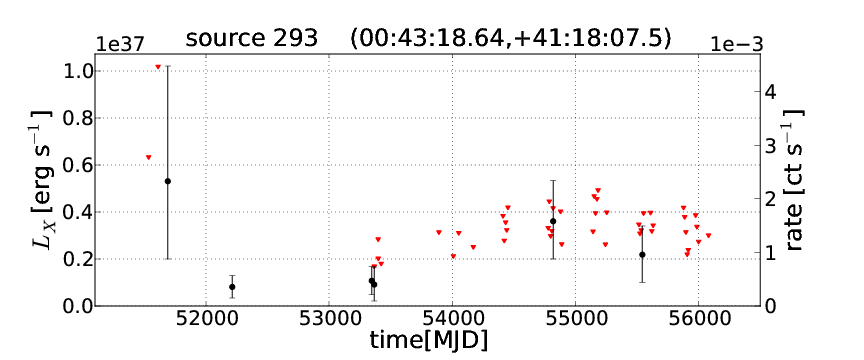}
\includegraphics[width=\linewidth]{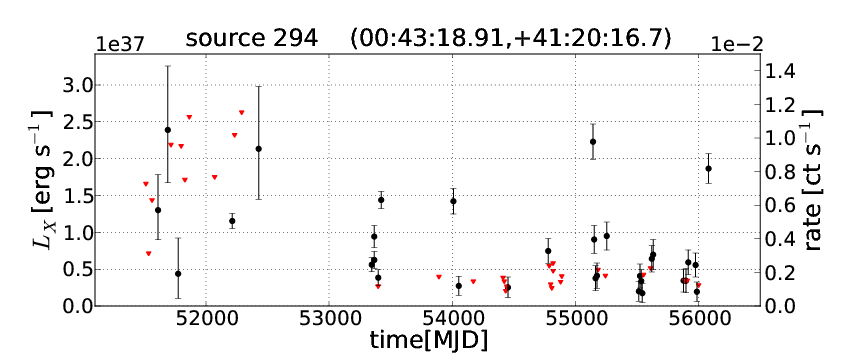}
\includegraphics[width=\linewidth]{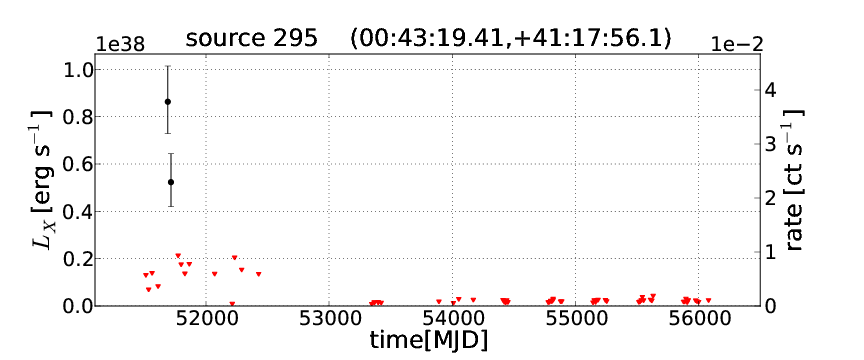}
\includegraphics[width=\linewidth]{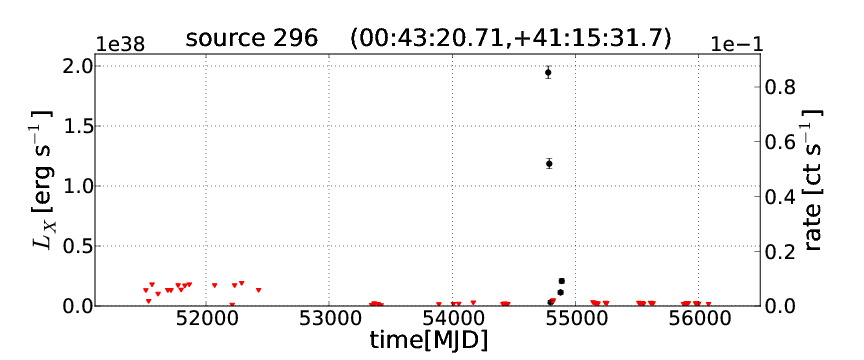}
\includegraphics[width=\linewidth]{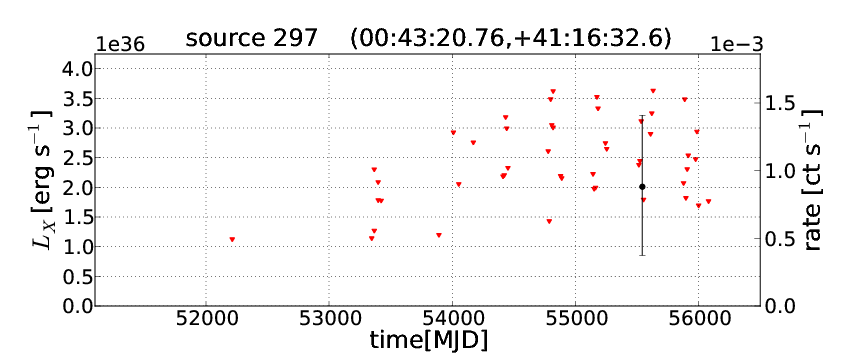}
\includegraphics[width=\linewidth]{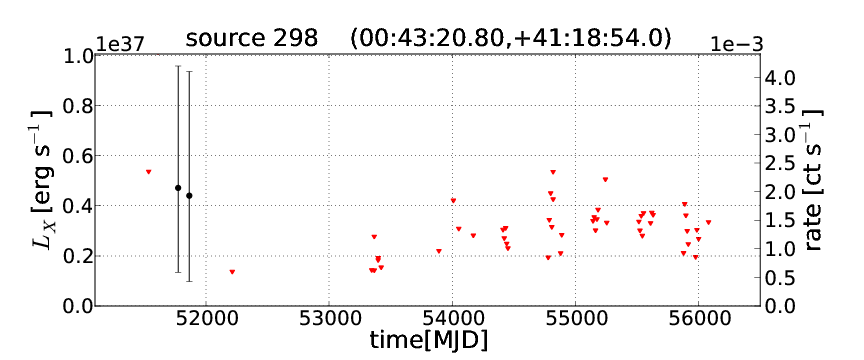}
\end{minipage}
\caption{continued.}
\label{fig:lc_all}
\end{figure*}

\addtocounter{figure}{-1} 

\begin{figure*}
\begin{minipage}{0.5\linewidth}
\includegraphics[width=\linewidth]{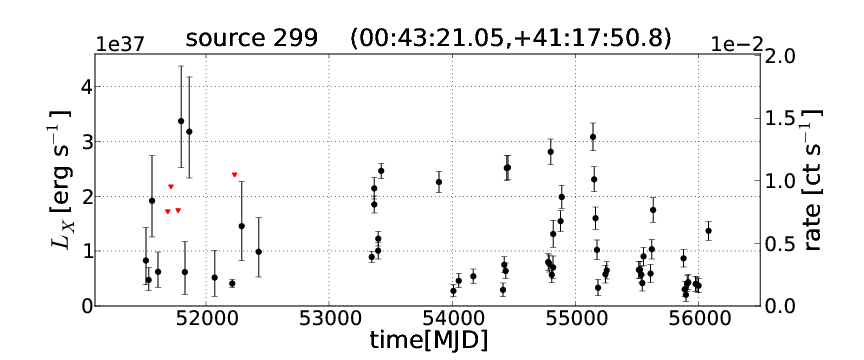}
\includegraphics[width=\linewidth]{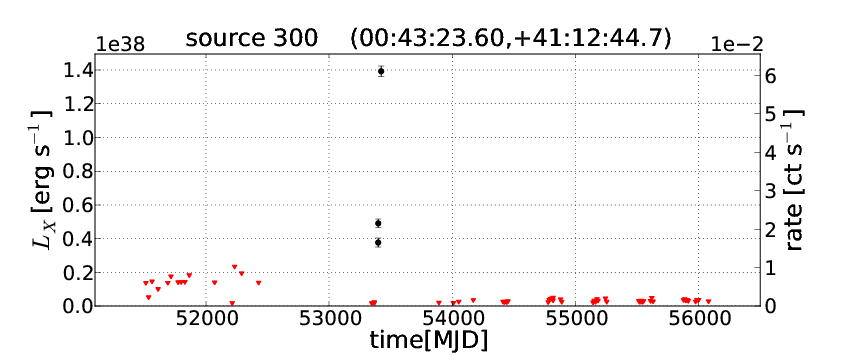}
\includegraphics[width=\linewidth]{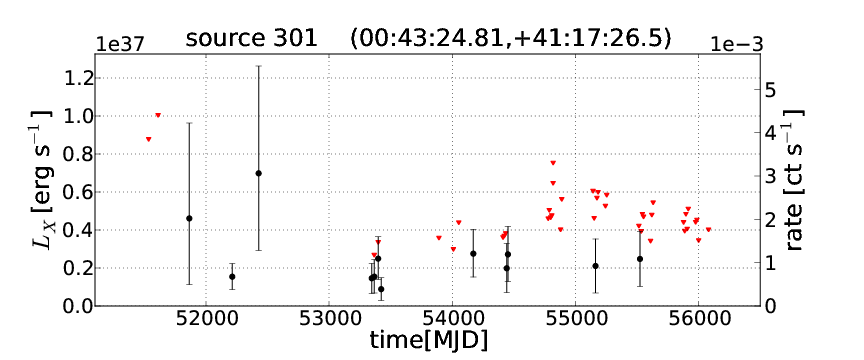}
\includegraphics[width=\linewidth]{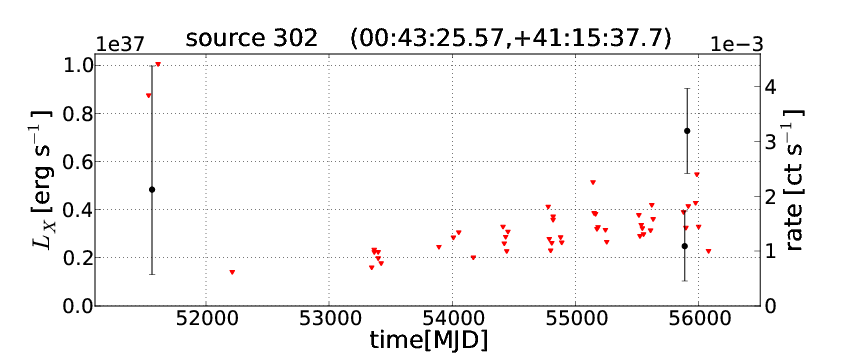}
\includegraphics[width=\linewidth]{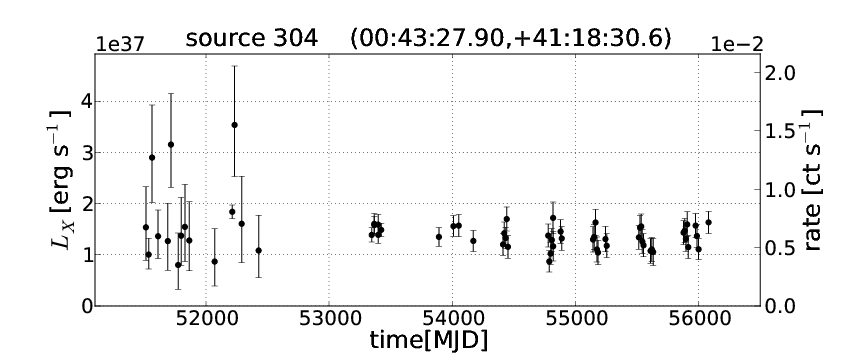}
\includegraphics[width=\linewidth]{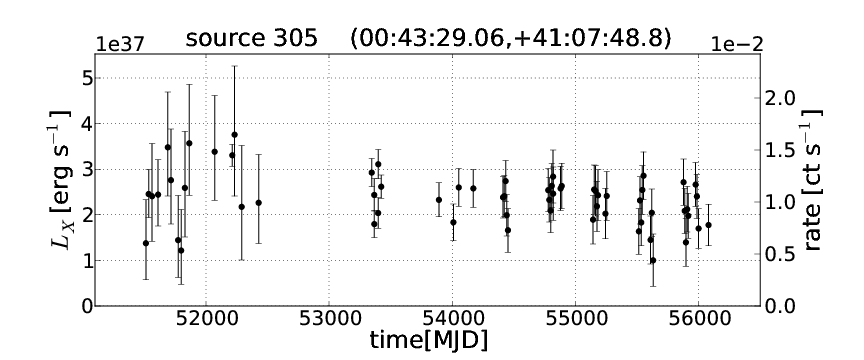}
\end{minipage}
\begin{minipage}{0.5\linewidth}
\includegraphics[width=\linewidth]{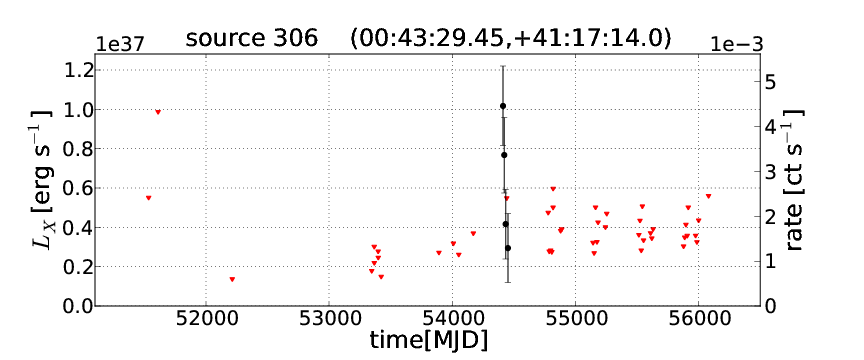}
\includegraphics[width=\linewidth]{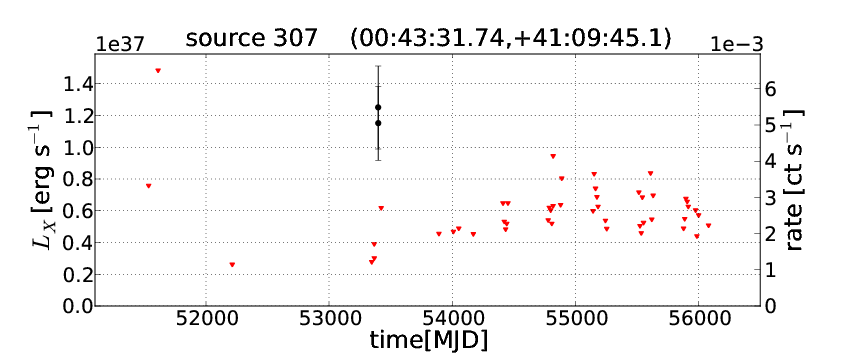}
\includegraphics[width=\linewidth]{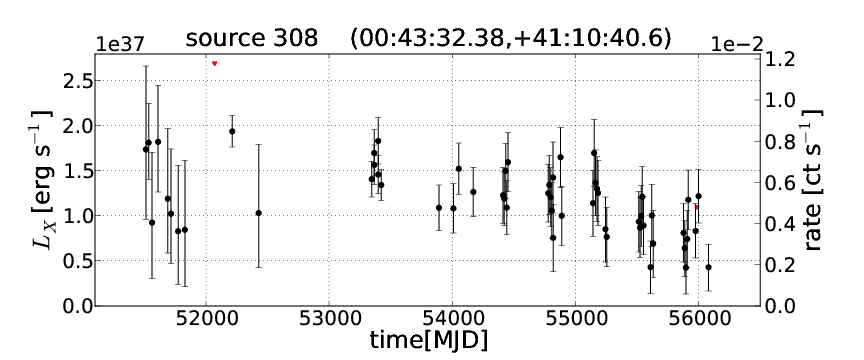}
\includegraphics[width=\linewidth]{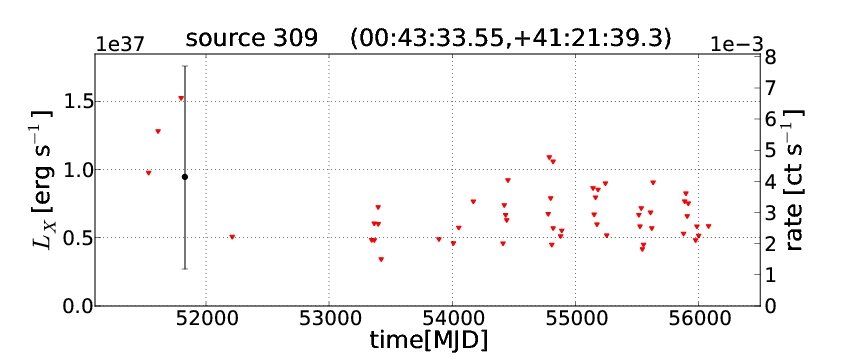}
\includegraphics[width=\linewidth]{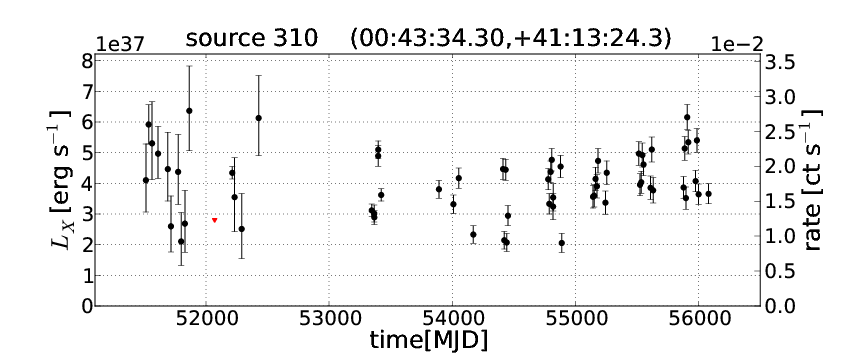}
\includegraphics[width=\linewidth]{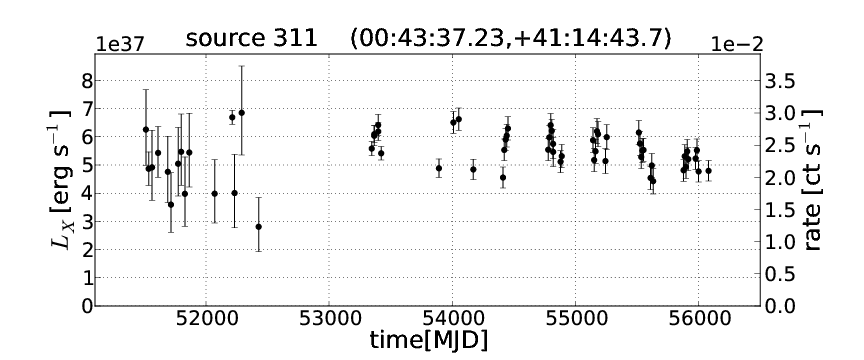}
\end{minipage}
\caption{continued.}
\label{fig:lc_all}
\end{figure*}

\addtocounter{figure}{-1} 

\begin{figure*}
\begin{minipage}{0.5\linewidth}
\includegraphics[width=\linewidth]{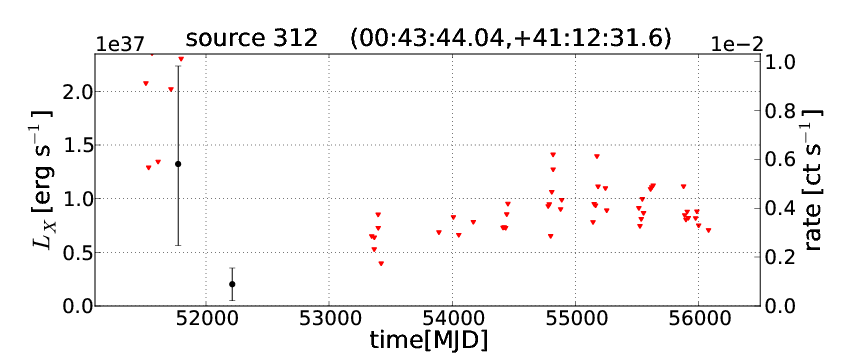}
\includegraphics[width=\linewidth]{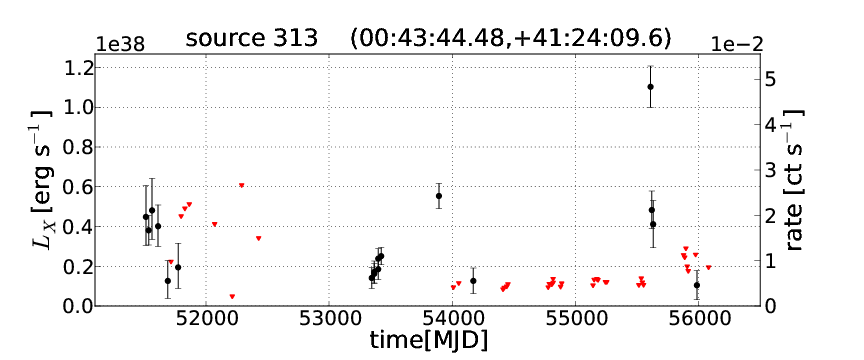}
\includegraphics[width=\linewidth]{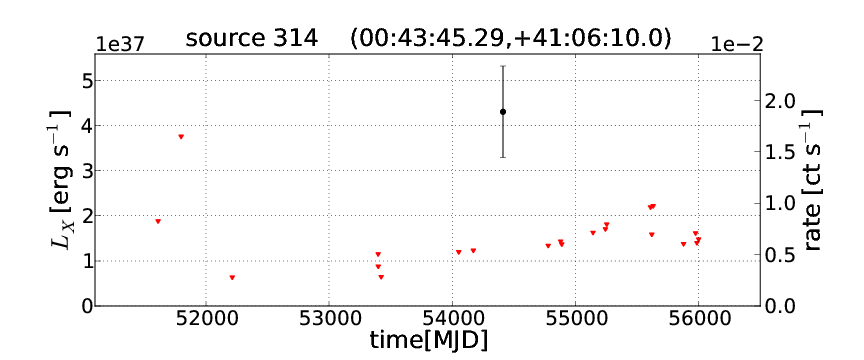}
\includegraphics[width=\linewidth]{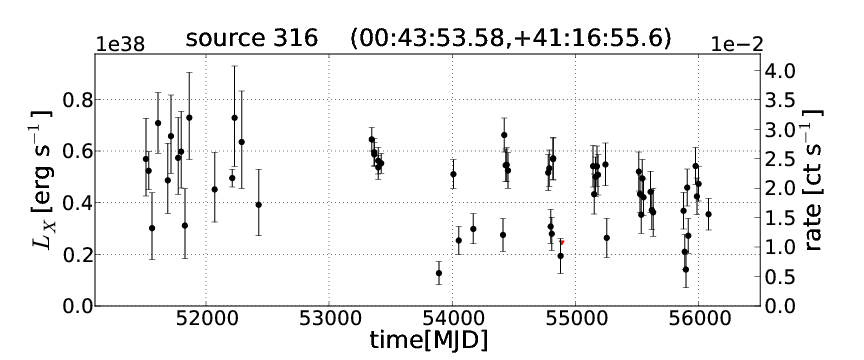}
\includegraphics[width=\linewidth]{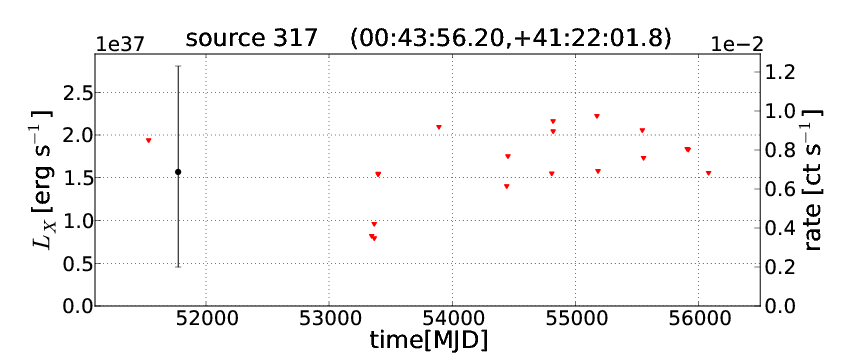}
\end{minipage}
\caption{continued.}
\label{fig:lc_all}
\end{figure*}

\end{appendix}

\end{document}